\documentclass[11pt,a4paper]{article}
\usepackage[dvipsnames]{xcolor}
\usepackage{jheppub}
\usepackage{orcidlink}
\hypersetup{colorlinks=true,linkcolor=MidnightBlue,citecolor=MidnightBlue,urlcolor=MidnightBlue}
\usepackage{mathtools,bm,mathrsfs}
\usepackage{booktabs,enumitem,array}
\usepackage{microtype,tikz}
\usetikzlibrary{arrows.meta,positioning}
\setlist{itemsep=2pt,topsep=3pt}
\newtheorem{theorem}{Theorem}
\newtheorem{proposition}[theorem]{Proposition}

\newcommand{\R}{\mathbb R}
\newcommand{\C}{\mathbb C}
\newcommand{\Z}{\mathbb Z}
\newcommand{\HH}{\mathbb H}
\newcommand{\tr}{\operatorname{tr}}
\newcommand{\ii}{\mathrm i}
\newcommand{\dd}{\mathrm d}
\newcommand{\spec}{\operatorname{spec}}
\newcommand{\arxiv}[1]{\href{https://arxiv.org/abs/#1}{arXiv:\nolinkurl{#1}}}
\usepackage{placeins}

\title{Instanton self-localization and conformality loss of 5d~non-Lorentzian Wilson lines}
\author{Abhishek Chowdhury\,\orcidlink{0000-0003-0948-4817}}
\affiliation{School of Basic Sciences, Indian Institute of Technology Bhubaneswar,\\Argul, Odisha 752050, India}
\emailAdd{achowdhury@iitbbs.ac.in}

\abstract{We study a spin-$s$ Wilson line with scalar coupling $\zeta$ in the one-instanton sector of five-dimensional non-Lorentzian $SU(2)$ gauge theory. At zero discrete theta angle this sector carries one unit of null momentum in the proposed light-cone description of the six-dimensional $A_1$ $(2,0)$ theory. Spatial self-duality reduces the classical dynamics of an isolated instanton exactly to motion of its centre, size and gauge orientation (moduli). The line is conformal when its Hamiltonian admits a scale-invariant self-adjoint realization bounded below. With bosonic moduli alone, instanton recoil protects every line with $s\le4$. For each $s\ge9/2$ a range of $\zeta$ gives fall to the centre at one discrete theta angle and a unique self-adjoint Hamiltonian at the other. Fermion zero modes change this conclusion. At the half-BPS endpoints $\zeta=\pm1$, reduction along the quaternionic Hopf fibration fixes the lowest angular eigenvalue at $-8$ for every nonzero spin, one unit above the Hardy threshold for collapse. The corresponding twistor-spinor states carry the same representations as the lowest Landau levels in the four-dimensional quantum Hall problem on $S^4$. At either theta angle, every line with $s\ge1$ loses conformality in a range containing $\zeta=0$, while the fundamental line has a unique self-adjoint, scale-invariant Hamiltonian throughout $|\zeta|\le1$. The curvature of the non-Abelian Berry connection acts as a magnetic field on the instanton centre. At large spin the Pauli energy of the centre fermion zero modes cancels the leading Landau zero-point energy, localizing the centre within a distance of order $\rho/\sqrt{s}$ of the line for an instanton of size $\rho$. The critical attraction $E_c=4s(s+1)(1-\zeta_c^2)$ obeys $E_c\ge1+3/(2s)-3/(4s^2)$ for $s\ge1$ and has the large-spin expansion $E_c=1+3/(2s)-9/(56s^2)+O(s^{-3})$.}

\begin{document}
\maketitle
\raggedbottom

\section{Introduction}
\label{sec:intro}

Line operators are natural probes of a gauge theory, and whether a given line defines a conformal defect is a dynamical question. In four-dimensional $\mathcal N=4$ super-Yang--Mills theory the Wilson line that couples to one adjoint scalar with strength $\zeta$ interpolates between the ordinary line at $\zeta=0$ and the half-BPS line at $\zeta=\pm1$. Its coupling runs at weak coupling, with the ordinary line in the ultraviolet and the supersymmetric lines in the infrared \cite{PolchinskiSully,BGT2017,BGT2022}, and lines in large representations can lose conformality altogether, through the merger of a pair of fixed points into complex ones \cite{KLSS,ACKMRletter}. Aharony, Cuomo, Komargodski, Mezei and Raviv-Moshe found this instability for a scalar-coupled line of spin $s$ at large $g^2s$ \cite{ACKMR}. In their double-scaling limit, with the colour treated classically, the dimensions of defect operators become complex once $(g^2s/2\pi)^2(1-\zeta^2)>1$, while the supersymmetric line remains stable for every representation. Related square-root mergers occur for spin impurities \cite{CKMR}. These results are obtained in weak-coupling or semiclassical regimes, where the threshold depends on the gauge coupling and the colour charge is treated classically.

This paper studies a setting in which the interaction between a line defect and a soliton follows exactly from a constraint, without a weak-coupling or semiclassical expansion, and in which the gauge coupling drops out of the stability problem. We use the maximally supersymmetric non-Lorentzian $SU(2)$ gauge theory of Lambert and Owen, proposed as a null reduction of the six-dimensional $(2,0)$ theory \cite{LambertRichmond,LO,LPS}. Because a Lagrange multiplier keeps the spatial gauge field self-dual at every time, the classical dynamics of a charge-one instanton reduces exactly to motion on its moduli space with no slow-motion approximation, and the resulting quantum mechanics describes, at zero discrete theta angle, the one-unit null-momentum sector in the discrete light-cone quantization (DLCQ) of the $A_1$ $(2,0)$ theory \cite{ABKSS,ABS,Seiberg1997,Sen1997}. A static Wilson line of spin $s$ and scalar coupling $\zeta$ introduces two interactions. Its colour is parallel transported as the instanton moves. The corresponding Berry connection is the universal instanton connection restricted to the position of the line, which we call the \emph{evaluation connection} \cite{TWADHM}. The second interaction is an attractive self-energy proportional to $1-\zeta^2$. Its derivation assumes the range condition on the source explained in Sec.~\ref{sec:selfenergy}. The covariant kinetic energy and the self-energy have the same scaling dimension. In five dimensions a charge-one $SU(2)$ instanton is a particle with a size modulus $\rho$, and its four centre coordinates and four size and orientation coordinates form an eight-dimensional moduli cone, on which the line's inverse-square attraction competes directly with the covariant kinetic energy as the instanton shrinks. The spectrum tests whether the instanton can lower its energy without bound by shrinking onto the defect. Section~\ref{sec:lightcone} considers the relation of this fixed-instanton-number result to a Wilson surface in discrete light-cone quantization.

The conformality criterion reduces to a single spectral number. Writing the eight-dimensional moduli cone in radial form turns the Hamiltonian into $-\partial_R^2-(7/R)\partial_R+L(\zeta)/R^2$, where $L(\zeta)$ is the \emph{link operator} acting on the seven-sphere linking the cone. The model admits a lower-bounded scale-invariant realization precisely when the lowest eigenvalue $\mu$ of $L(\zeta)$ stays at or above the eight-dimensional Hardy value $-9$. Below that value the radial scaling exponents $\tfrac12\pm\ii\sqrt{-(\mu+9)}$ of $R^{7/2}\psi$ turn complex and the unregulated Hamiltonian has no lower-bounded realization. A short-distance cutoff then gives an Efimov tower with Berezinskii--Kosterlitz--Thouless scaling at onset. Between the Hardy value and $-8$ the short-distance boundary condition acts as a contact coupling of the line at the cone point, and its two scale-invariant values, the infrared and ultraviolet fixed points of that coupling, merge at the Hardy value (Sec.~\ref{sec:radial}). The Hamiltonian in a harmonic trap, the compact generator of the conformal algebra, has lowest weight $k_+=\frac12(1+\sqrt{\mu+9})$ in the Friedrichs realization, and under the operator--state map \cite{DFF,NishidaSon} $2k_+$ would be the scaling dimension of a line operator, a reading that requires the matching of Sec.~\ref{sec:lightcone}. The loss of a real exponent at the Hardy bound is the moduli-space counterpart of the complex defect dimensions recalled above (Sec.~\ref{sec:radial}). The physical Hamiltonian is proportional to $g_5^2$, and the criterion is independent of this overall factor.

The bosonic computation isolates the role of recoil. Hold the instanton centre on the line and choose the five-dimensional discrete theta angle $\theta_{5d}=0$. Collapse then requires an attraction $E=4s(s+1)(1-\zeta^2)$ greater than $1$ for integer spin or $4$ for half-integer spin. A freely moving centre has four extra coordinates. They raise the critical attraction to at least $99/4$, and the magnetic coupling to the colour raises it to $16s+O(1)$ at large spin. The leading term is the Landau zero-point energy of the centre near the line. Every bosonic line with $s\le4$ is protected for all $|\zeta|\le1$ at either theta angle (Sec.~\ref{sec:freephases}). Localizing a scale-invariant system lowers its effective dimension and with it the critical attraction, as in the confinement-induced Efimov effect of mixed-dimensional atomic gases \cite{NishidaTan2008,NishidaTan2009,NishidaTan2011}. The five-dimensional theory, however, is maximally supersymmetric, and the instanton carries sixteen fermion zero modes. The field strength of the Berry connection on moduli space, the \emph{evaluation curvature}, couples their spins to the line's colour at the same order, $R^{-2}$, as the self-energy. Their Pauli interaction can therefore compete with the attraction even at arbitrarily small instanton size.

Our results are the following.
\begin{enumerate}
\item \emph{The lowest angular eigenvalue at the supersymmetric endpoints.} At $\zeta=\pm1$ the Hamiltonian is the square of a colour-twisted Dirac operator, and its link operator obeys $L(\pm1)\ge-8$ for every spin, with equality attained for $s\ge\tfrac12$ (Proposition~\ref{thm:floor}). This lower edge of the angular spectrum, which we call the \emph{endpoint floor}, follows from a gap about zero for the twisted Dirac operator on $S^7$. Its proof uses the fact that the evaluation connection is the pull-back of an instanton on $S^4$ along the Hopf fibration $S^7\to S^4$. In the endpoint superconformal algebra $OSp(4^*|2)$ the floor reads as a unitarity bound saturated by short multiplets (Sec.~\ref{sec:floor}).
\item \emph{Fermions overturn the bosonic protection.} For every $s\ge1$, in the supersymmetric quantum mechanics with a freely moving centre, the line loses conformality in a supercritical range that includes $\zeta=0$. We bound its outer boundary $\zeta_c(s)$ explicitly. Because the floor lies one unit above the Hardy value, a state at the floor of both supersymmetric halves collapses once its attraction $E\langle W\rangle$ exceeds one, where $W\le1$ is the polar profile of the self-energy $-EW/R^2$, equal to one in the pinned direction. Such \emph{doubly-BPS} states keep the supersymmetric link energy fixed at $-8$ while the self-energy drives the instability (Sec.~\ref{sec:instability}). At $s=1$ and $\theta_{5d}=\pi$ no doubly-BPS state exists, and the instability follows instead from mixing of endpoint floor states (Sec.~\ref{sec:bps}). By contrast, the fundamental line $s=\tfrac12$ is conformal for all $|\zeta|\le1$ and both theta sectors, with a unique self-adjoint realization. Its colour spin cannot pair with both fermion sets at once, so no state lies at the floor of both supersymmetric halves, and a three-spin recoupling bound with a scalar comparison keeps its link spectrum above $-8$ for $|\zeta|<1$ (Sec.~\ref{sec:fundamental} and Appendix~\ref{app:fb:s12}).
\item \emph{A two-sided bound on the threshold.} Introducing the dimensionless critical attraction $E_c(s)=4s(s+1)\bigl(1-\zeta_c(s)^2\bigr)$ (the supersymmetric pinned threshold is $E=1$), every integer $s\ge1$ at $\theta_{5d}=0$ obeys
\begin{equation}
 1+\frac3{2s}-\frac3{4s^2}\le E_c(s)\le1+\frac3{2s},
 \label{eq:law}
\end{equation}
and the same lower bound holds for every half-integer $s\ge\tfrac32$ with the analytic upper bound $E_{3\times3}(s)=1+3/(2s)-9/(56s^2)+O(s^{-3})$ obtained by mixing three endpoint floor states (Sec.~\ref{sec:bps}). At $\theta_{5d}=\pi$, the integer and half-integer families swap, and at $s=1$ mixing two such states in block $(4,5)$ bounds the threshold from above (Secs.~\ref{sec:instability} and~\ref{sec:bps}). These brackets hold at every spin $s\ge1$, the lower side being a computer-assisted proof in exact arithmetic and the upper side analytic. For integer $s$ the upper side sharpens to the bound $E_{2\times2}(s)$ from two endpoint floor states. As $s\to\infty$ the threshold is fixed analytically through two further orders,
\begin{equation}
 E_c(s)=1+\frac3{2s}-\frac9{56s^2}+\frac{c_3}{s^3}+O(s^{-7/2}),
 \label{eq:lawintro}
\end{equation}
where $c_3=-3447/43904$ for integer $s$ at $\theta_{5d}=0$ and for half-integer $s$ at $\theta_{5d}=\pi$, and $c_3=-1143/43904$ in the two converse cases, with remainder constants that are not explicit (Eqs.~\eqref{eq:lawcor2} and~\eqref{eq:lawcor3}).
\item \emph{The mechanism is self-localization.} Near the line the centre moves in the Berry curvature of the line's colour, which is the instanton's own field strength at the line. For the highest colour weight it is a magnetic field $B=8s/\rho^2$ in two orthogonal planes of the isotropic coordinate $Z=X/\sqrt2$. The Pauli energy of the anti-aligned centre fermions cancels the Landau zero-point energy $2B$, so at large $s$ the centre $X$ localizes at zero magnetic cost in a lowest Landau level of width $\rho/\sqrt{2s}$ in $Z$, that is within $\rho/\sqrt s$ of the line at instanton size $\rho$. The free instanton then behaves like a pinned one, approaching the supersymmetric pinned threshold $\zeta_{\rm pin}(s)=\sqrt{1-1/(4C_2)}$ (with $C_2=s(s+1)$), while the residual Landau-level spread sets the coefficient $3/2$, the same for integer and half-integer spin. Globally, the BPS floor states are products of twistor spinors on $S^4$, isomorphic as $Sp(2)$ representations to the lowest Landau levels of the four-dimensional quantum Hall problem \cite{ZhangHu}.
\item \emph{The discrete theta angle selects the channel.} The theories with $\theta_{5d}=0$ and $\theta_{5d}=\pi$, of which the first has the proposed untwisted $A_1$ light-cone interpretation, admit different physical channels through the parity rule of Eq.~\eqref{eq:physical}. At spin one a window of attractions, certified by a computer-assisted proof, separates them, with the free line collapsing at $\theta_{5d}=0$ and admitting a conformal, though not unique, realization at $\theta_{5d}=\pi$ (Proposition~\ref{prop:theta-window}). At large spin their thresholds first differ at order $s^{-3}$, by $18/(343s^3)$. With bosonic moduli alone the angle decides much more. For every $s\ge\tfrac92$ there is a scalar coupling at which one theory falls to the centre in a single angular multiplet while the other keeps a unique self-adjoint Hamiltonian, because the projection changes the orbital cost of the cheapest channel by at least three units while contact and collapse are only about one unit apart (Proposition~\ref{thm:bosphase}). The fermion zero modes, which cancel that orbital cost, shrink the difference to order $s^{-3}$.
\end{enumerate}
Two further results complete the picture. Near $|\zeta|=1$ several radial branches of one symmetry channel enter the window $-9\le\mu<-8$ together, so that the short-distance boundary conditions of that channel form at least a $U(2)$ family for $s=\tfrac32$ and integer $s\ge2$ and at least a $U(3)$ family for half-integer $s\ge\tfrac52$ (Sec.~\ref{sec:bps}). These count the polar branches of one channel and ordered block, and the magnetic copies of the channel repeat the family without adding transitions. Confining the instanton centre in a supersymmetric $\Omega$-trap tests the mechanism (Sec.~\ref{sec:windows}). In at least one fixed sector of the trap rotation, the free-conformal range $1<E\le E_c(s)$ carries an Efimov tower of shallow bound states, the supersymmetric counterpart of the confinement-induced Efimov effect, and the edge $E=1$ is sharp. Summed over sectors, the infinitely degenerate lowest Landau level of the trap lets centres displaced far from the line bind, and for $s\ge1$ the number of levels below $-\eta$ grows as $\eta^{-2}$ throughout $0<E\le E_c(s)$, an exponent set by the degeneracy of the trap rather than by the line.

The non-Lorentzian theory plays three enabling roles. First, its self-duality constraint makes the classical moduli dynamics of an isolated instanton exact at every size, whereas in a Lorentzian completion the large-spin source estimate requires $\rho\gg g_5^2s$ for weak backreaction, and the quantum prescription of Sec.~\ref{sec:models} then fixes the short-distance behaviour. Second, the constrained action uniquely fixes the fermion zero-mode couplings to the line (Appendix~\ref{app:fermions}), which drive the instability. Third, the null reduction identifies the instanton number with longitudinal momentum in the probe-free theory, motivating the conditional Wilson-surface interpretation developed in Sec.~\ref{sec:lightcone}. We quantize canonically with the flat moduli measure, $O(1)$ orbifold prescription, and covariant point-source subtraction (Secs.~\ref{sec:selfenergy} and~\ref{sec:models}), keeping the domain of the model explicit throughout.

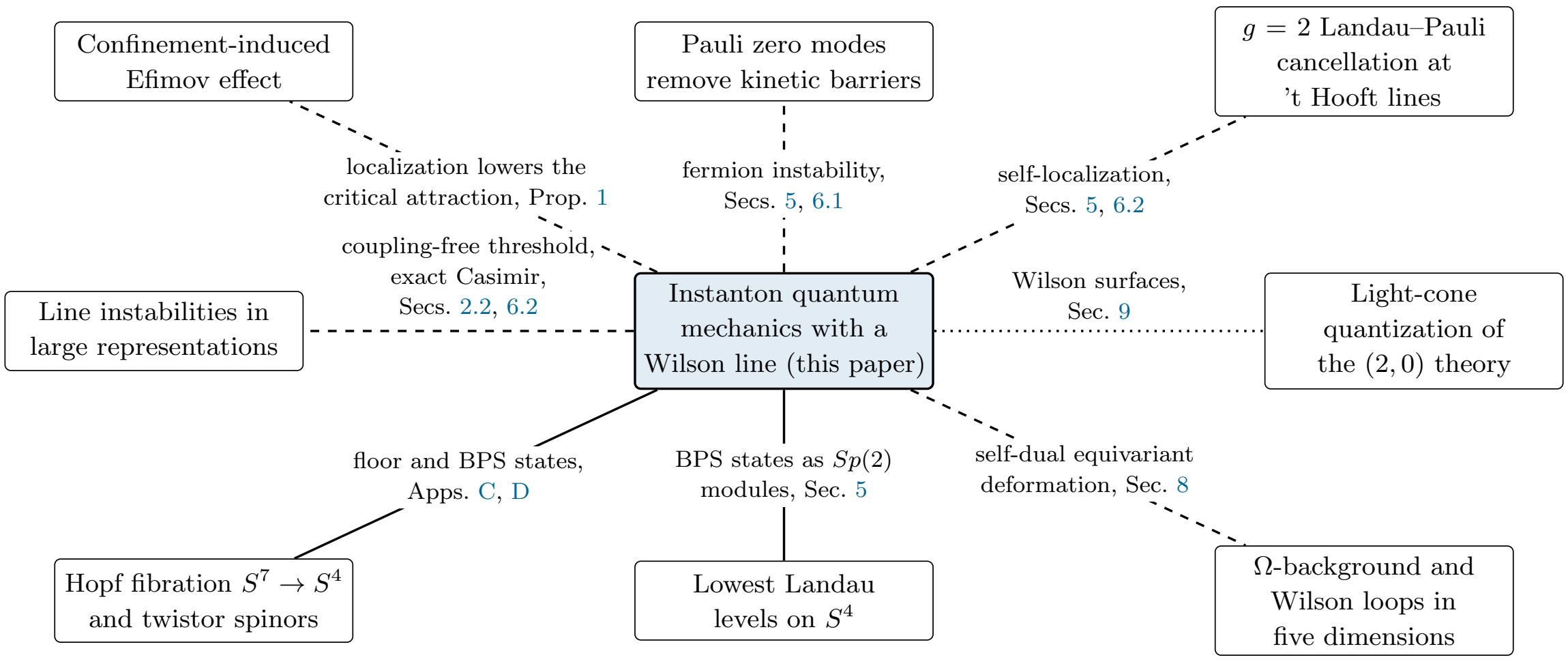
\begin{figure}[t]
\centering
\resizebox{\textwidth}{!}{%
\begin{tikzpicture}[font=\footnotesize,
  box/.style={draw,rounded corners=2pt,align=center,inner sep=3.5pt,text width=3.35cm,minimum height=0.95cm,
    execute at begin node={\hyphenpenalty=10000\exhyphenpenalty=10000}},
  core/.style={box,thick,text width=3.35cm,fill=MidnightBlue!8},
  lab/.style={font=\scriptsize,fill=white,inner sep=1.8pt,align=center},
  derived/.style={thick},
  mech/.style={thick,dashed},
  interp/.style={thick,dotted}]
 \node[core] (c) at (0,0) {Instanton quantum mechanics with a Wilson line (this paper)};
 \node[box] (efimov) at (-7.0,3.25) {Confinement-induced Efimov effect};
 \node[box] (pauli)  at (0,3.25)    {Pauli zero modes remove kinetic barriers};
 \node[box] (gtwo)   at (7.0,3.25)  {$g=2$ Landau--Pauli cancellation at 't~Hooft lines};
 \node[box] (large)  at (-7.6,0)   {Line instabilities in large representations};
 \node[box] (dlcq)   at (7.6,0)    {Light-cone quantization of the $(2,0)$ theory};
 \node[box] (hopf)   at (-7.0,-3.25) {Hopf fibration $S^7\to S^4$\\ and twistor spinors};
 \node[box] (qhe)    at (0,-3.25)   {Lowest Landau levels on $S^4$};
 \node[box] (omega)  at (7.0,-3.25) {$\Omega$-background and Wilson loops in five dimensions};
 \draw[mech] (c) -- node[lab,pos=0.52] {localization lowers the\\ critical attraction, Prop.~\ref{thm:L}} (efimov);
 \draw[mech] (c) -- node[lab,pos=0.50] {fermion instability,\\ Secs.~\ref{sec:pauli}, \ref{sec:instability}} (pauli);
 \draw[mech] (c) -- node[lab,pos=0.52] {self-localization,\\ Secs.~\ref{sec:pauli}, \ref{sec:mechanism}} (gtwo);
 \draw[mech] (c) -- node[lab,above=2pt] {coupling-free threshold,\\ exact Casimir,\\ Secs.~\ref{sec:selfenergy}, \ref{sec:mechanism}} (large);
 \draw[interp] (c) -- node[lab,above=2pt] {Wilson surfaces,\\ Sec.~\ref{sec:lightcone}} (dlcq);
 \draw[derived] (c) -- node[lab,pos=0.52] {floor and BPS states,\\ Apps.~\ref{app:floor}, \ref{app:fb:bps}} (hopf);
 \draw[derived] (c) -- node[lab,pos=0.50] {BPS states as $Sp(2)$\\ modules, Sec.~\ref{sec:bpsstates}} (qhe);
 \draw[mech] (c) -- node[lab,pos=0.52] {self-dual equivariant\\ deformation, Sec.~\ref{sec:windows}} (omega);
\end{tikzpicture}%
}
\caption{The settings this paper connects. Solid lines mark results used or derived here, dashed lines a shared mechanism or a comparison, and the dotted line an interpretation. The labels give the section or appendix that makes each link.}
\label{fig:connections}
\end{figure}

The semiclassical Wilson-line instability in four-dimensional $\mathcal N=4$ Yang--Mills theory \cite{ACKMRletter,ACKMR} illustrates how the onset of collapse depends on the dynamics of the probe. The four-dimensional threshold depends on $g^2s$ in a double-scaling limit. In the one-instanton Hamiltonian $g_5^2$ multiplies every term, so it sets the energy scale and drops out of the stability criterion. The scalar attraction depends on the exact Casimir $s(s+1)$, and the width of the centre wavefunction fixes the first large-$s$ correction. The fermion mechanism is a soliton-moduli realization of Pauli zero modes removing kinetic barriers against a scale-covariant attraction \cite{FLL1986,LossYau1986,FrankLoss}, and the Landau--Pauli cancellation plays a role analogous to the $g_m=2$ cancellation for charged particles at 't~Hooft lines. Figure~\ref{fig:connections} places the paper among these settings.

The first step is to quantize the constrained moduli dynamics and identify the angular eigenvalue that controls collapse (Secs.~\ref{sec:system} and~\ref{sec:spectral}). The bosonic theory isolates the energy cost of localizing a moving centre. The fermion zero modes then reveal how that cost can be cancelled, and why the supersymmetric endpoints attain $-8$ (Secs.~\ref{sec:bosonic} and~\ref{sec:susy}). Section~\ref{sec:selfloc} uses those floor states to find the free threshold, show that a single onset separates conformal and supercritical couplings, and quantify its theta dependence. The lowest endpoint eigenspaces also explain the exceptional stability of the fundamental line and distinguish the two theta sectors at spin one (Sec.~\ref{sec:bps}). We then test localization with a supersymmetric trap, distinguish its fixed-charge and global spectra, and examine the conditions for a six-dimensional interpretation (Secs.~\ref{sec:windows}--\ref{sec:discussion}). Appendices~\ref{app:frames}--\ref{app:localization} contain the complete bundle, fermion-coupling, Hopf--Dirac floor, twistor/BPS, exact-arithmetic lower-bound, and localization proofs, while Appendix~\ref{app:un:coverage} bounds the angular range at onset and Appendix~\ref{app:numerics} explains the radial enclosures and numerical scans.

\section{Instanton quantum mechanics with a Wilson line}
\label{sec:system}

 This section builds the quantum mechanics whose spectrum decides conformality. The constrained five-dimensional theory fixes the exact moduli dynamics (Sec.~\ref{sec:theory}); the line adds the evaluation connection (Sec.~\ref{sec:connection}) and a self-energy (Sec.~\ref{sec:selfenergy}); supersymmetry fixes the coupling of the fermion zero modes (Sec.~\ref{sec:fermions}); and Sec.~\ref{sec:models} states the quantization and separates the models that the later sections compare.

\subsection{The constrained gauge theory and the Wilson-line Berry connection}
\label{sec:theory}

 The five-dimensional theory of Lambert and collaborators is proposed as the null reduction of the six-dimensional $(2,0)$ theory \cite{LambertRichmond,LO,LPS}. It is the case of a null vector field in the $(2,0)$ tensor-multiplet system of Lambert and Papageorgakis \cite{LambertPapageorgakis}. It is also the scale-invariant fixed point of a non-Lorentzian rescaling of maximally supersymmetric Yang--Mills theory, with twenty-four supersymmetries, ordinary and superconformal \cite{LambertMouland}. Its fields live on $\R_t\times\R^4$ and comprise a gauge field $(A_t,A_i)$, five adjoint scalars $X^I$ ($I=6,\dots,10$), adjoint fermions $\Psi$ with sixteen real components, and an anti-self-dual two-form Lagrange multiplier $G_{ij}$ that imposes self-duality on each Euclidean spatial slice $\R^4$. The instanton is a particle whose collective coordinates evolve in $t$. Because of the constraint, the classical field equations reduce exactly to geodesic motion on the instanton moduli space, supplemented by the forces that background scalars and sources produce. The instanton number $k$ is the momentum along the null circle. In the normalization of Ref.~\cite{LPS}, $g_5^2=4\pi^2R_{\rm null}$, with $R_{\rm null}$ the radius of the null circle (written $R_+$ there, since that reference interchanges the light-cone labels $+$ and $-$ relative to Sec.~\ref{sec:lightcone}), and a configuration with instanton number $k$ carries light-cone energy $\tfrac1{2g_5^2}\,g_{\alpha\beta}\dot m^\alpha\dot m^\beta$, where $m^\alpha$ are the moduli and $g_{\alpha\beta}$ is the metric that the field-space inner product induces on the moduli space. We normalize the action as $\frac1{g_5^2}\int\dd t\,\dd^4x\,\tr\bigl(\tfrac12F_{ti}F_{ti}-\tfrac12D_iX^ID_iX^I+\dots\bigr)$ with $\tr(T_aT_b)=\tfrac12\delta_{ab}$, as in Appendix~\ref{app:fermions}. The moduli metric $g_{\alpha\beta}$ and the Gauss-law energy of Sec.~\ref{sec:selfenergy} both follow from it. We take gauge group $SU(2)$ and instanton number $k=1$, for which the moduli metric is flat and the angular problem has a smooth equivariant lift to $S^7$ (higher rank and $k>1$ are discussed in Sec.~\ref{sec:discussion}). A related family of five-dimensional theories, with an $\Omega$-deformed constraint and $SU(1,3)$ symmetry, arises from the null reduction of six-dimensional superconformal theories on conformally compactified Minkowski space \cite{LLMR2019,LLMR2020}. We use the undeformed theory.

The probe is a Wilson line of spin $s$ at the origin of space,
\begin{equation}
 \mathcal W_s[\zeta]=\tr_{V_s}\,\mathrm P\exp\!\left(\ii\!\int\!\dd t\,\bigl(A_t+\zeta\,X^6\bigr)\right),
 \label{eq:line}
\end{equation}
where $V_s$ is the spin-$s$ representation, with Casimir $C_2=s(s+1)$, and we align the scalar coupling along $X^6$ without loss of generality by $SO(5)_R$ invariance. The line with $\zeta=0$ is the ordinary electric probe, and $\zeta=\pm1$ are the two supersymmetric lines. In the instanton background the source in Eq.~\eqref{eq:line} produces line-sourced components $A_t^{\rm line}$ through the Gauss constraint and $X^{6,\rm line}$ through the scalar equation. The two equations involve the same covariant Laplacian with opposite signs, so these line-sourced fields obey $X^{6,\rm line}=-\zeta A_t^{\rm line}$ (while velocity and fermion-bilinear terms enter separately, Appendix~\ref{app:fermions}). We treat the line as a static heavy probe and $\zeta$ as a fixed parameter with $|\zeta|\le1$, the regime in which the static self-energy is attractive and the supersymmetric weights $c_\pm=(1\pm\zeta)/2$ in Eq.~\eqref{eq:Hzeta} below are nonnegative (for $|\zeta|>1$ one weight turns negative, flipping the sign of the corresponding Pauli term while the self-energy turns repulsive; we do not analyse that regime). The fields $A_t$ and $X^I$ appear in the constrained action without time derivatives, so their exchange along the line is instantaneous. The line self-energy has a power divergence removed by the covariant vacuum subtraction in Sec.~\ref{sec:selfenergy}. The reduced operator treats $\zeta$ as a fixed parameter. In four-dimensional gauge theory it can run \cite{PolchinskiSully,BGT2022}, and along such line-defect flows the defect entropy decreases \cite{CKR2022}. Inside the constrained action, the nondynamical fermion component also acts as a Lagrange multiplier projecting out all non-zero-mode fermions (Appendix~\ref{app:fermions}). Any running of $\zeta$ from non-BPS path-integral determinants or from modes exchanging null momentum in a six-dimensional $(2,0)$ parent or a Lorentzian five-dimensional completion lies outside the constrained moduli model (Sec.~\ref{sec:lightcone}).

The five-dimensional $SU(2)$ theory has a discrete theta angle $\theta_{5d}\in\{0,\pi\}$ associated with $\pi_4(SU(2))=\Z_2$ \cite{DKV,MorrisonSeiberg,BRZ}. In the one-instanton quantum mechanics it selects the character $e^{\ii\theta_{5d}}$ under the lifted redundancy $\lambda\to-\lambda$, where $\lambda$ is the size and orientation quaternion. The lift acts on the size fermions as well as on the moduli (Appendix~\ref{app:frames}). We use $\theta_{5d}=0$ as the reference sector when the angle is unspecified, and compare both characters in Sec.~\ref{sec:selfloc}. The spacelike reductions identify the untwisted $A_1$ theory with $\theta_{5d}=0$ and the $A_2$ theory with an outer-automorphism twist with $\theta_{5d}=\pi$ \cite{Tachikawa}. The proposed untwisted null reduction uses the first assignment. At $\theta_{5d}=\pi$, a light-cone interpretation of our Hamiltonian would require identifying the charge-one sector and its Wilson line in the null limit of the twisted $A_2$ compactification. Our comparison of their spectra uses the two five-dimensional theories directly.\footnote{\label{fn:theta}At $\theta_{5d}=\pi$ the angular projection in Eq.~\eqref{eq:physical} exchanges the integer and half-integer common-floor families. The spin-one sector then loses conformality through mixing without a common endpoint floor state (Sec.~\ref{sec:bps} and Appendix~\ref{app:fb:block45}).}

\paragraph*{The moduli space and the evaluation connection.}\phantomsection\label{sec:connection}

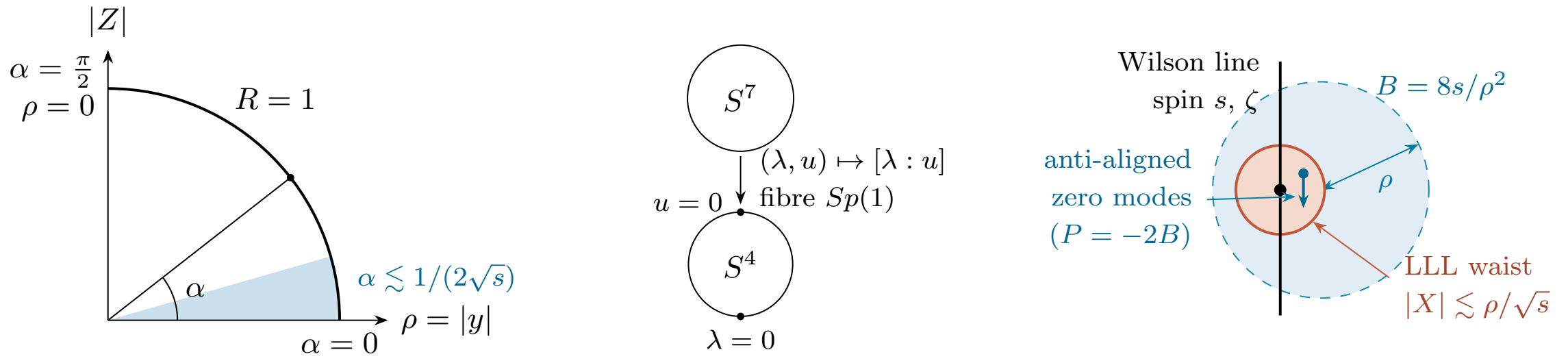
\begin{figure}[t]
\centering
\resizebox{\textwidth}{!}{%
\begin{tikzpicture}[baseline={(0,0)},font=\footnotesize,>={Stealth[length=4pt]}]
 \fill[MidnightBlue!14] (0,0) -- (2.4,0) arc[start angle=0,end angle=16,radius=2.4] -- cycle;
 \draw[->] (0,0) -- (2.9,0) node[right] {$\rho=|y|$};
 \draw[->] (0,0) -- (0,2.8) node[above] {$|Z|$};
 \draw[thick] (2.4,0) arc[start angle=0,end angle=90,radius=2.4];
 \node[above right=-1pt] at (60:2.4) {$R=1$};
 \draw (0,0) -- (38:2.4);
 \fill (38:2.4) circle (1.1pt);
 \draw (0.72,0) arc[start angle=0,end angle=38,radius=0.72];
 \node at (19:0.96) {$\alpha$};
 \node[below] at (2.4,0) {$\alpha=0$};
 \node[left,align=right] at (0,2.4) {$\alpha=\tfrac\pi2$\\[-1pt]$\rho=0$};
 \node[MidnightBlue,anchor=west] at (2.45,0.44) {\scriptsize$\alpha\lesssim1/(2\sqrt s)$};
 \node[anchor=north] at (1.25,-0.55) {(a) Moduli quadrant};
\end{tikzpicture}\hspace{0.75cm}%
\begin{tikzpicture}[baseline={(0,0)},font=\footnotesize,>={Stealth[length=4pt]}]
  \draw (0,2.30) circle (0.55);
  \node at (0,2.30) {$S^7$};
  \draw[->] (0,1.70) -- (0,1.20);
  \node[right,align=left] at (0.06,1.45) {\scriptsize$(\lambda,u)\mapsto[\lambda:u]$\\[-1pt]\scriptsize fibre $Sp(1)$};
  \draw (0,0.58) circle (0.54);
  \node at (0,0.58) {$S^4$};
  \fill (0,1.12) circle (1.1pt);
  \node[left] at (-0.05,1.22) {\scriptsize$u=0$};
  \fill (0,0.04) circle (1.1pt);
  \node[below] at (0,0.04) {\scriptsize$\lambda=0$};
  \node[anchor=north] at (0,-0.55) {(b) Hopf pull-back};
\end{tikzpicture}\hspace{0.75cm}%
\begin{tikzpicture}[baseline={(0,0)},font=\footnotesize,>={Stealth[length=4pt]}]
  \fill[MidnightBlue!8] (0.42,1.35) circle (1.12);
  \draw[dashed,MidnightBlue!70] (0.42,1.35) circle (1.12);
  \draw[<->,MidnightBlue!80] (0.42,1.35) -- ++(25:1.12) node[midway,below right=-2pt] {\scriptsize$\rho$};
  \fill[BrickRed!14] (0,1.35) circle (0.46);
  \draw[BrickRed!80,thick] (0,1.35) circle (0.46);
  \draw[thick,black] (0,0.05) -- (0,2.68);
  \fill[black] (0,1.35) circle (1.8pt);
  \node[left,align=right] at (-0.08,2.46) {\scriptsize Wilson line\\[-1pt]\scriptsize spin $s$, $\zeta$};
  \fill[MidnightBlue] (0.24,1.52) circle (1.5pt);
  \draw[->,MidnightBlue,thick] (0.24,1.52) -- (0.24,1.16);
  \node[MidnightBlue,anchor=west] at (0.85,2.42) {\scriptsize$B=8s/\rho^2$};
  \node[BrickRed!85!black,align=left,anchor=west] at (1.15,0.34) {\scriptsize LLL waist\\[-1pt]\scriptsize$|X|\lesssim\rho/\sqrt s$};
  \draw[->,BrickRed!80,thin] (1.13,0.44) -- (0.34,1.02);
  \node[MidnightBlue,align=right,anchor=east] at (-0.78,1.25) {\scriptsize anti-aligned\\[-1pt]\scriptsize zero modes\\[-1pt]\scriptsize($P=-2B$)};
  \draw[->,MidnightBlue!80,thin] (-0.76,1.25) -- (0.16,1.28);
  \node[anchor=north] at (0.35,-0.55) {(c) Self-localization in $\R^4$};
\end{tikzpicture}%
}
\caption{Geometry and physical mechanism. (a)~The quadrant of instanton size $\rho$ and centre displacement $|Z|$ on the moduli cone $\R^8/O(1)$; the arc $R=1$ is the orbit-space image of the link $S^7$, $\alpha=0$ is the pinned direction, $\alpha=\pi/2$ is the zero-size stratum, and at large $s$ the shaded wedge reaches the $1/e$ contour of the BPS density near $\alpha=0$ (Fig.~\ref{fig:profile} and Sec.~\ref{sec:mechanism}). (b)~ The quaternionic Hopf map $S^7\to S^4=\HH P^1$, along which the evaluation connection pulls back the basic instanton and collapses the pinned slice $u=0$ to one pole (Sec.~\ref{sec:connection}). (c)~Physical self-localization in $\R^4$; a static Wilson line generates the moduli-space Berry field $B=8s/\rho^2$ from the instanton's own field strength, while anti-aligned centre fermion zero modes cancel the Landau zero-point energy ($P=-2B$) and confine the instanton centre within $|X|\lesssim\rho/\sqrt s$ of the line (Secs.~\ref{sec:pauli} and~\ref{sec:mechanism}).}
\label{fig:geometry}
\end{figure}

A charge-one Belavin--Polyakov--Schwartz--Tyupkin (BPST) instanton \cite{BPST} is described by the Atiyah--Drinfeld--Hitchin--Manin (ADHM) data $\Delta(x)=(\lambda,\,x-X)^{\rm T}\in\HH^2$, with centre $X\in\R^4\simeq\HH$ and a quaternion $\lambda$ whose norm $\rho=|\lambda|$ is the size and whose phase is the global gauge orientation \cite{ADHM}. The data $\lambda$ and $-\lambda$ describe the same field. This is the residual $O(1)$ redundancy. The moduli metric is flat. Including the factor $1/g_5^2$ of the light-cone energy, the kinetic metric is
\begin{equation}
 \dd s^2=\frac{4\pi^2}{g_5^2}\bigl(|\dd X|^2+2|\dd\lambda|^2\bigr)
 =\frac{8\pi^2}{g_5^2}\bigl(|\dd Z|^2+|\dd y|^2\bigr),
 \label{eq:metric}
\end{equation}
where the isotropic coordinates $(Z,y)=(X/\sqrt2,\lambda)$ make the eight-dimensional moduli cone $\R^8/O(1)$ round, with $O(1)=\{\pm1\}$ acting by $\lambda\to-\lambda$. The centre-to-size inertia ratio $m_c/m_s=1/2$ fixes the relative scaling $\sqrt2$ between $X$ and $\lambda$ and will matter in Sec.~\ref{sec:bosonic} and in the conformal geometry of Sec.~\ref{sec:susy}.\footnote{Evaluating the background-gauge tangent norms $\int\dd^4x\,\tr(\delta_aA_i\,\delta_bA_i)$ on the BPST instanton gives $4\pi^2\delta_{\mu\nu}$ for translations $\delta_{X_\mu}A_i=F_{\mu i}$ against $8\pi^2\delta_{\alpha\beta}$ for size and orientation variations $\delta_{\lambda_\alpha}A_i$, fixing $m_c/m_s=1/2$. The same $\sqrt2$ ratio sets both the bosonic pinning criterion $m_c/m_s\le2$ (Sec.~\ref{sec:theoremL}) and the conformal dilation $\bar N$ of the instanton on $S^4$ (Sec.~\ref{sec:bpsstates}).} In these coordinates the free Hamiltonian is $-(g_5^2/16\pi^2)\nabla^2_{\R^8}$, and from now on we factor $g_5^2/(16\pi^2)$ out of the Hamiltonian. Its reduced eigenvalues have dimensions of inverse length squared. Write $R^2=|Z|^2+|y|^2$, $|Z|=R\sin\alpha$ and $\rho=|y|=R\cos\alpha$ with $\alpha\in(0,\pi/2)$. The direction $\alpha=0$ is the pinned direction, in which the instanton is centred on the line. Figure~\ref{fig:geometry} shows these coordinates.

The line's colour lives in the fibre, at the origin, of the instanton's gauge bundle. In the ADHM construction this fibre is the kernel of $\Delta(0)^\dagger=(\bar\lambda,\bar u)$, with $u=-X$; it is the quaternionic line $q^\perp\subset\HH^2$ orthogonal to $q=(\lambda,u)$ \cite{ADHM}. Choosing a unit vector $v$ spanning $q^\perp$ over the quaternions defines the evaluation connection $\mathcal A=v^\dagger\dd v$ on the moduli space. Dynamically, it enters the moduli Lagrangian through the compensating gauge potential $A_t^{\rm v}$ evaluated at the probe (Appendix~\ref{app:fermions}), and it transports the line's colour as the instanton moves \cite{TWADHM}. The same construction couples Wilson lines to monopoles \cite{TWmono}. The connection therefore fixes the two effects that compete below. Its covariant derivative sets the orbital energy a centre pays to approach the line (Sec.~\ref{sec:bosonic}), and its curvature couples to the spins of the fermion zero modes (Sec.~\ref{sec:pauli}). A convenient frame, regular in the pinned direction, is
\begin{equation}
 v=\frac1{\sqrt\varphi}\begin{pmatrix}\lambda\bar u\bar\lambda/|\lambda|^2\\ -\bar\lambda\end{pmatrix},
 \qquad \varphi=|\lambda|^2+|u|^2 .
 \label{eq:frame}
\end{equation}
Its curvature has a simple closed form, which we use throughout. For any unit frame $v$ of $q^\perp$,
\begin{equation}
 \mathcal F=\frac1\varphi\,(v^\dagger\dd q)\wedge\overline{(v^\dagger\dd q)},\qquad
 \mathcal F(\xi,\eta)=\frac2\varphi\,\operatorname{Im}\bigl[(v^\dagger\xi)\,\overline{(v^\dagger\eta)}\bigr].
 \label{eq:curvature}
\end{equation}
For the induced connection, $\mathcal F=\dd\mathcal A+\mathcal A\wedge\mathcal A=\dd v^\dagger\wedge(1-vv^\dagger)\dd v=\dd v^\dagger\wedge P_q\,\dd v$, where $P_q=qq^\dagger/\varphi$ projects onto $q\HH$, and $q^\dagger\dd v=-\dd q^\dagger v$ because $q^\dagger v=0$. Together these give Eq.~\eqref{eq:curvature}. Three properties of $\mathcal F$ follow from this form, and the later sections use each of them.
(i) Right multiplication by a unit quaternion $r$ multiplies both $v^\dagger\xi$ and $v^\dagger\eta$ on the right by $r$, which drops out of $\operatorname{Im}[a\bar b]$. Hence $\mathcal F(\xi r,\eta r)=\mathcal F(\xi,\eta)$ for every unit quaternion $r$, so $\mathcal F$ is of type $(1,1)$ for the three complex structures given by right multiplication and takes values in $sp(2)\otimes su(2)$. Here $Sp(2)\subset SO(8)$ is the group of quaternion-linear isometries of $\HH^2\simeq\R^8$. The form part of $\mathcal F$ lies in $sp(2)\subset so(8)\simeq\Lambda^2\R^8$, and the colour part in $su(2)$.
(ii) Since $v^\dagger\xi$ depends only on the component of $\xi$ orthogonal to the quaternionic line $q\HH$, $\mathcal F$ is homogeneous of degree $-2$ and is pulled back from $\HH P^1\simeq S^4$ along $(\lambda,u)\mapsto[\lambda:u]$.
(iii) On the pinned slice $u=0$, $q^\perp=(0,1)\HH$ and $v^\dagger\xi$ vanishes for $\xi=(\xi_\lambda,0)$, so $\mathcal F$ vanishes on every size direction, while its centre--centre components do not.

The Hopf map identifies $(\lambda,u)$ up to right quaternionic rescaling and sends the pinned slice $u=0$ to one pole of $S^4=\HH P^1$. The complement of the tautological quaternionic line carries the basic instanton connection on $S^4$. This connection is the restriction to the unit sphere of the $SU(2)$ Yang monopole on an auxiliary Euclidean $\R^5\setminus\{0\}$ \cite{AHS,Atiyah1979,Trautman1977,Yang1978,Minami1980}. Pulling the connection on $S^4$ back along the Hopf map gives the Berry connection on instanton moduli space. In the isotropic moduli coordinates $(Z,y)$ the base connection includes a conformal dilation. This geometry determines the endpoint spectrum in Sec.~\ref{sec:susy}. Its type $(1,1)$ property follows from the $\mathcal N=(0,4)$ supersymmetry of the instanton quantum mechanics with a supersymmetric line. It is the hyperholomorphic condition of Ref.~\cite{Verbitsky}, and it implies the Yang--Mills equations. On the curve $(\lambda,u)=(\cos\alpha,-\sqrt2\sin\alpha)$ through each point of the link modulo symmetry, Eq.~\eqref{eq:curvature} reduces to
\begin{equation}
 \mathcal F^c_{AB}=\frac4{D^2}\bigl[e_c\,\hat w\hat w^{\rm T}\bigr]_{AB},
 \label{eq:basecurve}
\end{equation}
in the $(Z,y)$ components, with $\hat w=(\cos\alpha,-\sin\alpha)$ and $D=1+\sin^2\alpha=\varphi/R^2$. Here $e_c$ is an imaginary unit quaternion, $e_c\,\hat w\hat w^{\rm T}$ projects $\HH^2$ onto the quaternionic line spanned by $\hat w$ and multiplies on the left by $e_c$, and $[\,\cdot\,]_{AB}$ is the real $8\times8$ matrix of such a real-linear map of $\HH^2\simeq\R^8$. At every point, $\mathcal F$ is the curvature of the basic instanton on the quaternionic line spanned by the direction $\hat w$. At the pinned direction it becomes $\mathcal F_{Z_iZ_j}=-4\eta^c_{ij}e_c/\rho^2$, a self-dual field acting on the centre. Here $Z=Z^0+Z^ce_c$, $\eta^c_{0a}=\delta_{ca}$, $\eta^c_{ab}=\epsilon_{cab}$ (so $[e_c]_{ij}=-\eta^c_{ij}$), and the orientation is $\dd Z^0\wedge\dd Z^1\wedge\dd Z^2\wedge\dd Z^3$. Appendix~\ref{app:floor} orients the Hopf base oppositely, so there the same field is anti-self-dual.

The redundancy $\lambda\to-\lambda$ acts on $\HH^2$ by $\mathsf T=\operatorname{diag}(1,-1)$, up to the overall sign of the ADHM vector, which leaves the connection unchanged, and the frame~\eqref{eq:frame} obeys $v(-\lambda)=\mathsf Tv(\lambda)$, so the redundancy acts trivially on the colour components in this frame. Appendix~\ref{app:frames} relates this frame to a second regular frame $v_R$, of the kind used for calorons \cite{KraanVanBaal}, in which $v_R(-\lambda)=-\mathsf Tv_R(\lambda)$ and spin-$s$ sections obey $\psi(-\lambda)=(-1)^{2s}\psi(\lambda)$, and shows that both describe the same physical bundle.

\subsection{The self-energy and the fermion couplings}
\label{sec:selfenergy}

The line sources $A_t$ and $X^6$ in the instanton background, and eliminating them produces the potential that competes with the kinetic energy, as eliminating $A_t$ in a scalar background produces the potential of dyonic instantons \cite{LambertTong}. We define the static source covariantly. A smooth form factor $f_\ell$ of radius $\ell$ about the position of the line, with the colour $Q=2J^aT_a$ transported along fixed radial paths by the Wilson parallel transporter $\mathcal U$, gives the adjoint density $ Q_\ell=f_\ell\,\mathcal U Q\mathcal U^\dagger$, where $\tr(T_aT_b)=\delta_{ab}/2$. With $\mathscr L=-D_iD_i$ the covariant Laplacian of the self-dual background, the Gauss constraint sets the electric potential to $-g_5^2\mathscr L^{-1} Q_\ell$, and eliminating it adds the energy $\tfrac{g_5^2}2\int\tr( Q_\ell\mathscr L^{-1} Q_\ell)\ge0$ to the moduli Hamiltonian. The scalar $X^6$ is sourced by $\zeta Q_\ell$ through the same operator and enters with the opposite sign. The static energy is therefore
\begin{equation}
 V_{\rm self}=(1-\zeta^2)\,\frac{g_5^2}2\int\tr\bigl( Q_\ell\,\mathscr L^{-1} Q_\ell\bigr)\,\dd^4x ,
 \label{eq:gausselimination}
\end{equation}
which vanishes on the supersymmetric lines, as it must for a static BPS configuration. The elimination assumes that the sourced configuration lifts to the constrained theory. On the background-gauge slice let $\mathscr K a=(D_{A}a)^-$ (the anti-self-dual part of the linearized field strength), whose kernel is the tangent space of the moduli space. The anti-self-dual multiplier $G_{ij}$ (carrying opposite four-dimensional duality to the self-dual field strength $F_{ij}$, matching $H_{ij+}$ against $H_{ij-}$ in Sec.~\ref{sec:lightcone}) satisfies $D_j G_{ji}=\mathcal R_i$, where $\mathcal R_i$ is the residual spatial force left after eliminating $A_t$ and $X^6$. Orthogonality of $\mathcal R_i$ to $\ker\mathscr K$ is the Euler--Lagrange equation on the moduli space, and $G_{ij}$ can be reconstructed when $\mathcal R_i$ lies in the range of $\mathscr K^*$, with the required decay. We assume this range condition for smooth sources. The argument leaves open whether the inverse stays bounded as $\rho\to0$ and whether a distributional point source admits a lift. The quantum model of Sec.~\ref{sec:models} uses a separate point-source prescription.

For a point source we subtract the free singularity covariantly. On flat $\R^4$ the coincident heat kernel of $\mathscr L$ has $a_0=\mathbf1$ and $a_1=0$, so
\begin{equation}
 \mathscr R_\tau=\int_0^\infty\Bigl[H_{\mathscr L}(t;x_0,x_0)-\frac{\mathbf 1}{(4\pi t)^2}\,\mathbf 1_{t<\tau}\Bigr]\dd t
 \label{eq:pointprescription}
\end{equation}
is finite. Its limit $\mathscr R_\infty$ as $\tau\to\infty$ is the vacuum-subtracted coincident Green function, and the point-source energy is $(1-\zeta^2)\tfrac{g_5^2}2\tr(Q\mathscr R_\infty Q)$. For one instanton $\mathscr R_\infty$ follows from the closed-form propagators \cite{BrownGreen,CorriganGreen,GreenKovacsProp}. With $v(x)$ the unit vector spanning the kernel of $\Delta(x)^\dagger$ and $M(x,x')=v(x)^\dagger v(x')$, the adjoint Green function of $\mathscr L$ is the tensor product $J\mapsto M(x,x')\,J\,M(x,x')^\dagger/(4\pi^2|x-x'|^2)$. The additional term of Ref.~\cite{CorriganGreen} is proportional to $\sum_\mu e_\mu Q\bar e_\mu=4\operatorname{Re}Q$ and vanishes for traceless sources, a cancellation that Ref.~\cite{GreenKovacsProp} also notes for $SU(2)$. Along any direction $n$, $M(x,x+\varepsilon n)$ differs from the parallel transporter by the factor $1-\tfrac12\varepsilon^2\rho^2/(\rho^2+|x-X|^2)^2+O(\varepsilon^3)$, and subtracting the free Green function gives
\begin{equation}
 \mathscr R_\infty=-\frac{\rho^2}{4\pi^2\bigl(\rho^2+|X|^2\bigr)^2}\,\mathbf 1_{\rm adj}
 =-\frac{\delta^{\mu\nu}q_{\mu\nu}}{16\pi^2}\,\mathbf 1_{\rm adj},
 \label{eq:BPSTfinite}
\end{equation}
where $q_{\mu\nu}=\rho^2\delta_{\mu\nu}/(\rho^2+|x-X|^2)^2$ is the pull-back to space, by the normalized ADHM kernel, of the round metric of the four-sphere of radius $\tfrac12$. The self-energy is therefore the conformal factor of the instanton's four-sphere at the position of the line. In the spin-$s$ representation, $Q=2J^aT_a$ and $\tr(T_aT_b)=\delta_{ab}/2$ give $\tr Q^2=2\sum_a(J^a)^2=2C_2\mathbf 1$, with quadratic Casimir $C_2=s(s+1)$, and in units of $g_5^2/16\pi^2$
\begin{equation}
 V_{\rm self}=-4C_2\,(1-\zeta^2)\,\frac{\rho^2}{\bigl(\rho^2+|X|^2\bigr)^2}=\frac{1-\zeta^2}{R^2}\,V(\alpha),\qquad
 V(\alpha)=-4C_2\,\frac{\cos^2\alpha}{(1+\sin^2\alpha)^2}.
 \label{eq:selfenergy}
\end{equation}
In terms of the attraction $E=4C_2(1-\zeta^2)$ the self-energy is $V_{\rm self}=-EW/R^2$, with the angular profile $W=\cos^2\alpha/(1+\sin^2\alpha)^2\le1$. The colour enters only through the Casimir, because the finite part is proportional to the identity in adjoint colour. The self-energy is homogeneous of degree $-2$, attractive for $|\zeta|<1$ and strongest in the pinned direction, where it equals $-4C_2(1-\zeta^2)/\rho^2$. The attraction arises physically because the non-flat instanton field screens the point probe relative to flat space. The unsubtracted Gauss energy is positive, and the instanton lowers it below the vacuum divergence.\footnote{In the heat-kernel representation~\eqref{eq:pointprescription}, the vector-bundle Kato inequality and heat-semigroup domination \cite{HessSchraderUhlenbrock1980} bound the norm of the coincident heat kernel by $(4\pi t)^{-2}$. Equation~\eqref{eq:BPSTfinite} gives strict negativity of the adjoint finite part at positive size. The local heat coefficient $a_1=0$ excludes a logarithmic divergence. Higher-dimension terms such as $c\ell^2\tr F^2$ vanish at fixed $\rho>0$ as $\ell\to0$. This observation gives no uniform estimate at $\rho\sim\ell$.} A source of finite size bounds it. In physical units, for a spherical shell of radius $\ell$ centred on the instanton, the vacuum-subtracted energy is $-(1-\zeta^2)\,\tfrac{g_5^2C_2}{12\pi^2}\,(2\ell^2+3\rho^2)/(\ell^2+\rho^2)^2$, which reduces to Eq.~\eqref{eq:selfenergy} for $\ell\ll\rho$ and tends to $-(1-\zeta^2)g_5^2C_2/(6\pi^2\ell^2)$ as $\rho\to0$.

\paragraph*{Fermion zero modes and their coupling to the line.}\phantomsection\label{sec:fermions}

A charge-one instanton of the maximally supersymmetric theory has sixteen real fermion zero modes, eight from the broken supersymmetries and eight from the broken superconformal symmetries of the constrained theory \cite{LambertMouland}; for $SU(2)$ these are all its zero modes \cite{DHKM}. The fermion coupling of the line follows from the Yukawa and kinetic terms of the constrained theory once the line-sourced relation $X^{6,\rm line}=-\zeta A_t^{\rm line}$ is inserted (with $\Gamma_\pm=(\Gamma_5\pm\Gamma_t)/\sqrt2$, Appendix~\ref{app:fermions}); it is
\begin{equation}
 \tfrac12\,\bar\Psi\,\Gamma_-\,(1+\zeta\Gamma^6)\,[A_t,\Psi] .
 \label{eq:fermioncoupling}
\end{equation}
Since $\Gamma_t\Gamma_-$ commutes with $\Gamma^6$, the bilinear is diagonal in $\Gamma^6$. The zero modes split into two sets $\chi_\pm$ with $\Gamma^6=\pm1$, eight real modes each, coupled to the line with weights $1\pm\zeta$. At $\zeta=1$ one set decouples. As recalled in Sec.~\ref{sec:connection}, the instanton quantum mechanics with a supersymmetric Wilson line admits an $\mathcal N=(0,4)$ supersymmetric completion, and the set that remains coupled carries that structure. In each set the eight modes are tangent vectors to the moduli space, four along the centre and four along the size, and their quantization gives a sixteen-dimensional Clifford module $S_\pm$ of $\R^8$.

The fields $A_t$ and $X^6$ enter the constrained action quadratically and without time derivatives, so eliminating them is exact. Their sources are the line, the moduli velocities and the fermion bilinears. Besides the kinetic term and the evaluation connection, the elimination produces three terms of order $R^{-2}$. The line--line term is the self-energy found above, with its factor $1-\zeta^2$. The line--bilinear term couples each set to the curvature of the evaluation connection, with the weight $1\pm\zeta$ of Eq.~\eqref{eq:fermioncoupling}. The bilinear--bilinear term does not involve the line, so it is that of the one-instanton quantum mechanics without a probe, which for gauge group $SU(2)$ is the free supersymmetric quantum mechanics on the flat moduli space \cite{ABS} (the supersymmetric moduli-space sigma models of these non-Lorentzian theories are derived in Ref.~\cite{Mouland2020}); it vanishes. Appendix~\ref{app:fermions} carries out the reduction. The line--bilinear term is the line's colour contracted with the field $A_t^{\rm f}(0)$ that the zero-mode bilinear sources at the line, and by an identity of the universal connection $\mathbb A$ of Atiyah and Singer \cite{AtiyahSinger1984} on $\R^4\times\mathcal M$ (in the Hermitian convention $D_i=\partial_i-\ii[\mathbb A_i,\cdot\,]$, $D_a=\partial_a-\ii[\mathbb A_a,\cdot\,]$ with mixed curvature $\mathbb F_{ia}=-\delta_aA_i$ in background gauge $D_i\delta_aA_i=0$ and $\mathbb F_{ab}(0)=\ii\mathcal F_{ab}$, Appendix~\ref{app:fermions}), the Bianchi identity $D_i\mathbb F_{ab}=D_a\mathbb F_{ib}-D_b\mathbb F_{ia}$ gives
\begin{equation}
 D_iD_i\,\mathbb F_{ab}=-2\ii\,[\delta_aA_i,\delta_bA_i],
 \label{eq:main:universal}
\end{equation}
so inverting $\mathscr L=-D_iD_i$ on the bilinear source $[\delta_aA_i,\delta_bA_i]$ returns $-\tfrac\ii2\mathbb F_{ab}(0)=\tfrac12\mathcal F_{ab}$ with no free parameters.  At $\zeta=\pm1$ the Hamiltonian therefore emerges as the square of the colour-twisted Dirac operator of the coupled set, confirming independently the $\mathcal N=(0,4)$ structure (Appendix~\ref{app:fermions} fixes the normalization of the Pauli term). With $c_\pm=(1\pm\zeta)/2$ the Hamiltonian is
\begin{equation}
 H(\zeta)=c_+H(1)+c_-H(-1)+\frac{1-\zeta^2}{R^2}\,V,\qquad
 H(\pm1)={\not{\!\mathbf D}}_\pm^{\dagger}{\not{\!\mathbf D}}_\pm\ \ge0,
 \label{eq:Hzeta}
\end{equation}
acting on sections of $E_s\otimes S_+\otimes S_-$ over the moduli space, where $E_s$ is the bundle with fibre $V_s$ associated with the evaluation connection. The fibre has dimension $256(2s+1)$. By the Lichnerowicz formula on flat $\R^8$, $H(\pm1)=-\nabla_{\mathcal A}^2+P^{(\pm)}$, where
\begin{equation}
 P^{(\pm)}=-\sum_{a<b}\gamma^{ab}_\pm\,\mathcal F_{ab}
 \label{eq:pauli}
\end{equation}
is the Pauli coupling of the set $\chi_\pm$ to the evaluation curvature, with $\gamma^{ab}_\pm$ the Clifford generators of that set. The decomposition~\eqref{eq:Hzeta} isolates the two ingredients whose competition decides the answer, the supersymmetric operators $H(\pm1)$, which are nonnegative, and the self-energy, which on states that saturate the floor at both endpoints (Sec.~\ref{sec:bpsstates}) is the only term that distinguishes a generic line from a supersymmetric one.
\subsection{Quantization and the models}
\label{sec:models}

\begin{table}[!htbp]
\centering
\small
\begin{tabular}{@{}llll>{\raggedright\arraybackslash}p{5.1cm}@{}}
\toprule
Model & Space & Measure & Fibre & Role and principal statements\\
\midrule
$\mathsf S$ (SUSY free centre) & $\R^8/O(1)$ & $\dd^8x$ & $V_s\otimes S_+\otimes S_-$ & principal model; floor, instability, two-sided threshold bound (Secs.~\ref{sec:susy}, \ref{sec:selfloc})\\\addlinespace[3pt]
$\mathsf B$ (bosonic free centre) & $\R^8/O(1)$ & $\dd^8x$ & $V_s$ & invariant sector of $\mathsf S$; unique Hamiltonian for $s\le4$; theta-selected collapse near $E=16s$ for $s\ge\tfrac92$ (Sec.~\ref{sec:freephases})\\\addlinespace[3pt]
$\mathsf B_{\rm pin}$ (bosonic pinned) & $\R^4/O(1)$ & $\rho^3\dd\rho\,\dd\Omega_3$ & $V_s$ & theta-dependent selection rule and collapse (Sec.~\ref{sec:pinned})\\\addlinespace[3pt]
$\mathsf S_{\rm pin}$ (SUSY pinned) & $\R^4/O(1)$ & $\rho^3\dd\rho\,\dd\Omega_3$ & $V_s\otimes F_y$ & endpoint floor $\beta=1$ and threshold $\zeta_{\rm pin}(s)$ for $s>0$; reproduced by $\mathsf S_\Omega$ at leading Born--Oppenheimer order (Sec.~\ref{sec:windows})\\\addlinespace[3pt]
$\mathsf S_\Omega$ ($\Omega$-trapped centre) & $\R^8/O(1)$ & $\dd^8x$ & as $\mathsf S$ & logarithmic fixed-charge and power-law global counts of shallow levels in the ranges of Sec.~\ref{sec:windows}\\\addlinespace[3pt]
$\mathsf B_{\rm trap}$ (bosonic trap) & $\R^8/O(1)$ & $\dd^8x$ & $V_s$ & bounded below at every finite stiffness if and only if $\mathsf B$ is conformal (Prop.~\ref{thm:L}(c)); stiff limit not identified with $\mathsf B_{\rm pin}$\\
\bottomrule
\end{tabular}
\caption{The models compared in this paper. All share the self-energy~\eqref{eq:selfenergy} and the evaluation connection, which the pinned models restrict to the slice $X=0$, where it is flat (Sec.~\ref{sec:connection}). All are measured from the energy at infinite separation. ``Pinned'' means the holonomic constraint $X=0$ imposed before quantization, and $F_y$ is the Fock space of the size fermions; $\Omega$ denotes the supersymmetric centre trap of Sec.~\ref{sec:windows}. Here $\beta$ is the radial inverse-square index of Sec.~\ref{sec:radial}, $\zeta_{\rm pin}(s)=\sqrt{1-1/(4C_2)}$ is the pinned threshold of Eq.~\eqref{eq:pinbound}.}
\label{tab:models}
\end{table}

We quantize the moduli dynamics canonically. The Hilbert space consists of square-integrable sections of $E_s\otimes S_+\otimes S_-$ over $\R^8$, with the flat measure of the metric~\eqref{eq:metric}, that transform by the chosen theta character under the lifted $O(1)$ redundancy (Sec.~\ref{sec:channels}). The Hamiltonian is Eq.~\eqref{eq:Hzeta}. At both supersymmetric endpoints the operator ordering is fixed by the supercharges, since $H(\pm1)$ is the square of a Dirac operator. Away from them $H(\zeta)$ combines the two endpoint operators with the weights $c_\pm$ and adds only the multiplication operator $V$. Lambert and Richmond quantize the reduced moduli Hamiltonian canonically \cite{LambertRichmond} and Lambert and Owen derive the constrained field equations from a five-dimensional action \cite{LO}. For the line defect we adopt the flat moduli measure and the point-source subtraction of Sec.~\ref{sec:selfenergy} as the explicit definition of the quantum model, leaving a first-principles derivation of the measure from the parent path integral as an open problem (Sec.~\ref{sec:lightcone}). 

The decisive region is the cone point $R\to0$, where the instanton shrinks to zero size at the position of the line, and there the moduli space is singular. Ref.~\cite{LambertRichmond} notes that the small-instanton singularities must either be removed or given a prescription for the wavefunction, and proposes wavefunctions even under the orbifold action. The even character at $\theta_{5d}=0$ extends that prescription to the fermions and the line. Aharony, Berkooz and Seiberg instead remove the singularities by a Fayet--Iliopoulos deformation of the ADHM quotient, whose parameter has mass dimension $-2$ and serves as an ultraviolet cutoff of the quantum mechanics \cite{ABS}. If a resolved operator converges to the same exterior differential expression, essential self-adjointness for $\beta\ge1$ fixes its exterior realization (Sec.~\ref{sec:radial}). This observation leaves convergence and possible additional resolved-core states undecided. For $0\le\beta<1$ the zero-cutoff limit, when it exists, selects a specific self-adjoint realization dictated by the resolved core. Determining that selection, and how the Fayet--Iliopoulos deformation acts on a Wilson line, is a concrete ultraviolet problem outside the orbifold model analysed here.\footnote{The Fayet--Iliopoulos term exists in the $U(1)$-quotient ADHM description of $U(2)$ instantons and has no counterpart in the $O(1)$ description used here. It describes instantons on noncommutative $\R^4$, and for $k=1$ it resolves the size cone $\R^4/O(1)$ into the Eguchi--Hanson space \cite{NekrasovSchwarz}. Because gauge transformations then act through the noncommutative product, a Wilson line at a point is not by itself gauge invariant, so the resolved theory must also specify what replaces the point-localized line and how it couples to the core. The commutative reading of the same data shows the obstruction. The deformed ADHM data describe framed torsion-free sheaves \cite{NakajimaYoshioka}, and the exceptional locus of the resolution consists of point-like instantons, at which the sheaf is not a vector bundle. For $k=1$ the middle cohomology after evaluating the commutative ADHM monad at the probe has dimension two except when a point-like instanton sits on the line, where a direct rank computation gives dimension three. The locus has real codimension six in the resolved moduli space. The fibre dimension of this pointwise commutative construction jumps there.} This choice does not affect the conformality criterion of Sec.~\ref{sec:radial}, which asks only whether a scale-invariant realization bounded below exists. It matters in the near-BPS window of Sec.~\ref{sec:bps}, where several radial branches enter $0\le\beta<1$. For $\beta^2<0$, by contrast, a short-distance cutoff that leaves a supercritical inverse-square tail to infinity produces asymptotically geometric shallow levels. Dilated test functions in the uncut region force the lower bound to diverge to $-\infty$ as the cutoff is removed, so no lower-bounded scale-invariant continuum limit exists. That divergence is the resolved counterpart of conformality loss.

The model has no intrinsic scale. The coupling multiplies the Hamiltonian, and the moduli are the only lengths. Scales enter only through deformations. A physical source of finite extent $\ell$ replaces the point-source subtraction at $\rho\lesssim\ell$, a trap of length $\ell_{\rm trap}$ confines the centre, and a periodic identification of space, as for calorons, sets an infrared length (Sec.~\ref{sec:windows}). The question of conformality concerns the regime in which none of these matters, between the source size and the infrared length. Throughout, $k=1$, the gauge group is $SU(2)$ with $\theta_{5d}=0$ unless stated otherwise, the probe is static and heavy, and $\zeta$ is fixed with $|\zeta|\le1$. The exact classical reduction applies at positive size and motivates taking the small-size limit in the prescribed cone Hamiltonian. Its singular core still requires the quantum domain specified above.\footnote{For a Lorentzian five-dimensional Yang--Mills completion, the ratio $g_5^2/\rho$ measures the coupling at the inverse size scale. At large probe spin the source estimate $A_t/A_i\sim g_5^2s/\rho$ gives the parametric requirement $\rho\gg g_5^2s$ for weak backreaction. These estimates do not control radiation out of the moduli sector.} 

Several models appear in the comparisons below, and they differ in their configuration spaces and Hilbert spaces as well as in their potentials. Table~\ref{tab:models} lists them. The principal model is the supersymmetric free centre~$\mathsf S$. Its bosonic sector, in which both fermion sets are in the $Sp(2)$-singlet states, is an exactly invariant subspace (Sec.~\ref{sec:sector}). The bosonic free-centre model $\mathsf B$ is that sector. The pinned models impose $X=0$ as a holonomic constraint before quantization. The supersymmetric pinned model $\mathsf S_{\rm pin}$ also removes the centre fermions, so its fibre is $V_s$ tensored with the Fock space of the size fermions, which are spectators there. A positive scalar trap on the centre neither creates nor cures a fall to the centre (Proposition~\ref{thm:L}(c)). The stiff limit of a generic bosonic trap is delicate, whereas the supersymmetric $\Omega$-trap of Sec.~\ref{sec:windows} cancels the centre Pauli energy at the pinned point and reproduces $\mathsf S_{\rm pin}$ at leading Born--Oppenheimer order. Each relation among the models of Table~\ref{tab:models} that we use is either proved by an explicit spectral inequality or stated as a leading Born--Oppenheimer result.

\section{Conformality as a spectral problem}
\label{sec:spectral}

Because every term in the Hamiltonian~\eqref{eq:Hzeta} scales homogeneously under dilations of the moduli cone, the stability of a Wilson line reduces to an eigenvalue problem on the compact seven-sphere $S^7$. Below we carry out the radial reduction, fix the angular coordinates on the link, and decompose the resulting spectral problem into physical symmetry channels.

\paragraph*{Radial reduction and the decisive number.}\phantomsection\label{sec:radial}

Every term of $H(\zeta)$ is homogeneous of degree $-2$ under the dilation $(Z,y)\to\Lambda(Z,y)$, and the connection is dilation invariant with no radial component (Sec.~\ref{sec:connection}). In polar form on the cone,
\begin{equation}
 H(\zeta)=-\partial_R^2-\frac7R\,\partial_R+\frac{L(\zeta)}{R^2},\qquad
 L(\zeta)=-\Delta_{S^7}^{\mathcal A}+(1-\zeta^2)V+c_+P^{(+)}+c_-P^{(-)},
 \label{eq:radial}
\end{equation}
where $L(\zeta)$ is the link operator, acting on sections over the seven-sphere $S^7$ that links the cone. Its Pauli terms denote the restrictions to $R=1$ of the homogeneous terms in the cone Hamiltonian. Here $-\Delta_{S^7}^{\mathcal A}$ is the colour-twisted Laplacian there. Writing $L_\pm=L(\pm1)$ for the supersymmetric endpoint link operators, we denote its supersymmetric part without the self-energy by $L_0(\zeta)=c_+L_++c_-L_-=L(\zeta)-(1-\zeta^2)V$.  The dilation generator and $K=R^2$ obey the conformal algebra with $H$ as differential expressions \cite{DFF}. A lower-bounded, scale-invariant quantization also requires a suitable common radial domain. Since $g_5^2$ multiplies the whole Hamiltonian, Pauli terms included, nothing in the spectral problem depends on the coupling.

Let $\mu$ be the lowest eigenvalue of $L(\zeta)$ in the physical sector, defined below, and set $\beta^2=\mu+9$ for that lowest channel. More generally, each link eigenvalue $\mu_k$ has $\beta_k^2=\mu_k+9$. Writing $\psi=R^{-7/2}\chi$ turns its radial equation into that of $-\partial_R^2+(\beta_k^2-\tfrac14)R^{-2}$ on $(0,\infty)$, and the theory of the inverse-square potential \cite{CaseSingular} gives three regimes.
\begin{itemize}
\item For $\beta\ge1$ ($\mu\ge-8$), the operator is essentially self-adjoint, with a unique, scale-invariant realization.
\item For $0\le\beta<1$ ($-9\le\mu<-8$), both independent radial solutions are square integrable near the origin, and the self-adjoint realizations form a $U(n)$ family, where $n$ is the total multiplicity of link eigenvalues in $[-9,-8)$. The scale-invariant choices in this space are specified below. We use the Friedrichs realization.\footnote{At $\beta=0$ the two linearly independent radial behaviours are $R^{1/2}$ and $R^{1/2}\ln R$. The Friedrichs realization is the canonical extension defined by closing the quadratic form on $C_c^\infty(\R^8\setminus0)$. It selects the more regular $R^{1/2+\beta}$ (or $R^{1/2}$) branch and preserves $SO(1,2)$ covariance across the full range $\beta\ge0$. Physically, the extension parameters are contact couplings of the line at the cone point. The two scale-invariant realizations are the infrared and ultraviolet fixed points of such a coupling, which merge at $\beta=0$ \cite{KLSS}. For scalar core matching at radius $R_0$ the logarithmic derivative $\gamma=R_0\chi'/\chi-\tfrac12$ gives $a/b=R_0^{2\beta}(\beta-\gamma)/(\beta+\gamma)$ for the coefficients of $\chi=aR^{1/2-\beta}+bR^{1/2+\beta}$. Moving the matching radius at fixed boundary data, the zero-energy equation gives $\dd\gamma/\dd\ln R_0=\beta^2-\gamma^2$. Its fixed points $\gamma=\pm\beta$ are the two scale-invariant realizations, they merge at $\beta=0$, and for $\beta^2=-\nu^2<0$ the flow has no fixed point and runs through every real value of $\gamma$ in each interval of $\ln R_0$ of length $\pi/\nu$, which is the discrete scale invariance of the supercritical regime. The flow concerns this contact coupling at fixed $\zeta$; it is not a beta function for $\zeta$ itself. Friedrichs is the untuned limit when the ratio of the two bracketed factors stays bounded. With $n$ branches in the window the $U(n)$ parameters form a matrix of such couplings.}
\item For $\beta^2<0$ ($\mu<-9$), the radial scaling exponents $\tfrac12\pm\ii\sqrt{-\beta^2}$ turn complex; no realization is bounded below, continuous scale invariance breaks to discrete Berezinskii--Kosterlitz--Thouless or Efimov scaling upon ultraviolet regulation, and the instanton collapses onto the line.
\end{itemize}
The dilation action separates distinct positive radial indices. For an eigenvalue with $0<\beta<1$, write the short-distance coefficients as $u=aR^{1/2-\beta}+bR^{1/2+\beta}$. Their dilation weights are $1-\beta$ and $1+\beta$. An invariant self-adjoint boundary plane has $a\in U_\beta$, $b\in U_\beta^\perp$ for a subspace $U_\beta$ of the equal-$\beta$ multiplicity space. This follows because the different weights separate the two coefficient spaces, and the boundary form pairs them nondegenerately. When the eigenspace has multiplicity one, scale invariance therefore requires either $a=0$ or $b=0$. Each choice retains just one of the powers $R^{1/2\pm\beta}$. At $\beta=0$, write $u=R^{1/2}(a+b\ln R)$. Dilation invariance would include $(b,0)$ in the boundary plane whenever $(a,b)$ is included. Boundary isotropy then forces $b=0$. These restrictions give the scale-invariant subset of the $U(n)$ family \cite{DerezinskiRichard2017}, and the contact conditions sit at the joint cone tip.\footnote{The stratum $y=0$ of zero-size instantons away from the tip needs no boundary condition. There the regular frame of Appendix~\ref{app:loc:guiding} gives smooth coefficients on the cover, and a cutoff in $\ln|y|$ between $|y|=\delta^2$ and $\delta$ approximates regular sections in graph norm, since the squared norm of its worst Laplacian commutator is $O(\int_{\delta^2}^{\delta}\dd r/(r|\ln\delta|^2))=O(|\ln\delta|^{-1})$. The finite quotient projection preserves this approximation.}

We call the line conformal when $\mu\ge-9$, so that a scale-invariant realization bounded below exists, and we say that it loses conformality otherwise. Since $L(\zeta)$ depends on $\zeta$ through bounded operators, $\mu$ is continuous in $\zeta$. We define the threshold $\zeta_c(s)$ as the supremum of the couplings $|\zeta|\le1$ at which the line of spin $s$ loses conformality, with $\zeta_c(s)=0$ if there are none. The line is then conformal for $\zeta_c(s)\le|\zeta|\le1$. Section~\ref{sec:instability} proves that for $s\ge1$ a single onset separates the two phases. The radial scaling exponents give a quantum-mechanical counterpart of the complex fixed-point merger in defect field theory \cite{BGT2017}, evaluated at fixed bare $\zeta$ within the moduli Hamiltonian (Sec.~\ref{sec:lightcone} examines its lift to the six-dimensional $(2,0)$ theory). The value $-9=-((8-2)/2)^2$ is the Hardy value of $\R^8$ (with Hardy constant $+9$), and $-8$ marks the boundary of essential self-adjointness. At $\zeta=\pm1$ the operator $H(\pm1)$ is nonnegative, so $\mu\ge-9$ on the physical domain. Proposition~\ref{thm:floor} shows that in fact $\mu=-8$ there for every nontrivial spin $s\ge\tfrac12$, placing the supersymmetric lines exactly at $\beta=1$.

In the conformal regime, the compact generator $J_0=(H+K)/4$ in oscillator units, with $K=R^2$, has lowest weight $k_+=\frac12(1+\beta)$ in each Friedrichs radial channel \cite{DFF,BMSV}. This uses a conformal trap on the entire cone to define the compact generator. Section~\ref{sec:windows} instead studies a deformation of the centre Hamiltonian. Under a defect operator--state map of the type in Ref.~\cite{NishidaSon}, the quantity $2k_+=1+\sqrt{\mu+9}$ would be the $z=2$ scaling dimension of a line-localized insertion carrying one unit of instanton number. For bulk instanton operators defined by prescribed flux, see Refs.~\cite{LPSS2015,Tachikawa2015}. Its interpretation as the dimension of an operator in the parent surface theory requires the matching discussed in Sec.~\ref{sec:lightcone}. For $\mu<-9$ the unregulated Hamiltonian has no bounded-below conformal realization, and the complex exponents formally continue the two radial powers. The present computation concerns the fixed one-instanton sector, the first sector with dynamical instanton moduli, and the probing line specified in Sec.~\ref{sec:models}.

To parametrize the link $S^7$, we write a point as $(\alpha,\hat Z,\hat y)$ with $\hat Z,\hat y\in S^3$ and volume element $\omega\,\dd\alpha\,\dd\Omega_3\dd\Omega_3$, where $\omega=\sin^3\alpha\cos^3\alpha$, together with the shorthand
\begin{equation}
 \sigma=\sin^2\alpha,\quad \kappa=\cos^2\alpha,\quad D=1+\sigma,\quad t=\tan^2\alpha,\quad W=\frac{\kappa}{D^2}=\frac{1+t}{(1+2t)^2},
 \label{eq:abbrev}
\end{equation}
so that $V=-4C_2W$, with $0\le W\le1$, $W=1$ in the pinned direction and $W$ decreasing in $t$. We call $\alpha$ the polar angle of the link and refer to densities, profiles and excitations in $\alpha$ as polar, keeping ``radial'' for the radius of a cone.

\paragraph*{Symmetry, fermion blocks and physical channels.}\phantomsection\label{sec:channels}

The link operator commutes with $G=SU(2)_L\times SU(2)_R\times SU(2)_G$, acting on the moduli by $(\lambda,u)\mapsto(h\lambda\bar q,\,pu\bar q)$ for $(p,q,h)\in G$, with $p\in SU(2)_L$, $q\in SU(2)_R$ and $h\in SU(2)_G$ written as unit quaternions. The factors $SU(2)_L\times SU(2)_R$ act on the moduli as the spatial rotations, under which the relative position transforms as a vector of $SO(4)$ and the size quaternion only under $SU(2)_R$. The factor $SU(2)_G$ is the global gauge rotation, which rotates the instanton's orientation. In the frame~\eqref{eq:frame} the group acts on the colour fibre at the points $\lambda\in\R_+$, $u\in\R_-$ through $D_s(h\bar qp)$, the spin-$s$ representation matrix of $h\bar qp\in SU(2)$, and on each fermion set through the left multiplication $\operatorname{diag}(h,p)\in Sp(2)$ on $\HH^2$. On the zero modes, which are tangent vectors, the spatial $SU(2)_R$ acts by right multiplication. Composing it with the commuting inverse right-$Sp(1)$ rotation on the fermions strips off the fermion action entirely, so $j_R$ labels purely the orbital rotation along the Hopf fibre (while the physical spatial spin differs only by the inert $Sp(1)$ multiplicity and leaves both the spectrum and Eq.~\eqref{eq:physical} unchanged). The group acts transitively on each slice $\alpha=\text{const}$, with the diagonal $SU(2)$ as stabilizer, so every section decomposes into channels $\varpi=(j_L,j_R,j_G)$ and the link operator becomes a matrix differential operator in $\alpha$ in each channel (Appendix~\ref{app:numerics}).

The Pauli operators lie in $sp(2)\otimes su(2)$ (property~(i) in Sec.~\ref{sec:connection}), so they commute with the $Sp(1)$ that acts on each fermion set by right quaternionic multiplication. Under $Sp(2)\times Sp(1)$ each sixteen-dimensional set decomposes as $(5,1)\oplus(1,3)\oplus(4,2)$\footnote{This is the branching of $\mathbf 8_s\oplus\mathbf 8_c$ under $Sp(2)\cdot Sp(1)\subset SO(8)$. The light-cone content of the $(2,0)$ tensor multiplet, five scalars, three tensor polarizations and eight fermions, has the same form under $Sp(2)_R\times SU(2)$; here, however, $Sp(2)$ contains $SU(2)_L\times SU(2)_G$ rather than the R-symmetry, and each set mixes centre and size modes, so the blocks are not tensor-multiplet polarizations.}, and the spectrum of $L(\zeta)$ is that of the blocks $(A_+,A_-)$ with $A\in\{5,1,4\}$, the $Sp(1)$ factor contributing only a degeneracy. Block $(1,1)$, in which both sets are $Sp(2)$ singlets, carries no Pauli coupling.

The physical theta character acts under the lifted redundancy $\lambda\to-\lambda$. The redundancy sends the size fermion zero modes to their negatives. In the convention of Appendix~\ref{app:frames}, the fermion operator in this lift is the central element $-1\in SU(2)_G$, which equals the chirality $\Gamma_y$ of the size fermion zero modes,
\begin{equation}
 \Gamma_{O(1)}=e^{2\pi\ii J^3_{G,F}}=(-1)^{2j_G^F}=\Gamma_y .
 \label{eq:O1lift}
\end{equation}
Although every dynamical field is in the adjoint representation, a large gauge transformation in $\pi_4(SU(2))=\Z_2$ weights the charge-one instanton sector by $e^{\ii\theta_{5d}}$. The central element $-1\in SU(2)_G$ acts on the moduli and zero modes in the same way as the redundancy and on physical states by $e^{\ii\theta_{5d}}(-1)^{2s}$, the sign that the discrete theta angle attaches to the one-instanton wavefunction on $S^3/\Z_2$ \cite{Tachikawa2015}. Because the representation of $SU(2)_G$ on the zero modes is unambiguous ($SU(2)$ generators admit no constant shift), fixing $\theta_{5d}=0$ selects $\Gamma_{O(1)}=U_F(-1)=(-1)^{2j_G^F}$, with $U_F$ the representation of $SU(2)_G$ on the fermion Fock space. The opposite lift, $-\Gamma_y$, describes the theory with $\theta_{5d}=\pi$, where the lowest allowed gauge representation at $s=0$ is a doublet, in agreement with the one-instanton gauge character of Ref.~\cite{BRZ} and the channel rule shifts to $j_G\equiv s+\tfrac12\pmod1$ (see footnote~\ref{fn:theta} in Sec.~\ref{sec:theory}).  The physical channels in every fermion block therefore obey
\begin{equation}
 j_G\equiv s+\frac{\theta_{5d}}{2\pi}\pmod 1 .
 \label{eq:physical}
\end{equation}
In the frame $v_R$ of Appendix~\ref{app:frames}, $SU(2)_G$ acts trivially on the colour fibre. On the bosonic pinned slice a size harmonic of degree $n$ has $j_G=n/2$, so Eq.~\eqref{eq:physical} becomes $n\equiv2s+\theta_{5d}/\pi\pmod2$. The size-fermion chirality shifts this parity in the fermionic blocks, $n\equiv2s+\theta_{5d}/\pi+(1-\Gamma_y)/2\pmod 2$, a fact that will matter for the pinned problem with fermions.

The channel decomposition reduces the spectral problem to countably many matrix problems. The proved results below need no truncation. The instability bounds use explicit trial states, and the lower bounds treat every channel at once. For $s\ge1$ and $E\le8/3$, the Hopf gap and the angular potential bound~\eqref{eq:un:Gbound} exclude all onset channels outside $j_R=0$, $j_L\le s+\tfrac32$ and $j_G\le s+2$ (Appendix~\ref{app:un:coverage}). The Galerkin scans sample subsets of those channels as specified in Appendix~\ref{app:numerics}.

\section{Instanton recoil and collapse in the bosonic model}
\label{sec:bosonic}

The bosonic moduli set the baseline for the fermionic instability of Sec.~\ref{sec:selfloc}. Physically, freeing the instanton centre lets the soliton recoil into four extra transverse dimensions and dilute the inverse-square attraction of the line. Section~\ref{sec:theoremL} formulates this recoil protection as a general dimensional-localization criterion (Proposition~\ref{thm:L}), Sec.~\ref{sec:pinned} contrasts the exact pinned collapse thresholds under the $O(1)$ bundle selection at both theta angles against the strong protection of the freely moving centre, and Sec.~\ref{sec:freephases} finds its limit. The colour twist of the evaluation connection overcomes the protection at a spin and a coupling set by the discrete theta angle, and at fixed spin and scalar coupling the theta angle can then decide between a unique Hamiltonian and fall to the centre.

\subsection{Localization changes the effective dimension}
\label{sec:theoremL}

A scale-invariant inverse-square attraction is much easier to destabilize in four dimensions than in eight, because the radial centrifugal barrier $((N-2)/2)^2$ jumps from $1$ at $N=4$ to $9$ at $N=8$. Holding the instanton centre on the line removes four transverse dimensions and drops that barrier by eight units. To isolate the dimensional effect, consider a configuration space $\R^d_X\times C$, where $C$ is a cone of dimension $d_C$ with compact link, a kinetic metric $m_c|\dd X|^2+m_s(\dd\rho^2+\rho^2g_\Sigma)$, a unitary bundle with a dilation-invariant connection $\mathcal A$ that has no radial component, and a potential $U$ homogeneous of degree $-2$, with smooth link coefficients in regular local frames (or smooth equivariant lifts on an orbifold cover). With $Z=\sqrt{m_c/m_s}\,X$ the problem lives on a cone of dimension $N=d+d_C$, and the pinned problem is its restriction to $X=0$. Suppose that the strongest attraction on the link is attained in the pinned direction.

\begin{proposition}[Localization]
\label{thm:L}
For the dilation-invariant kinetic connection and degree-$(-2)$ scalar potential just specified, assume the potential on the link reaches its minimum at $X=0$. Then 
\begin{enumerate}
\item[(a)] the pinned operator admits a realization bounded below if and only if its lowest link eigenvalue in the physical sector obeys $\mu_{\rm pin}\ge-((d_C-2)/2)^2$, and otherwise no self-adjoint realization is bounded below;
\item[(b)] the free operator admits a realization bounded below if and only if $\mu_{\rm free}\ge-((N-2)/2)^2$, and it is conformal whenever $\sup(-\hat U_{\rm pin})\le((N-2)/2)^2$, where $U=R^{-2}\hat U$ and $\hat U_{\rm pin}$ is the restriction of $\hat U$ to the pinned link, the Hardy constants of the free and the pinned problems differing by $d(d+2d_C-4)/4$;
\item[(c)] a trap $\varkappa^2|Z|^2$ on the centre, with frequency $\varkappa$ (a symbol distinct from $\kappa=\cos^2\alpha$), gives an operator that is bounded below if and only if the free problem is conformal. For a scale-invariant short-distance domain of the free operator, dilation multiplies the trapped spectrum at frequency $\varkappa$ by $\varkappa$ relative to unit frequency. If the free problem is conformal while the pinned problem is supercritical in some channel, the Friedrichs trapped operator has infinitely many eigenvalues below the zero-point energy of the trap.
\end{enumerate}
\end{proposition}

Appendix~\ref{app:localization} proves Proposition~\ref{thm:L} using the Hardy and diamagnetic inequalities, the limit-point criterion \cite{ReedSimonII,LiebLoss}, dilation covariance and the min--max principle. The connection with atomic gases concerns the radial equation. In the resonantly interacting mixtures studied by Nishida and Tan, confining one species changes the scale-invariant three-body problem and can drive its hyperradial channel past the inverse-square instability \cite{NishidaTan2008,NishidaTan2009,NishidaTan2011,MohapatraBraaten}. Here the radial coordinate belongs to the instanton moduli, and the attraction comes from the Wilson line. Confining the centre removes transverse kinetic energy from the effective radial problem and can drive the same change from real to complex scaling exponents. Section~\ref{sec:windows} examines this mechanism in a supersymmetric trap. For the instanton $d=d_C=4$, $N=8$, the inertia ratio is $m_c/m_s=1/2$, and the profile $W=\kappa/D^2$ is maximal in the pinned direction. The hypothesis on the maximum holds whenever $m_c/m_s\le2$, so it holds here with room to spare\footnote{At $R=1$ the profile is $W=(1+m\xi)/(1+\xi)^2$, with $\xi=|X|^2/\rho^2$ and $m=m_c/m_s$. Its derivative in $\xi$ has the sign of $m-2-m\xi$. For $m\le2$ the maximum $W=1$ lies on the pinned slice, and for $m>2$ it moves to $\xi=1-2/m$, where $W=m^2/[4(m-1)]$. Since $m$ also enters the kinetic operator, this locates the deepest attraction without fixing a threshold.}. Physically, a heavier centre moves less in physical $X$ per unit angle $\alpha$, so only for $m_c/m_s>2$ does $(\rho^2+|X|^2)^2$ dilute more slowly than $\rho^2=R^2\cos^2\alpha$ shrinks, pushing the deepest attraction off the pinned slice. Pinning therefore lowers the Hardy bound on the critical attraction from $9$ to $1$. We show below that the bundle selection raises the pinned value to $4$ for half-integer $s$ at $\theta_{5d}=0$, and for integer $s$ at $\theta_{5d}=\pi$, while transverse recoil and angular dilution raise the free value to at least $99/4$, and the colour twist raises it further, to $16s+O(1)$ at large spin (Sec.~\ref{sec:freephases}).

\paragraph*{Pinned collapse versus free-centre protection.}\phantomsection\label{sec:pinned}\label{sec:freecentre}

On the pinned slice $X=0$ the configuration space is the size cone $\R^4/O(1)$ with measure $\rho^3\dd\rho\,\dd\Omega_3$, the restriction of the evaluation connection to the slice is flat (property~(iii) in Sec.~\ref{sec:connection}), and the self-energy is $-4C_2(1-\zeta^2)/\rho^2$. The centre components of the curvature do not vanish there; they act only when the centre moves. The only memory of the colour bundle is its holonomy around the identification $\lambda\to-\lambda$. In the frame of Appendix~\ref{app:frames}, where the connection vanishes on the slice, a spin-$s$ section obeys $\psi(-\lambda)=e^{\ii\theta_{5d}}(-1)^{2s}\psi(\lambda)$, so only spherical harmonics of degree $n\equiv2s+\theta_{5d}/\pi\pmod2$ on the size sphere are allowed. A harmonic of degree $n$ has eigenvalue $n(n+2)$ and carries $j_G=n/2$, so Eq.~\eqref{eq:physical} is recovered. With $u=\rho^{3/2}\psi_n$ the radial operator in such a channel is
\begin{equation}
 -\frac{\dd^2}{\dd\rho^2}+\frac{\beta_n^2-\tfrac14}{\rho^2},\qquad
 \beta_n^2=(n+1)^2-4C_2(1-\zeta^2).
 \label{eq:pinnedradial}
\end{equation}
Applied to this operator with $d_C=4$, Proposition~\ref{thm:L}(a) shows that in the bosonic pinned model the channel $j_G=n/2$ is supercritical if and only if $(n+1)^2<4C_2(1-\zeta^2)$, and that in a supercritical channel no self-adjoint realization is bounded below. Since $n=2j_G$, the condition $(2j_G+1)^2<4s(s+1)\le(2s+1)^2-1$ excludes $j_G\ge s$. The lowest allowed channel is $n=0$ for integer $s$ and $n=1$ for half-integer $s$ at $\theta_{5d}=0$, and the reverse at $\theta_{5d}=\pi$, so at $\theta_{5d}=0$ the pinned line of spin $s$ loses conformality for
\begin{equation}
 \zeta^2<1-\frac1{4C_2}\quad(s\in\Z,\ s\ge1),\qquad
 \zeta^2<1-\frac1{C_2}\quad(s\in\Z+\tfrac12,\ s\ge\tfrac32),
 \label{eq:pinnedthresholds}
\end{equation}
while at $\theta_{5d}=0$ the bosonic pinned fundamental line $s=\tfrac12$ is conformal at every $\zeta$, since its lowest allowed channel has $n=1$ and $4C_2=3<4$. At $\theta_{5d}=\pi$ the two cases of Eq.~\eqref{eq:pinnedthresholds} exchange, and the fundamental line has $n=0$. Its pinned problem then loses essential self-adjointness at every $|\zeta|<1$ and conformality for $\zeta^2<\tfrac23$. The freely moving fundamental line is conformal (below), so Proposition~\ref{thm:L}(c) gives a trapped centre infinitely many levels below the zero-point energy of the trap for $1<E\le3$, the smallest example of the confinement-induced tower. At $\theta_{5d}=0$ the bosonic pinned floor is therefore $\beta=1$ for integer spin (threshold $\sqrt{7/8}$ at $s=1$) and $\beta=2$ for half-integer spin (threshold $\sqrt{11/15}$ at $s=\tfrac32$), a shift produced entirely by the $\Z_2$ centre holonomy of the bundle. At $\theta_{5d}=\pi$ the theta character reverses the sign of the holonomy and the two floors exchange.\footnote{With $n_0\in\{0,1\}$, $n_0\equiv2s+\theta_{5d}/\pi\pmod2$, the lowest allowed degree, a pinned channel of degree $n$ needs a boundary condition at the tip when $(n+1)^2<E+1$, and it falls to the centre when $(n+1)^2<E$. Each degree carries $(n+1)^2$ harmonics on the size sphere and the flat colour fibre $2s+1$ states, so the deficiency indices of the pinned bosonic Hamiltonian are both $(2s+1)\sum(n+1)^2$, summed over $n\equiv n_0\pmod2$ with $(n+1)^2<E+1$.} Below the Hardy value, a Dirichlet wall at $r_0$ gives a geometric bound-state tower with deepest energy of order $-r_0^{-2}\exp(-2\pi/\sqrt{-\beta_n^2})$ near onset. This is the Berezinskii--Kosterlitz--Thouless scaling of conformality loss in that regulator.

Once the centre moves freely, Proposition~\ref{thm:L}(b) with $N=8$ reverses this picture. By the diamagnetic inequality \cite{LiebLoss} and the eight-dimensional scalar Hardy inequality, the covariant kinetic energy obeys $\int|\nabla_{\mathcal A}\Psi|^2\ge9\int|\Psi|^2/R^2$ for every unitary connection \cite{CazacuKrejcirik}, where the bundle case reduces to the scalar one through the pointwise inequality $|\nabla|\Psi||\le|\nabla_{\mathcal A}\Psi|$ that follows from metric compatibility and the Cauchy--Schwarz inequality (so $E=4C_2(1-\zeta^2)\le9$ is automatically conformal because $W\le1$), and the colour-twisted link operator is bounded below by the scalar operator $L_{\rm scal}=-\Delta_{S^7}-EW$. Barta's inequality \cite{Barta} raises this barrier using the angular fall-off of $W=\kappa/D^2$. For any positive test function $\psi$ on the link, $\inf\spec L_{\rm scal}\ge\inf(L_{\rm scal}\psi/\psi)$. Choosing $\psi=D^{-9/4}$ (matching the link denominator $D=1+\sigma$) gives
\begin{equation}
 D^2\Bigl(\frac{L_{\rm scal}\psi}{\psi}+9\Bigr)=\tfrac14\,\kappa\,(108-4E-9\sigma),
 \label{eq:barta}
\end{equation}
which is nonnegative on the whole link $\sigma\in[0,1]$ precisely when $E\le99/4$. The threshold of the scalar operator, computed numerically from the Heun-type eigenvalue problem, is $E\approx26.4576$ (the case $s=0$ of Sec.~\ref{sec:freephases}, read with $E$ as a free parameter, which also places the contact value of this operator exactly at $E=24$), so the analytic bound is within seven percent of the numerical threshold. Every bosonic free line with $s\le2$ is therefore conformal for every $\zeta$ by dimensional dilution and the scalar Hardy barrier alone. A single positive test function needs no representation theory, and for the same reason it cannot see what decides the larger spins, the colour twist and the theta character.\footnote{Written in $\sigma=\sin^2\alpha$, the scalar link equation has four regular singular points at $\sigma\in\{0,1,-1,\infty\}$, placing it in the Heun class. Since $4C_2=4s(s+1)$ reaches $24<99/4$ at $s=2$ and $35>26.46$ at $s=\tfrac52$, dimensional dilution alone guarantees conformality through $s=2$. Proposition~\ref{thm:bosphase} extends this to $s\le4$. In the coordinate $\sigma/(1+\sigma)$ the four singular points sit at $0$, $\tfrac12$, $1$ and $\infty$. Connection problems of such Heun equations are solved by the Nekrasov--Shatashvili limit of the instanton partition function of four-dimensional $\mathcal N=2$ $SU(2)$ gauge theory with four flavours, in which the position of the fourth point plays the role of the instanton counting parameter \cite{NekrasovShatashvili,BILT2022}. That would turn each threshold into a root of an explicit quantization condition, a route we do not pursue here.} For the lines proved conformal, a positive trap on the centre preserves boundedness from below (Proposition~\ref{thm:L}(c)) while inducing Efimov towers at $\rho\gg\ell_{\rm trap}$ whenever the pinned problem is supercritical.

\subsection{The theta angle and the bosonic phases of a moving instanton}\label{sec:freephases}

The scalar comparison discards the colour twist, and with it everything that distinguishes the representations of the line and the two theta angles. Restoring it, we follow two events as the attraction grows. Once the lowest link eigenvalue falls below $\mu=-8$ the radial Hamiltonian is no longer essentially self-adjoint and the line needs a contact coupling at the cone tip, and once it falls below $\mu=-9$ the line loses conformality. At $\mu=-8$ itself the Hamiltonian is still essentially self-adjoint, and at $\mu=-9$ it still has a scale-invariant realization bounded below (Sec.~\ref{sec:radial}). For a channel $\varpi=(j_L,j_R,j_G)$ we write $E_{\rm sa}^\varpi$ and $E_c^\varpi$ for the attractions at which its lowest polar state reaches these two values. Both are set by the orbital energy a centre pays near the line, and that energy depends on how the colour of the line is shared among the three rotations of $G$.

\paragraph*{An exact contact state.}The cheapest channel puts the whole colour into the orbital rotation of the centre, $\varpi=(s,0,0)$, and leaves the gauge orientation a singlet. There the bosonic link operator $L=-\Delta^{\mathcal A}_{S^7}-EW$ is the operator~\eqref{eq:fb:bosonic} with $C_L=C_2$ and $C_G=0$, and the positive function $T^{\rm bos}_s=\kappa^sD^{-(s+2)}$ obeys
\begin{equation}
 D^2\,\frac{(L+8)\,T^{\rm bos}_s}{T^{\rm bos}_s}=\kappa\,(16s+24-E)
 \label{eq:stretched}
\end{equation}
identically in $s$, as direct differentiation shows. The function has the regular behaviour at both ends of the polar interval, so it is the ground state of the channel at $E=16s+24$, and Barta's inequality, applied in both directions, gives
\begin{equation}
 E_{\rm sa}^{(s,0,0)}=16s+24
 \label{eq:Esa}
\end{equation}
for every spin, integer or half-integer. The leading term has a simple origin. Near the pinned direction the evaluation connection acts on the centre as the field $B=8s/\rho^2$ of Sec.~\ref{sec:pauli}, and $16s/\rho^2=2B$ is the zero-point energy of its lowest Landau level. A bosonic centre must pay this energy to stay near the line, and it is exactly the energy that the Pauli coupling of the fermion zero modes removes in Sec.~\ref{sec:pauli}.

Dividing by $T^{\rm bos}_s$ and changing variable to $v^2$, with $v=(1-\sigma)/(1+\sigma)$ as in Sec.~\ref{sec:instability}, turns the equation $LT=-8T$ in this channel into the Jacobi equation on $0<v^2<1$. Its polynomial solutions give every polar state of the channel that reaches $\mu=-8$,
\begin{equation}
 E_{\rm sa}^{(n)}=16(n+1)\Bigl(s+n+\frac32\Bigr),\qquad n=0,1,2,\ldots,
 \label{eq:ladder}
\end{equation}
where $n$ counts polar nodes. At $s=0$ this is the scalar sequence $24,80,168,\ldots$, so the scalar operator above, with $E$ as a free parameter, needs a contact condition exactly for $E>24$. The second state of the channel reaches $\mu=-8$ only $16s+56$ above the first, so between contact and collapse a single polar state of the channel lies below $\mu=-8$.

\paragraph*{Contact and collapse are about one unit apart.}In the ground state the profile has the mean $\langle W\rangle=(2s+2)/(2s+5)$, from two Beta integrals in $v$ as for the doubly-BPS state $T_s$ of Sec.~\ref{sec:instability}, whose mean is $2s/(2s+3)$. Lowering the polar eigenvalue from $-8$ to $-9$ costs at least one unit of attraction, because $W<1$ inside the link, and at most $1/\langle W\rangle$, by the variational principle with $T^{\rm bos}_s$ as trial state, so that
\begin{equation}
 16s+25<E_c^{(s,0,0)}<16s+25+\frac3{2s+2} .
 \label{eq:stretchbracket}
\end{equation}
Both thresholds grow like $16s$, but the interval between them, in which the line needs a contact coupling yet stays conformal, has width $1+O(1/s)$.\footnote{The expansion in powers of $1/s$ gives $E_c^{(s,0,0)}=16s+25+\frac3{2s}-\frac3{2s^2}+\frac{183}{128s^3}+O(s^{-4})$, which we checked against direct eigenvalues up to $s=160$. The cubic coefficient lies $\tfrac9{128}$ below that of the one-state estimate in Eq.~\eqref{eq:stretchbracket}. Mixing with the first excited state of Eq.~\eqref{eq:ladder} alone produces this shift. Let $T_0,T_1$ denote the ground and first excited polar states in Eq.~\eqref{eq:ladder}, normalized in $\langle f,g\rangle_W=\int\bar f g\,W\omega\,\dd\alpha$. The perturbation shifting the link eigenvalue from $-8$ to $-9$ is multiplication by $W^{-1}$ in this weighted eigenvalue problem. These two states have squared transition matrix element $|\langle T_1,W^{-1}T_0\rangle_W|^2=9(p+1)(p+7)/[2p^2(p+2)^2]$, with $p=2s+2$, and spacing $16s+56$, so second-order mixing lowers the one-state estimate by $9/(128s^3)+O(s^{-4})$, while the squared matrix elements from the ground state to the remaining polar states sum to $O(s^{-4})$ and shift the threshold only at $O(s^{-5})$. The same correction, $-9/(128s^3)$, lowers the Pauli thresholds of Sec.~\ref{sec:uniform}, where it also comes from virtual excitation of the next polar state.}

\paragraph*{Every channel and both theta angles.}Other channels pay more orbital energy. How much more depends on the representation, and the theta character decides which representations exist. In a general channel the bosonic orbital potential reads (Appendix~\ref{app:bosonic})
\begin{equation}
 M_\varpi=\frac{4\kappa\,C_{RG}}{D\sigma}+\frac{8\sigma\,C_{LR}}{D\kappa}+\frac{8C_L+4C_G}{D}-\frac{8C_2}{D^2}-EW ,
 \label{eq:Mbos}
\end{equation}
where $C_{RG}$ and $C_{LR}$ are the Casimir operators of the pairs $(j_R,j_G)$ and $(j_L,j_R)$ coupled inside the channel. At $j_R=0$ they reduce to $C_G$ and $C_L$, and Eq.~\eqref{eq:Mbos} to Eq.~\eqref{eq:fb:bosonic}. Both enter with positive coefficients, and each is bounded below by the smallest spin that its two Clebsch--Gordan triangles allow. Comparison functions $\sigma^a\kappa^bD^{-s-2}$, of which $T^{\rm bos}_s$ is the case $a=0$, $b=s$, then exclude every representation except one or two, and explicit vector trial functions settle those. The outcome is the following.

\begin{proposition}[Bosonic phases]
\label{thm:bosphase}
In the bosonic model~$\mathsf B$ call a theta sector integral when Eq.~\eqref{eq:physical} makes $j_G$ an integer, that is $\theta_{5d}=0$ for integer $s$ and $\theta_{5d}=\pi$ for half-integer $s$, and call the other sector complementary.
\begin{enumerate}
\item[(a)] For $s\le4$ the lowest link eigenvalue obeys $\mu>-8$ at every $|\zeta|\le1$ in both sectors, so the line has a unique self-adjoint Hamiltonian and is conformal.
\item[(b)] For $s\ge\tfrac92$, in the integral sector the first polar state to reach any value $\mu=\beta^2-9$ with $0\le\beta\le1$ is the ground state of the stretched channel $(s,0,0)$, a multiplet of $2s+1$ states. The sector loses essential self-adjointness exactly for $E>16s+24$ and conformality for $E>E_c^{(s,0,0)}$, with $E_c^{(s,0,0)}$ of Eq.~\eqref{eq:stretchbracket}, and for $E\le16s+27$ every other state has $\mu>-8$.
\item[(c)] For $s\ge\tfrac92$, in the complementary sector every state has $\mu>-8$ for $E\le16s+27$. At $s=\tfrac92$ this covers every physical coupling, up to the ordinary line $E=99$.
\end{enumerate}
\end{proposition}

Appendix~\ref{app:bosonic} proves Proposition~\ref{thm:bosphase} for all the spins stated. An independent numerical check at several couplings for each spin from $\tfrac52$ to $6$ excluded channels by pointwise and scalar lower bounds, then diagonalized the remaining channels by finite differences. It agrees with the predicted separation for every tested $s\ge\tfrac92$. At $E=16s+26$, a physical coupling for $s\ge\tfrac92$, the integral sector has fallen to the centre in a single angular multiplet while the complementary sector keeps a unique self-adjoint Hamiltonian. The integral threshold lies below $16s+26$, and a singlet trial in the channel $(s,\tfrac12,\tfrac12)$ places the complementary one below $16s+28+3/(2s+2)+12/(2s+1)$, so both sectors have lost conformality at the ordinary line $\zeta=0$ once $s\ge5$.

\paragraph*{Why the theta angle matters.}The two sectors differ only in which representations of the gauge orientation the instanton may carry. In the integral sector the orientation can be a singlet, and the whole colour rides on the orbital motion of the centre. In the complementary sector the orientation carries at least $j_G=\tfrac12$. The cheapest way to remove the resulting centrifugal pole at the pinned end couples it with the Hopf-fibre spin $j_R=\tfrac12$ to a singlet, at a price $3/\kappa$ in the orbital energy, which is at least $4j_R(j_R+1)=3$ and equals it at the pinned end. This offset exceeds the width of the window between contact and collapse, about one unit, so for every $s\ge\tfrac92$ there is an interval of attraction in which one sector has fallen to the centre while the other has not yet needed a contact coupling. By the bounds of Appendix~\ref{app:bosonic} its width in $E$ lies between $2-3/(2s+2)$ and $2+12/(2s+1)$, about $1/(4s^2)$ in the scalar coupling near $|\zeta|=1-2/s$ at large spin, which is why the two sectors still differ at a tuned coupling after both have collapsed at the ordinary line. The global topology of the gauge group reaches this local inverse-square problem only through this selection of representations.

\begin{figure}[t]
\centering
\includegraphics[width=\textwidth]{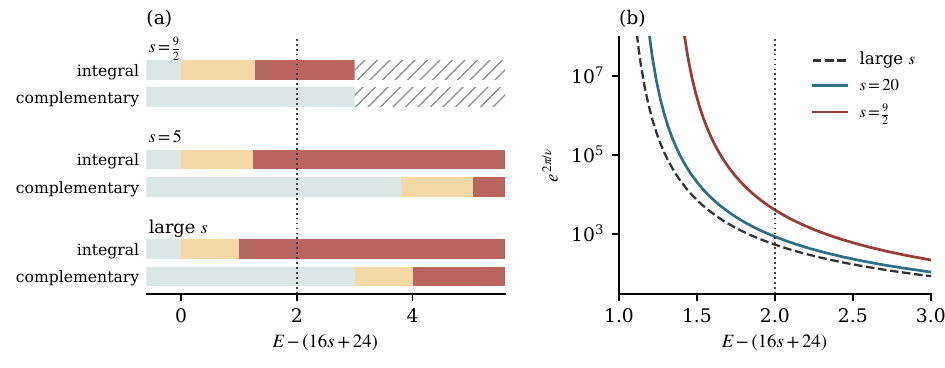}
\caption[Short-distance phases of the freely moving bosonic line]{Short-distance phases of the freely moving bosonic line, at fixed spin, against the attraction measured from the stretched contact value, $E-(16s+24)$. (a)~Each pair of bars compares the integral and complementary theta sectors of Proposition~\ref{thm:bosphase} at $s=\tfrac92$, at $s=5$ and at large spin. The complementary boundaries at $s=5$ are those of the channel $(5,\tfrac12,\tfrac12)$, computed numerically beyond the range $E\le16s+27$ in which Proposition~\ref{thm:bosphase}(c) proves uniqueness, and the large-spin bars show the leading offsets. The line has a unique self-adjoint Hamiltonian in the light region, needs a contact coupling but stays conformal in the middle region and falls to the centre in the dark region. Hatching marks attractions beyond the ordinary line $\zeta=0$, and the dotted line is $E=16s+26$. (b)~Ratio of successive binding energies, larger to smaller, $e^{2\pi/\nu}$, of the supercritical multiplet of the integral sector at $s=\tfrac92$, at $s=20$ and at large spin, for a short-distance boundary condition whose phase tends to a constant for shallow levels, slowly enough that each winding gives one level.}
\label{fig:thetawindow}
\end{figure}

\paragraph*{The first spin.}At $s=\tfrac92$ the ordinary line has $E=99$. At $\theta_{5d}=0$ the sector is complementary, and the line keeps a unique self-adjoint Hamiltonian at every scalar coupling. At $\theta_{5d}=\pi$ it loses essential self-adjointness for $E>96$, that is for $|\zeta|<1/\sqrt{33}$, and conformality above the lowest stretched-channel Hardy crossing. Equation~\eqref{eq:stretchbracket} bounds this spectral root by $97<E_c^{(s,0,0)}<97+3/11$. Numerically it is $97.2723426$, corresponding to $|\zeta_c|\simeq0.13210$. At $E=98$ the two quantizations of the same line, with the same local forces, therefore differ qualitatively. The $\theta_{5d}=0$ sector has a unique scale-invariant Hamiltonian. In the $\pi$ sector a single multiplet of ten states has radial exponents $\tfrac12\pm\ii\nu$, with $\nu^2\simeq0.5722$ from numerical diagonalization. Figure~\ref{fig:thetawindow} shows the intervals at $s=\tfrac92$, at $s=5$ and at large spin. 

\paragraph*{What the exterior predicts.}Proposition~\ref{thm:bosphase}(b) makes the count of boundary data exact. For $16s+24<E\le16s+27$ the deficiency indices of the bosonic Hamiltonian at the cone tip are $(2s+1,2s+1)$ in the integral sector and $(0,0)$ in the complementary one. Rotations commute with the Hamiltonian, so a boundary condition that respects them acts on the $2s+1$ states of the multiplet as a single phase, by Schur's lemma. Above $E_c^{(s,0,0)}$ the radial exponents become $\tfrac12\pm\ii\nu$. If the phase then tends to a constant for shallow levels, slowly enough that each winding of the radial solution gives one level, the bound states below zero energy form a geometric sequence whose successive binding energies, larger to smaller, have the ratio $e^{2\pi/\nu}$, as for the inverse-square potential \cite{CaseSingular} and in the Efimov effect \cite{Efimov1970}, and only the absolute scale depends on the core.\footnote{For a resolved core of size $R_0$ the logarithm of the ratio of two successive binding energies differs from $2\pi/\nu$ by at most $2/\nu$ times the change of the matching phase between the two levels, which is $O((kR_0)^2)$ at momentum $k$ for a regular interior. A core that breaks the rotations acts on the multiplet by a unitary matrix and can produce up to $2s+1$ interleaved sequences, each with the same ratio.} At $s=\tfrac92$ the ratio is about $4.0\times10^3$ at $E=98$ and $219$ at the ordinary line, and at large spin it tends to $e^{2\pi/\sqrt{E-16s-25}}$ for $16s+25<E\le16s+27$. The complementary quantization has no such sequence, since its Hamiltonian is unique.

In the bosonic model, then, localization decides conformality at small spin, and for $s\ge\tfrac92$ the discrete theta angle decides, at a given coupling, whether the line keeps a unique Hamiltonian or falls to the centre. In both sectors the critical attraction grows like $16s$, the Landau cost of holding the centre near the line. Section~\ref{sec:susy} shows that this answer holds only in one sector of the supersymmetric theory, and Sec.~\ref{sec:selfloc} shows that in the other sectors the line localizes the centre by itself, at an attraction of order one.

\section{Supersymmetry and the lowest angular eigenvalue}
\label{sec:susy}

Including the sixteen fermion zero modes opens new blocks in which the Pauli coupling to the evaluation curvature competes with the bosonic Landau barrier. We isolate the invariant bosonic sector, evaluate the Pauli operator and its Landau-level cancellation, prove that supersymmetry fixes the endpoint link floor at $-8$ (attained for every $s\ge\tfrac12$), and identify the floor states as twistor-spinor lowest Landau levels on $S^4$. In the bosonic sector this barrier is the Landau zero-point energy that places the first bosonic threshold of each sector near $16s$ (Sec.~\ref{sec:freephases}). The question is how much of it survives once the fermion spins can align with the curvature.

\paragraph*{The invariant bosonic sector and Landau--Pauli cancellation.}\phantomsection\label{sec:sector}\label{sec:pauli}

 The bosonic problem of Sec.~\ref{sec:bosonic} survives inside the supersymmetric theory as a single exact invariant subspace, namely the $(1,1)$ fermion block $V_s\otimes(1,3)_+\otimes(1,3)_-$ in which both fermion sets are $Sp(2)$ singlets. By property~(i) of Eq.~\eqref{eq:curvature}, $\mathcal F_{ab}$ lies in $sp(2)\otimes su(2)$, so $\sum\gamma^{ab}_\pm\mathcal F_{ab}$ acts through the spin representation of $sp(2)\subset so(8)$ and annihilates the $Sp(2)$ singlets; hence $P^{(\pm)}=0$ on this block. Because $SU(2)_G\subset Sp(2)$ acts trivially on $(1,3)$, the lift~\eqref{eq:O1lift} is the identity on the fermions and the physical sections coincide with those of model $\mathsf B$, tensored with a nine-dimensional spectator space. The same restriction applies to the pinned and positively trapped models, so every bosonic instability carries over to the supersymmetric theory, whereas bosonic protection holds in the singlet sector alone.

 In the remaining blocks, the Pauli couplings $P^{(\pm)}$ act against the bosonic Landau barrier. On the base curve, Eq.~\eqref{eq:basecurve} gives the Pauli operator of one set in closed form,
\begin{equation}
 P=\frac{32}{D^2}\,\mathbf S_{\hat w}\cdot\mathbf J,\qquad
 \spec P=\frac{32}{D^2}\Bigl\{\frac s2,\,0,\,-\frac{s+1}2\Bigr\}\quad(s\ge\tfrac12),
 \label{eq:pauliclosed}
\end{equation}
where $\mathbf J$ is the colour spin and $\mathbf S_{\hat w}$ is a spin-$\tfrac12$ operator on the fermion states built along the direction $\hat w$. This follows from Eq.~\eqref{eq:basecurve} since the curvature is the instanton curvature on the quaternionic line spanned by $\hat w$, and by the symmetry of Sec.~\ref{sec:channels} it holds up to conjugation at every point of the link.\footnote{In the convention $e_c=-2\ii T_c$, $[\nabla_a,\nabla_b]=-\ii J^c\mathbb F_{ab}^c(0)=-2\ii J^c\mathcal F_{ab}^c$ contributes a factor of $2$, while contracting $-\gamma^{ab}$ with the self-dual 't~Hooft tensor $\eta^c_{ij}$ gives $-\ii\sum_{i<j}\eta^c_{ij}\gamma^{ij}=4S_{\hat w}^c$, turning $4/D^2$ in Eq.~\eqref{eq:basecurve} into the Hermitian Pauli operator $(32/D^2)\mathbf S_{\hat w}\cdot\mathbf J$ of Eq.~\eqref{eq:pauliclosed}. At the pinned point the exact quantum minimum $-16(s+1)/\rho^2$ (spin $s$ coupled with fermion spin $\tfrac12$ to $s-\tfrac12$) agrees at leading order in $1/s$ with the highest-weight semiclassical Landau energy $-16s/\rho^2$ below.} For $s\ge\tfrac12$, at the pinned direction the most negative value is $-16(s+1)/\rho^2$, which is $4/s$ times the self-energy $-4C_2/\rho^2$ at $\zeta=0$, so for $s<4$ the Pauli energy is the larger of the two. In the combination $c_+P^{(+)}+c_-P^{(-)}$ the minimum is $-16(s+1)/D^2$ for every $\zeta$ when $s\ge1$. Each term is bounded below by its minimum, and both minima are attained together on the states in which the two fermion spins and the colour couple to total spin $s-1$, because there each fermion spin couples with the colour to $s-\tfrac12$. For $s=\tfrac12$ no such state exists, and we prove below analytically that no doubly-BPS state exists in any fermion block, so two-set Pauli frustration lifts $L_0(\zeta)$ above $-8$ whenever $|\zeta|<1$.

The physical content of this minimum is a cancellation. At the pinned direction the centre sees the self-dual field $\mathcal F_{Z_iZ_j}=-4\eta^c_{ij}e_c/\rho^2$. In the colour state of highest weight this is a magnetic field $B=8s/\rho^2$ in each of two orthogonal planes of the coordinate $Z$, with Landau zero-point energy $2B=16s/\rho^2$. The centre fermions anti-aligned with the field contribute the Pauli energy $-16s/\rho^2$, which cancels the zero point exactly, as for a particle of gyromagnetic ratio $2$ in a constant field of one chirality, the mechanism of the planar zero modes of Ref.~\cite{AharonovCasher}. The anti-aligned modes are the superpartners of the centre coordinates, generated by the supersymmetries that the instanton breaks (Appendix~\ref{app:fermions}), which is why their gyromagnetic ratio is exactly two. The centre can therefore occupy a lowest Landau level near the line at no magnetic cost, realizing on instanton moduli space the general mechanism by which Pauli zero modes remove kinetic barriers against critical attractions \cite{FLL1986,LossYau1986,FrankLoss}. In the bosonic sector this Landau energy opposes localization of the free centre and sets the scale $16s$ of the first bosonic threshold in each sector (Sec.~\ref{sec:freephases}), although the protection of Sec.~\ref{sec:freecentre} at small spin needs only the eight-dimensional Hardy barrier. Replacing the Pauli terms by their pointwise minimum and discarding the Landau energy, as a diamagnetic estimate would, gives bounds below $-9$ even at $\zeta=1$, where $H\ge0$ is guaranteed. The cancellation is the moduli-space counterpart of a result for monopoles. For a Dirac particle at a fixed magnetic monopole the centrifugal barrier vanishes in the lowest partial wave, and the Hamiltonian then needs a boundary condition at the monopole, which an infinitesimal extra magnetic moment supplies \cite{KazamaYangGoldhaber1977}. Ref.~\cite{ACKMR} found the same vanishing barrier for fermions of magnetic gyromagnetic ratio $g_m=2$ at 't~Hooft lines.

\paragraph*{The floor at the supersymmetric endpoints.}\phantomsection\label{sec:floor}

Positivity of the supersymmetric endpoints $H(\pm1)={\not{\!\mathbf D}}_\pm^{\dagger}{\not{\!\mathbf D}}_\pm\ge0$ guarantees $\mu\ge-9$ automatically, placing the supersymmetric lines on the conformal side of the Hardy barrier. A deeper physical question is how close the Pauli attraction comes to $-9$, because that gap decides both whether the supersymmetric endpoint dynamics is unique (essentially self-adjoint, $\beta\ge1$) and how much self-energy is needed to trigger collapse once $|\zeta|<1$. The Pauli coupling stops at least one full unit above $-9$ for every spin and attains $-8$ for every nontrivial spin $s\ge\tfrac12$, placing every nontrivial supersymmetric line on the uniqueness boundary $\beta=1$ and setting the pinned attraction scale at $E=1$.

\begin{proposition}[Floor]
\label{thm:floor}
For every $s$, in every fermion block and every channel, $L(\pm1)\ge-8$. For $s\ge\tfrac12$ the value $-8$ is attained. More precisely, on the subspace of Hopf-fibre spin $j_R$ the bound sharpens to $L(\pm1)\ge4j_R(j_R+1)-8$, and the bound holds for the pull-back along the Hopf fibration of any anti-self-dual connection on $S^4$ (Appendix~\ref{app:floor}). Equivalently (Appendix~\ref{app:fb:dirac}), the spherical Dirac operator $\mathcal D$ on $S^7$ defined by ${\not{\!\mathbf D}}=\Gamma(\hat x)\bigl(\partial_R+(7/2-\mathcal D)/R\bigr)$, with $\Gamma(\hat x)=\hat x_\mu\Gamma^\mu$ radial Clifford multiplication, has no spectrum in $(-\tfrac32,\tfrac32)$, or
\begin{equation}
 \int_{\R^8}|{\not{\!\mathbf D}}\psi|^2\ \ge\ \int_{\R^8}\frac{|\psi|^2}{R^2}\qquad\bigl(\psi\in C_c^\infty(\R^8\setminus0)\bigr).
 \label{eq:hardydirac}
\end{equation}
\end{proposition}

Three geometric steps explain the origin of this one-unit gap. Appendix~\ref{app:floor} gives the complete proof. First, the Lichnerowicz formula and radial decomposition give $L(\pm1)=(\mathcal D-\tfrac12)^2-9$, while the chirality involution $\Gamma(\hat x)\mathcal D\Gamma(\hat x)=-\mathcal D$ makes $\spec\mathcal D$ symmetric about zero. Excluding $(-\tfrac12,\tfrac32)$ from $\spec\mathcal D$ therefore excludes the full interval $(-\tfrac32,\tfrac32)$ and turns the floor into the twisted Dirac--Hardy inequality~\eqref{eq:hardydirac}, with constant $1$ in place of the scalar value $9$. Second, because the evaluation connection is pulled back from $S^4$ along the Riemannian Hopf submersion $\pi:S^7(1)\to S^4(\tfrac12)$ (property~(ii) in Sec.~\ref{sec:connection}), every channel splits orthogonally into the fibre Casimir and a horizontal Pauli operator on the base,
\begin{equation}
 L(\pm1)=4j_R(j_R+1)+\mathcal H_A,\qquad \mathcal H_A\ \ge\ \begin{cases}-8,& A\in\{4,5\},\\ 0,& A=1,\end{cases}
 \label{eq:main:hopfsplit}
\end{equation}
where $j_R$ is the $SU(2)_R$ spin along the Hopf fibre and $A$ labels the active fermion block. The floor and $W\le1$ give $L(\zeta)+9\ge1+4j_R(j_R+1)-E$. Thus $j_R\ge\tfrac12$ stays above the Hardy threshold for $E<4$, which includes the onset bracketed in Sec.~\ref{sec:instability}. Third, the round base $S^4(\tfrac12)$ has scalar curvature $4\cdot3/(\tfrac12)^2=48$. The Lichnerowicz formula for the squared twisted Dirac operator therefore contributes the scalar term $\tfrac14\mathrm{scal}=12$. The flat fermion connection of $\R^8$ descends to the Killing-spinor connection $\nabla+\tau X\cdot$ (with Killing sign $\tau=\pm1$ and Clifford multiplication $X\cdot$) in block~$4$ and the second-fundamental-form connection $\nabla+\mathcal B$ of $S^4(\tfrac12)\subset\R^5$ in block~$5$, shifting the square by $-4$ and completing $\mathcal H_A+8$ into a manifestly nonnegative sum of squares ($12-4=8$). Equality at $j_R=0$ occurs precisely on the zero modes $\mathcal K_4=\ker\mathsf D_-$ of the instanton-twisted Dirac operator on $S^4(\tfrac12)$ in block~$4$, and on the space $\mathcal K_5$ of twisted one-forms obeying $\dd^*\gamma=\dd^+\gamma=0$ (zero codifferential and zero self-dual exterior derivative) in block~$5$. By the Atiyah--Singer index theorem, the single-set block-$4$ floor space has exact dimension $I(s)=\frac23s(s+1)(2s+1)=\binom{2s+2}{3}$ \cite{JackiwRebbi1977,BCL1977}, ensuring that $-8$ is attained on physical sections for every $s\ge\tfrac12$. Here $I(s)$ counts active block-$4$ multiplets, one per inert $Sp(1)$ doublet, before the spectator tensor product and the physical parity restriction.

The endpoint operators carry a superconformal structure that organizes this gap. Because the connection is dilation invariant and has no radial component, the colour-twisted Dirac operator and Clifford multiplication by the position obey $\{{\not{\!\mathbf D}},\Gamma(x)\}=2R\partial_R+8$ and $[{\not{\!\mathbf D}},\Gamma(x)]=2\mathcal D+1$, so $-\ii{\not{\!\mathbf D}}$ and $\Gamma(x)$, whose squares are $H(\pm1)$ and $K=R^2$, are odd generators that extend the $SO(1,2)$ of Sec.~\ref{sec:radial} to $OSp(1|2)$. The spherical Dirac operator $\mathcal D$ is its Scasimir \cite{Lesniewski1995,ABF1997}, anticommuting with both odd generators, and each multiplet pairs two link channels with opposite values of $\mathcal D$, with lowest weights $\tfrac12|\mathcal D|+\tfrac14$ and $\tfrac12|\mathcal D|+\tfrac34$, so the floor $\mathcal D=\tfrac32$ is the multiplet with $k_+=1$. Since unitary multiplets of $OSp(1|2)$ exist for every $k>0$, the superalgebra converts the spectral gap of $\mathcal D$ into the bound $k_+\ge1$ but does not by itself fix the position of the floor.

The four supercharges of the $(0,4)$ structure organize the floor into superconformal multiplets. Because the curvature $\mathcal F$ of the Berry connection on instanton moduli space is of type $(1,1)$ for the three right complex structures and annihilates the quaternionic Hopf fibre (properties~(i) and~(ii) after Eq.~\eqref{eq:curvature}), the operators $Q_A=-\ii\Gamma(I_A\nabla_{\mathcal A})$ and $S_A=\Gamma(I_Ax)$, with $I_0=1$ and $I_a$ right multiplication by $i,j,k$, close with $H(\pm1)$, $R^2$ and the dilatation into $OSp(4^*|2)$, the real form of $D(2,1;\alpha)$ at $\alpha=-2$, as in superconformal quantum mechanics on hyperk\"ahler cones \cite{BMSV}. Its bosonic part is $SL(2,\R)\times SU(2)_R\times Sp(1)_{\rm F}$, with $SU(2)_R$ the rotation along the Hopf fibre (spin $j_R$) and $Sp(1)_{\rm F}$ the right rotation of the active fermion set (spin $j_{\rm F}=0,\tfrac12,1$ in blocks $5$, $4$ and $1$). Unitarity \cite{GimenezLiendo2020} gives $k\ge j_R+1$, which is the bound $L(\pm1)\ge4j_R(j_R+1)-8$ of Eq.~\eqref{eq:main:hopfsplit}. At $j_R=0$ it gives the universal floor, while the sharper statements $\mathcal H_1\ge0$ in block~1 and the gap $\mathsf G_4^{\rm F}=8$ implied by Friedrich's Dirac-eigenvalue inequality \cite{Friedrich1980} in block~4 (Appendix~\ref{app:fb:consequences}) follow from the geometric arguments, not from unitarity alone, and the floor states form a short multiplet in block~4 and a multiplet at threshold in block~5. The algebra interprets the floor but does not replace the proof of Proposition~\ref{thm:floor}, since the odd generators act within the realization only in channels with $|\mathcal D|\ge\tfrac12$ and do not show that the floor is attained. The closure relations were checked symbolically on flat $\R^4$ and $\R^8$ and at sample points for the evaluation connection, and the norm conditions with exact Gram determinants for the charges that occur at $j_R=0$. The quaternionic structure suggests that the relations persist in every dimension divisible by four.

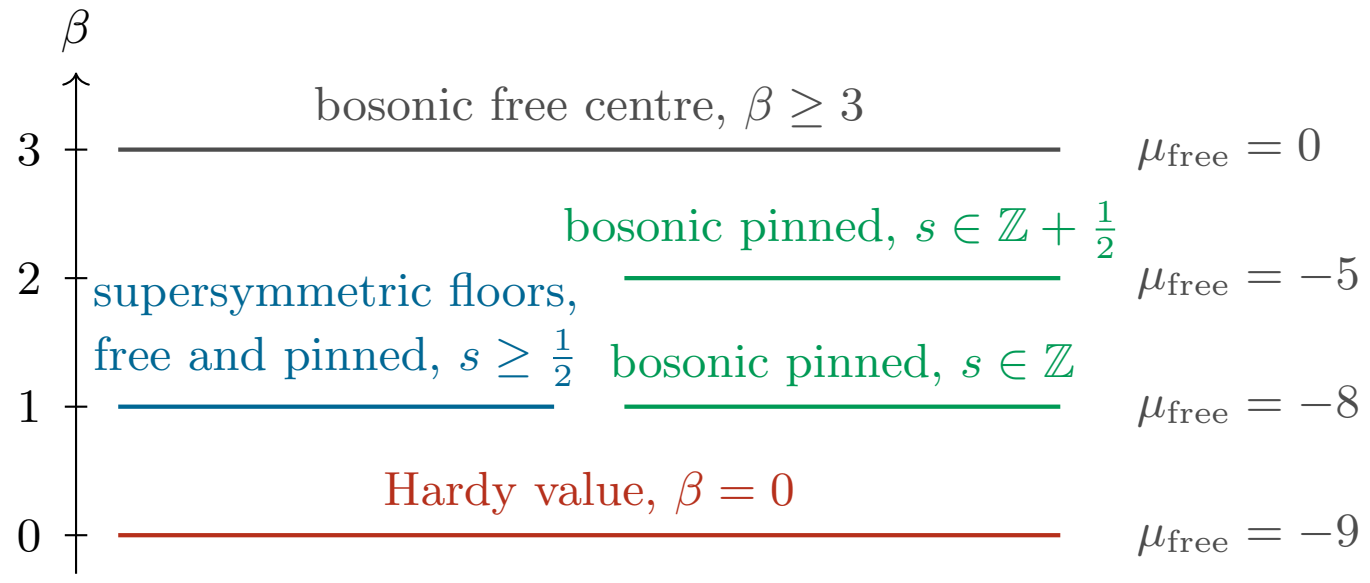
\begin{figure}[t]
\centering
\begin{tikzpicture}[x=0.93cm,y=0.85cm,font=\footnotesize]
 \draw[->] (0,-0.3) -- (0,3.6) node[above] {$\beta$};
 \foreach \b/\m in {0/-9,1/-8,2/-5,3/0}{
  \draw (-0.08,\b) -- (0.08,\b);
  \node[left] at (-0.1,\b) {$\b$};
  \node[right,black!70] at (7.4,\b) {$\mu_{\rm free}=\m$};}
 \draw[thick,BrickRed] (0.3,0) -- (7,0);
 \node[above,BrickRed] at (3.65,0) {Hardy value, $\beta=0$};
 \draw[thick,MidnightBlue] (0.3,1) -- (3.4,1);
 \node[above,MidnightBlue,align=center] at (1.85,1) {supersymmetric floors,\\ free and pinned, $s\ge\frac12$};
 \draw[thick,ForestGreen] (3.9,1) -- (7,1);
 \node[above,ForestGreen] at (5.45,1) {bosonic pinned, $s\in\Z$};
 \draw[thick,ForestGreen] (3.9,2) -- (7,2);
 \node[above,ForestGreen] at (5.45,2) {bosonic pinned, $s\in\Z+\tfrac12$};
 \draw[thick,black!70] (0.3,3) -- (7,3);
 \node[above,black!70] at (3.65,3) {bosonic free centre, $\beta\ge3$};
\end{tikzpicture}
\caption{Floors of the radial problem, in $\beta$ ($\beta=\sqrt{\mu_{\rm free}+9}$ for the free-centre models, with the right-hand ticks showing $\mu_{\rm free}$, and $\beta=\sqrt{\mu_{\rm pin}+1}$ for the pinned models), before the self-energy is switched on. The self-energy lowers $\beta^2$ by at most $4C_2(1-\zeta^2)$, so a floor at $\beta=1$ requires an attraction of strength greater than $1$ for collapse, while the bosonic free centre, at $\beta=3$, needs at least $9$ and, for the actual self-energy profile, at least $99/4$, and $16s+O(1)$ at large spin (Sec.~\ref{sec:freephases}). Supersymmetry places the free centre at the pinned value $\beta=1$ for each $s\ge\tfrac12$ (Proposition~\ref{thm:floor}). For half-integer spin the bosonic pinned problem sits higher, at $\beta=2$, because of the bundle selection. The bosonic pinned labels refer to $\theta_{5d}=0$, and at $\theta_{5d}=\pi$ the integer and half-integer floors exchange. The supersymmetric pinned floor is that of the model $\mathsf S_{\rm pin}$ of Table~\ref{tab:models}, which the $\Omega$-trap of Sec.~\ref{sec:windows} reproduces at leading Born--Oppenheimer order.}
\label{fig:ladder}
\end{figure}

Figure~\ref{fig:ladder} summarizes these floors. The floor bounds the full link operator at every coupling. For every $s$ and $|\zeta|\le1$, $L(\zeta)\ge-8+(1-\zeta^2)V\ge-8-4C_2(1-\zeta^2)$, because $L(\zeta)-(1-\zeta^2)V=c_+L(1)+c_-L(-1)\ge-8$ by Proposition~\ref{thm:floor} and $V\ge-4C_2$. The line is therefore conformal whenever $4C_2(1-\zeta^2)\le1$, and in terms of $E_c(s)=4C_2\bigl(1-\zeta_c(s)^2\bigr)$ the resulting bound on the free threshold for $s\ge1$ is
\begin{equation}
 E_c(s)\ \ge\ 1,\qquad \zeta_c(s)\ \le\ \zeta_{\rm pin}(s)=\sqrt{1-1/(4C_2)} .
 \label{eq:pinbound}
\end{equation}

\paragraph*{States at the floor.}\phantomsection\label{sec:bpsstates}

 The states saturating the supersymmetric endpoint bound $L(\pm1)\ge-8$ are BPS states. For $s=1$ they include the supersymmetric zero modes of the evaluation connection, since the Bianchi identity and the Yang--Mills equations give ${\not{\!\mathbf D}}\,(\gamma^{AB}\mathcal F_{AB}\,\epsilon)=0$ for a constant spinor $\epsilon$ of the appropriate chirality \cite{NSVZ,DHKM}, and the degree $-2$ homogeneity of $\mathcal F$ places these states at $\mu=-8$. In the critical $(0,0,0)$ channel of block $(4,4)$, decomposing along the centre and size polarizations $e_Z=(1,0)$ and $e_y=(0,1)$ of the spectator fermion set makes the $\zeta=+1$ floor space two-dimensional, spanned by $\mathcal F_b=-\hat w_b\,\hat w\otimes e_b/D^2$ ($b\in\{Z,y\}$). Section~\ref{sec:bps} constructs the corresponding pair for every integer spin $s\ge1$.

Globally, the conformal geometry of $S^4$ generates the entire $j_R=0$ floor space in closed algebraic form (Appendix~\ref{app:fb:bps}). On the Hopf base $S^4(\tfrac12)$ with round metric $\mathsf g_1$, the inertia rescaling $N(Z,y)=(y,-\sqrt2 Z)$ induces a conformal diffeomorphism $\bar N$ of $S^4$. The colour connection $a_N=\bar N^*a$ is the pull-back of the basic instanton. Passing to the conformally equivalent round metric $\mathsf g_2=\bar N^*\mathsf g_1$ identifies the spin-$s$ colour bundle with $E_s=\mathrm{Sym}^{2s}S^-_{\mathsf g_2}$, where $S^\pm$ denote the two-dimensional chiral spinor bundles of $S^4$ (distinct from the sixteen-dimensional $\R^8$ modules $S_\pm$ of Sec.~\ref{sec:fermions}), with unprimed indices $A,B$ on $S^+$ and primed indices $A',B'$ on $S^-$. Writing $\epsilon\odot\omega$ for contraction of a fermion index with a colour index by $\epsilon_{A'C'}$ followed by symmetrization over the $2s$ colour indices, the four-dimensional space $\mathbb T$ of twistor spinors on $S^4$, the conformal Killing spinors obeying $\nabla_X\psi=-\tfrac14X\cdot{\not\!\nabla}\psi$ for every vector $X$ \cite{PenroseRindler}, builds the complete single-set block-$4$ floor space as $\mathcal K_4=\epsilon\odot\mathcal T_{2s-1}$. Here $\mathcal T_k=\mathrm{Sym}^k\mathbb T$ has dimension $\binom{k+3}{3}$, which reproduces $I(s)=\binom{2s+2}{3}$. Since $\mathbb T=(\tfrac12,0)\oplus(0,\tfrac12)$ under $SU(2)_L\times SU(2)_G$, the space $\mathcal T_k$ is the sum of the \emph{twistor types} $(\tfrac a2,\tfrac b2)$ with $a+b=k$.

Requiring a state to saturate the floor of both supersymmetric halves simultaneously, $(L_++8)T=(L_-+8)T=0$, defines the doubly-BPS subspace. On any doubly-BPS state, $L_0(\zeta)=-8$ identically for all $\zeta\in[-1,1]$, leaving the self-energy as the sole term distinguishing a generic line from a supersymmetric one. In block $(4,4)$, a state is doubly BPS if and only if it is a symmetric twisted anti-self-dual Maxwell field $\phi_{A'B'}\in C^\infty(\mathrm{Sym}^2S^-\otimes E_s)$ obeying $\nabla_A{}^{A'}\phi_{A'B'}=0$, and Appendix~\ref{app:fb:twistor} proves that the twistor family
\begin{equation}
 \mathcal H_{\rm dBPS}^{(4,4)}(s)=\epsilon\odot\epsilon\odot\mathcal T_{2s-2},\qquad
 \dim\mathcal H_{\rm dBPS}^{(4,4)}(s)=\binom{2s+1}{3},
 \label{eq:main:dbps}
\end{equation}
exhausts the block-$(4,4)$ doubly-BPS space for every $s\ge\tfrac12$. Two immediate consequences follow. First, at $s=\tfrac12$, $\binom{2}{3}=0$ combines with $\dim\mathcal K_5=2I(\tfrac12)-2=0$ (Appendix~\ref{app:fb:block5}) to prove analytically that no doubly-BPS state exists in any fermion block for the fundamental line. This is the physical reason to expect stability, and a pointwise bound on the two Pauli couplings, a three-spin recoupling and a scalar comparison prove analytically that $s=\tfrac12$ stays conformal for every $|\zeta|\le1$ and both $\theta_{5d}\in\{0,\pi\}$ (Sec.~\ref{sec:bps} and Appendix~\ref{app:fb:s12}). Second, because block-$5$ floor states live in $\Lambda^1\otimes\C=S^+\otimes S^-$, the equivariant map $\iota_\epsilon:\phi_{A'B'}\mapsto\epsilon_{AB}\phi_{A'B'}$ embeds the entire $(4,4)$ core, isometrically up to a constant factor, into block $(5,5)$ (Appendix~\ref{app:fb:core}), preserving both $L_0(\zeta)+8$ and $\langle W\rangle$. This embedding explains why blocks $(4,4)$ and $(5,5)$ tie at first order with weight average $w_s=2s/(2s+3)$ (Sec.~\ref{sec:instability}). Numerical scans resolve a small splitting away from the common trial-state bound.

This twistor construction places our instanton problem inside a broader Hopf-fibration hierarchy of lowest Landau levels (LLLs) \cite{Hasebe}.\footnote{Because the inertia factor $\sqrt2$ deforms the round metric $\mathsf g_1$ to $\mathsf g_2=\bar N^*\mathsf g_1$, it breaks isometric $Sp(2)$ invariance to $SU(2)_L\times SU(2)_G$. Conformal invariance of the twisted Dirac and Maxwell equations preserves the conjugate action $\bar N^{-1}Sp(2)\bar N$ on $\mathcal K_4$, $\mathcal K_5$ and $\mathcal H_{\rm dBPS}^{(4,4)}$, displacing the Zhang--Hu lowest Landau level conformally without changing its representation content.} As $Sp(2)$ representations and as explicit polynomials, the single-set floor states $\epsilon\odot\mathcal T_{2s-1}$ and the doubly-BPS family $\epsilon\odot\epsilon\odot\mathcal T_{2s-2}$ are isomorphic to the LLLs of the four-dimensional quantum Hall problem of Zhang and Hu on $S^4$ \cite{ZhangHu} at levels $p=2s-1$ and $p=2s-2$ (Appendix~\ref{app:fb:twistor}), and the Zhang--Hu degeneracy $\binom{p+3}{3}$ at $p=2s-1$ reproduces the index $I(s)$. The correspondence is canonical, since both spaces are the holomorphic sections $H^0(\mathbb{CP}^3,\mathcal O(p))$ on the twistor space $\mathbb{CP}^3$ fibred over $S^4$, through the Penrose transform for the twistor spinors and through the holomorphy of lowest-Landau-level wavefunctions on $S^7\subset\C^4$; physically, each anti-aligned fermion spin-$\tfrac12$ lowers the effective Yang-monopole isospin by $\tfrac12$, reducing $2s$ to $2s-1$ for one active set and $2s-2$ for both, while the inertia ratio $m_c/m_s=\tfrac12$ enters purely as the conformal dilation $\bar N$ weighting the LLL around the pinned pole. The identification concerns the representation content and the spatial profile of the floor states. A Hall response or an edge would need a different Hamiltonian and boundary. This sits between Haldane's complex Hopf map $S^3\to S^2$ \cite{Haldane1983} (also the coadjoint-orbit phase space of spin-$s$ impurities \cite{CKMR}) below and the octonionic Hopf map $S^{15}\to S^8$ of eight-dimensional $SO(8)$ instantons and quantum Hall LLLs \cite{GrossmanKephartStasheff1984,Bernevig2003} above (Sec.~\ref{sec:discussion}). Although doubly-BPS states are eigenstates of $L(\pm1)$ and of $c_+L(1)+c_-L(-1)$ at $-8$, they are not normalizable zero-energy states of the full quantum mechanics, and at generic $\zeta$ they are not supersymmetric. Their role is to furnish the exact variational trial states that drive the instability of Sec.~\ref{sec:instability}.

The superconformal algebra also distinguishes the two floor modules by a protected trace, an index of the endpoint quantum mechanics in which paired states cancel \cite{KMMR2007}. Its first two terms are $-q\,\chi_{\mathcal K_4}-q^2\chi_{\mathcal K_5}$, where $\chi_{\mathcal K_A}$ is the $SU(2)_L\times SU(2)_G$ character of the floor space of the active block $A$, before the spectator factors and the theta projection. The block-4 count $I(s)$ and its Zhang--Hu character are therefore protected against deformations that preserve the superconformal algebra, the symmetries that grade the trace and the discreteness of the spectrum of its compact generator. The index counts floor states with signs. The gap that isolates them comes from the Dirac bound behind Proposition~\ref{thm:floor}, not from the index. Appendix~\ref{app:fb:index} gives the grading, the spectator characters and the effect of the theta projection.

\section{Loss of conformality and self-localization}
\label{sec:selfloc}

 The floor of Sec.~\ref{sec:susy} reduces the question to a competition. On a doubly-BPS state the supersymmetric part of $L(\zeta)$ equals $-8$ at every $\zeta$, and the self-energy $(1-\zeta^2)V$ decides whether the state falls below $-9$. The endpoint trial states first give an upper bound and identify the localization mechanism. The all-channel lower bound then isolates the competitors through cubic order, and a separate spin-one comparison resolves the two theta thresholds at finite spin. Section~\ref{sec:windows} tests the same mechanism with a confining rotation, in a different spectral regime, where the size can grow or the centre can drift to infinity.

\subsection{The critical channel and the instability}
\label{sec:instability}

At $\theta_{5d}=0$ and integer $s\ge1$, the numerical scans favour the channel $(j_L,j_R,j_G)=(s-1,0,0)$ of the fermion block $(4,4)$. The scans cover the numerically delimited boxes for $s\le3$ and cover the $j_R=0$ channels in the stated scan boxes for integer $s\le6$ (Appendix~\ref{app:numerics}). The lower bound of Sec.~\ref{sec:uniform} applies to every channel. In this channel the two fermion sets are coupled to total spin $1$ and then, with the colour, to $j=s-1$. The multiplicity space is four-dimensional for every $s$, with basis $T_{ab}$, $a,b\in\{Z,y\}$, labelling the doublet of each set. The link operator becomes an exact $4\times4$ system,
\begin{equation}
\begin{aligned}
 L_s(\zeta)&=-\omega^{-1}\partial_\alpha\,\omega\,\partial_\alpha+K_s+(1-\zeta^2)V+\Pi_s(\zeta),\\
 \Pi_s(\zeta)&=-\frac{16(s+1)}{D^2}\bigl[c_+P_{\hat w}\otimes1+c_-1\otimes P_{\hat w}\bigr],
\end{aligned}
 \label{eq:4x4}
\end{equation}
with $P_{\hat w}=\hat w\hat w^{\rm T}$ and $\hat w=(\cos\alpha,-\sin\alpha)$ as in Eq.~\eqref{eq:basecurve}. The angular term is diagonal, $K_s=\operatorname{diag}(k_1,k_2,k_2,k_4)$ in the order $(ZZ,Zy,yZ,yy)$, with
\begin{equation}
\begin{gathered}
 k_1=\frac{16C_2\sigma}{D^2\kappa},\qquad
 k_2=\frac3{\sigma\kappa}+\frac{8(s+1)\bigl((2s-1)\sigma-1\bigr)}{D^2\kappa},\\
 k_4=\frac4{\sigma\kappa}\Bigl(\frac{4C_2\sigma^2}{D^2}-\frac{4(s+1)\sigma}{D}+2\Bigr).
\end{gathered}
 \label{eq:Ks}
\end{equation}
The derivation uses the closed-form curvature~\eqref{eq:basecurve}, stabilizer equivariance along the two orbit directions and the Wigner--Eckart theorem for the Pauli term. Equivalently, Eq.~\eqref{eq:4x4} is the specialization of the general channel form~\eqref{eq:un:Mform}--\eqref{eq:un:orbit} of Appendix~\ref{app:un:coverage} to block $(4,4)$ and the channel $(s-1,0,0)$. Appendix~\ref{app:numerics} describes the independent numerical reduction, whose matrix potential agrees with that of Eq.~\eqref{eq:4x4} pointwise to relative accuracy $7\times10^{-15}$ for $s\le5$.

The channel contains a doubly-BPS state in closed form. For every $s\ge1$, integer or half-integer, the state
\begin{equation}
 T_s(\alpha)=\cos^{2s-2}\!\alpha\,\bigl(1+\sin^2\alpha\bigr)^{-(s+1)}\,\hat w\otimes\hat w
 \label{eq:Ts}
\end{equation}
obeys $\bigl(c_+L_s(1)+c_-L_s(-1)+8\bigr)T_s=0$ for every $\zeta\in[-1,1]$, is regular at both ends of the interval, and lies in the form domain. Since $(P_{\hat w}\otimes1)T_s=(1\otimes P_{\hat w})T_s=T_s$, the Pauli part acts on $T_s$ as $-16(s+1)D^{-2}$ for every $\zeta$, so it suffices to verify the equation at $\zeta=1$. Writing $T_s=A\,(\kappa,-\sqrt{\sigma\kappa},-\sqrt{\sigma\kappa},\sigma)$ with $A=\kappa^{s-1}D^{-(s+1)}$, the equation becomes three identities, for the $ZZ$, $Zy=yZ$ and $yy$ components, each rational in $\sigma$ and $s$. Each vanishes identically. We verified this symbolically and checked it independently by high-precision finite differences. The identity also follows without computer algebra from Appendix~\ref{app:fb:bps}, where $T_s$ is the member of type $(s-1,0)$ of the doubly-BPS family built from twistor spinors (Sec.~\ref{app:fb:twistor}) and Eq.~\eqref{eq:fb:Wb} gives its weight average $w_s$. The indicial exponents of $T_s$ are the regular roots of the system, $(0,1,2)$ for $(ZZ,Zy,yy)$ at $\alpha\to0$ and $(2s,2s-1,2s-2)$ at $\alpha\to\pi/2$, which gives the regularity at both ends and places $T_s$ in the form domain. Since the equation holds at $\zeta=\pm1$, $T_s$ is supersymmetric for both halves separately, a BPS state of $L(1)$ and of $L(-1)$ at once.

Its self-energy average reduces to two Beta integrals. With the measure $\omega\,\dd\alpha$, $\|T_s\|^2=\bigl(16s(2s+1)\bigr)^{-1}$ and
\begin{equation}
 w_s=\frac{\langle T_s,W\,T_s\rangle}{\|T_s\|^2}=\frac{2s}{2s+3}.
 \label{eq:vs}
\end{equation}
Changing variable to $v=(1-\sigma)/(1+\sigma)=1/(1+2t)\in(0,1)$ turns $A^2\omega\,\dd\alpha$ into $\tfrac18v^{2s-1}(1-v)\,\dd v$ and $W$ into $\tfrac12v(1+v)$, giving the two Beta integrals $\tfrac18B(2s,2)=\bigl(16s(2s+1)\bigr)^{-1}$ and $\tfrac1{16}\int_0^1v^{2s}(1-v^2)\,\dd v=\bigl(8(2s+1)(2s+3)\bigr)^{-1}$, whose ratio is $w_s$.

These two results give the instability. For every integer $s\ge1$ the free-centre line of model~$\mathsf S$ loses conformality whenever
\begin{equation}
 \zeta^2<1-\frac{2s+3}{8s^2(s+1)},
 \label{eq:thmC}
\end{equation}
so that $E_c(s)\le1+3/(2s)$. For $s=1$ this reads $\zeta_c\ge\sqrt{11}/4$. Because $T_s$ is doubly BPS, $L_s(\zeta)T_s=-8T_s+(1-\zeta^2)VT_s$, and Eq.~\eqref{eq:vs} gives
\begin{equation}
 \frac{\langle T_s,L_s(\zeta)T_s\rangle}{\|T_s\|^2}=-8-(1-\zeta^2)\,4C_2\,w_s=-8-(1-\zeta^2)\frac{8s^2(s+1)}{2s+3},
\end{equation}
which lies below $-9$ exactly under condition~\eqref{eq:thmC}. The channel is an invariant subspace of the link operator, so by min--max the link spectrum lies below $-9$.

At $\theta_{5d}=0$, half-integer spins $s\ge\tfrac32$ probe the $\Z_2$ centre of $SU(2)$ and carry nontrivial orbifold holonomy around $\lambda\to-\lambda$, so the selection rule~\eqref{eq:physical} forbids $j_G=0$ and excludes the state $T_s$, whose active twistor type $(\frac a2,\frac b2)$ (Sec.~\ref{sec:bpsstates}) has $SU(2)_G$ spin $\frac b2=0$. Staying in block $(4,4)$ forces $b=1$ ($j_G=\tfrac12$), whose extra centrifugal factor $t$ pushes the instanton away from the line and weakens $\langle W\rangle$; instead, numerical scans place the threshold in channel $(s-1,0,\tfrac12)$ of the mixed block $(4,5)$, where placing one fermion set in block~5 carries the half-unit of $j_G$ while preserving $b=0$ polar localization. Its doubly-BPS trial states $X_\mu\cdot\chi$ contract the quartet of gradient conformal Killing fields $X_\mu$ of $S^4$ (carrying $j_G=\tfrac12$) into $b=0$ twisted massless spin-$\tfrac32$ fields $\chi$ built from twistor spinors. Summing over the quartet gives the trial-state bound $E_c(s)\le1/w_{\rm qrt}(s)$ for every half-integer $s\ge\tfrac32$, where
\begin{equation}
 w_{\rm qrt}(s)=\frac{\int_0^\infty W\varrho_1\,\dd t}{\int_0^\infty\varrho_1\,\dd t},\qquad
 \varrho_1(t)=\frac{t\,(1+t+t^2)}{(1+t)\,(t+\tfrac12)^{2s+2}} .
 \label{eq:vpp}
\end{equation}
Isolating the unique highest-weight vector $T'_s=X_{++}\cdot\chi_{(2s-3,0)}^{\rm hw}$, with $X_{++}$ the top-weight member of the quartet $X_\mu$, inside channel $(s-1,0,\tfrac12)$ sharpens this bound. Within the one-parameter density family $\varrho_\vartheta(t)=t(1+t+\vartheta t^2)/\bigl[(1+t)(t+\tfrac12)^{2s+2}\bigr]$ ($\varrho_1$ being $\vartheta=1$), Appendix~\ref{app:fb:halfint} proves that $T'_s$ has exact profile $\varrho_{\rm crit}=\varrho_{\vartheta_s}$ with $\vartheta_s=(2s-1)/(4s-4)$ for $s\ge\tfrac52$ ($\vartheta_{3/2}=1$, the dimension ratio partitioning the large-$t$ quartet weight between $(s-1,\tfrac12)$ and $(s-2,\tfrac12)$), giving $E_c(s)\le1/w_{\rm crit}(s)\le1/w_{\rm qrt}(s)$ (strictly tighter for all $s\ge\tfrac52$, e.g.\ $1/w_{\rm crit}(\tfrac52)=\tfrac79(113-160\ln2)$ against $1/w_{\rm qrt}(\tfrac52)=\tfrac{28}{73}(337-480\ln2)$). Because the $(1+t)$ factors in $W(t)$ and $\varrho_\vartheta(t)$ cancel identically, the numerator integral is purely rational in $s$, and alternating geometric bounds on $(1+t)^{-1}$ prove analytically through $O(s^{-4})$ that both reciprocals have vanishing $s^{-2}$ terms,
\begin{equation}
 \frac1{w_{\rm qrt}(s)}=1+\frac3{2s}+\frac9{16s^3}+\frac{15}{32s^4}+O(s^{-5}),\qquad
 \frac1{w_{\rm crit}(s)}=1+\frac3{2s}+\frac9{32s^3}+\frac3{8s^4}+O(s^{-5}).
 \label{eq:vpp-asymp}
\end{equation}
Enlarging the trial space from one doubly-BPS state to the full $\zeta=1$ BPS subspace of the critical channel gives sharper proved bounds $E_{2\times2}(s)$ (integer $s$) and $E_{3\times3}(s)$ (half-integer $s$) whose expansions share the term $-\tfrac{9}{56s^2}$ for both spin families (Sec.~\ref{sec:bps}, Tables~\ref{tab:thresholds} and~\ref{tab:nearbps}).

\begin{figure}[t]
\centering
\includegraphics[width=0.72\textwidth]{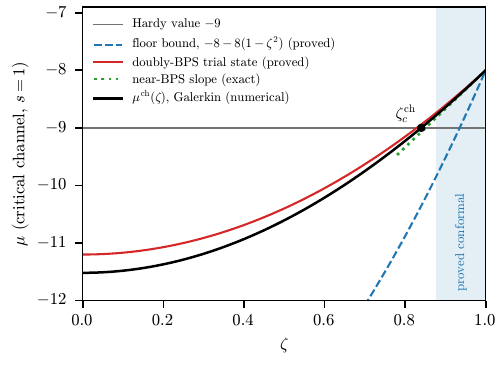}
\caption{Lowest eigenvalue $\mu^{\rm ch}$ of the $(4,4)$ channel $(0,0,0)$ for $s=1$ at $\theta_{5d}=0$ (black, Galerkin solution, numerical), against the Hardy value $-9$ below which the line loses conformality. The doubly-BPS trial state gives the proved upper bound $\mu^{\rm ch}\le-8-\frac{16}5(1-\zeta^2)$ (red), which crosses $-9$ at $\sqrt{11}/4$. The floor gives the proved lower bound $\mu^{\rm ch}\ge-8-8(1-\zeta^2)$ (dashed), and the ground-state representation of Eq.~\eqref{eq:channellaw} proves $\mu^{\rm ch}\ge-9$ in the shaded region $\zeta\ge\sqrt{70-16\sqrt{13}}/4$. For all channels together Eq.~\eqref{eq:O2prime} gives $\zeta\ge5\sqrt2/8$. The dotted line is the exact near-BPS slope $\mu_1^{(1)}(1)$ of Eq.~\eqref{eq:slope}. The numerical curve lies between the proved bounds and crosses $-9$ at $\zeta_c^{\rm ch}\simeq0.83996$.}
\label{fig:mu}
\end{figure}

\paragraph*{Lower bound in the critical channel.}
This bound concerns the threshold $E_c^{\rm ch}$ of one channel. Appendix~\ref{app:fb:bps} uses its ground-state representation to identify $T_s$, and Eq.~\eqref{eq:O2prime} below bounds the threshold over all channels. Because $T_s$ is positive after a constant change of sign, it also controls the channel from below. For an integer $s\ge1$ let $\delta_*(s)$ be the positive root of $16\delta^2+(32s+24)\delta=3$. In the critical channel $c_+L_s(1)+c_-L_s(-1)\ge-8$ for every $|\zeta|\le1$, and $L_s(\zeta)\ge-9$ whenever $4C_2(1-\zeta^2)\le1+16\delta_*(s)$. Together with the instability bound~\eqref{eq:thmC}, whose trial state lies in this channel, the channel threshold therefore obeys
\begin{equation}
 1+16\delta_*(s)\ \le\ E_c^{\rm ch}(s)\ \le\ 1+\frac3{2s},\qquad
 16\delta_*(s)=\frac3{2s}-\frac9{8s^2}+\frac{99}{128s^3}+O(s^{-4}).
 \label{eq:channellaw}
\end{equation}
Figure~\ref{fig:mu} shows these bounds for $s=1$ together with the numerical eigenvalue of the channel. Both lower bounds follow from a cooperative-system Picone identity \cite{Allegretto1985}. In the sign gauge $\mathcal G=\operatorname{diag}(1,-1,-1,1)$, every off-diagonal entry of $L_0=c_+L_s(1)+c_-L_s(-1)$ is nonpositive ($-16(s+1)c_\pm\sqrt{\sigma\kappa}/D^2\le0$) and $\mathcal GT_s>0$ on $(0,\pi/2)$. Consequently, for any positive vector $\Phi$ and $\varphi_i=\psi_i/\Phi_i$, the quadratic form $\langle\psi,\mathcal L\psi\rangle$ splits into $\sum_i\int\omega\,\Phi_i^2[(\varphi_i')^2+q_i\varphi_i^2]$ plus nonnegative off-diagonal squares $\sum_{i<j}\int\omega(-M_{ij})\Phi_i\Phi_j(\varphi_i-\varphi_j)^2$, where $q_i=(\mathcal L\Phi)_i/\Phi_i$. Choosing $\Phi=\mathcal GT_s$ in $\mathcal L=L_0+8$ makes every $q_i=0$, proving the channel floor; deforming to $\Phi=D^{-2\delta}\mathcal GT_s$ in $\mathcal L=L_s(\zeta)+9$ gives $q_{yy}>q_{Zy}=q_{yZ}>q_{ZZ}$, with $D^2q_{ZZ}=(1-E+16\delta)+(2+E-40\delta-16\delta^2-32s\delta)\sigma+(1-4\delta)^2\sigma^2$. At $\delta=\delta_*$ and $E\le E_*=1+16\delta_*$, $D^2q_{ZZ}=(E_*-E)(1-\sigma)+(1-4\delta_*)^2\sigma^2\ge0$ everywhere on $[0,1]$, closing the bound on the Friedrichs form core.

\paragraph*{A single onset of instability.} The trial-state bounds also identify a single transition between stable and supercritical couplings for $s\ge1$. Physically, lowering $|\zeta|$ raises the weight $1-\zeta^2$ of the attraction linearly in $E$, while the supersymmetric part changes only through $c_-$, which stays below $\tfrac18$ in the relevant range. Once a state has crossed $-9$, the attraction outgrows any loss of supersymmetric binding and the state cannot return. To see this, exchange the two fermion sets if necessary so that $\zeta\ge0$, and use $E=4C_2(1-\zeta^2)$. For a unit vector in the link form domain write $a=\langle L_++8\rangle\ge0$, $d=\langle L_--L_+\rangle$, and $w=\langle W\rangle\le1$. Its shifted quadratic form is $f(E)=1+a+c_-(E)d-Ew$. At a zero of $f$, using $Ec_-'(E)-c_-(E)=c_-^2/\zeta$, we have

\begin{equation}
 Ef'(E)=-1-a+\frac{c_-^2}{\zeta}\,d .
 \label{eq:onset}
\end{equation}

If $d\le0$ this is negative. If $d>0$, the zero gives $c_-d\le E-1$, and for $E\le7/2$ and $s\ge1$ we have $\zeta\ge3/4$ and $c_-\le1/8$. Hence the derivative is again strictly negative. A negative test form therefore stays negative up to $E=7/2$, and compactness of the link ensures downward crossing at its first zero eigenvalue. For $E\ge7/2$, the common-floor trials above are negative. The half-integer bound follows from $1/w_{\rm qrt}(3/2)<7/2$ and from the fact that increasing $s$ tilts its density towards smaller $t$, where $W$ is larger (the likelihood-ratio ordering used in Appendix~\ref{app:un:uniform}). In the exceptional $s=1,\theta_{5d}=\pi$ channel the vector $(1,1)$ in Eq.~\eqref{eq:E2x2} gives $5/2+6c_-(E)-E$, which is convex and negative at both $7/2$ and $8$. Thus $\mu<-9$ throughout $E>E_c$ for the stated spin families, and $\mu>-9$ for $E<E_c$. At $\theta_{5d}=\pi$ the argument is unchanged with the two families exchanged. Half-integer $s\ge\tfrac32$ then have the block-$(4,4)$ state $T_s$ in channel $(s-1,0,0)$, with $1/w_s\le2$, and integer $s\ge2$ the block-$(4,5)$ states of Eq.~\eqref{eq:vpp}, with $1/w_{\rm qrt}(s)\le1/w_{\rm qrt}(\tfrac32)<\tfrac72$, so a single onset separates the phases at both values of the theta angle.

 The same argument with a target $-8-\Delta\mu$, $0<\Delta\mu\le1$, in place of $-9$ (so that $1$ becomes $\Delta\mu$ in $f$, and a zero gives $c_-d\le E-\Delta\mu$) shows more. At $E=\Delta\mu$ the form is positive, since $a+c_-d=\langle c_+(L_++8)+c_-(L_-+8)\rangle\ge0$ and $W<1$. Every zero below $E=7\Delta\mu/2$ is a downward crossing, since $c_-/\zeta\le\tfrac16$ there gives $Ef'\le-\tfrac7{12}\Delta\mu$. For the block-$(4,4)$ state $T_s$ of Eq.~\eqref{eq:Ts} and the normalized mixed-block quartet average in Eq.~\eqref{eq:vpp}, the shifted expectation is $\Delta\mu-Ew$, with $w=w_s$ and $w=w_{\rm qrt}$ respectively. It vanishes at $E=\Delta\mu/w\le7\Delta\mu/2$ and is strictly negative above this value. A negative quartet average implies a negative expectation on at least one of its states. At $s=1$, $\theta_{5d}=\pi$, where no doubly-BPS state exists, the trial wavefunction is $\phi_0$, with coordinates $(1,0)$ in the orthogonal endpoint basis of Sec.~\ref{sec:floorspace}. Since $\mathsf N_0=60\ln2-40<\tfrac{13}8$ and $8c_-\le\tfrac27E$ for $E\le\tfrac72$, its expectation $\Delta\mu\,\mathsf N_0+8c_--\tfrac34E$ is negative for $7\Delta\mu/2\le E\le\tfrac72$, and above $\tfrac72$ the trial $\phi_0+\phi_1$, with coordinates $(1,1)$, gives $\tfrac52\Delta\mu+6c_--E<0$. For all $s\ge1$ at both theta angles, each value of the lowest link eigenvalue between $-8$ and $-9$ is therefore reached exactly once, at an attraction between $\Delta\mu$ and $7\Delta\mu/2$. Together with the negative near-endpoint slopes of Sec.~\ref{sec:bps}, the radial index $\beta$ of the lowest channel falls monotonically from $1$ at the supersymmetric endpoint to $0$ at $E_c$, so any departure from the endpoint already requires a contact condition at the cone tip.

\subsection{Self-localization}
\label{sec:mechanism}

\begin{figure}[t]
\centering
\includegraphics[width=0.72\textwidth]{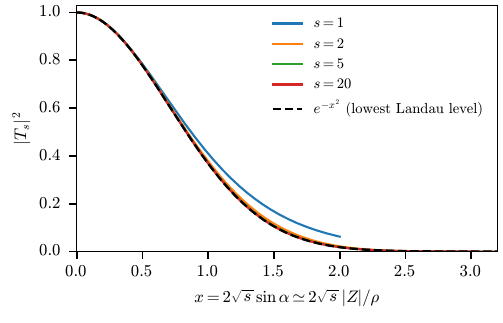}
\caption{Exact profile $|T_s|^2=\kappa^{2s-2}D^{-2s-2}$ of the doubly-BPS state, Eq.~\eqref{eq:LLL}, against $x=2\sqrt s\,\sin\alpha$ for $s=1,2,5,20$, with the lowest-Landau-level Gaussian $e^{-x^2}$ (dashed). Near the pinned direction $\sin\alpha\simeq|Z|/\rho$, so $x\in[0,2\sqrt s]$ measures the displacement of the centre in units of $\rho/(2\sqrt s)$ (matching the $1/e$ wedge $\alpha\lesssim 1/(2\sqrt s)$ of Fig.~\ref{fig:geometry}, while the root-mean-square spread lies at $x_{\rm rms}\simeq\sqrt2$). As $s$ grows the profiles collapse onto the Gaussian, which is the self-localization of the centre.}
\label{fig:profile}
\end{figure}

 The doubly-BPS state explains the instability. Near the pinned direction,
\begin{equation}
 |T_s|^2=\kappa^{2s-2}D^{-2s-2}=e^{-4s\sin^2\alpha}\bigl(1+O(s\sin^4\alpha)\bigr),
 \label{eq:LLL}
\end{equation}
and since $\sin^2\alpha\simeq|Z|^2/\rho^2$ this is the Gaussian $e^{-B|Z|^2/2}$ of the lowest Landau level in the field $B=8s/\rho^2$ of Sec.~\ref{sec:pauli}. Nothing holds the free centre in place, yet it is localized in the lowest Landau level of the Berry curvature of the line's colour. Unlike a flat lowest Landau level, which is infinitely degenerate, the floor on the compact Hopf base $S^4$ is finite-dimensional. The inertia dilation concentrates its $b=0$ member at the pinned pole, and the attraction selects that member. Each of the two orthogonal two-planes of $Z$ carries $\langle|z_a|^2\rangle=2/B=\rho^2/(4s)$, summing to $\langle|Z|^2\rangle=\rho^2/(2s)$, that is $\langle|X|^2\rangle=\rho^2/s$, because the Pauli energy of the centre fermions removes the Landau cost of doing so. A freely moving instanton therefore behaves like a pinned one, up to the residual spread of its centre, and its threshold approaches the pinned threshold $\zeta_{\rm pin}(s)$. This Landau-level description is the large-$s$ limit of the exact results of Sec.~\ref{sec:instability}, and Fig.~\ref{fig:profile} shows the exact profile collapsing onto the Gaussian. The cancellation of Sec.~\ref{sec:pauli} is exact for the highest colour weight, while $T_s$ carries total spin $s-1$, where the Pauli minimum is $-16(s+1)/\rho^2$. The two pictures agree at leading order in $1/s$.

Two features distinguish this localization from the Pauli--Coulomb problem \cite{FLL1986,LossYau1986,FrankLoss}, where the magnetic field is varied together with the wavefunction and its field energy enters the variational functional, and from the $g_m=2$ cancellation at a magnetic line. Here the line is electric, and the field of the Landau problem is the curvature of the colour connection on moduli space, which is the instanton's own field strength evaluated at the line. The field and the fermions that cancel its zero-point energy both belong to the instanton.

The spread fixes the first correction. The self-energy profile is $W=\kappa/D^2=1-3\sin^2\alpha+O(\sin^4\alpha)$, so on the localized state $\langle W\rangle=1-3\langle\sin^2\alpha\rangle+\dots=1-3/(2s)+O(s^{-2})$, in agreement with $w_s=2s/(2s+3)$. For $T_s$ itself the polar mean is exact, $\langle|Z|^2/\rho^2\rangle=\langle\tan^2\alpha\rangle=1/(2s-1)$ for $s\ge1$, so the centre stays within a distance of order $\rho/\sqrt s$ of the line. Self-localization refers to this polar profile at a fixed ratio of centre distance to size, not to a normalizable bound state of the scale-invariant free Hamiltonian. The trial expectation reaches $-9$ at $E=1/\langle W\rangle$, giving an upper bound on $E_c$. The uniform lower bound in Sec.~\ref{sec:uniform} fixes the same $3/(2s)$ coefficient for the full threshold. The product $4C_2(1-\zeta^2)\langle W\rangle$ thus plays the role that $(g^2s/2\pi)^2(1-\zeta^2)$ plays in the double-scaling limit recalled in Sec.~\ref{sec:intro}, with the exact Casimir in place of $s^2$ and no gauge coupling. The coefficient $3/2$ is the product of the quadratic fall-off of the self-energy away from the pinned direction, the $3$ in $W=1-3\sin^2\alpha+\dots$, and the mean-square angular spread $s\langle\sin^2\alpha\rangle=\tfrac12$ of a lowest Landau level of width $\rho/\sqrt{2s}$.\footnote{For a general centre-to-size inertia ratio $m=m_c/m_s$, the isotropic profile on the link is $W_m(\alpha)=\cos^2\alpha/(\cos^2\alpha+m^{-1}\sin^2\alpha)^2=1-(2m^{-1}-1)\sin^2\alpha+O(\sin^4\alpha)$, so $m=\tfrac12$ fixes the coefficient $2m^{-1}-1=3$. Notice that the pinned maximum criterion $m\le2$ of Sec.~\ref{sec:theoremL} is precisely the condition $2m^{-1}-1\ge0$ that keeps this quadratic slope nonnegative.} The pinned limit $\langle W\rangle\to1$ recovers $E_c=1$, the pinned bound~\eqref{eq:pinbound}.

Within the endpoint floor space, the same picture explains the subleading improvement of the variational bound. The common floor state $T_s=\phi_0+\phi_1$ has relative weight $\|\phi_1\|^2/\|T_s\|^2\sim1/(2s)$ in the size component. The orthogonal floor state $\widetilde T_s$ is supported predominantly there. At $E\simeq1$, the weight $c_-\simeq1/(16s^2)$ lifts this state by about $1/s$, and its weaker self-energy adds a splitting of about $3/(4s)$. The off-diagonal self-energy in this basis is of order $3/(4\sqrt{2}\,s^{3/2})$. Thus the ratio of the squared mixing matrix element to the splitting gives the estimate
\begin{equation}
 \Delta E_{\rm var}^{(2)}=-\frac{[3/(4\sqrt2\,s^{3/2})]^2}{7/(4s)}=-\frac{9}{56s^2}.
 \label{eq:main:c2phys}
\end{equation}
Section~\ref{sec:uniform} derives the coefficient for the full problem, and Sec.~\ref{sec:bps} expands the two variational upper bounds. In the full integer-spin channel, computations in a mapped Legendre basis give a rescaled difference close to $-9/56$ at $s=400$. Appendix~\ref{app:fb:chgap} proves that allowing excited link states changes either of these critical-channel thresholds by at most $O(s^{-3})$, on both sides of the channel. For half-integer spin the side $\zeta<0$, block $(5,4)$, uses a block-5 gap that grows linearly with $s$ and the Pauli lift of the block-4 set on its endpoint floor. The proof uses the linear growth of the excited-state gap on the round Hopf base and bounds both the Pauli and self-energy couplings to its complement.

The same structure holds globally, since the BPS states are the lowest Landau levels of the instanton on $S^4$ (Sec.~\ref{sec:bpsstates}) and $T_s$ is the global completion of the local Gaussian~\eqref{eq:LLL}, whose exact profile $\kappa^{2s-2}D^{-2s-2}$ replaces the naive extrapolation $D^{-4s}$. The next step is a lower bound on the full link. It excludes all but a finite list of large-spin competitors, for which the polar correction determines the first theta splitting.

\subsection{A uniform lower bound and the large-spin threshold}
\label{sec:uniform}

Variational trial states bound the threshold only from above; without a global lower bound holding across all nine fermion blocks and infinitely many angular-momentum channels, an unexamined channel could collapse earlier, invalidating both the Landau-level asymptotic coefficient $\tfrac32$ and the openness of the $\Omega$-trap localization windows in Sec.~\ref{sec:windows}. Let $\hat E(s)=1+\tfrac3{2s}-\tfrac3{4s^2}$, whose $O(s^{-2})$ coefficient $\tfrac34$ is chosen above the asymptotic deficit $c_*=\tfrac{33+3\sqrt3}{64}$ of the Schur test derived below, leaving room for the finite-spin checks. For every spin $s\ge1$, integer or half-integer, the single-set inequality $L(1)+9-\hat E(s)\,W\ge0$ holds on the full state space of the link operator (Appendix~\ref{app:un:uniform}), implying conformality whenever $4C_2(1-\zeta^2)\le\hat E(s)$ and hence
\begin{equation}
 E_c(s)\ \ge\ \hat E(s)=1+\frac3{2s}-\frac3{4s^2}.
 \label{eq:O2prime}
\end{equation}
Exchanging the two fermion sets carries the single-set bound to the other endpoint. Since $c_++c_-=1$, $L(\zeta)+9=c_+(L(1)+8)+c_-(L(-1)+8)+1-EW$. The exchange preserves $W$, hence $L(\pm1)+8\ge\hat EW-1$ gives $L(\zeta)+9\ge(\hat E-E)\,W\ge0$ for every $|\zeta|\le1$ whenever $E\le\hat E$.

Four physical and structural features turn the single-set inequality into a finite exact-arithmetic verification (Appendix~\ref{app:un:uniform}) without losing $O(1/s)$ precision. First, passing to a single active fermion set preserves the leading Landau-level waist. The tightest single-set floor profile $g_0\propto t(1+t)^{-1}(t+\tfrac12)^{-(2s+2)}$ shares $\langle\sin^2\alpha\rangle=\tfrac{1}{2s}+O(s^{-2})$ with the doubly-BPS profile $h_0$ of $T_s$, matching $\langle W\rangle=1-\tfrac{3}{2s}+O(s^{-2})$. Second, channels with $j_R\ge\tfrac12$ clear immediately via the vertical Casimir $4j_R(j_R+1)\ge3$ in Eq.~\eqref{eq:main:hopfsplit}, while in each $j_R=0$ channel containing a floor state $\chi$ (polar density $\varrho$) separated from excited states by an $O(1)$ gap $\mathsf G$, bounding the inverse in the exact Schur complement gives the sufficient test for $\mathcal H_A+9-\hat EW\ge0$
\begin{equation}
 1-\hat E\,\langle W\rangle_\varrho-k(\mathsf G)\,\mathrm{Var}_\varrho W\ \ge\ 0,\qquad k(\mathsf G)=\frac{\hat E^2}{\mathsf G+1-\hat E},
 \label{eq:main:schur}
\end{equation}
where mixing across the gap $\mathsf G$ lowers the energy by at most $k(\mathsf G)\,\mathrm{Var}_\varrho W$. For fixed $b$ this term is $O(s^{-2})$. Third, higher gauge angular momentum $b=2j_G$ multiplies the polar density by a centrifugal factor $t^b$, so apart from $g_0$ and $\sigma g_0$ (the $b=0$ dilation mode of the first block-$5$ floor family $\mathcal K_5^{(1)}$ of Appendix~\ref{app:fb:block5}, protected by gap $16$), every floor-state profile dominates $g_1$ in the likelihood-ratio order \cite{ShakedShanthikumar}; combining this ordering with $\mathrm{Var}_\varrho W\le w(1-w)$ removes all channel-label dependence for $s\ge\tfrac52$ and leaves only three explicit rational inequalities. Fourth, in the tightest channel (block~4, $b=0$), the $O(s^{-2})$ curvature of $\langle W\rangle_{g_0}$ and Feshbach mixing across the Lichnerowicz gap $\mathsf G_4^{\rm Lich}=12-4\sqrt3$ (Eq.~\eqref{eq:fb:gap4}) combine to the asymptotic deficit $c_*=\tfrac38+\tfrac{9}{8\mathsf G_4^{\rm Lich}}=\tfrac{33+3\sqrt3}{64}$ of the test (Friedrich's bound $8$ in place of $\mathsf G_4^{\rm Lich}$ would give $\tfrac{33}{64}$, Appendix~\ref{sec:un:scope}); choosing $\tfrac34>c_*$ builds in an asymptotic buffer $\tfrac{3(5-\sqrt3)}{64s^2}$ that clears both shifted-polynomial positivity for all $s\ge\tfrac52$ and exact rational interval checks for $1\le s\le5$ \cite{MooreKearfottCloud}, with no floating-point number entering any decision.

Combined with the instability bounds of Secs.~\ref{sec:instability} and~\ref{sec:bps}, the uniform lower bound brackets the threshold at every spin. At $\theta_{5d}=0$, for every integer $s\ge1$, $1+\tfrac3{2s}-\tfrac3{4s^2}\le E_c(s)\le E_{2\times2}(s)<1+\tfrac3{2s}$, and for every half-integer $s\ge\tfrac32$, $1+\tfrac3{2s}-\tfrac3{4s^2}\le E_c(s)\le E_{3\times3}(s)<1/w_{\rm crit}(s)\le1/w_{\rm qrt}(s)$ (the inequalities against the one-state bounds are strict because $T_s$ and $T_s'$ are not null vectors of the pencil~\eqref{eq:E2x2} at the one-state thresholds). At $\theta_{5d}=\pi$ the two families exchange, with the bound $E_{2\times2}^{(4,5)}(1)$ at $s=1$. Since $E_{3\times3}(s)=1+3/(2s)-9/(56s^2)+O(s^{-3})$ (Sec.~\ref{sec:bps}), both brackets close at order $1/s$, proving the large-spin threshold formula
\begin{equation}
 4s(s+1)\bigl(1-\zeta_c(s)^2\bigr)=1+\frac3{2s}+O(s^{-2}),\qquad
 \lim_{s\to\infty}s\,\bigl(E_c(s)-1\bigr)=\frac32 ,
 \label{eq:lawcor}
\end{equation}
with a computer-assisted lower side and an analytic upper side. Appendix~\ref{app:fb:chgap} also proves Eq.~\eqref{eq:lawcor} analytically as $s\to\infty$, with remainder constants that are not made explicit, and it resolves the next two orders. Exchanging the two fermion sets maps $\zeta$ to $-\zeta$, so the threshold depends only on $|\zeta|$, and the competitors are organized by \emph{sides}, each an ordered pair of fermion blocks, the first active at $\zeta=1$, together with a channel $(j_L,0,j_G)$. Only sides whose endpoint floor contains the narrowest profile $g_0$ can compete at order $s^{-2}$. There are seventeen of them (Table~\ref{tab:competing-floors}), and on all but five the Pauli lift of the second fermion set grows like $s$ and raises the onset by at least $1/s$. The five survivors are the sides that carry a doubly-BPS state, namely block $(4,4)$ in channel $(s-1,0,0)$, the mixed blocks $(4,5)$ and $(5,4)$ in channel $(s-1,0,\tfrac12)$, and block $(5,5)$ in channels $(s-1,0,0)$ and $(s-1,0,1)$. In a given theta sector Eq.~\eqref{eq:physical} keeps either the three with integer $j_G$ or the two with half-integer $j_G$. On each survivor a two-set estimate that keeps the Pauli lift of the second fermion set gives the coefficient. The narrow floor state has the inverse weight $1+\frac3{2s}-\frac3{8s^2}$, its second-set Pauli lift costs $\frac1{2s^2}$, and mixing with its partners returns $\frac2{7s^2}$, so that $-\frac38+\frac12-\frac27=-\frac9{56}$ and
\begin{equation}
 E_c(s)=1+\frac3{2s}-\frac9{56s^2}+O(s^{-3}).
 \label{eq:lawcor2}
\end{equation}
In the basis $T_s,\tilde T_s$ of Sec.~\ref{sec:mechanism} the same coefficient is the mixing of the doubly-BPS state with its orthogonal partner. The narrow-state decomposition separates the width correction, the Pauli cost and the mixing, and it is the form that extends to every competing side. This order does not distinguish the two theta sectors. They separate at the next order, where the narrow state can also make a virtual excitation into the first excited polar state above the floor. That gap grows like $16s$ while the variance of $W$ on the narrow state is $\frac9{8s^2}$, so the excitation lowers every competing side by the same amount, $\frac9{128s^3}$, and the ordering at order $s^{-3}$ is decided by the floor pencils alone.

\begin{proposition}[Cubic order and the discrete theta angle]\label{prop:cubic}
At either theta angle let $p$ be the parity of $j_G$ selected by $j_G\equiv s+\theta_{5d}/(2\pi)\pmod{1}$. For sufficiently large spin the first threshold lies in block $(4,4)$, channel $(s-1,0,0)$, when $p$ is an integer, and in the mixed blocks, channel $(s-1,0,\tfrac12)$, when $p$ is a half-integer. In both cases $E_c^p=E_{\rm var}^p-9/(128s^3)+O(s^{-7/2})$, where $E_{\rm var}^p$ is the pencil root $E_{2\times2}$ or $E_{3\times3}$ of Eq.~\eqref{eq:E2x2}, so that
\begin{equation}
E_c^{p}(s)=1+\frac{3}{2s}-\frac{9}{56s^2}
  +\frac{1}{43904s^3}
    \begin{cases}-3447,&p\text{ integer},\\-1143,&p\text{ half-integer},\end{cases}
  +O(s^{-7/2}).
\label{eq:lawcor3}
\end{equation}
\end{proposition}

 At fixed $s$, changing $\theta_{5d}$ switches the allowed gauge angular momentum $j_G$ by a half-unit. The $(4,4)$ state can localize at the pole with $j_G=0$. In the mixed block the spectator carries $j_G=\tfrac12$, so the polar profile remains equally narrow to leading order. The two endpoint spaces nevertheless have different mixing partners, with two floor vectors in $(4,4)$ and three in $(4,5)$. Appendix~\ref{app:fb:cubic} gives the proof. The virtual polar excitation shifts all five sides equally. The first distinction between $j_G$ parities consequently comes from the variational $s^{-3}$ coefficients, $18/(343s^3)+O(s^{-7/2})$, which moves $|\zeta_c|$ by $-9/(1372s^5)+O(s^{-11/2})$. The discrete angle thus changes the short-distance phase in a very narrow window. At the midpoint between the two thresholds the two theories have $\beta^2=\mp9/(343s^3)$ to leading order, since the slope $-\partial_E\mu$ of both critical branches tends to one, and even one cycle of the flow of the contact coupling in Sec.~\ref{sec:radial}, $\ln(R_{\rm out}/R_0)=\pi/\nu=(7\pi\sqrt7/3)\,s^{3/2}$, requires a separation between a core of size $R_0$ and an outer scale $R_{\rm out}$ that grows exponentially in $s^{3/2}$.\footnote{The next term of the splitting, $1266/(16807s^4)$, follows from the perturbation series of the two critical channels in powers of $1/s$. Against the thresholds computed directly, it reduces the error of the leading term at $s=\tfrac52$ from about $38\%$ to $2\%$. In the two critical channels, for a general target $-8-\Delta\mu$, $0<\Delta\mu\le1$, of the lowest link eigenvalue, each attraction is $\Delta\mu\,[1+3/(2s)-9/(56s^2)+O(s^{-3})]$, and the same series gives their difference as $\Delta\mu\,[18/(343s^3)+1266/(16807s^4)+9(91552-108045\,\Delta\mu)/(105413504\,s^5)+O(s^{-6})]$, both as formal series in $1/s$. Through fourth order in $1/s$ the curves $E(\Delta\mu)$ of the two theta angles are therefore rescaled copies of each other, and the fifth-order coefficient only records the first, small departure from that. A direct computation of the two critical channels at $\Delta\mu=\tfrac12$ and $1$, for spins from $6$ to $32$, confirms the expansion (Appendix~\ref{app:numerics}). It reproduces $18/343$ and $1266/16807$ to about $10^{-4}$ and $10^{-3}$ of their values, and the scaled differences at the two targets agree through order $s^{-4}$ and separate at order $s^{-5}$ by $405/(87808\,s^5)$, the value implied by the term $-108045\,\Delta\mu$, to better than $1\%$.} The two values belong to the two theories $\theta_{5d}=0$ and $\pi$ of Sec.~\ref{sec:theory}, and the channel selection holds beyond a spin that the non-explicit remainders leave undetermined.

\textit{Resummation.} Beyond the orders controlled in Eq.~\eqref{eq:lawcor3}, perturbation theory in the critical channel generates a formal expansion in $1/s$. Near the pinned direction, in the variable $4s\sigma$, the endpoint operator of the critical channel becomes a Landau problem whose levels are spaced by $16s$, the cyclotron energy $2B$ in link units, so each order of the series in $1/s$ is a finite computation in exact arithmetic. In the critical channel of block $(4,4)$ the next two coefficients are $486261/2151296$ and $-635135379/3373232128$. We extrapolate to finite spin by Borel--Pad\'e approximation \cite{AnicetoBasarSchiappa}, replacing its Borel transform $\sum_{n\ge1}E_n\mathsf t^{n-1}/(n-1)!$ by a Pad\'e approximant and integrating against $e^{-s\mathsf t}$ along the positive axis. The $[6/5]$ approximant built from the first twelve coefficients has no pole on that axis and reproduces the thresholds of this channel, computed to twelve digits (Appendix~\ref{app:numerics}), with absolute errors of about $2\times10^{-6}$ at $s=1$, $8\times10^{-8}$ at $s=2$ and $2\times10^{-9}$ at $s=3$. At $s=1$, the absolute errors of the $[5/5]$ and $[4/6]$ approximants, both regular on the integration path, are $1.08\times10^{-4}$ and $1.09\times10^{-4}$, respectively. Here the expansion parameter $1/s$ is unity. The smallest absolute error among the partial sums tested is $0.0164$, after $s^{-2}$. The twelfth-order sum gives $2.08434$ against the independently computed channel threshold $2.35574$, an error of $11.5\%$. The spin-one phase comparison therefore uses the independent spectral bounds of Proposition~\ref{prop:theta-window}. All three approximants have a complex-conjugate pair of poles near $\mathsf t\approx\pm5.2\ii$, slightly to the left of the imaginary axis.\footnote{The $[5/6]$ approximant has an uncancelled pole at $\mathsf t\simeq0.682673$ on the integration path and would require a separate prescription.} The pole locations and the nonmonotonic truncation errors are finite-order observations. Proving divergence or Borel summability of the full threshold series would require large-order control.  For the shape response of the bosonic line, whose Borel transform is known exactly, Appendix~\ref{app:bosonic} locates its singularities on the imaginary axis.

\textit{The merger of the radial weights.} Near the threshold the weights $k_\pm=\tfrac12(1\pm\beta)$ of Sec.~\ref{sec:radial} coalesce at $\tfrac12$. On the conformal side they describe two scale-invariant boundary conditions. Across the threshold they become a complex-conjugate pair, and a real self-adjoint boundary condition runs periodically with the logarithm of the short-distance cutoff. For a simple critical channel $\beta^2=\mu_c'(E_c-E)+O((E_c-E)^2)$, with $\mu_c'=-\dd\mu/\dd E$ at $E_c$ the mean of $W$ in the critical state corrected by the derivative of the Pauli weights, so that $k_\pm=\tfrac12\pm[2C_2\zeta_c\mu_c'(|\zeta|-\zeta_c)]^{1/2}+O(|\zeta|-\zeta_c)$. The coefficient of $(|\zeta|-\zeta_c)^{1/2}$ is $1.2033$ at $s=1$ and $2.6060$ at $s=2$ in the critical channel at $\theta_{5d}=0$, and it grows as $\sqrt2\,s\,(1-1/(4s))$ at large spin, where $\mu_c'\to1-3/(2s)$. The fixed-point merger of Sec.~\ref{sec:radial} thus proceeds at a rate set by the spin of the line.

\begin{figure}[t]
\centering
\includegraphics[width=\textwidth]{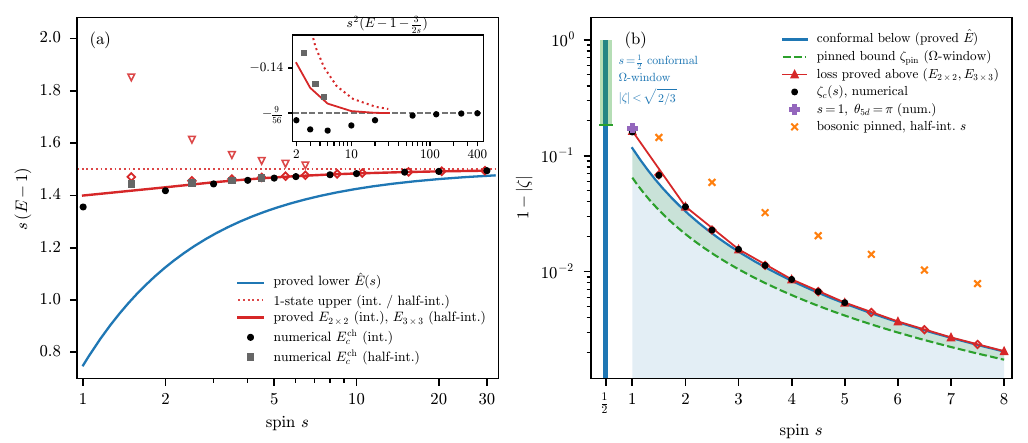}\par
\includegraphics[width=0.83\textwidth]{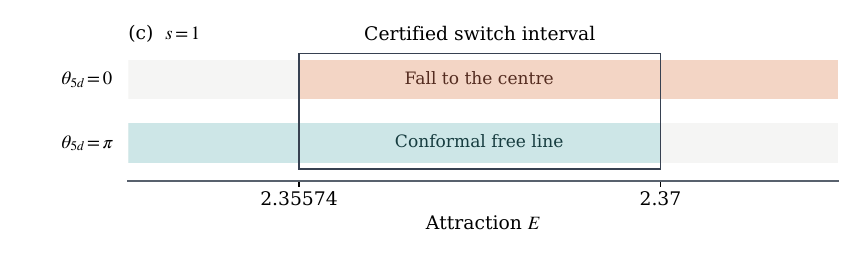}
\caption{Bounds and numerical thresholds. (a)~The curves show $s(E-1)$ for the proved lower bound $\hat E$ and the one-state and endpoint-floor upper bounds. Filled symbols show Galerkin values in the candidate channels at $\theta_{5d}=0$. The inset shows the rescaled $s^{-2}$ residual for the variational bounds and critical-channel numerics. The integer numerical series reaches $s=400$, while the half-integer numerical series ends at $s=9/2$. The coefficient $-9/56$ is common to both physical sectors. Eq.~\eqref{eq:lawcor3} resolves their first difference at the next order. (b)~The vertical coordinate is $1-|\zeta|$. For $s\ge1$, the free model is conformal below the blue boundary $E=\hat E$ and supercritical above the corresponding red variational bound for $\theta_{5d}=0$. The shaded green strip marks the explicit part $1<E<\hat E$ of the trap window, which extends to the threshold $E_c$, and the dashed line $E=1$ is its proved upper edge in $|\zeta|$, sector by sector. The dashed point itself has finitely many negative levels per sector. At $s=\tfrac12$ the blue bar denotes free conformality and the green segment denotes the proved trap towers for $|\zeta|<\sqrt{2/3}$. The purple symbol is the numerical mixed-channel threshold at $s=1$, $\theta_{5d}=\pi$. Orange crosses are the bosonic pinned half-integer thresholds at $\theta_{5d}=0$. (c)~Certified spin-one window from Eq.~\eqref{eq:theta-window}, with $E=8(1-|\zeta|^2)$. The outlined interval lies inside the supercritical range at $\theta_{5d}=0$ and inside the free-conformal range at $\theta_{5d}=\pi$. The light grey segments are undecided by the two bounds plotted in this panel.}
\label{fig:law}
\end{figure}

Table~\ref{tab:thresholds} collects the brackets at $\theta_{5d}=0$ for $1\le s\le5$. Figure~\ref{fig:law} compares these bounds and numerical thresholds, with the $s=\tfrac12$ result and the exceptional $s=1$ point at $\theta_{5d}=\pi$. For $s=1$ at $\theta_{5d}=0$ the analytic bracket is $0.83669<\zeta_c\le5\sqrt2/8$, and the certified crossing of Appendix~\ref{app:radial-certificate} gives the stronger lower bound $\zeta_c>8399601/10^7=0.8399601$. The analytic lower end comes from the endpoint-mixing bound $E_{2\times2}(1)$ of Sec.~\ref{sec:bps} (the single trial state gives $\sqrt{11}/4$), and the upper end already improves the pinned bound $\sqrt{7/8}$. In the three common normalizations of the colour charge, $s(s+1)$, $(s+\tfrac12)^2$ and $s^2$, the expansion reads $4s(s+1)(1-\zeta_c^2)=1+3/(2s)+O(s^{-2})$, equivalently $4(s+\tfrac12)^2(1-\zeta_c^2)=1+3/(2s)+O(s^{-2})$ and $4s^2(1-\zeta_c^2)=1+1/(2s)+O(s^{-2})$. Relative to the classical colour $s^2$ of that limit, the Casimir contributes $-1/s$ and self-localization $+3/(2s)$. In the scalar coupling the threshold sits at $1-|\zeta_c|=1/(8s^2)+O(s^{-3})$, whereas Proposition~\ref{thm:bosphase} keeps both bosonic sectors essentially self-adjoint for $E<16s+24$. Throughout $s^{-2}\ll1-|\zeta|\ll s^{-1}$, where $1\ll E\ll s$, the line with fermion zero modes has therefore lost conformality while the same line with bosonic moduli alone still has a unique Hamiltonian at either theta angle. The Pauli coupling lowers the scale of the instability from $16s$ to $1$.

\begin{table}[!htbp]
\centering\small
\setlength{\tabcolsep}{4.5pt}
\begin{tabular}{@{}lcccccc@{}}
\toprule
& \multicolumn{4}{c}{$E_c$} & & \\
\cmidrule(lr){2-5}
$s$ & lower $\hat E$ & numerical & sharpened upper & 1-state upper & $\zeta_c$ (numerical) & $\zeta_{\rm pin}$\\
\midrule
$1$ & $1.75000$ & $2.35574^\dagger$ & $2.39954$ & $2.50000$ & $0.83996^\dagger$ & $0.93542$\\
$\tfrac32$ & $1.66666$ & $1.96276$ & $1.98053$ & $2.23443$ & $0.93228$ & $0.96610$\\
$2$ & $1.56250$ & $1.70897$ & $1.71554$ & $1.75000$ & $0.96374$ & $0.97895$\\
$\tfrac52$ & $1.48000$ & $1.57866$ & $1.58239$ & $1.64524$ & $0.97719$ & $0.98562$\\
$3$ & $1.41666$ & $1.48130$ & $1.48342$ & $1.50000$ & $0.98445$ & $0.98953$\\
$\tfrac72$ & $1.36734$ & $1.41652$ & $1.41793$ & $1.44424$ & $0.98869$ & $0.99204$\\
$4$ & $1.32812$ & $1.36443$ & $1.36538$ & $1.37500$ & $0.99144$ & $0.99374$\\
$\tfrac92$ & $1.29629$ & $1.32575$ & $1.32643$ & $1.34048$ & $0.99328$ & $0.99494$\\
$5$ & $1.27000$ & $1.29325$ & $1.29375$ & $1.30000$ & $0.99460$ & $0.99583$\\
\bottomrule
\end{tabular}
\caption{Proved brackets for $E_c(s)=4s(s+1)(1-\zeta_c^2)$ and corresponding critical-channel estimates at $\theta_{5d}=0$ for $1\le s\le5$. The lower bound $\hat E(s)$ in Eq.~\eqref{eq:O2prime} is rounded down. Upper bounds, including the pinned coupling, are rounded up. The sharpened upper bound is the endpoint-pencil root $E_{2\times2}(s)$ for integer spin and $E_{3\times3}(s)$ for half-integer spin. The one-state column uses $1/w_s$ at integer spin and the quartet bound $1/w_{\rm qrt}$ at half-integer spin. Their exact forms are given in Eqs.~\eqref{eq:vs} and~\eqref{eq:vpp} and Table~\ref{tab:vpp}. The numerical columns list converged critical-channel Galerkin thresholds (compared with competing channels in the numerically delimited boxes for $s\le3$ and $j_R=0$ scans for integer $s\le6$, Appendix~\ref{app:numerics}). The last column gives the pinned bound $\zeta_{\rm pin}=\sqrt{1-1/(4C_2)}$. Entries marked $\dagger$ are certified channel values (Appendix~\ref{app:radial-certificate}).}
\label{tab:thresholds}
\end{table}

The two thresholds can also be separated at spin one, where the large-spin expansion gives no error bound. The comparison uses the certified crossing of a single integer-$j_G$ channel to bound the onset at $\theta_{5d}=0$ from above. At $\theta_{5d}=\pi$ the positivity test must cover the whole physical link. The endpoint gap reduces it to eight angular channels, including the effect of polar states outside their endpoint floors.

\begin{proposition}[A spin-one window controlled by the discrete theta angle]\label{prop:theta-window}
At $s=1$, write $E_{c,0}$ and $E_{c,\pi}$ for the global critical attractions at the two discrete theta angles. Then
\begin{equation}
 E_{c,0}<2.35574<\frac{237}{100}<E_{c,\pi}.
 \label{eq:theta-window}
\end{equation}
Throughout $2.35574\le E\le237/100$, the free quadratic form is unbounded below at $\theta_{5d}=0$. At $\theta_{5d}=\pi$ it admits a nonnegative scale-invariant Friedrichs realization.
\end{proposition}
\emph{Proof.} At $\theta_{5d}=0$ the critical-channel crossing in Appendix~\ref{app:radial-certificate} implies $E_{c,0}\le8[1-(8399601/10^7)^2]=2.3557362\ldots$. At $\theta_{5d}=\pi$ the global positivity estimate of Appendix~\ref{app:fb:block45} places $237/100$ strictly below the onset. Throughout the stated interval, the lowest angular eigenvalue therefore lies below the Hardy value $-9$ at $\theta_{5d}=0$ and above it at $\theta_{5d}=\pi$. The radial Hamiltonian is unbounded below in the first theory and has a nonnegative Friedrichs realization in the second.\hfill\ensuremath{\square}

At $\theta_{5d}=\pi$ that realization is not the only one. The vector $(1,1)$ of Eq.~\eqref{eq:E2x2}, of norm squared $\tfrac52$, gives $\langle L+8\rangle=(6c_-(E)-E)/\tfrac52<0$ throughout the window, so the lowest channel there already needs a contact condition, and the two theta angles differ between fall to the centre and a conformal line with a free boundary parameter.

Figure~\ref{fig:law}(c) illustrates the change at fixed $|\zeta|$. The discrete angle changes the physical colour parity while leaving the local differential operator unchanged. At $\theta_{5d}=\pi$ the first endpoint-floor state gives the stronger inequality $\mathsf N_0\langle L+8\rangle=8c_--3E/4<0$ for every $E>0$ on the positive-$\zeta$ branch, and exchange of the fermion sets gives the negative branch. The conformal realization in the switch interval uses the specified boundary condition. Section~\ref{sec:windows} uses the same spin-one window to compare the trapped spectra.

\section{Fundamental-line conformality and the near-BPS regime}
\label{sec:bps}

 At $\zeta=\pm1$ the floor $\mu=-8$ places every nontrivial spin $s\ge\tfrac12$ at $\beta=1$, the boundary of essential self-adjointness where the dynamics is unique and insensitive to short-distance data. For the fundamental line no state reaches the floor of both supersymmetric halves, and we show below that its link spectrum stays above $-8$ for $|\zeta|<1$ and reaches $-8$ only at the endpoints, which keeps its realization unique. For $s\ge1$, moving $|\zeta|<1$ raises two physical questions that the single-state variational bounds of Sec.~\ref{sec:instability} do not answer. First, even inside a single irreducible $G$-symmetry channel (where symmetry forbids inter-channel mixing), how many independent radial branches cross below $-8$ into the multi-extension strip $0\le\beta<1$, forcing any ultraviolet resolution of the small-instanton core to pick a $U(n_{\rm ch})$ matrix of boundary conditions before conformality is lost at $\beta=0$? Second, does level repulsion inside the degenerate $\zeta=1$ BPS floor subspace explain the term $-\tfrac{9}{56s^2}$ that the variational thresholds of both spin families share, and what happens at $\theta_{5d}=\pi$, $s=1$ where the decisive channel has no doubly-BPS state?

\paragraph*{The fundamental line.}\phantomsection\label{sec:fundamental}
For the fundamental line the outcome is decided at every coupling, and the reason is a frustration among three spins. At $s=\tfrac12$ the colour is itself a spin $\tfrac12$, and the Pauli term of each fermion set lowers the energy only when the spin of that set pairs with the colour into a singlet. A spin $\tfrac12$ cannot form a singlet with two partners at once, so the two Pauli energies cannot reach their minima together. Recoupling the three spins bounds their sum below by $-(8+16\sqrt{1-E/4})/D^2$, a bound that Appendix~\ref{app:fb:s12} extends to fermion sets carrying no spin, whereas a single set at an endpoint reaches $-24/D^2$ and cancels exactly the magnetic barrier $-\Delta_{S^7}^{\mathcal A}+8\ge24/D^2$ of the colour-twisted scalar Laplacian on the link, which two explicit positive solutions of the scalar link equation establish (Appendix~\ref{app:fb:s12}). At zero Hopf-fibre spin the two estimates combine into
\begin{equation}
 L(\zeta)+8\ \ge\ \frac{E}{D}.
 \label{eq:s12bound}
\end{equation}
The attraction raises this bound, since as $|\zeta|$ decreases the Pauli bound improves by $2E/D^2$ while the self-energy costs only $E\kappa/D^2$. Nonzero Hopf-fibre spin is controlled by the vertical Casimir. The lowest link eigenvalue therefore lies strictly above $-8$ for $|\zeta|<1$ and equals $-8$ only at the supersymmetric endpoints, so every radial channel is limit-point at the origin and the fundamental line has a unique self-adjoint realization at every coupling and in both theta sectors. Near the endpoints the lowest branch leaves $-8$ with the slope $(3-4\ln2)^{-1}$, because the Pauli operator of the second set has no matrix element on the single block-$4$ floor state at $s=\tfrac12$.\footnote{On that floor state the Pauli operator of the active set equals $-24/D^2$, so first-order perturbation theory gives the slope $6\langle D^{-1}\rangle$ in $1-|\zeta|$. In $t=\tan^2\alpha$ the floor density is proportional to $t/[(1+t)(t+\tfrac12)^3]$, whose integral is $6-8\ln2$ and whose $D^{-1}$-weighted integral is $\tfrac13$.}

\paragraph*{The endpoint floor subspace for $s\ge1$.}\phantomsection\label{sec:floorspace}
For every spin $s\ge1$ (integer $s\ge1$ in block $(4,4)$ channel $(s-1,0,0)$, half-integer $s\ge\tfrac32$ in block $(4,5)$ channel $(s-1,0,\tfrac12)$, and the $\theta_{5d}=\pi$ edge case $s=1$ in block $(4,5)$ channel $(0,0,\tfrac12)$), the $\zeta=1$ floor subspace $\ker(L_++8)$ in the indicated channel has the same tridiagonal graph-Laplacian structure. Under $SU(2)_L\times SU(2)_G$, tensoring the complete single-set block-$4$ floor space $\mathcal K_4=\bigoplus_{b=0}^{2s-1}(s-\tfrac12-\tfrac b2,\tfrac b2)$ (Appendix~\ref{app:fb:twistor}) with the flat spectator $\mathbb F_4=(\tfrac12,0)\oplus(0,\tfrac12)$ hits $(s-1,0)$ with exact multiplicity $2$ ($b=0,1$), while tensoring with $\mathbb F_5=(0,0)\oplus(\tfrac12,\tfrac12)$ hits $(s-1,\tfrac12)$ with exact multiplicity $3$ for all $s\ge\tfrac32$ ($b=0,1,2$) and multiplicity $2$ at $s=1$ (where $b=2$ is absent). In block $(4,4)$, decomposing $T_s=A\,(\hat w\otimes\hat w)$ ($A=\kappa^{s-1}D^{-(s+1)}$) along the spectator polarizations $e_Z,e_y$ gives the two orthogonal floor states explicitly,
\begin{equation}
 \phi_0=A\,(\kappa,\,0,\,-\sqrt{\sigma\kappa},\,0),\qquad
 \phi_1=A\,(0,\,-\sqrt{\sigma\kappa},\,0,\,\sigma),
 \label{eq:bpsbasis}
\end{equation}
which reduce to $\mathcal F_Z,\mathcal F_y$ at $s=1$ and satisfy $T_s=\phi_0+\phi_1$. Substituting either vector into Eq.~\eqref{eq:4x4} gives $(L_s(1)+8)\phi_b=0$ in all four components for every allowed $s$. In block $(4,5)$ at $s\ge\tfrac32$, the corresponding orthogonal basis is $\phi_0=\psi_0$, $\phi_1=\psi_1$ and $\phi_2=\sqrt{\vartheta_s}\,\psi_2$ (with $\vartheta_s=(2s-1)/(4s-4)$ from Eq.~\eqref{eq:fb:rhocrit}), so the critical doubly-BPS state is $T'_s=\phi_0+\phi_1+\phi_2$ with profile $\varrho_{\rm crit}=g_0+g_1+\vartheta_s g_2$ (Appendix~\ref{app:fb:block45}). At $s=1$ in the theta-$\pi$ channel the basis contains only $\phi_0,\phi_1$ and has no doubly-BPS state.

Because distinct $\mathcal K_4$ types carry profiles $g_b(t)=t^{b+1}(1+t)^{-1}(t+\tfrac12)^{-(2s+2)}$ and are mutually orthogonal under both $L^2$ and the $SO(4)$-invariant weight $W(\alpha)$, normalizing uniformly by $2(2s+1)(2s+3)2^{-(2s+1)}$ puts the Gram matrix $\mathsf N$, the self-energy matrix $\mathsf W$, and the Pauli perturbation matrix $\mathsf M$ of $L_-+8=P^{(-)}-P^{(+)}$ on $\ker(L_++8)$ into one single family across both blocks (Appendix~\ref{app:fb:block45}),
\begin{equation}
\begin{gathered}
 \mathsf N=\operatorname{diag}\bigl(\mathsf N_0,\,\mathsf N_1\bigl[,\,\mathsf N_2\bigr]\bigr),\qquad
 \mathsf W=\operatorname{diag}\!\Bigl(\frac{2s+1}{2s+2},\,\frac{1}{2s+2}\Bigl[,\,\frac{3\vartheta_s}{8s(s+1)}\Bigr]\Bigr),\\
 \mathsf M^{(4,4)}=8\,(\hat e_0-\hat e_1)(\hat e_0-\hat e_1)^{\rm T},\qquad
 \mathsf M^{(4,5)}(s)=\mathsf M^{(4,4)}+\frac{6}{s}\,(\hat e_1-\hat e_2)(\hat e_1-\hat e_2)^{\rm T},
\end{gathered}
 \label{eq:bpsmoments}
\end{equation}
where $\hat e_0,\hat e_1,\hat e_2$ are the standard basis vectors in the order of the basis $\phi_0,\phi_1,\phi_2$ and the bracketed third entry appears in block $(4,5)$ for $s\ge\tfrac32$ (with $\mathsf M^{(4,4)}$ embedded in the top-left $2\times2$ block) and is absent in block $(4,4)$ and at $s=1,\theta_{5d}=\pi$ (truncating $\mathsf M^{(4,5)}(1)$ to its top-left $2\times2$ block $\bigl(\begin{smallmatrix}8&-8\\-8&14\end{smallmatrix}\bigr)$). Explicitly, $\mathsf N_b=2(2s+1)(2s+3)n_b(s)$ for $b=0,1$ and $\mathsf N_2=\vartheta_s\cdot 2(2s+1)(2s+3)n_2(s)$ with $n_b(s)=2^{-(2s+1)}\int_0^\infty g_b(t)\,\dd t$. For integer $s\ge1$, changing variable to $v=(1-\sigma)/(1+\sigma)=1/(1+2t)$ evaluates $\mathsf N_0$ and $\mathsf N_1=\frac{2s+3}{2s}-\mathsf N_0$ in elementary closed form,
\begin{equation}
 \mathsf N_0(s)=(2s+1)(2s+3)\!\left[4\!\left(\ln2-\sum_{k=1}^{2s}\frac{(-1)^{k-1}}{k}\right)-\frac{2}{2s+1}\right].
 \label{eq:allslope}
\end{equation}
Expanding $L(\zeta)+8=(L_++8)+\tfrac{1-\zeta}{2}(L_-+8)-8C_2(1-\zeta)W+O((1-\zeta)^2)$ on $\ker(L_++8)$ determines all near-BPS slopes $\mu^{(k)}(\zeta)=-8+\mu_1^{(k)}(1-\zeta)+O((1-\zeta)^2)$ from the generalized eigenvalue equation
\begin{equation}
 \det\!\bigl[\tfrac12\mathsf M-8C_2\mathsf W-\mu_1\mathsf N\bigr]=0,
 \label{eq:secular-slope}
\end{equation}
while applying the min--max principle to $\ker(L_++8)$ at arbitrary $|\zeta|<1$ gives the exact proved variational upper bounds $E_c(s)\le E_c^{\rm ch}(s)\le E_{\rm var}(s)$ ($E_{2\times2}$ in block $(4,4)$, $E_{3\times3}$ in block $(4,5)$, and $E_{2\times2}^{(4,5)}(1)$ at $\theta_{5d}=\pi,s=1$) from the first loss of positivity along the compressed matrix pencil, at a root of
\begin{equation}
 \det\!\bigl[c_-(E)\,\mathsf M+\mathsf N-E\,\mathsf W\bigr]=0,\qquad
 c_-(E)=\frac{1-\sqrt{1-E/(4C_2)}}{2}.
 \label{eq:E2x2}
\end{equation}

\begin{figure}[t]
\centering
\includegraphics[width=\textwidth]{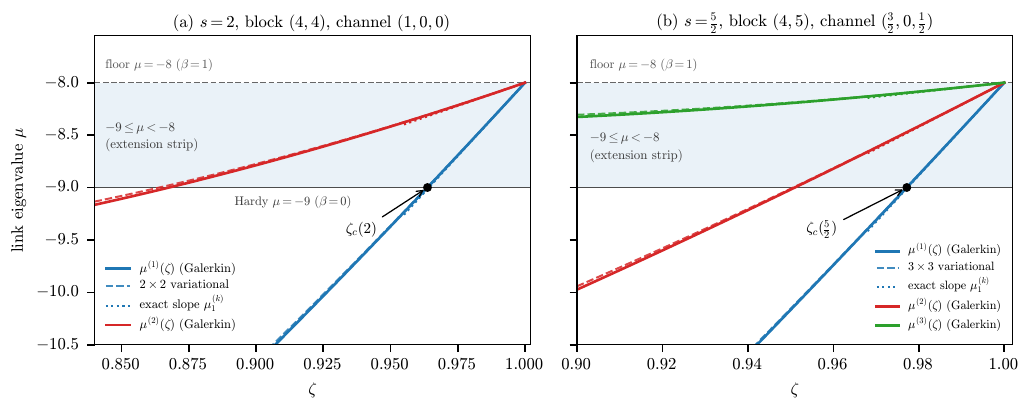}
\caption{Near-BPS spectral flow in the designated critical channels. (a)~Integer spin $s=2$ in block $(4,4)$, channel $(1,0,0)$. (b)~Half-integer spin $s=\tfrac52$ in block $(4,5)$, channel $(\tfrac32,0,\tfrac12)$. The exact slopes give two and three descending branches, respectively, near $\zeta=1$. The Galerkin curves (solid) track them in the extension strip $-9\le\mu<-8$, giving a $U(p)$ radial extension family in the indicated ordered block, with $p\ge2$ or $p\ge3$, when spectator symmetries are preserved. The dashed curves solve the endpoint-floor variational problems of Eq.~\eqref{eq:E2x2}. The dotted curves are the tangents from Eq.~\eqref{eq:secular-slope}. The arrows mark estimated channel crossings of the Hardy value at $\zeta_c^{\rm ch}\simeq0.96374$ and $0.97719$, respectively, at $\theta_{5d}=0$.}
\label{fig:nearbps}
\end{figure}

At $s=1$ in block $(4,4)$, where $\mathsf N_0=60\ln2-40$, $\mathsf N_1=\tfrac{85}{2}-60\ln2$ and the $(2,2)$ entry of $\tfrac12\mathsf M^{(4,4)}-8C_2\mathsf W$ vanishes identically, Eq.~\eqref{eq:secular-slope} collapses to closed-form slopes for both branches,
\begin{equation}
 \mu_1^{(1,2)}(1)=-\frac{1\pm(17-24\ln2)^{-1/2}}{15\ln2-10}\simeq-6.68774,\ +1.65259\,.
 \label{eq:slope}
\end{equation}
\paragraph*{Extension families and the second-order term.}\phantomsection\label{sec:consequences}
For general $s\ge1$, Eqs.~\eqref{eq:bpsmoments}--\eqref{eq:E2x2} produce three immediate physical and quantitative consequences, summarized in Table~\ref{tab:nearbps} and Fig.~\ref{fig:nearbps}.

\begin{enumerate}
\item \emph{Single-channel growth of the self-adjoint extension space to $U(2)$ and $U(3)$.} While the total multiplicity of diving channels across all $G$-sectors already grows as $O(s^3)$ from Eq.~\eqref{eq:main:dbps}, every strictly negative root $\mu_1^{(k)}(s)<0$ of Eq.~\eqref{eq:secular-slope} sends an independent radial branch into $-9\le\mu<-8$ within the single irreducible critical channel, enlarging its symmetry-preserving extension family. In block $(4,4)$, $\det(\tfrac12\mathsf M^{(4,4)}-8C_2\mathsf W)$ is proportional to $s(2s^2-s-2)$. It is negative at $s=1$ (one diving branch, $U(1)$), whereas for every integer $s\ge2$ both branches dive simultaneously ($\ge U(2)$). In block $(4,5)$, two branches already dive at $s=\tfrac32$ ($\ge U(2)$), and for every half-integer $s\ge\tfrac52$ all three dive simultaneously, enlarging the single-channel extension family to at least $U(3)$ (Fig.~\ref{fig:nearbps}).
\item \emph{The common term $-\tfrac{9}{56s^2}$ of the two variational bounds.} The first two entries of $\mathsf N$ and $\mathsf W$ agree in both blocks. The Pauli matrix adds an edge of weight $6/s$ to the third node, with norm $\mathsf N_2=O(s^{-2})$. Expanding Eq.~\eqref{eq:E2x2} gives the same $-9/(56s^2)$ correction, derived for the full problem in Sec.~\ref{sec:uniform}, to both variational upper bounds. Their $s^{-3}$ coefficients differ by $18/343$,
\begin{equation}
\begin{aligned}
 E_c^{\rm ch}(s)\le E_{2\times2}(s)&=1+\frac{3}{2s}-\frac{9}{56s^2}-\frac{45}{5488s^3}+O(s^{-4})=\frac{8C_2}{|\mu_1^{(1)}(s)|}+O(s^{-4}),\\
 E_c^{\rm ch}(s)\le E_{3\times3}(s)&=1+\frac{3}{2s}-\frac{9}{56s^2}+\frac{243}{5488s^3}+O(s^{-4})=\frac{8C_2}{|\mu_1^{(1)}(s)|}+O(s^{-4}).
\end{aligned}
 \label{eq:E2x2-asymp}
\end{equation}
Two-state mixing closes $70\%$ of the gap between $1+\tfrac3{2s}$ and $E_c^{\rm ch}(s)$ already at $s=1$ and $93\%$ by $s=5$, while three-state mixing closes $93\%$--$94\%$ of the gap between $1/w_{\rm crit}(s)$ and $E_c^{\rm ch}(s)$ for $s=\tfrac32,\dots,\tfrac92$ (Table~\ref{tab:nearbps}). The asymptotic relation $E_{\rm var}=8C_2/|\mu_1^{(1)}|+O(s^{-4})$ also explains why the linear tangent estimate $1-\zeta_c\simeq1/|\mu_1^{(1)}|$ underestimates $1-\zeta_c$ by only $6.6\%$ down to $0.23\%$ for $s=1,\dots,5$.\footnote{Since $1-\zeta_c=\tfrac12(1-\zeta_c^2)\bigl[1+\tfrac12(1-\zeta_c)+\dots\bigr]$, replacing $1-\zeta_c$ by $1/|\mu_1^{(1)}(s)|$ omits the quadratic kinematic factor $\tfrac12(1-\zeta_c)\simeq\tfrac1{16s^2}$, which equals $\tfrac{1}{16},\tfrac{1}{64},\tfrac{1}{144},\tfrac{1}{256},\tfrac{1}{400}$ ($6.25\%$ down to $0.25\%$) for $s=1,\dots,5$. This accounts for much of the difference in the finite-spin examples.} Finite-spin channel computations measure the difference between the variational threshold and the channel threshold (Sec.~\ref{sec:mechanism}).
\item \emph{An unstable channel without doubly-BPS states at $\theta_{5d}=\pi$, $s=1$.} Setting $s=1$ removes $\phi_2$ and truncates $\mathsf M^{(4,5)}(1)$ to $\bigl(\begin{smallmatrix}8&-8\\-8&14\end{smallmatrix}\bigr)$ with $\det\mathsf M^{(4,5)}(1)=48>0$. The full endpoint floor census in Appendix~\ref{app:fb:block45} excludes doubly-BPS states in this theta sector and finds one negative slope in each ordered mixed block in channel $(0,0,\tfrac12)$. In the $(4,5)$ block the lowest slope $\mu_1^{(1)}(1)=-6.19951<0$ (exact closed form in $\ln2$ from Eq.~\eqref{eq:secular-slope}) still dives immediately below $-8$ and gives the proved instability bound $E_c^{\theta=\pi}(1)\le E_{2\times2}^{(4,5)}(1)\simeq2.62270$ (lower bound $\zeta_c^{\theta=\pi}(1)\ge0.81985$, with mixed-channel estimates $E_c^{\rm ch}\simeq2.52924$ and $|\zeta_c^{\rm ch}|\simeq0.82695$). A channel can therefore become supercritical through endpoint mixing even when it lacks a common floor state.
\end{enumerate}
The large-spin expansions of the slopes, whose coefficients of $s$ form an arithmetic progression after the common $-8s^2$, are given in Appendix~\ref{app:fb:block45}.

\begin{table}[!htbp]
\centering\small
\setlength{\tabcolsep}{4.5pt}
\begin{tabular}{@{}lcccccc@{}}
\toprule
$s$ (block) & $\mu_1^{(1)}(s)$ & $\mu_1^{(2)}(s)$ & $\mu_1^{(3)}(s)$ & $E_{\rm var}(s)$ (proved) & $E_c^{\rm ch}$ (num.) & $E_{\rm var}-E_c^{\rm ch}$\\
\midrule
$1$ $(4,4)$ & $-6.68774$ & $+1.65259$ & --- & $2.39954$ & $2.35574$ & $0.04380$\\
$2$ $(4,4)$ & $-27.98684$ & $-8.85177$ & --- & $1.71554$ & $1.70897$ & $0.00656$\\
$3$ $(4,4)$ & $-64.71914$ & $-33.37631$ & --- & $1.48342$ & $1.48130$ & $0.00212$\\
$4$ $(4,4)$ & $-117.18640$ & $-72.98576$ & --- & $1.36538$ & $1.36443$ & $0.00094$\\
$5$ $(4,4)$ & $-185.51001$ & $-128.10014$ & --- & $1.29375$ & $1.29325$ & $0.00050$\\
\midrule
$\tfrac32$ $(4,5)$ & $-15.17180$ & $-4.56289$ & $+3.83203$ & $1.98053$ & $1.96276$ & $0.01777$\\
$\tfrac52$ $(4,5)$ & $-44.24444$ & $-21.04605$ & $-4.52504$ & $1.58239$ & $1.57866$ & $0.00373$\\
$\tfrac72$ $(4,5)$ & $-88.86593$ & $-52.72364$ & $-25.49012$ & $1.41793$ & $1.41652$ & $0.00140$\\
$\tfrac92$ $(4,5)$ & $-149.27533$ & $-99.89501$ & $-60.78369$ & $1.32643$ & $1.32575$ & $0.00068$\\
\midrule
$1_{\theta=\pi}$ $(4,5)$ & $-6.19951$ & $+4.45683$ & --- & $2.62270$ & $2.52924$ & $0.09345$\\
\bottomrule
\end{tabular}
\caption{Near-BPS slopes as $\zeta\to+1^-$, rounded to five decimal places from the exact pencil in Eq.~\eqref{eq:secular-slope} and proved endpoint-BPS variational upper bounds $E_{\rm var}(s)$ from Eq.~\eqref{eq:E2x2} (rounded upward) across representative spins ($E_{2\times2}$ for integer $s$ in block $(4,4)$, $E_{3\times3}$ for half-integer $s$ in block $(4,5)$, and the $2\times2$ truncation $E_{2\times2}^{(4,5)}(1)$ at $\theta_{5d}=\pi,s=1$), compared with converged Galerkin thresholds $E_c^{\rm ch}(s)$. Negative slopes count independent radial branches entering $-9\le\mu<-8$ near $\zeta=1$ within the single irreducible critical channel, enlarging its symmetry-preserving extension family to $U(1)$ at $s=1$, at least $U(2)$ for integer $s\ge2$ and $s=\tfrac32$, and at least $U(3)$ for every half-integer $s\ge\tfrac52$. A dash in the third-slope column means that the endpoint space has dimension two, so a third eigenvalue does not exist. }
\label{tab:nearbps}
\end{table}

\section{Localization windows under a supersymmetric trap}
\label{sec:windows}

 This section leaves the free problem of Secs.~\ref{sec:selfloc} and~\ref{sec:bps} for the trapped model $\mathsf S_\Omega$ of Table~\ref{tab:models}. Self-localization makes the free centre resemble a pinned one, and confining the centre in a trap tests whether dimensional localization can switch on Efimov towers inside the free conformal phase. The trap acts at two different scales. Near the tip of the cone a smooth trap is of lower order than $R^{-2}$ and cannot cure a supercritical free problem, while at large instanton size it can hold the centre close enough to the line to make a size channel supercritical in the infrared. It can therefore add shallow bound states to a line whose free Hamiltonian is semibounded, so accumulation of levels at zero binding energy is a different question from the loss of a lower bound. A naive harmonic trap $\varkappa^2|Z|^2$ fails to isolate the pinned threshold cleanly, because it leaves behind the uncompensated centre-fermion Pauli energy $-16(s+1)/\rho^2$ and produces Efimov towers even at $\zeta=\pm1$. A supersymmetric deformation resolves this obstruction by pairing harmonic confinement with a self-dual rotation whose colour moment map cancels the centre Pauli energy at the pinned point, inside the square of each deformed supercharge (Eq.~\eqref{eq:omegaQ}), while the trapped Hamiltonian $\sum_\pm c_\pm Q_\pm^2-EW/R^2$ is itself such a square only at the endpoints. Concretely, the centre rotation generated by $\xi=e_1Z$ (left multiplication by $e_1$; any unit imaginary quaternion is equivalent by $SU(2)_L$ invariance) is triholomorphic and preserves the $SU(2)_R$ Hopf symmetry, corresponding to the self-dual $\Omega$-background $\epsilon_1+\epsilon_2=0$ (whereas a generic deformation $\epsilon_1+\epsilon_2\ne0$ breaks triholomorphicity and spoils the pinned-point cancellation below). With trap frequency $\varkappa$ and trap length $\ell_{\rm trap}=(2/\varkappa)^{1/2}$, the trapped supercharge of each fermion set and its square are
\begin{equation}
 Q_\pm=\ii{\not{\!\mathbf D}}_\pm+\varkappa\,\mathsf c_\pm(\xi),\qquad
 Q_\pm^2=-(\nabla_{\mathcal A}-\ii\varkappa\xi)^2+P^{(\pm)}+2\varkappa\,\mathcal S_\pm ,
 \label{eq:omegaQ}
\end{equation}
where $\mathsf c_\pm(\xi)$ is Clifford multiplication by $\xi$, and $Q_\pm^2$ combines the harmonic potential $\varkappa^2|Z|^2$, the covariant rotation $2\ii\varkappa\,\xi\cdot\nabla_{\mathcal A}$ along $\xi$, and the fermion Zeeman generator $\mathcal S_\pm$. The trapped model $\mathsf S_\Omega$ of Table~\ref{tab:models} has $H_\Omega(\zeta)=c_+Q_+^2+c_-Q_-^2+(1-\zeta^2)R^{-2}V$. Without the instanton connection, $Q_\pm^{(0)2}=-(\partial-\ii\varkappa\xi)^2+2\varkappa\mathcal S_\pm$ acts as a constant self-dual $U(1)$ magnetic field on $\R^4_Z$ whose lowest Landau level $z_1^{m_1}z_2^{m_2}e^{-\varkappa|Z|^2/2}\varepsilon$ ($\mathcal S_\pm\varepsilon=-2\varepsilon$), in complex coordinates $z_1,z_2$ on which left multiplication by $e_1$ acts as multiplication by $\ii$, so that the rotation charge of a Landau monomial depends only on its total degree, confines $|Z|$ within each fixed eigensector of the lifted $U(1)_L$ rotation (Appendix~\ref{app:localization}(d)). On the lowest-weight Gaussian $g(Z)\propto e^{-\varkappa|Z|^2/2}$, the oscillator zero-point energy $4\varkappa$ cancels $2\varkappa\mathcal S_\pm=-4\varkappa$, so $Q_\pm^{(0)}$ annihilates $g\varepsilon$ and the linear terms in its ground-state expectation cancel; explicitly, the radial-$Z$ gauge expansion $\mathcal A_{Z_i}=\tfrac12 Z_j\mathcal F_{Z_j Z_i}+O(|Z|^2)$ inside $2\ii\varkappa\,\xi\cdot\mathcal A$ produces an orbital moment-map coupling $-16J'_1/\rho^2$ (with $J'_1=\operatorname{Ad}(\hat\lambda)e_1\cdot\mathbf J$, independent of $\varkappa$ because $\varkappa\langle|Z|^2\rangle_{\ell_{\rm trap}}=2$) that cancels the Pauli energy $+16J'_1/\rho^2$ of the polarized centre fermions for every colour weight.

The trapped operator connects with three other settings. First, the trapped supercharge is the Dirac operator of the combined curvature of $\mathcal A$ and the Abelian rotation connection, realizing the Witten--Bismut deformation of a Dirac operator by a Killing field (following the localization tradition of Refs.~\cite{Witten1982,Bismut1986}, with the Killing-field square in Refs.~\cite{ProkhorenkovRichardson,WuZhang}), augmented here by the exact pinned-point cancellation between the Pauli term and the orbital moment map. Second, $Q_\pm$ uses the Killing field of the self-dual ($\epsilon_1+\epsilon_2=0$) equivariant action in the Nekrasov $\Omega$-background \cite{Nekrasov2002} with the line at its fixed point, the torus action under which five-dimensional supersymmetric Wilson loops are computed \cite{GaiottoKim,Nekrasov2015,HCKim2016}, with its colour lift encoding the equivariant structure of the universal bundle. Third, in a fixed rotation sector this equivariant deformation can switch on a tower of shallow levels absent in the free-conformal regime, casting the self-dual $\Omega$-torus action as a supersymmetric realization of the Nishida--Tan confinement switch \cite{NishidaTan2011}.

At leading Born--Oppenheimer order, projecting onto the polarized centre ground state leaves the pinned model $\mathsf S_{\rm pin}$ of Table~\ref{tab:models}. The two size-fermion chiralities remain available and implement the two bundle parities. The rule becomes $n\equiv2s+\theta_{5d}/\pi+(1-\Gamma_y)/2\pmod2$, so the channel $n=0$ is available for either discrete theta angle. The supersymmetric pinned floor is $\beta=1$, and for $s>0$ the pinned model has threshold $\zeta_{\rm pin}(s)$, with $\zeta_{\rm pin}(s)^2=1-1/(4C_2)$. At $\theta_{5d}=0$ and integer $s>0$ this also equals the bosonic pinned threshold~\eqref{eq:pinnedthresholds}. For $s\ge1$, Eq.~\eqref{eq:pinbound} gives $\zeta_c\le\zeta_{\rm pin}$, leaving an interval in which the free problem is conformal and the pinned size channel is supercritical. The free fundamental line is conformal throughout its physical coupling range. In the supercritical $n=0$ pinned channel it predicts the larger-to-smaller binding-energy ratio $e^{2\pi/\nu}$, where $\nu=\sqrt{4C_2(1-\zeta^2)-1}$. The full trapped result below makes this switch precise.\footnote{Three features distinguish this realization from atomic mixed-dimensional gases, namely (i) the free threshold never exceeds the pinned one ($\zeta_c\le\zeta_{\rm pin}$), (ii) for half-integer spin, size-fermion chirality reverses the bosonic pinned shift $\beta=2$ back to $\beta=1$, and (iii) at large $s$ the window narrows as $\Delta\zeta=\zeta_{\rm pin}-\zeta_c\sim 3/(16s^3)$, with a non-uniform $\varkappa\to0$ limit (at any fixed $\varkappa>0$ the Efimov tower lives in the far-size region $\rho\gg\ell_{\rm trap}$, whereas taking $\varkappa\to0$ at fixed $\rho$ recovers the free conformal problem without towers).}

\paragraph*{The rotation-sector threshold.}\label{prop:omega-sector}
For $s\ge1$ and either discrete theta angle, the $\Omega$-trapped form is semibounded for $0\le E\le E_c(s)$ and unbounded below for $E>E_c(s)$, where $E_c(s)$ is the critical attraction of the chosen theta sector. In its Friedrichs realization each fixed lifted-rotation sector has finitely many negative levels for $0\le E\le1$. For $1<E\le E_c(s)$ at least one fixed sector has infinitely many negative levels accumulating at zero. For $s=\tfrac12$ the form is semibounded throughout the physical range $0\le E\le3$. Each fixed sector has finitely many negative levels for $E\le1$, and at least one has an accumulating sequence for $1<E\le3$. For $s=0$ the form is nonnegative. Appendix~\ref{app:localization}(d) gives the proof. A logarithmic Hardy remainder absorbs the residual Pauli terms at $E=1$, where the leading inverse-square terms cancel. For $s=1$ the tower interval in $|\zeta|$ is $[\zeta_c(1),\sqrt{7/8})$ with the corresponding theta-dependent $\zeta_c(1)$.

At fixed lifted rotation charge $q$, the large-size limit retains a finite-dimensional polarized Landau space. Allowing the rotation charge to vary also admits magnetic translations that take the centre far from the line. The size degrees of freedom then see a BPST connection with scale set by the centre displacement. The number of shallow levels can therefore grow at different rates according to whether the rotation charge is resolved.

\paragraph*{Counting the shallow levels.}\label{prop:omega-count}
For the Friedrichs realization, let $N_q(\eta)$ count eigenvalues below $-\eta$ in charge $q$, and $N(\eta)$ the count over all charges. Write $r_q$ for the rank of the polarized Landau space at charge $q$, including colour and the constant size-spinor states that pass the physical projection. For $1<E\le E_c(s)$, or $1<E\le3$ at $s=\tfrac12$, the fixed-charge count is
\begin{equation}
N_q(\eta)=\frac{r_q\sqrt{E-1}}{2\pi}\ln\frac{\varkappa}{\eta}+O(1)
\quad (\eta\downarrow0).
\label{eq:omega-count-sector}
\end{equation}
For $s\ge1$ and $0<E\le E_c(s)$, and also for $s=\tfrac12$ and $1<E\le3$, the global count satisfies
\begin{equation}
0<\liminf_{\eta\downarrow0}\left(\frac{\eta}{\varkappa}\right)^2N(\eta)
\le\limsup_{\eta\downarrow0}\left(\frac{\eta}{\varkappa}\right)^2N(\eta)<\infty .
\label{eq:omega-count-global}
\end{equation}
For $s=\tfrac12$ and $0<E\le1$ the global upper bound still holds, and in the $\theta_{5d}=\pi$ sector the count then stays bounded as $\eta\downarrow0$.

The \hyperref[app:loc:count]{fixed-charge proof in Appendix~\ref*{app:localization}} reduces the problem to the size channel with zero angular momentum. Its coefficient comes from the radial Hardy deficit $E-1$, multiplied by the allowed Landau rank. The bounded remainder leaves the phases and adjacent level ratios undetermined. For the fundamental line below the sector threshold, the global finiteness at $\theta_{5d}=\pi$ follows from a separate chirality comparison on the displaced size slice (\hyperref[app:loc:fundamental]{fundamental-line estimate}). The estimate in a fixed charge is nonuniform as the occupied Landau degree grows. The Landau states make this quantitative. A centre state built from a holomorphic polynomial of degree $h$ on the Gaussian $e^{-\varkappa|Z|^2/2}$ has mean squared radius $\langle|Z|^2\rangle=(h+2)/\varkappa$, the degrees available at fixed charge grow linearly with $|q|$ along the populated tail $q\to-\infty$, and so does the rank $r_q$. The pinned expansion behind the fixed-charge count needs $\varkappa\rho^2\gg h$, and at the size $\rho\sim\eta^{-1/2}$ of the shallow levels this fails once the degree reaches order $\varkappa/\eta$. This explains why summing the logarithmic count over charges would not recover Eq.~\eqref{eq:omega-count-global}. Physically, the two counts measure different things. The fixed-charge count is the Efimov tower of the size coordinate, with the logarithmic density of any supercritical inverse-square channel, as in the confinement-induced Efimov effect. The global exponent $2$ is the density of guiding centres. Because the self-dual trap rotates critically, its lowest Landau level is infinitely degenerate in four dimensions, and a centre displaced by $L$ costs no energy while it binds its size modes at a depth of order $L^{-2}$ through a BPST zero mode of the size variables near the supersymmetric endpoints (the \hyperref[app:loc:guiding]{global counting argument}), so that the packets displaced by up to $\eta^{-1/2}$ number of order $\eta^{-2}$. The exponent therefore reflects the degeneracy of the trap, and the line enters through the existence of the bound size mode.

In the certified spin-one interval, the same $\Omega$-trap preserves the fall to the centre at $\theta_{5d}=0$. At $\theta_{5d}=\pi$ the trap is semibounded and exhibits both the fixed-sector logarithmic tower in sectors with $r_q>0$ and the global power-law accumulation. This gives two distinct outcomes at the same attraction and trap strength by changing the discrete projection alone.

\paragraph*{An infrared cutoff from a compact direction.} On $\R^4$ the inverse-square tail in the effective pinned channel persists to arbitrarily large size. In the trapped model the magnetic length controls the onset of this tail. The following finite-annulus estimate concerns this size channel. Compactifying one spatial direction on a circle of circumference $\ell_{\rm IR}$ gives a spatial manifold $\R^3\times S^1$. Its periodic instanton is a charge-one $SU(2)$ caloron \cite{KraanVanBaal,LeeLu}. We fix the asymptotic holonomy away from the centre, for example at $\omega_{\rm hol}=\tfrac14$. Its centralizer $U(1)_G\subset SU(2)_G$ is the asymptotic gauge symmetry at every constituent separation. The ordinary and Kaluza--Klein monopole constituents carry opposite magnetic charges. In either central-holonomy limit, $\omega_{\rm hol}\to0$ or $\tfrac12$, one constituent becomes massless and ceases to be separately localized. At the fixed generic holonomy their separation is $d=\pi\rho^2/\ell_{\rm IR}$ and their relative moduli space is the positive-mass Taub--NUT quotient $\mathrm{TN}/\Z_2$ \cite{KraanVanBaalTN}. For $\rho\ll\ell_{\rm IR}$ the metric near the nut reduces to the size cone $\R^4/O(1)$ and windings around $S^1$ are suppressed in the coincident Green function, preserving the inverse-square tail. At $\rho\sim\ell_{\rm IR}$ the constituent cores become distinguishable ($d\gtrsim\ell_{\rm IR}$), and the scale-invariant approximation underlying the log-periodic estimate reaches its infrared cutoff. In a supercritical pinned channel of Eq.~\eqref{eq:pinnedradial} ($\nu^2=-\beta_n^2=E-(n+1)^2>0$), the zero-energy radial solution $u=\rho^{1/2}\cos\bigl(\nu\ln(\rho/\rho_0)\bigr)$ has consecutive zeros spaced by $\Delta\ln\rho=\pi/\nu$, so Sturm oscillation truncates the tower to the logarithmic Efimov count \cite{Efimov1970}
\begin{equation}
 N_{\rm Ef}=\frac{\nu}{\pi}\ln\frac{\ell_{\rm IR}}{\ell_{\rm UV}}+O(1),
 \label{eq:main:efimov}
\end{equation}
for an inverse-square interval with $\ell_{\rm UV}\ll\ell_{\rm IR}$ at fixed $\nu$. Its binding scales range from order $\ell_{\rm IR}^{-2}$ to order $\ell_{\rm UV}^{-2}$ in the stated Hamiltonian units. Here $\ell_{\rm UV}=\ell$ is the finite source size for the pinned models (the $O(1)$ offset depends on the Taub--NUT crossover and any separated-constituent bound states at $\rho\gg\ell_{\rm IR}$), while $\ell_{\rm UV}=\ell_{\rm trap}$ gives the corresponding leading-order estimate at $n=0$ for $\mathsf S_\Omega$ (the rotation $\xi=e_1Z$ mixes all four spatial coordinates and does not itself descend to $\R^3\times S^1$).

Embedding this centre-only trap into a six-dimensional $(2,0)$ parent remains open. The null reduction of six-dimensional superconformal theories on conformally compactified Minkowski space produces five-dimensional theories whose anti-self-duality constraint is itself $\Omega$-deformed, so that instanton sizes and positions evolve along worldlines \cite{LLMR2020,LLMR2021} instead of remaining on a fixed moduli space with a centre-only potential; similarly, the string and M-theory fluxtraps of Hellerman, Orlando and Reffert \cite{HOR2011,HOR2012,LOR2014} trap transverse brane coordinates without realizing a D0--D4 instanton sector with a probe line, and spacelike-circle $\Omega$-backgrounds do not descend directly to a null $x^-$ reduction. The localization windows are therefore statements about the $\Omega$-deformed instanton quantum mechanics itself, with their sector dependence stated above.

\section{Light-cone interpretation}
\label{sec:lightcone}

 Because the constrained five-dimensional theory is proposed as the null reduction of the six-dimensional $(2,0)$ theory and the charge-one instanton carries one unit of null momentum, the thresholds above naturally invite interpretation as statements about Wilson surface operators \cite{Ganor1996} of the $A_1$ $(2,0)$ theory wrapping the null circle. This six-dimensional interpretation concerns the sector $\theta_{5d}=0$, the untwisted reduction of the $A_1$ theory (Sec.~\ref{sec:theory}). For the twisted $A_2$ parent at $\theta_{5d}=\pi$, the identification of the wrapped defect and its null-momentum sector remains to be derived. Three observations motivate the identification in the untwisted sector. First, the discrete light-cone quantization of the $(2,0)$ theory is proposed to be described by the quantum mechanics on instanton moduli space, with null momentum equal to instanton number \cite{ABKSS}. In Seiberg's derivation the null circle is the limit of a boosted small spacelike circle and the momentum modes become D0-branes \cite{Seiberg1997}. Second, a boost rescales the null radius $R_{\rm null}=g_5^2/(4\pi^2)$ of Sec.~\ref{sec:theory}. The dimensionless thresholds at fixed instanton number are independent of this radius in the scale-free model, although dimensionful energies keep their factor $g_5^2$. Third, at the abelian level a surface operator that wraps the null circle reduces to the line~\eqref{eq:line}. For fields independent of the null coordinate $x^-$, the self-dual three-form $H=\dd B$ of the $(2,0)$ theory gives $F_{ij}=2\pi R_{\rm null}H_{ij-}$ for $A_\mu=2\pi R_{\rm null}B_{\mu-}$. This curvature is self-dual and obeys the constraint of the five-dimensional theory. The anti-self-dual component $H_{ij+}$ corresponds to its multiplier. For a surface $\Sigma$ spanned by $(x^+,x^-)$ at the origin of space, with chiral two-form $B$ reducing via $A_t=2\pi R_{\rm null}B_{t-}$ and the scalar rescaled in the same way, 
\begin{equation}
 \exp\!\left(\ii\!\int_\Sigma\bigl(B+\zeta\,n\cdot\Phi\,\mathrm{vol}_\Sigma\bigr)\right)\ \longmapsto\ \exp\!\left(\ii\!\int\!\dd t\,\bigl(A_t+\zeta\,n\cdot X\bigr)\right),
 \label{eq:nulldictionary}
\end{equation}
where $n$ is a unit vector in the space of the five scalars $\Phi$, so that $n\cdot X=X^6$ in Eq.~\eqref{eq:line}. Only the zero mode in $x^-$ couples, so the surface preserves every momentum sector. The symmetry common to the null reduction and the surface is $sl(2,\R)\oplus u(1)\oplus so(4)$, the symmetry of our quantum mechanics, together with the $so(4)_R$ that fixes $n$, and for $\zeta=\pm1$ the supersymmetry projector of the surface reduces to that of the line under the null-reduced supersymmetry rules of the constrained theory. In the light-cone frame of Seiberg and Sen, applying the M-theory/type IIA brane dictionary suggests a fundamental string ending on D4-branes in a D0--D4 background \cite{Seiberg1997,Sen1997}, and its moduli dynamics is expected to be the bosonic-impurity variant of Tong and Wong's quantum mechanics of instantons with Wilson lines \cite{TWADHM}.  Using bosonic impurity degrees of freedom to carry the spin-$s$ colour space, that quantum mechanics describes every $s$.\footnote{By contrast, the D4$'$-brane realization of that quantum mechanics uses fermionic impurity fields, which produce antisymmetric representations and therefore only the fundamental $s=\tfrac12$ line for gauge group $SU(2)$; realizing higher-spin $SU(2)$ Wilson lines microscopically requires symmetric-representation brane constructions (bosonic impurities) analogous to those of Ref.~\cite{AKKS}.}

Promoting this null-reduced moduli computation to a full non-abelian theorem in the six-dimensional $(2,0)$ theory requires five concrete advances, starting with the non-abelian surface dictionary. For spacelike circles, Agarwal, Kim, Kim and Sciarappa compute Wilson-surface observables on $\Omega$-deformed $\R^4\times T^2$ in two ways, from five-dimensional instanton partition functions with Wilson lines and from elliptic genera of two-dimensional $(0,4)$ theories of self-dual strings. The tested minuscule examples agree and non-minuscule examples agree in part \cite{AKKS}. Our lines with $s\ge1$ are non-minuscule, and no corresponding statement is available for null circles or for $|\zeta|<1$. The other steps are the treatment of the small-instanton singularity with a line present (Sec.~\ref{sec:models}), a derivation of the quantum measure from the parent theory (Sec.~\ref{sec:models}), control of any running of $\zeta$ from non-BPS determinants or null-momentum-exchanging modes of the parent theory (beyond the zero-mode projection of Appendix~\ref{app:fermions}), and a source of finite thickness in place of the point source. The results therefore support a precise conditional statement. In the one-momentum light-cone sector of the $A_1$ theory, if the dictionary extends to non-minuscule representations and to $|\zeta|<1$ and if the reduced Hamiltonian with this measure and point-source prescription applies, the surface inherits the corresponding one-sector supercritical instability for the couplings proved in Sec.~\ref{sec:instability}. With an explicitly specified core of scale $\ell_{\rm UV}$ the collapse becomes a tower of bound states through the radial matching of Sec.~\ref{sec:radial}, and the tower stays inside the sector of one unit of null momentum, because a surface wrapping the null circle conserves that momentum and cannot by itself move the system to another sector.

The quantum numbers of the collapsing states suggest what the instability does to the line. At large $s$, and in the finite-spin scans, the critical channels carry the smallest gauge charge that Eq.~\eqref{eq:physical} allows, $j_G=0$ for integer $p$ and $j_G=\tfrac12$ for half-integer $p$, the physical charge being the $SU(2)_G$ representation (Appendix~\ref{app:frames}), while the colour spin of the line reappears as orbital spin $j_L=s-1$ of the bound instanton, as isospin becomes spin for a particle bound to a monopole \cite{JackiwRebbi1976,HasenfratzTHooft1976}. On this interpretation the instanton that falls onto the line screens its charge as far as the centre of $SU(2)$ permits, which is the fate that the large-charge analysis of Ref.~\cite{ACKMR} finds for supercritical Wilson lines. Turning the interpretation into a statement about the endpoint of the instability requires defect dynamics beyond the constrained moduli model.

Drukker, Giombi, Tseytlin and Zhou study the surface in the fundamental representation of the $(2,0)$ theory holographically at large $N$, and briefly discuss a renormalization-group flow from the non-supersymmetric surface to the half-BPS one, consistent with the defect $b$-theorem \cite{JensenOBannon2016} (see also Ref.~\cite{DGTZ}). The holographic flow concerns a running coupling on the fundamental surface in the decompactified theory. Our reduced $A_1$ model fixes the bare $\zeta$ and one unit of null momentum. Its fundamental line stays conformal, while lines with $s\ge1$ can become supercritical. A computation that interpolates between them would need both larger momentum and running $\zeta$. Graham and Witten computed the surface Weyl anomaly holographically from a membrane in $AdS_7\times S^4$ \cite{GrahamWitten1999}. Henningson and Skenderis derived the anomaly of a Wilson surface coupled to a free abelian two-form gauge field in six dimensions \cite{HenningsonSkenderis1999}. Local insertions into the half-BPS surface and its anomaly coefficients have since been studied with defect conformal field theory and bootstrap methods \cite{DPT2020,DPT2021,MT2023}. Under the proposed light-cone identification, $2k_+=1+\beta$ would be the dimension of an insertion carrying one unit of null momentum, and matching it with such defect data would test the dictionary directly.
\section{Discussion}
\label{sec:discussion}

In the charge-one sector of five-dimensional non-Lorentzian $SU(2)$ gauge theory, the spatial self-duality constraint leaves the instanton moduli as bosonic dynamical variables. In their canonically quantized Hamiltonian, conformality depends on the competition between attraction and motion of the instanton centre. The self-energy draws the instanton towards the line, while the free motion of the centre spreads that attraction over an eight-dimensional cone, whose Hardy barrier is nine units deep. With bosonic moduli alone the spreading wins for every line with $s\le4$, and from $s=\tfrac92$ onwards the two theta sectors can differ in phase at the same coupling. The fermion zero modes change the outcome. The curvature of the colour bundle acts on the centre as a magnetic field, the Pauli energy of the centre zero modes cancels its Landau zero-point energy, and the centre localizes in a lowest Landau level of width of order $\rho/\sqrt s$ at no magnetic cost. The free instanton then behaves like a pinned one, and the barrier left by supersymmetry is one unit. Every line with $s\ge1$ therefore loses conformality in a range of couplings that contains the ordinary line, while the fundamental line, for which no state reaches the floor of both supersymmetric halves, stays conformal at both values of the discrete theta angle. Compared with the four-dimensional instabilities recalled in the introduction, the product $4s(s+1)(1-\zeta^2)\langle W\rangle$ plays the role of $(g^2s/2\pi)^2(1-\zeta^2)$, with the exact Casimir in place of $s^2$, the moduli geometry in place of the gauge coupling, and the fermion zero modes deciding whether small spins collapse at all. The spin-one comparison isolates a change of phase caused by the discrete projection at fixed attraction, between the two five-dimensional theories $\theta_{5d}=0$ and $\pi$. The large-spin expansion explains why the colour parity first enters at cubic order. The residual width raises the pinned threshold by $3/(2s)$, and mixing within the degenerate endpoint space lowers it by $9/(56s^2)$. A virtual polar excitation shifts each competing threshold by the same $-9/(128s^3)$. Their ordering therefore measures the different mixing allowed by the two gauge-spin parities. These statements concern the Hamiltonian obtained from the self-duality constraint, with the canonical quantization, point-source prescription and domain stated in Sec.~\ref{sec:models}.

\paragraph*{Structures of wider scope.}
The Hopf reduction links the floor states to four-dimensional quantum Hall wavefunctions at the level of representations and explicit polynomials \cite{Hasebe}. The inertia ratio changes their concentration about the pinned pole through a conformal dilation. The same concentration controls the matrix elements of the inverse-square attraction and makes centrifugal ordering useful in the Schur bound. An external $\Omega$-trap uses a different aspect of the geometry. Its Killing-field deformation cancels the projected centre Pauli term and lets the far-size channel retain the pinned inverse-square tail. This gives confinement-induced towers in at least one fixed rotation sector of the free-conformal phase. At fixed rotation charge the far-size Hardy deficit gives the logarithmic coefficient. Globally the four centre directions permit a power-law count, with a BPST size zero mode allowing binding even below the fixed-sector onset for $s\ge1$. Some ingredients of the analysis have a wider scope. Proposition~\ref{thm:L} uses a homogeneous connection on a cone with a compact angular cover and smooth link data. Its comparison applies when the attractive potential is strongest on the pinned slice. The Hopf submersion and its pulled-back colour connection turn the Dirac--Hardy bound into a base estimate and a fibre Casimir. The floor also uses the curvature of the round base and the induced fermion connections. A Berry curvature of one chirality and gyromagnetic ratio two give the local Landau--Pauli cancellation. The same cancellation should localize other supersymmetric solitons near probes that couple to their orientation, such as monopoles near Wilson or 't~Hooft lines, once the global spectrum of their moduli space is controlled as here.

The radial inverse-square problem explains the relation to atomic collapse. In a simple channel with $\beta\ge1$ the radial Hamiltonian has a unique self-adjoint realization. For $0<\beta<1$ either of the two radial powers defines a scale-invariant contact condition. Degenerate branches admit choices of the subspaces assigned to each power, subject to the symmetries imposed. Hardy--Dirac inequalities and distinguished domains give a related analysis for the relativistic Coulomb problem \cite{EstebanLoss2007}. Supercritical Coulomb fields give vacuum instabilities for heavy nuclei and log-periodic resonances for graphene impurities \cite{ZeldovichPopov1972,SKL2007}. Here the inverse-square attraction and the change of its coefficient by the Pauli interaction are properties of the Wilson-coupled instanton Hamiltonian. The two scale-invariant choices for a simple branch are analogous to the two quantizations between the Breitenlohner--Freedman and unitarity bounds in anti-de~Sitter space. They coalesce at $\beta=0$, where the Friedrichs contact condition remains scale invariant. For $\beta^2<0$ the free quadratic form is unbounded below and the regulated tower acquires a short-distance scale \cite{KLSS}. The near-BPS window of Sec.~\ref{sec:bps}, where two or three radial branches enter together, is then a quantum-mechanical model of a defect renormalization group with several couplings whose fixed points merge branch by branch.

\paragraph*{Evidence summary.}
All results govern the Hamiltonian~\eqref{eq:Hzeta}, or its pinned and trapped versions of Table~\ref{tab:models}, with the canonical quantization of Sec.~\ref{sec:models} and the couplings derived in Appendix~\ref{app:fermions}. Table~\ref{tab:evidence} separates the evidence across five tiers, with detailed scope and verification notes in Appendices~\ref{app:un:uniform} and~\ref{app:numerics}. 

\begin{table}[!htbp]
\centering\small
\setlength{\tabcolsep}{5pt}
\begin{tabular}{@{}>{\raggedright\arraybackslash}p{3.3cm}p{10.9cm}@{}}
\toprule
Evidence tier & Covered statements and locations\\
\midrule
Analytic derivations and proofs & Covariant Hardy criterion (Proposition~\ref{thm:L}) and the endpoint floor in its fibre-resolved form (Proposition~\ref{thm:floor}); bosonic pinned thresholds at both theta angles and free-centre protection for $E\le99/4$ (Sec.~\ref{sec:pinned}); the exact stretched contact ladder and the bosonic phases at every spin and both theta angles (Proposition~\ref{thm:bosphase}, Appendix~\ref{app:bosonic}); instability of every line with $s\ge1$, $\zeta=0$ included (Eq.~\eqref{eq:thmC}), and its half-integer and $\theta_{5d}=\pi$ counterparts (Sec.~\ref{sec:bps}); fundamental-line conformality with a unique realization, Eq.~\eqref{eq:s12bound} (Appendix~\ref{app:fb:s12}); endpoint BPS system, near-BPS slopes, the bounds $E_{2\times2}$ and $E_{3\times3}$, the single-channel $U(2)/U(3)$ multiplicity (Sec.~\ref{sec:bps}); the single onset (Sec.~\ref{sec:instability}); the large-spin expansion~\eqref{eq:lawcor2} and its theta-dependent cubic term~\eqref{eq:lawcor3} as $s\to\infty$, with remainder constants that are not explicit and the $s^{-3}$ pencil coefficients expanded symbolically (Appendix~\ref{app:fb:chgap}); endpoint symmetry and contact domains (Secs.~\ref{sec:radial} and~\ref{sec:floor}); the sector threshold and the logarithmic and power-law counts of the $\Omega$-trap (Sec.~\ref{sec:windows}); angular exclusion at onset (Appendix~\ref{app:un:coverage}).\\\addlinespace[3pt]
Symbolic computation & General-spin floor polynomials, compressed cubic coefficients and the $s=1$ slope census at $\theta_{5d}=\pi$ (Appendix~\ref{app:fb:bps}); the $s^{-3}$ terms of $E_{2\times2}$ and $E_{3\times3}$ and the slope expansions (Sec.~\ref{sec:bps}); the higher coefficients of the large-spin series of the critical channel and of the theta splitting (Sec.~\ref{sec:uniform}); endpoint protected characters (Appendix~\ref{app:fb:index}).\\\addlinespace[3pt]
Exact-arithmetic proof & Uniform lower bound $\hat E(s)$ for every $s\ge1$, the lower side of Eq.~\eqref{eq:lawcor} (Appendix~\ref{app:un:uniform}); spin-one theta window~\eqref{eq:theta-window} (Appendix~\ref{app:fb:block45}); certified crossing $0.8399601<|\zeta_c^{\rm ch}|<0.8399602$ of the $s=1$ critical channel (Appendix~\ref{app:radial-certificate}); outward rounding of the ten variational upper bounds (Appendix~\ref{app:numerics}).\\\addlinespace[3pt]
Effective-channel estimates and physical interpretations & Caloron Efimov count~\eqref{eq:main:efimov} in a finite inverse-square interval and the leading Born--Oppenheimer reduction of $\mathsf S_\Omega$ to $\mathsf S_{\rm pin}$ (Sec.~\ref{sec:windows}); the perturbative estimate~\eqref{eq:main:c2phys} of the coefficient $-9/(56s^2)$ (Sec.~\ref{sec:mechanism}); the superconformal interpretation of the floor (Sec.~\ref{sec:floor}); the proposed light-cone, Wilson-surface and screening interpretations (Sec.~\ref{sec:lightcone}); the predicted density ratio $N_0/N_\pi\to8/5$ (Sec.~\ref{sec:discussion}); the ratio $e^{2\pi/\nu}$ of bosonic binding energies for a core whose matching phase tends slowly to a constant (Sec.~\ref{sec:freephases}).\\\addlinespace[3pt]
Galerkin scans & Finite-spin channel estimates and ordering within the sampled set; all-channel scans confirming Proposition~\ref{thm:bosphase} for $\tfrac52\le s\le6$, with pointwise and scalar exclusions and finite differences for the remaining channels; the large-spin series of the stretched bosonic threshold checked to $s=160$; Borel--Pad\'e resummation of the critical-channel series, and the theta splitting through order $s^{-5}$ from direct thresholds of both critical channels up to $s=32$ (Sec.~\ref{sec:uniform}); complementary polar resolvent check through $s=1600$ in $(4,4)$ (Appendix~\ref{app:numerics}).\\
\bottomrule
\end{tabular}
\caption{Summary of evidence tiers across the manuscript. The analytic bounds and the exact-arithmetic lower-bound proof use no scanned eigenvalues.}
\label{tab:evidence}
\end{table}

\paragraph*{Remaining questions within the charge-one model.}

The results above leave seven well-posed problems inside the charge-one model, each of which would sharpen a statement of this paper.
\begin{enumerate}
\item \emph{Effective large-spin control.} The expansions~\eqref{eq:lawcor2} and~\eqref{eq:lawcor3} select the critical channel and the theta splitting beyond a spin that their implicit remainders leave undetermined. Explicit remainder constants would turn them into statements at a computable spin, and the angular rectangle of Appendix~\ref{app:un:coverage} with the polar enclosure criterion of Appendix~\ref{app:numerics} reduces each finite spin to finitely many certified signs, as already carried out at $s=1$.
\item \emph{The density of trapped levels.} The global count of the $\Omega$-trap has a proved exponent and an unknown coefficient. Freezing the centre gives negative eigenvalues $-b_j(E)$ of the unit-scale BPST size Hamiltonian, which a displacement $Z$ rescales to $-b_j(E)/(2|Z|^2)$ (Appendix~\ref{app:loc:guiding}). If the projection onto the polarized lowest Landau level controls the full operator, summing over orbital degrees gives $N_\theta(\eta)\sim(\varkappa/\eta)^2\sum_jb_j(E)^2/8$, and at $s=1$ near the supersymmetric endpoint the zero-mode slopes $-\frac15$ once and $-\frac1{10}$ four times at $\theta_{5d}=0$, against $-\frac1{20}$ and $-\frac1{10}$ four times each at $\theta_{5d}=\pi$, predict $N_0(\eta)/N_\pi(\eta)\to\frac85$ as $\eta\to0$ and then $E\to0$, the ratio of the sums of squared slopes, $\tfrac2{25}$ and $\tfrac1{20}$. This is a conjecture whose test requires control of the size continuum and of centre derivatives of the size states. The fundamental line at $\theta_{5d}=0$ and $0<E\le1$, where the frozen size problem has an exact zero mode, and the ratio of consecutive shallow levels, left open by the bounded phase error of the fixed-charge count, belong to the same problem.
\item \emph{Ultraviolet selection.} The scale-invariant contact conditions of Sec.~\ref{sec:radial} classify the choices visible in the exterior problem. The Fayet--Iliopoulos resolution of the instanton core, in which the size cone becomes the Eguchi--Hanson space and the colour space of the line jumps in rank at point-like instantons (Sec.~\ref{sec:models}), would select one of them, once a gauge-invariant replacement of the point-localized line and a convergence theorem for the resolved operator are available. In the bosonic window of Sec.~\ref{sec:freephases} the matching phase at the core is the only datum the exterior leaves free. The exterior fixes the ratio $e^{2\pi/\nu}$ of successive binding energies, so a resolution of the core would be tested by their absolute scale.
\item \emph{Superconformal origin of the floor.} Excluding the window $|\mathcal D|<\tfrac12$ algebraically and establishing the full unitarity region of the multiplets that occur would derive Proposition~\ref{thm:floor} from the endpoint $OSp(4^*|2)$ alone and fix the floor multiplicities without the Hopf reduction (Sec.~\ref{sec:floor}).
\item \emph{Five-dimensional comparison.} The refined endpoint trace of Appendix~\ref{app:fb:index} and the one-instanton contribution to five-dimensional Wilson loops computed from the instanton quantum mechanics with Wilson lines \cite{HCKim2016,AsselSciarappa2018} differ in their fugacities and defect degrees of freedom. A map between them would test the floor content independently, and the partial non-minuscule agreement of Sec.~\ref{sec:lightcone} marks where the distinction matters.
\item \emph{The number of states that reach the core.} Proposition~\ref{thm:bosphase} counts the bosonic states that need boundary data only up to $E=16s+27$. At larger attraction further polar states of the stretched channel reach $\mu=-8$, about $s(\sqrt{2-\zeta^2}-1)/2$ of them at fixed $\zeta$ to leading order in $s$ by Eq.~\eqref{eq:ladder}, and other channels follow. How many states of each representation reach the core over all channels is open, as is its counterpart with the fermion zero modes. These multiplicities fix the contact couplings of the line, since $n$ states below $-8$ give deficiency indices $(n,n)$ and a $U(n)$ family of boundary conditions, which rotation invariance reduces to one $U(m)$ factor for each representation that reaches the core $m$ times.
\item \emph{Thresholds as a connection problem.} A threshold is a coupling at which the solution regular at one end of the polar interval is regular at the other. For the scalar link equation, a Heun equation, such connection problems are solved by Nekrasov--Shatashvili functions (Sec.~\ref{sec:freecentre}), and the exact singularities of the bosonic shape response at $\mathsf t=\pm2\pi\ii(2k+1)$ (Appendix~\ref{app:bosonic}) fix its Stokes phenomenon exactly. Solving the connection problem for the colour-twisted and Pauli-coupled equations would give the thresholds at every spin in closed form and decide whether they contain terms beyond all orders in $1/s$, which neither the present coefficients nor the numerical resummation of Sec.~\ref{sec:uniform} can resolve.
\end{enumerate}

\paragraph*{Long-term directions beyond charge-one $SU(2)$.}
Three farther-reaching problems connect the mechanism to other settings.
\begin{enumerate}
\item \emph{Multi-instanton ($k>1$) cooperative binding.} For $k>1$ the moduli space is curved, the evaluation connection is no longer a pull-back from $S^4$, and cluster strata on which one instanton collapses onto the line ($\rho_1\to0$) while $k-1$ remain separated create internal inverse-square singularities on the link itself (violating the smoothness hypotheses of Proposition~\ref{thm:L}). This turns $k>1$ into a coupled multi-body Faddeev spectral problem testing whether cooperative multi-instanton binding can overcome the $s=\tfrac12$ Pauli barrier or cluster instantons around the probe.
\item \emph{Higher-rank, octonionic and monopole moduli spaces.} Higher-rank gauge groups split cleanly by topology; $USp(2n)$ retains $\pi_4=\Z_2$ and $\theta_{5d}$, whereas $SU(N)$ ($N\ge3$) has $\pi_4(SU(N))=0$ and even at $k=1$ carries a curved $4N$-dimensional moduli cone with $U(1)$ quotient and $8N$ fermion zero modes, tracking how self-localization scales with $N$. One step up the Hopf hierarchy of Sec.~\ref{sec:bpsstates} lies a sharper mathematical question. The octonionic map $S^{15}\to S^8$ carries the instantons that solve quartic self-duality $F\wedge F=\pm{*(F\wedge F)}$ \cite{GrossmanKephartStasheff1984} and the lowest Landau levels of an eight-dimensional quantum Hall problem \cite{Bernevig2003}. Whether the corresponding twisted Dirac operator on $S^{15}$ has a sharp floor attained by those levels, as the quaternionic operator on $S^7$ does here, is a well-posed spectral problem, although no known line defect realizes it. For magnetic-monopole moduli spaces \cite{TWmono}, the Higgs scale breaks the homogeneity of Proposition~\ref{thm:L}, and the smooth one-monopole moduli space has no cone tip, so an analogous short-distance inverse-square problem would need a singular monopole or a line--monopole system. The semiclassical framed BPS states of 't~Hooft lines, which are $L^2$ kernels of Dirac operators on singular-monopole moduli spaces \cite{MRV2016}, are a magnetic counterpart of the instanton states that saturate the supersymmetric bound $L(\pm1)\ge-8$.
\item \emph{The six-dimensional $(2,0)$ Wilson-surface lift and infrared completions.} Bridging our fixed-$\zeta$, $k=1$ instability to the holographic large-$N$ defect flows of Ref.~\cite{DGTZ} requires the five field-theoretic advances identified in Sec.~\ref{sec:lightcone}, above all extending the non-abelian Wilson-surface dictionary of Ref.~\cite{AKKS} to non-minuscule representations and $|\zeta|<1$ on a null circle, computing any running of $\zeta$ from non-BPS determinants and null-momentum exchange, and testing whether separated caloron constituents at $\rho\gg\ell_{\rm IR}$ bind additional states beyond Eq.~\eqref{eq:main:efimov}. The same programme would decide whether the screening suggested by the critical quantum numbers is the endpoint of the instability. Taking the null limit of the twisted $A_2$ compactification would test whether the $\theta_{5d}=\pi$ threshold also has a fixed-null-momentum interpretation.
\end{enumerate}

 The instanton--Wilson-line system thus brings defect conformality, soliton moduli geometry, four-dimensional lowest Landau levels and Efimov scaling into one Hamiltonian, derived without weak-coupling or classical-charge approximations and analysed largely in closed form. Deciding whether multi-instanton clustering can break fundamental-line conformality at $k>1$, and how the single-channel $U(2)/U(3)$ ultraviolet boundary conditions lift to resolved six-dimensional surface defects, will test how far soliton self-localization reaches beyond the one-instanton sector of the light-cone $(2,0)$ theory.

\acknowledgments
A.C. thanks Ashoke Sen and Amihay Hanany for discussions, and Suhas Mahesh for sharing a paper on non-Lorentzian spacetimes. A.C. acknowledges the support and hospitality of the International Centre for Theoretical Sciences (ICTS), Bengaluru, during participation in the programme \emph{Monopole Moduli 2026} (ICTS/MM2026/09).

\section*{Use of large language models}
The authors acknowledge the use of the free versions of ChatGPT (GPT-5 series), Claude (5 series) and Gemini (3 series) for assistance with organizing the literature, secondary numerical cross-checking, and LaTeX formatting. All primary research, derivations, and conclusions remain the sole work of the authors, who retain full responsibility for the manuscript.

\section*{Data availability}
The exact-arithmetic verification scripts and output logs for Appendices~\ref{app:fb:block45}, \ref{app:un:uniform}, and~\ref{app:numerics}, and the datasets underlying the numerical tables and figures, are available upon reasonable request from the author.

\FloatBarrier
\appendix
\section{Gauge frames and the discrete theta projection on instanton states}
\label{app:frames}

 This appendix fixes the conventions behind the physical channels of Sec.~\ref{sec:channels}. All identities below were verified in exact arithmetic.

\emph{The canonical lift.} The ADHM data of Sec.~\ref{sec:connection} transform as $\Delta\mapsto\operatorname{diag}(h,\pm1)\,\Delta\,(\pm1)$, with $h\in Sp(1)$ the global gauge rotation and the common sign in $O(1)=\{\pm1\}$ the redundancy \cite{ADHM}. The element $-1\in O(1)$ maps $(\lambda,u)$ to $(-\lambda,u)$ and carries the kernel of $\Delta^\dagger$ to itself by $v\mapsto\mathsf Tv$ with $\mathsf T=\operatorname{diag}(1,-1)$. The element $-\mathsf T=\operatorname{diag}(-1,1)$ also maps $\lambda$ to $-\lambda$, but it is the central element $h=-1$ of the global gauge group acting alone, a physical symmetry, distinct from a gauge redundancy. A physical section $\psi$, written in a frame $v$, therefore obeys
\begin{equation}
 v(-\lambda)\,\psi(-\lambda)=\mathsf T\,v(\lambda)\,\psi(\lambda).
 \label{eq:app:lift}
\end{equation}

\emph{Two frames.} First take a bosonic section at $\theta_{5d}=0$. The frame~\eqref{eq:frame} obeys $v_F(-\lambda)=\mathsf Tv_F(\lambda)$, so in it Eq.~\eqref{eq:app:lift} reads $\psi_F(-\lambda)=\psi_F(\lambda)$. The frame
\begin{equation}
 v_R=\frac1{\sqrt\varphi}\begin{pmatrix}\hat\lambda\bar u\\-|\lambda|\end{pmatrix},\qquad\hat\lambda=\lambda/|\lambda|,
\end{equation}
is regular at the pinned direction too, and it obeys $v_R(-\lambda)=-\mathsf Tv_R(\lambda)$, as does the frame used for the caloron in Ref.~\cite{KraanVanBaal}. In it $\psi_R(-\lambda)=(-1)^{2s}\psi_R(\lambda)$. The two frames are related by $v_F=v_R\,\bar{\hat\lambda}$, an odd gauge function, which accounts for the sign. Under $(p,q,h)\in G$ the frames transform as
\begin{equation}
\begin{aligned}
 v_F(h\lambda\bar q,pu\bar q)&=\operatorname{diag}(h,p)\,v_F(\lambda,u)\,\hat\lambda\bar pq\bar{\hat\lambda}\bar h,\\
 v_R(h\lambda\bar q,pu\bar q)&=\operatorname{diag}(h,p)\,v_R(\lambda,u)\,\bar p .
\end{aligned}
\end{equation}
In the frame $v_R$ on bosonic colour sections, the global gauge rotation acts through the moduli. At $\theta_{5d}=0$ its $U(1)$ subgroup is generated by $-\ii\partial_\chi$ along the orbit $\lambda(\chi)=e^{\ii\chi T_3}\lambda$, and the weights $m$ of this generator obey $m\in s+\Z$ and $|m|\le n/2$ on harmonics of degree $n$. In the frame $v_F$ the same generator includes a rotation of the colour, and a change of frame $\psi'=\mathcal U^\dagger\psi$ maps it to $\mathcal U^\dagger(-\ii\partial_\chi)\mathcal U$. At $\theta_{5d}=0$ the central element $h=-1$ acts on physical bosonic sections in both frames by $(-1)^{2s}$, in $v_R$ through the moduli and in $v_F$ through the colour factor $D_s(-1)$. The physical charge is the $SU(2)_G$ representation $j_G$, and the rule~\eqref{eq:physical} is frame independent.

\emph{The fermion lift and the theta angle.} The redundancy acts on the size zero modes by $-1$, and its lift to the Fock space is determined up to sign by that action, $\Gamma_{O(1)}=\pm\Gamma_y$. The central element $-1\in SU(2)_G$ acts on the moduli and on the zero modes exactly as the redundancy does, and on the line by $(-1)^{2s}$. With all dynamical fields in the adjoint representation and a large gauge transformation in $\pi_4(SU(2))=\Z_2$ weighted by $e^{\ii\theta_{5d}}$, this element acts on physical states by $e^{\ii\theta_{5d}}(-1)^{2s}$; setting $\theta_{5d}=0$ therefore selects $\Gamma_{O(1)}=U_F(-1)=(-1)^{2j_G^F}$. In the Clifford conventions of our numerical code, $U_F(-1)=\exp\bigl(\pi\,dS(\operatorname{diag}([e_a],0))\bigr)$ (on each fermion set, with $dS$ the differential of the spin representation of $\mathfrak{so}(8)$ on its sixteen-dimensional Clifford module), built from the $SU(2)_G$ generators acting on the size slot $\lambda$, is the same for $a=1,2,3$, squares to one, anticommutes with the four size-direction generators and commutes with the others, equals the product of the size-direction generators with sign $+1$, and equals $(-1)^{2j_G}$ on every $SU(2)_G$ multiplet. On the blocks of one set it acts as $+1$ on all of $(1,3)$, as $+1$ on one and $-1$ on four states of $(5,1)$, and as $+1$ and $-1$ on four states each of $(4,2)$. At $s=0$, the lift $+\Gamma_y$ gives only integer $j_G$, with $j_G\in\{0,1\}$ at the lowest level, while $-\Gamma_y$ gives only half-integer $j_G$, with a gauge doublet at the lowest level. This is the distinction between the two values of the discrete theta angle found in Ref.~\cite{BRZ} for the pure $SU(2)$ theory with eight supercharges, where the choice of projection $\tfrac12(1\pm(-1)^P)$, with $P$ the element $-1\in O(1)$, is identified with $\theta_{5d}$ and the ground state of the instanton particle is gauge-charged only for $\theta_{5d}=\pi$. Tachikawa shows that the reduction of the $SO(2n+2)$ theory with a $\Z_2$ twist gives $USp(2n)$ with $\theta_{5d}=0$, and that $\theta_{5d}=\pi$ arises from the twisted $SU(2n+1)$ theory \cite{Tachikawa}. For $n=1$, the twisted $SO(4)$ theory is the $A_1$ theory on a circle of twice the length, so the reduction of the $A_1$ theory without twist gives $\theta_{5d}=0$ by the exchange of the two $A_1$ factors. The double-circle argument is our inference from the twisted reduction. The main text takes $\theta_{5d}=0$ as the reference sector, while Sec.~\ref{sec:bps} and Appendices~\ref{app:fb:s12} and~\ref{app:fb:block45} extend the complete phase classification ($s=\tfrac12$ conformality and an instability range for $s\ge1$) to $\theta_{5d}=\pi$.

\emph{The curvature on the base curve.} At $(\lambda,u)=(\cos\alpha,-\sqrt2\sin\alpha)$, $\varphi=1+\sin^2\alpha=D$ and $v_F=(-\sqrt2\sin\alpha,-\cos\alpha)/\sqrt D$. With $\xi_\lambda=\xi_y$ and $\xi_u=-\sqrt2\,\xi_Z$, $v_F^\dagger\xi=(2/D)^{1/2}(\cos\alpha\,\xi_Z-\sin\alpha\,\xi_y)$, and Eq.~\eqref{eq:curvature} gives Eq.~\eqref{eq:basecurve}. The general formula was checked against the curvature of the connection at five rational points, the base-curve form symbolically in $\alpha$, and the right-$Sp(1)$ invariance exactly.

\section{The fermion couplings of the line}
\label{app:fermions}

 This appendix derives Eq.~\eqref{eq:Hzeta} from the action of Ref.~\cite{LO}. The derivation fixes the sign and the normalization of the line--bilinear term and shows that eliminating $A_t$ and $X^6$ produces no other term that depends on the line. We use its conventions, taking a real representation of the Clifford algebra of $Spin(1,10)$, $\bar\Psi=\Psi^{\rm T}\Gamma_t$, $\Gamma_{t12345}\Psi=-\Psi$, $\Gamma_\pm=(\Gamma_5\pm\Gamma_t)/\sqrt2$, and the fermionic terms $\frac1{g_5^2}\tr\bigl(\frac\ii2\bar\Psi\Gamma_-D_t\Psi+\frac\ii2\bar\Psi\Gamma_iD_i\Psi-\frac12\bar\Psi[X^I,\Gamma_-\Gamma^I\Psi]\bigr)$. The fields are Hermitian, $D_\mu=\partial_\mu-\ii[A_\mu,\cdot\,]$, and the line contributes $\int\dd t\,\tr\,Q\,(A_t+\zeta X^6)$ at the origin, with $Q=2J^aT_a$ as in Sec.~\ref{sec:selfenergy}.

\emph{The dynamical fermions are the zero modes.} Let $P_1=\frac12\Gamma_+\Gamma_-=\frac12(1+\Gamma_t\Gamma_5)$ and $P_2=1-P_1$. Since $\Gamma_t\Gamma_-=\sqrt2P_1$, the time derivative enters only through $\bar\Psi\Gamma_-\partial_t\Psi=\sqrt2\,(P_1\Psi)^{\rm T}\partial_t(P_1\Psi)$, and the Yukawa terms also involve $P_1\Psi$ alone. The matrices $\Gamma_i$ and $\Gamma^I$ commute with $P_1$, while $\Gamma_t$ exchanges $P_1$ and $P_2$, so $P_2\Psi$ enters the action only through the spatial Dirac term, and linearly. Like the multiplier $G_{ij}$ for the gauge field, it imposes $\Gamma_iD_iP_1\Psi=0$, which restricts the dynamical fermions exactly to the zero modes of the Dirac operator in the instanton background. The line does not enter that operator.

\emph{The zero modes.} Let $m^a$ be the coordinates $(Z,y)$ of Eq.~\eqref{eq:metric} and $\delta_aA_i$ the zero modes in background gauge, $D_i\delta_aA_i=0$, so that $\int\tr(\delta_aA_i\delta_bA_i)\,\dd^4x=8\pi^2\delta_{ab}$. Let $\mathcal E_{\rm sp}$ be the space of constant spinors with $\Gamma_+\varepsilon=0$ and $\Gamma_{t12345}\varepsilon=\varepsilon$. It has dimension eight, $\Gamma_{1234}=1$ on it, and the self-dual combinations of the $\Gamma_{ij}$ annihilate it. For $\varepsilon\in\mathcal E_{\rm sp}$ the spinors $\chi_a=\Gamma^i\delta_aA_i\,\varepsilon$ lie in $P_1\Psi$ and are zero modes, because $\Gamma^j\Gamma^iD_j\delta_aA_i=D_i\delta_aA_i+\frac12\Gamma^{ji}\delta_aF_{ji}$ and $\delta_aF$ is self-dual. Since $\Gamma^6$ anticommutes with $\Gamma^i$, the four-dimensional subspaces $\mathcal E_{\rm sp,\mp}$ with $\Gamma^6\varepsilon=\mp\varepsilon$ give zero modes with $\Gamma^6=\pm1$, and $\Gamma^7$ exchanges the two eigenspaces, so each holds eight of the sixteen zero modes. The spinors $\Gamma^i\varepsilon$ are orthogonal with common norm $|\varepsilon|$, so for a fixed $\varepsilon\in\mathcal E_{\rm sp,\mp}$ the eight $\chi_a$ are linearly independent (matching the eight real background-gauge moduli $\delta_aA_i$ and the eight-dimensional $\Gamma^6=\pm1$ half of the sixteen instanton zero modes) and span the set $\chi_\pm$. This is the identification of each set with the tangent space. The centre directions give the supersymmetric modes and the size directions the superconformal ones.

\emph{Bilinears.} For $\varepsilon\in\mathcal E_{\rm sp}$,
\begin{equation}
 (\Gamma^i\varepsilon)^{\rm T}\,\Gamma_t\Gamma_-\,\Gamma^j\varepsilon=\sqrt2\,|\varepsilon|^2\delta_{ij},
 \label{eq:ff:bilinear}
\end{equation}
because $\Gamma^i$ is symmetric and commutes with $P_1$, and $\varepsilon^{\rm T}\Gamma^{ij}\varepsilon=0$ for a real spinor. Write the fermions of the set $\chi_+$ as $\Psi=\sum_a\psi^a\chi_a$ with a fixed $\varepsilon\in\mathcal E_{\rm sp,-}$. The kinetic term becomes $\frac\ii2\,(8\sqrt2\pi^2|\varepsilon|^2/g_5^2)\sum_a\psi^a\dot\psi^a$, together with fermion--velocity terms from the moduli dependence of $\chi_a$, so the canonically normalized modes $\theta^a=(8\sqrt2\pi^2|\varepsilon|^2/g_5^2)^{1/2}\psi^a$ quantize as $\theta^a=\gamma^a_+/\sqrt2$, with $\gamma^a_+$ the Clifford generators of the set. Bilinears that pair the two sets vanish in every term that contains $\Gamma_-(1+\zeta\Gamma^6)$, because $\Gamma^6$ is symmetric and commutes with $\Gamma^i\Gamma^j$, and $1+\zeta\Gamma^6$ acts on $\chi_\pm$ as $1\pm\zeta$.

\emph{Elimination.} Write $\mathscr L=-D_iD_i$ and $\langle\mathcal X,\mathcal Y\rangle=\int\tr\mathcal X\mathcal Y\,\dd^4x$. The terms with $A_t$ are $\frac1{2g_5^2}\langle A_t,\mathscr LA_t\rangle+\langle A_t,S\rangle$, with $S=g_5^{-2}(D_i\partial_tA_i+\mathcal J)+Q\delta_0$ and $\tr(\mathcal X\mathcal J)=\frac12\tr(\bar\Psi\Gamma_-[\mathcal X,\Psi])$, and eliminating $A_t$ gives $-\frac{g_5^2}2\langle S,\mathscr L^{-1}S\rangle$. The quadratic term of $X^6$ has the opposite sign, and its source is $g_5^{-2}\mathcal J_6+\zeta Q\delta_0$ with $\tr(\mathcal X\mathcal J_6)=-\frac12\tr(\bar\Psi\Gamma_-\Gamma^6[\mathcal X,\Psi])$, so eliminating $X^6$ gives $+\frac{g_5^2}2\langle S_6,\mathscr L^{-1}S_6\rangle$. The terms that contain $Q$ are therefore the self-energy of Sec.~\ref{sec:selfenergy}, the line--velocity term $\tr\,Q\,A_t^{\rm v}(0)$ with $A_t^{\rm v}=-\mathscr L^{-1}D_i\partial_tA_i$, and the line--bilinear term $\tr\,Q\,A_t^{\rm f}(0)$, where
\begin{equation}
 A_t^{\rm f}=-\mathscr L^{-1}\mathcal J_\zeta,\qquad
 \tr(\mathcal X\mathcal J_\zeta)=\tfrac12\tr\bigl(\bar\Psi\Gamma_-(1+\zeta\Gamma^6)[\mathcal X,\Psi]\bigr),
 \label{eq:ff:Af}
\end{equation}
which is Eq.~\eqref{eq:fermioncoupling} evaluated on the field of the line. No other term contains $Q$, since $X^I$ with $I\ne6$ has no line source and $X^6$ has no velocity source.

\emph{The universal connection.} Following Atiyah and Singer \cite{AtiyahSinger1984}, extend $A_i$ to a connection $\mathbb A$ on $\R^4\times\mathcal M$ whose moduli components are fixed by the background gauge, $\partial_aA_i=\delta_aA_i+D_i\mathbb A_a$. Its mixed curvature is $\mathbb F_{ia}=D_i\mathbb A_a-\partial_aA_i=-\delta_aA_i$. Three properties of $\mathbb A$ give the result. First, $D_i\partial_tA_i=\dot m^aD_iD_i\mathbb A_a$, so in the decaying gauge $\mathbb A_a(\infty)=0$ where $\mathscr L$ is invertible, $A_t^{\rm v}=\dot m^a\mathbb A_a$. This is the solution for which $F_{ti}=\dot m^a\delta_aA_i$ is a background-gauge zero mode, and hence the one whose kinetic energy is that of Eq.~\eqref{eq:metric}. Second, the Bianchi identity $D_i\mathbb F_{ab}=D_a\mathbb F_{ib}-D_b\mathbb F_{ia}$ and the background gauge $D_i\mathbb F_{ia}=0$ give
\begin{equation}
 D_iD_i\mathbb F_{ab}=[D_i,D_a]\mathbb F_{ib}-[D_i,D_b]\mathbb F_{ia}=-2\ii\,[\delta_aA_i,\delta_bA_i],
 \label{eq:ff:universal}
\end{equation}
and since $\mathbb F_{ab}$ vanishes at infinity, $\mathscr L^{-1}[\delta_aA_i,\delta_bA_i]=-\frac\ii2\mathbb F_{ab}$. Third, in the ADHM construction $\mathbb A=\ii U^\dagger\dd U$, with $U(x)$ the normalized kernel of $\Delta(x)^\dagger$ and $\dd$ taken on $\R^4\times\mathcal M$. At the origin $U=v$, so $\mathbb A_a(0)=\ii\mathcal A_a$ and $\mathbb F_{ab}(0)=\ii\mathcal F_{ab}$, with $\mathcal A$ and $\mathcal F$ the evaluation connection and its curvature, Eq.~\eqref{eq:curvature} (under $e_c=-2\ii T_c$, their components obey $\mathbb A_a^c(0)=2\mathcal A_a^c$ and $\mathbb F_{ab}^c(0)=2\mathcal F_{ab}^c$). We have checked the background gauge of $\mathbb F_{ia}$, Eq.~\eqref{eq:ff:universal} for all pairs of moduli directions, and the value of $\mathbb F_{ab}$ at the origin in exact arithmetic at random points.

\emph{The result.} On the set $\chi_+$, Eq.~\eqref{eq:ff:bilinear} gives $\mathcal J_\zeta=-\frac{1+\zeta}{\sqrt2}\,|\varepsilon|^2\sum_{a,b}\psi^a\psi^b[\delta_aA_i,\delta_bA_i]$, and therefore
\begin{equation}
 A_t^{\rm f}(0)=-\ii\,c_+\,\frac{g_5^2}{16\pi^2}\sum_{a,b}\theta^a\theta^b\,\mathbb F_{ab}(0).
\end{equation}
The line--velocity term adds $J^c\mathbb A^c_a(0)\,\dot m^a$ to the Lagrangian, so the kinetic operator becomes $-\nabla_a\nabla_a$, in units of $g_5^2/16\pi^2$, with $\nabla_a=\partial_a-\ii J^c\mathbb A^c_a(0)$ and $[\nabla_a,\nabla_b]=-\ii J^c\mathbb F^c_{ab}(0)$. The line--bilinear term of the Hamiltonian, $-\tr\,QA_t^{\rm f}(0)$, is then, in the same units,
\begin{equation}
 \ii\,c_+\sum_{a<b}\gamma^{ab}_+J^c\,\mathbb F^c_{ab}(0)=-c_+\sum_{a<b}\gamma^{ab}_+[\nabla_a,\nabla_b]=c_+P^{(+)},
 \label{eq:ff:result}
\end{equation}
and the set $\chi_-$ gives $c_-P^{(-)}$ in the same way. The factors $c_\pm$ and the unit $g_5^2/16\pi^2$ come out of the computation, and the normalization of $\varepsilon$ drops out. At $\zeta=\pm1$ Eq.~\eqref{eq:ff:result} is the Lichnerowicz term of ${\not{\!\mathbf D}}_\pm^\dagger{\not{\!\mathbf D}}_\pm$, so the $\mathcal N=(0,4)$ supersymmetry of the endpoints is an independent check on Eq.~\eqref{eq:Hzeta}. At fixed nonzero moduli radius $R$, this term is finite for a point source because $\mathbb F_{ab}$ is smooth at the spatial position of the line.

\emph{Terms without the line.} The remaining terms are those of the one-instanton quantum mechanics without a probe. Within one set the exchanges of $A_t$ and $X^6$ between bilinears cancel identically, since $\mathcal J_6=\mp\mathcal J$ on the set $\chi_\pm$ and the two eliminations enter with opposite signs. The exchanges between the sets, of $A_t$, $X^6$ and $X^7,\dots,X^{10}$, and the fermion--velocity terms belong to the reduction of the probe-free theory, which is the supersymmetric sigma model on the moduli space. For $k=1$ the moduli space is flat and they vanish. Equation~\eqref{eq:Hzeta} therefore follows from the constrained action, under the range condition of Sec.~\ref{sec:selfenergy} and the quantization of Sec.~\ref{sec:models}.

\section{Hopf reduction and the lower bound on the angular spectrum}
\label{app:floor}

 This appendix proves Proposition~\ref{thm:floor} in the sharper fibre-resolved form $L(\pm1)\ge4j_R(j_R+1)-8$ (Eq.~\eqref{eq:fb:H4}), characterizes equality by twisted Dirac kernels and counts the lowest states at $j_R=0$, and derives the resulting threshold and spectral-gap bounds.

\subsection{The Dirac spectral gap and the quaternionic Hopf fibration}
\label{app:fb:conventions}\label{app:fb:dirac}

We retain the link coordinates, profile $W=\kappa/D^2$, attraction $E=4C_2(1-\zeta^2)$, link operator $L(\zeta)$ (Eq.~\eqref{eq:radial}), and physical channels $(j_L,j_R,j_G)$ of Secs.~\ref{sec:system}--\ref{sec:spectral}, writing $L_\pm=L(\pm1)$ and $L_0(\zeta)=c_+L_++c_-L_-=L(\zeta)+EW$, with the symmetry group $G=SU(2)_L\times SU(2)_R\times SU(2)_G$ of Sec.~\ref{sec:channels}. Each fermion set is an $\R^8$ Clifford module $S=S_{\R^8}^+\oplus S_{\R^8}^-$ decomposing under $Sp(2)\times Sp(1)_{\rm F}$ ($Sp(1)_{\rm F}$ the right rotation of the set) into $S_{\R^8}^+=(5,1)\oplus(1,3)$ and $S_{\R^8}^-=(4,2)$, with $P_A$ denoting the Pauli term in block $A\in\{5,1,4\}$ ($P_1=0$). At $\zeta=+1$ only the $(+)$ set has a Pauli coupling, while the $(-)$ set is a flat spectator, and vice versa at $\zeta=-1$. On any physical doubly-BPS state $T$ obeying $(L_++8)T=(L_-+8)T=0$,
\begin{equation}\label{eq:fb:trial}
\langle T,L(\zeta)T\rangle=-8\|T\|^2-E\,\langle T,WT\rangle ,
\end{equation}
so min--max proves loss of conformality whenever $E\langle W\rangle_T>1$ ($\langle W\rangle_T=\langle T,WT\rangle/\|T\|^2$).

Fix $\zeta=+1$. The case $\zeta=-1$ follows by exchanging the two sets, and the spectator is suppressed from the notation.
Let $\Gamma^\mu$ ($\mu=1,\dots,8$) be Hermitian generators of the Clifford algebra of $\R^8$ acting on $S$ (the generators $\gamma^a_+$ of Appendix~\ref{app:fermions}, the spectator suppressed), write $\Gamma(v)=v_\mu\Gamma^\mu$ and $\hat x=x/R$, and let ${\not{\!\mathbf D}}=\Gamma^\mu\nabla_\mu$, $\nabla=\dd+\mathcal A$, be the colour-twisted Dirac operator on $\R^8\setminus0$.
It is anti-Hermitian, and ${\not{\!\mathbf D}}^\dagger{\not{\!\mathbf D}}$ is the supersymmetric Hamiltonian at $\zeta=1$, whose link operator is $L(1)$ (Eq.~\eqref{eq:Hzeta}),
\begin{equation}\label{eq:fb:Hsusy}
{\not{\!\mathbf D}}^\dagger{\not{\!\mathbf D}}=-\partial_R^2-\frac7R\,\partial_R+\frac{L(1)}{R^2}.
\end{equation}
We use the frame~\eqref{eq:frame}, which is homogeneous of degree zero, so that $\iota_{x\cdot\partial}\mathcal A=0$ and $\hat x\cdot\nabla=\partial_R$.
Define the spherical Dirac operator $\mathcal D$ on $S^7$ by
\begin{equation}\label{eq:fb:Dsph}
{\not{\!\mathbf D}}=\Gamma(\hat x)\Big(\partial_R+\frac{7/2-\mathcal D}{R}\Big).
\end{equation}
We show in turn that $\mathcal D$ is self-adjoint with a chirality symmetry, that it determines $L(1)$, and that the floor is a spectral gap of $\mathcal D$.

The operator $\mathcal D$ is self-adjoint on $L^2(S^7;S\otimes V_s)$, preserves $S_{\R^8}^\pm$, and satisfies $\Gamma(\hat x)\,\mathcal D\,\Gamma(\hat x)=-\mathcal D$, so that $\spec(\mathcal D|_{S_{\R^8}^-})=-\spec(\mathcal D|_{S_{\R^8}^+})$.
To see this, let $\{e_i\}_{i=1}^7$ be an orthonormal frame tangent to the sphere of radius $R$. Since $\Gamma(\hat x)^2=1$,
$\Gamma(\hat x){\not{\!\mathbf D}}=\hat x\cdot\nabla+\sum_i\Gamma(\hat x)\Gamma(e_i)\nabla_{e_i}$, so $\mathcal D=\frac72+\mathcal D_0$ with $\mathcal D_0=-R\sum_i\Gamma(\hat x)\Gamma(e_i)\nabla_{e_i}$.
The operator $\mathcal D_0$ contains only tangential derivatives and is even in the Clifford algebra, so it acts on $S^7$ and preserves $S_{\R^8}^\pm$.
Using $\nabla_{e_i}\hat x=e_i/R$ and $\Gamma(e_i)\Gamma(\hat x)=-\Gamma(\hat x)\Gamma(e_i)$,
\begin{equation}
\Gamma(\hat x)\mathcal D_0\Gamma(\hat x)=-R\sum_i\Gamma(e_i)\Big(\frac{\Gamma(e_i)}{R}+\Gamma(\hat x)\nabla_{e_i}\Big)=-7-\mathcal D_0 ,
\end{equation}
which is $\Gamma(\hat x)\mathcal D\Gamma(\hat x)=-\mathcal D$.
On $L^2(\R^8)=L^2(\R_+,R^7\dd R)\otimes L^2(S^7)$ the adjoint of the radial derivative is $\partial_R^\dagger=-\partial_R-7/R$; taking the adjoint of \eqref{eq:fb:Dsph} and comparing with ${\not{\!\mathbf D}}^\dagger=-{\not{\!\mathbf D}}$ gives $\Gamma(\hat x)\mathcal D^\dagger\Gamma(\hat x)=-\mathcal D$, hence $\mathcal D^\dagger=\mathcal D$ on smooth sections.
Since $\mathcal D$ is a first-order elliptic operator on the compact manifold $S^7$, it is essentially self-adjoint.
The unitary $\Gamma(\hat x)$ exchanges $S_{\R^8}^+$ and $S_{\R^8}^-$ and intertwines $\mathcal D|_{S_{\R^8}^+}$ with $-\mathcal D|_{S_{\R^8}^-}$.

The link operator is a function of $\mathcal D$, namely $L(1)=(\mathcal D-\frac12)^2-9$. In particular $L(1)\ge-9$.
For $\psi=R^{-7/2}\tilde\psi$, \eqref{eq:fb:Dsph} gives $R^7|{\not{\!\mathbf D}}\psi|^2=|\partial_R\tilde\psi-R^{-1}\mathcal D\tilde\psi|^2$.
Because $\mathcal D$ is self-adjoint and independent of $R$, the cross term integrated over $S^7$ is $-R^{-1}\partial_R\langle\tilde\psi,\mathcal D\tilde\psi\rangle$, and an integration by parts in $R$ gives
\begin{equation}\label{eq:fb:radialform}
\int_{\R^8}|{\not{\!\mathbf D}}\psi|^2\,\dd^8x=\int_0^\infty\!\dd R\,\Big(\|\partial_R\tilde\psi\|^2+\frac{\langle\tilde\psi,(\mathcal D^2-\mathcal D)\tilde\psi\rangle}{R^2}\Big),
\end{equation}
with norms in $L^2(S^7)$.
The same substitution in \eqref{eq:fb:Hsusy} gives $\langle\psi,{\not{\!\mathbf D}}^\dagger{\not{\!\mathbf D}}\psi\rangle=\int_0^\infty\dd R\,\big(\|\partial_R\tilde\psi\|^2+R^{-2}\langle\tilde\psi,(L(1)+\frac{35}4)\tilde\psi\rangle\big)$.
Choosing $\tilde\psi=f(R)\chi$ with arbitrary $f\in C^\infty_c(0,\infty)$ and $\chi\in C^\infty(S^7)$ shows that $L(1)+\frac{35}4=\mathcal D^2-\mathcal D$, which is the stated identity.

The floor is therefore a spectral gap of $\mathcal D$, and equivalently a Hardy inequality for ${\not{\!\mathbf D}}$.
The following three conditions are equivalent, namely (i) $L(1)\ge-8$ in blocks 4 and 5; (ii) $\mathcal D^2\ge\frac94$; (iii) for every $\psi\in C^\infty_c(\R^8\setminus0;S\otimes V_s)$,
\begin{equation}\label{eq:fb:hardy}
\int_{\R^8}|{\not{\!\mathbf D}}\psi|^2\,\dd^8x\ \ge\ \int_{\R^8}\frac{|\psi|^2}{R^2}\,\dd^8x .
\end{equation}
The Pauli term vanishes on block 1, where $L(1)=-\Delta_{S^7}^{\mathcal A}\ge0$, and $L(1)$ preserves the blocks. So (i) holds if and only if $(\mathcal D-\frac12)^2\ge1$ on all of $S\otimes V_s$, that is, if and only if neither $\spec(\mathcal D|_{S_{\R^8}^+})$ nor $\spec(\mathcal D|_{S_{\R^8}^-})$ meets $(-\frac12,\frac32)$. By the chirality symmetry this says that $\spec(\mathcal D|_{S_{\R^8}^-})$ meets neither $(-\frac32,\frac12)$ nor $(-\frac12,\frac32)$, that is, $\mathcal D^2\ge\frac94$ on $S_{\R^8}^-$, and, by the same symmetry, on $S_{\R^8}^+$. This is (ii).
For (iii), write $f=R^{1/2}g$ for $f\in C^\infty_c(0,\infty)$; then $\int_0^\infty|f'|^2\dd R=\int_0^\infty R|g'|^2\dd R+\frac14\int_0^\infty R^{-2}|f|^2\dd R$, so the one-dimensional Hardy inequality $\int|f'|^2\ge\frac14\int R^{-2}|f|^2$ holds with a sharp constant, approached by slowly varying $g$.
By \eqref{eq:fb:radialform}, (iii) therefore holds if and only if $\mathcal D^2-\mathcal D-1+\frac14\ge0$, that is, $(\mathcal D-\frac12)^2\ge1$, which is (i).
When these conditions hold, $L(1)\chi=-8\chi$ if and only if $\mathcal D\chi=\frac32\chi$. Indeed, by the identity $L(1)=(\mathcal D-\frac12)^2-9$, $L(1)\chi=-8\chi$ means $(\mathcal D-\frac12)^2\chi=\chi$, so $\chi$ is a combination of eigenvectors with eigenvalues $\frac32$ and $-\frac12$, and the second is excluded by (ii).

The floor states at $\zeta=1$ are thus the eigenvectors of $\mathcal D$ with eigenvalue $\frac32$, in either chirality, and $R^{-2}\chi$ is then a zero mode of ${\not{\!\mathbf D}}$ on $\R^8\setminus0$. By the chirality symmetry, $\Gamma(\hat x)$ maps them to eigenvectors with eigenvalue $-\frac32$ of the opposite chirality, on which $L(1)=-5$.
Conformality at $\zeta=\pm1$ is automatic from $L(1)\ge-9$.

\paragraph*{Hopf structure and horizontal splitting.}\phantomsection\label{app:fb:hopf}

Let $N:\HH^2\to\HH^2$, $N(Z,y)=(y,-\sqrt2\,Z)$.
The colour connection is $\mathcal A=\tilde v^\dagger\dd\tilde v$, where $\tilde v(x)\in\HH^2$ is a unit vector with $\tilde v^\dagger N(x)=0$. This is the evaluation connection of Sec.~\ref{sec:connection}, with $(\lambda,u)=N(Z,y)$.
Let $\pi:\HH^2\setminus0\to\HH P^1\cong S^4$, $x\mapsto[x]=x\HH$, be the quaternionic Hopf projection, and let $a$ be the basic instanton on $\HH P^1$, the connection induced by orthogonal projection on the quaternionic line bundle whose fibre at $[v]$ is $(v\HH)^\perp$~\cite{Atiyah1979,AHS}.

The map $N$ induces a conformal diffeomorphism $\bar N$ of the round $\HH P^1=S^4$ with $\pi\circ N=\bar N\circ\pi$, and, up to gauge,
\begin{equation}\label{eq:fb:pullback}
\mathcal A=(\pi\circ N)^*a=\pi^*a_N,\qquad a_N=\bar N^*a .
\end{equation}
The connection $a_N$ is an anti-self-dual instanton of charge one on the round $S^4$.
The conditions $U^\dagger N(x)=0$ and $|U|=1$ determine $\tilde v$ up to $\tilde v\mapsto\tilde v\,g$ with $g\in Sp(1)$, under which $\mathcal A$ undergoes a gauge transformation. They depend on $x$ only through the line $N(x)\HH=N(x\HH)$, because $N$ commutes with right multiplication. Choosing $\tilde v$ locally as a function of $[N(x)]$ gives $\mathcal A=(\pi\circ N)^*a$.
Since $N\in GL(2,\HH)$ maps right $\HH$-lines to right $\HH$-lines, it induces $\bar N$ with $\pi\circ N=\bar N\circ\pi$. The group $GL(2,\HH)$ acts on $\HH P^1=S^4$ by orientation-preserving conformal transformations, and the anti-self-duality equation is conformally invariant, so $a_N$ is again an anti-self-dual instanton.

Consequently the curvature $\mathcal F$ of $\mathcal A$ obeys $\iota_X\mathcal F=0$ for every $X$ tangent to the quaternionic line $x\HH$, which contains the dilation and the $SU(2)_R$ orbit through $x$. On the horizontal four-plane $(x\HH)^\perp$ it is anti-self-dual, in the orientation in which the K\"ahler forms of right multiplication by $i,j,k$ are self-dual.
In particular $\mathcal F$ takes values in $sp(2)\otimes su(2)$.
This holds because $\mathcal F=\pi^*\mathcal F_{a_N}$ vanishes on $\ker\dd\pi_x=x\HH$, and $\dd\pi_x$ maps $(x\HH)^\perp$ conformally onto $T_{[x]}S^4$. An anti-self-dual form on $(x\HH)^\perp$ commutes with the right complex structures and acts on $(x\HH)^\perp$ by left multiplication, so it lies in $sp(2)$.

The factor $\sqrt2$ is the ratio of the centre and size normalizations of the moduli metric. It enters only through $\bar N$, which moves the instanton by a conformal map of $S^4$ (for the Hopf structure of the BPST instanton see Refs.~\cite{Trautman1977}).
The Hopf base carries the round metric $\mathsf g_1$ of $S^4(\frac12)$ (written in sans-serif to distinguish it from the polar profiles $g_b$ of Sec.~\ref{sec:floorspace}), which is $Sp(2)$-homogeneous, whereas $a_N$ is homogeneous under the conjugate group $\bar N^{-1}Sp(2)\bar N$. The two groups share the subgroup $SU(2)_L\times SU(2)_G$, since $N\,{\rm diag}(p,h)={\rm diag}(h,p)\,N$. The factor $\sqrt2$ is therefore the only obstruction to $Sp(2)$ homogeneity.
As an independent check, at six random points the vertical contractions $\iota_X\mathcal F$ vanish to $4.1\times10^{-16}$ and the self-dual part of $\mathcal F$ on the horizontal plane vanishes to $6.2\times10^{-16}$ (both below $7\times10^{-16}$). %

To decompose each channel into vertical (fibre) and horizontal (base) kinetic terms, let $\xi_{L,a}(x)=(e_aZ,0)$, $\xi_{G,a}(x)=(0,e_ay)$ and $\xi_{R,a}(x)=-(Ze_a,ye_a)$, $a=1,2,3$, be the vector fields generated by the imaginary units $e_a$ in the three factors of $G$.
The fields $\xi_{R,a}$ are orthonormal at every point of $S^7$ and span the tangent spaces of the Hopf fibres $x\,Sp(1)$.
At a base point $x_\alpha$, where $Z=\sin\alpha$ and $y=\cos\alpha$ are real, the fields $\xi_{L,a}$ and $\xi_{G,a}$ are mutually orthogonal, $|\xi_{L,a}|=\sin\alpha$, $|\xi_{G,a}|=\cos\alpha$, and $\xi_{R,a}=-(\xi_{L,a}+\xi_{G,a})$.
The unit vectors
\begin{equation}\label{eq:fb:VH}
\nu_a=\xi_{L,a}+\xi_{G,a},\qquad  Y_a=\cot\alpha\,\xi_{L,a}-\tan\alpha\,\xi_{G,a}
\end{equation}
are respectively vertical and horizontal, and $\{Y_a,\partial_\alpha\}$ is an orthonormal basis of the horizontal four-plane at $x_\alpha$.

In this frame every channel splits into a vertical Casimir and a horizontal operator.
On the subspace of $SU(2)_R$ spin $j_R$, in every channel, block and parity,
\begin{equation}\label{eq:fb:H3}
L(\pm1)=4j_R(j_R+1)+\mathcal H_A ,
\end{equation}
where $A$ is the block of the active set and the horizontal Pauli operator $\mathcal H_A$ is the operator of the quadratic form $\int_{S^7}\big(|\nabla^{h}\psi|^2+\langle\psi,P_A\psi\rangle\big)$, $\nabla^h$ being the covariant derivative along the horizontal distribution. In the coordinates of a channel,
$\mathcal H_A=-(\sin\alpha\cos\alpha)^{-3}\partial_\alpha\big(\sin^3\!\alpha\cos^3\!\alpha\,\partial_\alpha\big)+\sum_a\nabla_{Y_a}^*\nabla_{Y_a}+P_A$.

To derive the splitting, note that the tangent bundle of $S^7$ splits orthogonally into the vertical bundle, spanned by the orthonormal fields $\xi_{R,a}$, and the horizontal bundle. Hence the Dirichlet form of $-\Delta_{S^7}^{\mathcal A}$ is $\sum_a\|\nabla_{\xi_{R,a}}\psi\|^2+\|\nabla^h\psi\|^2$, and the Pauli term is unaffected.
In the gauge of Eq.~\eqref{eq:fb:pullback}, $\iota_{\xi_{R,a}}\mathcal A=0$, and $\mathcal A$ is invariant under right multiplication because $\pi$ is. Two infinitesimal lifts of a group action that preserve an irreducible connection differ by a parallel section of the adjoint bundle, hence coincide; so the $SU(2)_R$ action of $G$, which preserves $\mathcal A$ and does not act on the fermions, is pull-back of the coefficient functions in this gauge.
Thus $\nabla_{\xi_{R,a}}$ is the infinitesimal generator of $SU(2)_R$, and $\sum_a\nabla_{\xi_{R,a}}^*\nabla_{\xi_{R,a}}=4j_R(j_R+1)$ on spin $j_R$, the normalization of the Laplacian of the unit three-sphere, whose eigenvalues are $l(l+2)=4j(j+1)$.
In the coordinates of a channel, the orbit directions at $x_\alpha$ have the orthonormal basis $\xi_{L,a}/\sin\alpha$, $\xi_{G,a}/\cos\alpha$; with $u_a=\nabla_{\xi_{L,a}}\psi$ and $v_a=\nabla_{\xi_{G,a}}\psi$ the orbit part of the Dirichlet density is $\sum_a(|u_a|^2/\sigma+|v_a|^2/\kappa)$. The algebraic identity
\begin{equation}\label{eq:fb:split}
\frac{|u|^2}{\sigma}+\frac{|v|^2}{\kappa}=|u+v|^2+\frac{|\kappa u-\sigma v|^2}{\sigma\kappa}\qquad(\sigma+\kappa=1)
\end{equation}
separates it into the vertical part $|u_a+v_a|^2=|\nabla_{\nu_a}\psi|^2$ and the horizontal part $|\kappa u_a-\sigma v_a|^2/(\sigma\kappa)=|\nabla_{Y_a}\psi|^2$. Together with $|\partial_\alpha\psi|^2$, whose operator is the polar term of $\mathcal H_A$ given above, the latter forms the horizontal Dirichlet density.
All operators involved commute with $G$, so \eqref{eq:fb:H3} holds on every channel and parity subspace.

\subsection{The sharp bound at fixed angular momentum along the Hopf fibre}
\label{app:fb:sharp}

On the round sphere $S^4(\frac12)$ of radius $\frac12$ (the Hopf base, with metric $\mathsf g_1$ and scalar curvature $48$) let $S^\pm$ be its two-dimensional chiral spinor bundles (distinct from the $\R^8$ Clifford half-spaces of Sec.~\ref{app:fb:conventions}).
For $j\in\frac12\Z_{\ge0}$ let $F_j={\rm Sym}^{2j}S^+$ with its Levi-Civita connection, whose curvature is self-dual, and let $E_s$ be the spin-$s$ bundle of $a_N$, whose curvature is anti-self-dual.
For a bundle $\mathcal E$ with connection, $\mathsf D_{\mathcal E}=\sum_ie_i\cdot\nabla_{e_i}$ denotes the $\mathcal E$-twisted Dirac operator (with $e\cdot e=-|e|^2$), $\mathsf D_{\mathcal E,+}:C^\infty(S^+\otimes\mathcal E)\to C^\infty(S^-\otimes\mathcal E)$ and $\mathsf D_{\mathcal E,-}=\mathsf D_{\mathcal E,+}^\dagger$ its chiral parts, and $\mathfrak F_{\mathcal E}=\sum_{i<j}e_i\cdot e_j\cdot\otimes F^{\mathcal E}(e_i,e_j)$ the Clifford action of its curvature.
The Lichnerowicz formula~\cite{Lichnerowicz} on $S^4(\frac12)$ reads
\begin{equation}\label{eq:fb:lich}
\mathsf D_{\mathcal E}^2=\nabla^*\nabla+12+\mathfrak F_{\mathcal E},
\end{equation}
and self-dual (anti-self-dual) two-forms act by Clifford multiplication on $S^+$ ($S^-$) only.

We now prove Proposition~\ref{thm:floor} in the following sharper form.
For every spin $s$, integer or half-integer, every channel and parity, and on the subspace of $SU(2)_R$ spin $j_R$,
\begin{equation}\label{eq:fb:H4}
L(\pm1)\ \ge\
\begin{cases}
4j_R(j_R+1)-8, & \text{blocks 4 and 5},\\
4j_R(j_R+1), & \text{block 1},
\end{cases}
\end{equation}
where the block is that of the active set.
The floor $L(\pm1)\ge-8$ of Proposition~\ref{thm:floor} is the case $j_R=0$.
The argument also determines the equality cases. Let $\mathcal E=E_s\otimes F_{j_R}\otimes\mathbb F_{\rm sp}$, where $\mathbb F_{\rm sp}$ is the flat fermion bundle of the spectator, and $\mathcal E'=S^+\otimes\mathcal E$. In block 4 equality holds exactly on the lift of $\ker\mathsf D_{\mathcal E,-}$, and in block 5 exactly on the lift of the kernel of $\mathsf D_{\mathcal E',-}$ acting on $S^-\otimes\mathcal E'$.
Only the fact that $\mathcal A$ is the Hopf pull-back of an anti-self-dual connection enters, so \eqref{eq:fb:H4} holds for every anti-self-dual instanton on $S^4$ in place of $a_N$.

\paragraph*{Proof of the fibre-resolved bound and its saturation.} Take $\zeta=+1$. By the splitting~\eqref{eq:fb:H3} it suffices to show $\mathcal H_A+8\ge0$ for $A=4,5$ and $\mathcal H_1\ge0$.

\emph{Reduction to $S^4$.}
The Hopf map $\pi:S^7(1)\to S^4(\frac12)$ is a Riemannian submersion with totally geodesic fibres, the unit three-spheres $x\,Sp(1)$, and a principal $Sp(1)$-bundle whose horizontal distribution is a connection.
In the gauge $\mathcal A=\pi^*a_N$ the fermion fibre of the active set is a constant vector space $\mathbb F_A$, the $Sp(2)$ representation of block $A$, the spectator's fibre $\mathbb F_{\rm sp}$ is another constant factor, and $SU(2)_R$ acts by pull-back alone (see the derivation of Eq.~\eqref{eq:fb:H3}).
The spin-$j$ subspace of sections over $S^7$ is therefore isomorphic to $\C^{2j+1}\otimes C^\infty(F_j\otimes E_s\otimes\mathbb F_A\otimes\mathbb F_{\rm sp})$ over $S^4(\frac12)$, where $F_j$ is the bundle associated with the Hopf bundle in spin $j$ with its induced connection~\cite{Trautman1977}.
Under this isomorphism the horizontal Dirichlet form of $S^7$ is the Dirichlet form of the Bochner Laplacian of $S^4(\frac12)$~\cite{BBB1982}, the fermion factors carrying the trivial connection.

Let $x\HH$ be the tautological quaternionic line at $[x]$ and $x^\perp$ its orthogonal complement.
In block 4, $\mathbb F_4=\HH^2=x\HH\oplus x^\perp$, and $x\HH=S^+$, $x^\perp=S^-$ are the spinor bundles of $S^4(\frac12)$, whose Levi-Civita connections are the projections of the trivial connection~\cite{Atiyah1979}. The off-diagonal part of the trivial connection is the second fundamental form of the splitting, which is Clifford multiplication. The trivial connection is the Killing-spinor connection $\nabla+\tau X\cdot$ of $S^4(\frac12)$, with a fixed sign $\tau\in\{\pm1\}$~\cite{Bar1993}. By the Hopf structure of Sec.~\ref{app:fb:hopf}, $\mathcal F$ acts on $\HH^2$ through $sp(2)$, preserving $x^\perp$ and annihilating $x\HH$; hence $P_4=\mathfrak F_{E_s}$ on $S^-$ and $P_4=0$ on $S^+$, both sides being the Clifford action of the same two-form.

In block 5, $\mathbb F_5=\Lambda^2_0\HH^2\cong\C^5$ is the complexified vector representation of $Sp(2)\cong Spin(5)$. Viewing $S^4(\frac12)$ as a sphere in $\R^5$, it splits at each point into the normal line and the tangent space, so sections are pairs $(f,\gamma)$ of a function and a complex one-form, $\gamma\in S^+\otimes S^-$. The trivial connection is $\nabla+\mathcal B$ with $\mathcal B_X(f,\gamma)=(-2\gamma(X),\,2fX^\flat)$, the second fundamental form of $S^4(\frac12)\subset\R^5$. The Pauli term annihilates $f$ and acts on $\gamma$ through its $S^-$ factor, as $\mathfrak F_{E_s}$.

In block 1, $\mathbb F_1=\C$ with the trivial connection and $P_1=0$.
The spectator is a flat factor of the twisting bundle $\mathcal E$ and contributes nothing to the curvature terms below.

\emph{Block 4.}
Write $\psi=(\psi_+,\psi_-)\in C^\infty((S^+\oplus S^-)\otimes\mathcal E)$ and $\mathsf D=\mathsf D_{\mathcal E}$. Clifford multiplication by a unit vector is unitary and skew-adjoint, so $\sum_i|e_i\cdot\psi|^2=4|\psi|^2$ and $\sum_i\langle\nabla_{e_i}\psi,e_i\cdot\psi\rangle=-\langle\mathsf D\psi,\psi\rangle$ pointwise. Since $\mathsf D$ is self-adjoint,
\begin{equation}
\|(\nabla+\tau X\cdot)\psi\|^2=\|\nabla\psi\|^2+4\|\psi\|^2-2\tau\langle\psi,\mathsf D\psi\rangle .
\end{equation}
The self-dual curvature of $F_{j_R}$ does not act on $S^-$, so \eqref{eq:fb:lich} gives $\mathsf D_+\mathsf D_-=\nabla^*\nabla+12+\mathfrak F_{E_s}=\nabla^*\nabla+12+P_4$ on $S^-\otimes\mathcal E$. With $\langle\psi,\mathsf D\psi\rangle=2\operatorname{Re}\langle\psi_+,\mathsf D_-\psi_-\rangle$,
\begin{equation}\label{eq:fb:block4sq}
\begin{aligned}
\langle\psi,(\mathcal H_4+8)\psi\rangle
&=\|\nabla\psi_+\|^2+12\|\psi_+\|^2+\|\mathsf D_-\psi_-\|^2-4\tau\operatorname{Re}\langle\psi_+,\mathsf D_-\psi_-\rangle\\
&=\|\nabla\psi_+\|^2+8\|\psi_+\|^2+\|\mathsf D_-\psi_--2\tau\psi_+\|^2\ \ge\ 0,
\end{aligned}
\end{equation}
with equality if and only if $\psi_+=0$ and $\mathsf D_-\psi_-=0$.

\emph{Block 5.}
Expanding the $B$-terms and integrating the cross terms by parts, with $\sum_i(\nabla_{e_i}\gamma)(e_i)=-\dd^*\gamma$,
\begin{equation}
\|(\nabla+\mathcal B)(f,\gamma)\|^2=\|\nabla f\|^2+\|\nabla\gamma\|^2+16\|f\|^2+4\|\gamma\|^2-8\operatorname{Re}\langle f,\dd^*\gamma\rangle .
\end{equation}
Regard $\gamma$ as a section of $S^-\otimes\mathcal E'$ and let $\mathsf D'=\mathsf D_{\mathcal E'}$. The bundles $S^+$ and $F_{j_R}$ have self-dual curvature, so \eqref{eq:fb:lich} gives $\|\mathsf D'\gamma\|^2=\|\nabla\gamma\|^2+12\|\gamma\|^2+\langle\gamma,P_5\gamma\rangle$. Under $S^+\otimes S^+=\Lambda^0\oplus\Lambda^2_+$ (complexified) the operator $\mathsf D'$ becomes $(\dd^*,\dd^+)$, the twisted codifferential and self-dual exterior derivative, and $\|\mathsf D'\gamma\|^2=\|\dd^*\gamma\|^2+2\|\dd^+\gamma\|^2$ for the norm on two-forms that sums the squared components over independent index pairs. Hence
\begin{equation}\label{eq:fb:block5sq}
\begin{aligned}
\langle(f,\gamma),(\mathcal H_5+8)(f,\gamma)\rangle
&=\|\nabla f\|^2+24\|f\|^2-8\operatorname{Re}\langle f,\dd^*\gamma\rangle+\|\dd^*\gamma\|^2+2\|\dd^+\gamma\|^2\\
&=\|\nabla f\|^2+8\|f\|^2+\|\dd^*\gamma-4f\|^2+2\|\dd^+\gamma\|^2\ \ge\ 0,
\end{aligned}
\end{equation}
with equality if and only if $f=0$ and $\dd^*\gamma=\dd^+\gamma=0$, that is, $f=0$ and $\mathsf D'\gamma=0$.

\emph{Block 1 and domains.}
On block 1, $\mathcal H_1=\nabla^*\nabla\ge0$.
The identities hold on smooth sections over $S^7$, which form a form core for $L(\pm1)$. Every channel and parity subspace is invariant, so \eqref{eq:fb:H4} holds on each of them.
This proves the fibre-resolved lower bound~\eqref{eq:fb:H4}. To prove attainment in Proposition~\ref{thm:floor}, it remains to exhibit physical states at the floor.

\emph{Equality cases and the index.}
For a single set, without the spectator factor, the number of floor states in the spin-$j_R$ sector of block 4 is $(2j_R+1)\dim\ker\mathsf D_{\mathcal E,-}$ with $\mathcal E=E_s\otimes F_{j_R}$, which is at least $(2j_R+1)$ times the Atiyah--Singer index $(2j_R+1)I(s)-(2s+1)I(j_R)$, $I(j)=\frac23j(j+1)(2j+1)$ (up to an orientation convention; $E_s$ has charge one and $F_{j}$ the opposite charge), with equality when $\ker\mathsf D_{\mathcal E,+}=0$.
At $j_R=0$ this vanishing holds, because $\mathsf D_-\mathsf D_+=\nabla^*\nabla+12$ on $S^+\otimes E_s$. The floor states of one set then number $I(s)$, the isospin-$s$ zero-mode count of the $k=1$ instanton~\cite{JackiwRebbi1977,BCL1977}. A single-set floor state becomes physical when tensored with a spectator component of suitable $SU(2)_G$ spin. The spectator can carry either integer or half-integer $SU(2)_G$ spin. For example, the active type $(0,0)$ at $s=\frac12$ needs a spectator of half-integer spin at $\theta_{5d}=0$ and integer spin at $\theta_{5d}=\pi$. Thus Eq.~\eqref{eq:physical} retains a state with energy $-8$ for each $s\ge\frac12$ and either angle.
In block 5 the twisting bundle $S^+\otimes F_j=F_{j+1/2}\oplus F_{j-1/2}$ gives the analogous count with two terms. At $j_R=0$ and $s\ge\tfrac12$ it is $2I(s)-(2s+1)$ (Sec.~\ref{app:fb:block5}). The untwisted case $s=0$ has no state at the floor.\unskip\nobreak\hfill\penalty50\hskip1.2em\hbox{}\nobreak\hfill\ensuremath{\square}

At $j_R=\tfrac12$ the opposite-chirality kernel can also be determined. Since $F_{1/2}=S^+$, the positive-chirality Dirac operator is the adjoint of the instanton deformation operator on one-forms, up to a constant normalization of its self-dual component. Its kernel is $H^0(E_s)\oplus H^2_+(E_s)$. For $s>0$, full $SU(2)$ holonomy excludes a parallel colour section and the positive Weitzenb\"ock term $16$ excludes a harmonic self-dual two-form, as shown in Appendix~\ref{app:fb:block5}. For $s=0$ the constant scalar survives. For $j_R\ge1$ the index alone fixes the difference of chiral multiplicities. In particular $j_R>s$ gives a negative index, which forces an opposite-chirality kernel. In block 5 the criterion applies to each twist $F_{j_R\pm1/2}$ separately. Thus at $j_R=\tfrac12$ the $F_0$ summand is covered by the untwisted positive-chirality estimate, while $F_1$ has a separate index problem. Attainment of the full floor uses $j_R=0$. The numbers of floor states found numerically, summed over $j_R$, are $14$, $330$, $2548$ in block 4 and $28$, $660$, $5096$ in block 5 for $s=1,2,3$, equal to the index sums $\sum_j(2j+1)[\operatorname{ind}(j)]_+$, with $\operatorname{ind}(j)=(2j+1)I(s)-(2s+1)I(j)$, and $\sum_j(2j+1)\bigl([\operatorname{ind}(j+\frac12)]_++[\operatorname{ind}(j-\frac12)]_+\bigr)$, which equals twice the first by reindexing. %

\emph{Independent checks.}
The following comparisons use numerical diagonalization of the channel operators and do not enter the proof. %
At $s=0$, \eqref{eq:fb:block4sq} at $j_R=0$ predicts $\mathcal H_4+8=\Lambda^2-2\tau\Lambda$ on the Dirac spectrum $\Lambda=\pm2(l+2)$, $l\ge0$, of $S^4(\frac12)$, with lowest values $8$ ($\times4$) and $24$ ($\times20$). The block-5 prediction is $8$ ($\times5$) and $24$ ($\times25$). Both are reproduced.
Galerkin scans of all channels with $j_L\le s+\frac32$, $j_R\le3$, $j_G\le s+2$ and both parities give $\min\big(\mu_0-4j_R(j_R+1)\big)=-7.99996$, $-8.00000$, $-8.00000$ for $s=1,2,3$, with $\mu_0$ the lowest eigenvalue of $L(\pm1)$ in a channel.
Replacing $\sqrt2$ in $N$ by $c\in\{\tfrac12,1,2,3\}$, which moves the instanton conformally, leaves the floor and its multiplicity unchanged ($s=1,2$).

\emph{Summary.}
The floor $-8$ is the Lichnerowicz constant $12$ of the Hopf base reduced by the $4$ from the induced Killing-spinor or second-fundamental-form connection. At $j_R=0$, the floor states are obtained from the instanton Dirac kernel on $S^4$ and counted using the Atiyah--Singer index. For general $j_R$ the lower bound retains the vertical Casimir, and the fibre spin costs exactly $4j_R(j_R+1)$, with no Pauli compensation.

\paragraph*{Consequences for the threshold.}\phantomsection\label{app:fb:consequences}

In the two-set problem the sharp floor bounds the whole link operator. For every spin $s$, every $\zeta\in[-1,1]$ and every channel of every block,
\begin{equation}\label{eq:fb:Lbound}
L_0(\zeta)\ge4j_R(j_R+1)-8,\qquad L(\zeta)\ge4j_R(j_R+1)-8-E .
\end{equation}
In particular $L(\pm1)\ge-8$ in every channel, every channel with $j_R\ge\frac12$ is conformal for $E\le4$, and every channel is conformal for $E\le1$.
Indeed, \eqref{eq:fb:H4} applies to $L_+$ with the set $-$ as spectator and to $L_-$ with the set $+$ as spectator. Because $SU(2)_R$ does not act on the fermions, both bounds refer to the same $j_R$, and $c_\pm\ge0$, $c_++c_-=1$ give the first inequality. Since $0<W\le1$, $(1-\zeta^2)V=-EW\ge-E$, which gives the second. The right-hand side is at least $-5-E$ for $j_R\ge\frac12$ and equals $-8-E$ for $j_R=0$.

At $\theta_{5d}=0$, the trial bounds~\eqref{eq:thmC} and~\eqref{eq:vpp} give $E_c\le5/2$ for $s\ge1$. At $\theta_{5d}=\pi$ the two trial families exchange their spin parities. The exceptional $s=1$ endpoint pencil gives $E_c<2.623$ (Sec.~\ref{sec:bps}). Thus every onset at $s\ge1$ lies below $E=4$, and only $j_R=0$ can set it.

For every spin $s\ge\frac12$ the free-centre line is therefore conformal whenever $\zeta^2\ge1-1/(4C_2)$, which is the case $E\le1$ of \eqref{eq:fb:Lbound}. For $s\ge1$ the threshold of Sec.~\ref{sec:radial} then obeys $\zeta_c(s)\le\zeta_{\rm pin}(s)=\sqrt{1-1/(4C_2)}$, the supersymmetric pinned threshold of Eq.~\eqref{eq:pinbound},
and
\begin{equation}\label{eq:fb:R2ge1}
E_c(s)=4C_2\big(1-\zeta_c(s)^2\big)\ \ge\ 1 .
\end{equation}
At $\theta_{5d}=0$, combining this with the trial-state bound $E_c\le1+\frac3{2s}$ for integer $s$ (Eq.~\eqref{eq:thmC})
and the bound $E_c\le1/w_{\rm qrt}(s)$ for half-integer $s\ge\frac32$ (Sec.~\ref{sec:instability}, proved in Sec.~\ref{app:fb:halfint}), gives
\begin{equation}\label{eq:fb:H5}
1\le E_c(s)\le 1+\frac{3}{2s}\quad(\text{integer }s\ge1),\quad
1\le E_c(s)\le \frac{1}{w_{\rm qrt}(s)}\quad(\text{half-integer }s\ge\tfrac32),
\end{equation}
with $1/w_{\rm qrt}(s)=1+\frac3{2s}+O(s^{-2})$ by \eqref{eq:fb:vpp-bound}, so that $E_c(s)\to1$ and $\zeta_c(s)^2=1-1/(4C_2)+O(s^{-3})$ as $s\to\infty$. The last statement follows from $1-\zeta_c^2=E_c/(4C_2)$ and $E_c=1+O(s^{-1})$.

The bound \eqref{eq:fb:R2ge1} holds for the full problem, in every channel and block. It does not use the identity of the critical channel. It fixes the leading order only. The lower side of the coefficient $\frac32$ of the $1/s$ term does not follow from \eqref{eq:fb:H4}. It follows from the uniform lower bound~\eqref{eq:O2prime}, whose proof is computer-assisted and uses exact arithmetic (Appendix~\ref{app:un:uniform}).

At $j_R=0$ the block-4 operator has an explicit gap above its kernel. Let $\mathsf D=\mathsf D_{E_s}$ and $\mathcal K_4=\ker\mathsf D_-$. Then $\mathcal H_4+8=\mathsf D^2-2\tau\mathsf D$, $\ker(\mathcal H_4+8)=\mathcal K_4$, and
\begin{equation}\label{eq:fb:gap4}
\mathcal H_4+8\ \ge\ 12-4\sqrt3\qquad\text{on }\mathcal K_4^\perp .
\end{equation}
At $j_R=0$ the bundle $F_0$ is trivial, so the curvature term of \eqref{eq:fb:lich} acts on $S^-$ only and equals $P_4$ there. By the block-4 computation leading to \eqref{eq:fb:block4sq}, $\langle\psi,(\mathcal H_4+8)\psi\rangle=\|(\nabla+\tau X\cdot)\psi\|^2+\langle\psi,P_4\psi\rangle+8\|\psi\|^2=\|\nabla\psi\|^2+12\|\psi\|^2+\langle\psi,P_4\psi\rangle-2\tau\langle\psi,\mathsf D\psi\rangle$, and by \eqref{eq:fb:lich} this is $\langle\psi,(\mathsf D^2-2\tau\mathsf D)\psi\rangle$.
On $S^+\otimes E_s$, $\mathsf D_-\mathsf D_+=\nabla^*\nabla+12\ge12$, so $\ker\mathsf D_+=0$, $\ker\mathsf D=\mathcal K_4$, and every eigenvector of $\mathsf D$ with eigenvalue $\Lambda\ne0$ has a nonzero $S^+$ component on which $\mathsf D_-\mathsf D_+=\Lambda^2$, so $\Lambda^2\ge12$. On $\mathcal K_4^\perp$ therefore $\mathsf D^2-2\tau\mathsf D\ge\min_{|\Lambda|\ge\sqrt{12}}(\Lambda^2-2|\Lambda|)=12-4\sqrt3$. Friedrich's twistor inequality \cite{Friedrich1980} sharpens $\mathsf G_4^{\rm Lich}=12-4\sqrt3$ immediately to the bound $\mathsf G_4^{\rm F}=8$ for every $s\ge0$. On $S^4(\frac12)$ (dimension $n=4$), the Penrose twistor operator $(\mathscr P_{\rm tw}\psi)_X=\nabla_X\psi+\tfrac14X\cdot\mathsf D_+\psi$ acting on $\psi\in C^\infty(S^+\otimes E_s)$ obeys $0\le\|\mathscr P_{\rm tw}\psi\|^2=\|\nabla\psi\|^2-\tfrac14\|\mathsf D_+\psi\|^2$, so $\|\nabla\psi\|^2\ge\tfrac14\|\mathsf D_+\psi\|^2$. Substituting into the Lichnerowicz identity $\|\mathsf D_+\psi\|^2=\|\nabla\psi\|^2+12\|\psi\|^2$ (where the anti-self-dual curvature of $E_s$ vanishes on $S^+$) gives $\tfrac34\|\mathsf D_+\psi\|^2\ge12\|\psi\|^2$, hence $|\Lambda|\ge4$ for every nonzero eigenvalue of $\mathsf D$ and
\begin{equation}\label{eq:fb:gap4sharp}
\mathcal H_4+8\ \ge\ 4^2-2\cdot4=8=\mathsf G_4^{\rm F}\qquad\text{on }\mathcal K_4^\perp\quad(\forall s\ge0),
\end{equation}
which is sharp, attained at $s=0$ and $s=\frac12$. At $s=0$ the Killing spinors attain it. At $s=\frac12$ the bundle $S^+\otimes E_{1/2}$ is the complexified cotangent bundle of the conformally related metric $\mathsf g_2$, and the gradients of the first spherical harmonics of $\mathsf g_2$ are twistor spinors. The twistor equation is conformally covariant, so they remain twistor spinors for $\mathsf g_1$, and for a twistor spinor the two relations above are equalities, $\|\mathsf D_+\psi\|^2=16\|\psi\|^2$. The spin-one theta comparison in Appendix~\ref{app:fb:block45} uses Eq.~\eqref{eq:fb:gap4sharp}. Appendix~\ref{app:un:uniform} retains $\mathsf G_4^{\rm Lich}=12-4\sqrt3$ because upgrading to $\mathsf G_4^{\rm F}=8$ lowers the asymptotic deficit $c_*$ only from $\tfrac{33+3\sqrt3}{64}$ to $\tfrac{33}{64}$, still well above the mixing coefficient $\tfrac{9}{56}$ that fixes $-\tfrac{9}{56s^2}$, so the weaker gap suffices there.

\section{BPS states and the threshold for instanton collapse}
\label{app:fb:bps}

 This appendix describes the floor and doubly-BPS states in closed form, derives the block-5 inputs for Appendix~\ref{app:un:uniform}, proves fundamental-line conformality analytically, and solves the exact block-$(4,5)$ endpoint system. By the sharp floor~\eqref{eq:fb:H4}, all floor states sit at $j_R=0$ and lift from $S^4(\frac12)$, on which we work throughout.

We use two-spinor notation on $S^4(\frac12)$, with unprimed indices $A,B,C,\dots$ on $S^+$, primed indices $A',B',\dots$ on $S^-$, and skew forms $\epsilon_{AB},\epsilon_{A'B'}$ raising and lowering indices ($\Lambda^1\otimes\C=S^+\otimes S^-$, $\Lambda^2_\pm\otimes\C={\rm Sym}^2S^\pm$, $(\mathsf D_-\psi)_A=\nabla_A{}^{A'}\psi_{A'}$, colour indices suppressed unless shown, and $\dd,\dd^*,\dd^+$ twisted by $a_N$). In the stereographic coordinate $x$ of $\mathsf g_1=|\dd x|^2/(1+t)^2$ centred at $Z=0$ ($|x|^2=t$), the conformal map $\bar N$ acts as $x\mapsto-\sqrt2\,x$, so $a_N$ is the basic instanton of the conformally equivalent round metric
\begin{equation}\label{eq:fb:g2}
\mathsf g_2=\bar N^*\mathsf g_1=\frac{2\,|\dd x|^2}{(1+2t)^2}=\frac{|\dd\tilde x|^2}{(1+|\tilde x|^2)^2},\qquad
\tilde x=\sqrt2\,x,\qquad \mathsf g_1=\Omega^2\mathsf g_2,\qquad \Omega=\frac{D}{\sqrt2},
\end{equation}
identifying $E_{1/2}\cong S^-_{\mathsf g_2}$ and $E_s={\rm Sym}^{2s}S^-_{\mathsf g_2}$, while $SU(2)_L\times SU(2)_G$ preserves $\mathsf g_1,\mathsf g_2,a_N$ and acts at $x=0$ on $S^+$ via $SU(2)_G$ and on $S^-$ via $SU(2)_L$. Under $\mathsf g_1=\Omega^2\mathsf g_2$, a twisted massless field with $n\in\{1,2,3\}$ primed indices ($\nabla^{AA'}\phi_{A'B'\cdots}=0$, namely Dirac, anti-self-dual Maxwell, and spin-$\frac32$) rescales by $\Omega^{-(n+2)/2}$ \cite{PenroseRindler}, changing its density by
\begin{equation}\label{eq:fb:weights}
|\phi|^2\,\dd{\rm vol}_{\mathsf g_1}=\Omega^{2-n}\,|\phi|^2\,\dd{\rm vol}_{\mathsf g_2}.
\end{equation}
The space $\mathcal T_k\subset\Gamma({\rm Sym}^kS^-_{\mathsf g_2})$ of valence-$k$ twistor spinors ($\nabla_{A(B'_1}\omega_{B'_2\cdots B'_{k+1})}=0$) satisfies $\mathcal T_k={\rm Sym}^k\mathbb T$ of dimension $\binom{k+3}{3}$, where $\mathbb T=\mathcal T_1$ ($\dim\mathbb T=4$) consists in flat $\tilde x$-spin-frame coordinates of $\omega^{B'}(\tilde x)=\pi^{B'}+\tilde x^{BB'}\lambda_B$ with constant $SU(2)_L$ doublet $\pi^{B'}$, $SU(2)_G$ doublet $\lambda_B$, and $|\omega|^2_{\mathsf g_2}=(1+|\tilde x|^2)^{-k}|\omega|^2_{\rm flat}$ \cite{PenroseRindler}.

\paragraph*{The $(4,4)$ core and its $(5,5)$ image.}\phantomsection\label{app:fb:core}

Let $s\ge1$ be an integer and work at $j_R=0$. Let $T_s$ be the doubly-BPS state of the channel $(s-1,0,0)$ of block $(4,4)$ and $w_s$ its weight average, from Eqs.~\eqref{eq:Ts} and~\eqref{eq:vs}. We show that the doubly-BPS core of block $(4,4)$ embeds isometrically, up to a constant factor, into block $(5,5)$, so that the image of $T_s$ is a doubly-BPS state of block $(5,5)$ with the same weight average.

A state of block $(4,4)$ is doubly BPS if and only if it is a twisted anti-self-dual Maxwell field, $\phi_{A'B'}=\phi_{(A'B')}\in\Gamma({\rm Sym}^2S^-\otimes E_s)$ with $\nabla_A{}^{A'}\phi_{A'B'}=0$, where $A'$ is the index of the set $+$ and $B'$ that of the set $-$. On the core $\Gamma({\rm Sym}^2S^-\otimes E_s)$ the quadratic forms of $L(1)+8$ and $L(-1)+8$ both equal $\|\nabla_C{}^{A'}\phi_{A'B'}\|^2$.
To see this, note first that for $|\zeta|<1$ both $c_\pm>0$ and $L(\pm1)+8\ge0$, so the kernel of $L_0(\zeta)+8$ is $\ker(L(1)+8)\cap\ker(L(-1)+8)$.
Apply the block-4 identity~\eqref{eq:fb:block4sq} to the set $+$, with the set $-$ as spectator. Written in the splitting $\HH^2=S^+\oplus S^-$ of the reduction to $S^4$ in Sec.~\ref{app:fb:sharp}, the trivial connection of the spectator's fibre is its Killing connection, so the Dirac operator of the set $+$ acquires the term $\tau\,\Xi\psi$, $\Xi=\sum_ie_i\cdot\otimes e_i\cdot$. Equality requires the set $+$ to lie in $S^-$, and, with the roles exchanged, the set $-$ as well, so $\psi\in\Gamma(S^-\otimes S^-\otimes E_s)$.
For such $\psi$ the derivative part of $\mathsf D_-\psi$ lies in $S^+\otimes S^-$ and $\Xi\psi$ in $S^+\otimes S^+$, so both vanish. By the completeness relation $\sum_i(e_i)_{AA'}(e_i)_{BB'}\propto\epsilon_{AB}\epsilon_{A'B'}$, $(\Xi\psi)_{AB}\propto\epsilon_{AB}\,\epsilon^{A'B'}\psi_{A'B'}$, which vanishes if and only if $\psi$ is symmetric. The derivative part is $\nabla_A{}^{A'}\psi_{A'B'}$, and the condition from the set $-$, $\nabla_B{}^{B'}\psi_{A'B'}=0$, is the same by symmetry. A symmetric $\phi_{A'B'}$ with $\nabla_A{}^{A'}\phi_{A'B'}=0$ is a twisted anti-self-dual Maxwell field. On the core $\psi_+=0$ and $\Xi\psi=0$, so \eqref{eq:fb:block4sq} reduces to $\|\nabla_C{}^{A'}\phi_{A'B'}\|^2$ for each set.

The map $\iota_\epsilon:\phi_{A'B'}\mapsto\psi_{AA'BB'}=\epsilon_{AB}\phi_{A'B'}$ embeds the core into block $(5,5)$, each set lying in $\Lambda^1$. It is $G$-equivariant, isometric up to the factor $|\epsilon|^2=2$, and its image is the $\Lambda^2_-$ part of $\Lambda^1\otimes\Lambda^1\otimes E_s$. For every $\zeta\in[-1,1]$ the quadratic forms of $L_0(\zeta)+8$ and of $W$ on the image are $|\epsilon|^2$ times those on the core.
Indeed, for $\psi=\epsilon_{AB}\phi_{A'B'}$ with $\phi$ symmetric, neither set has a scalar component. Apply the block-5 identity~\eqref{eq:fb:block5sq} to the set $+$, with the set $-$ as spectator, whose trivial connection in the splitting $\Lambda^0\oplus\Lambda^1$ is $\nabla+\mathcal B$. Since $f=0$, the square \eqref{eq:fb:block5sq} equals $\|\mathsf D'\psi\|^2$. The spectator's $B$-term maps the one-form of the set $-$ into its scalar slot and contributes to $\mathsf D'\psi$ a multiple of $\psi_{AA'C}{}^{A'}=\epsilon_{AC}\phi_{A'}{}^{A'}=0$, the trace of a symmetric spinor. The derivative part is $\nabla_C{}^{A'}(\epsilon_{AB}\phi_{A'B'})=\epsilon_{AB}\nabla_C{}^{A'}\phi_{A'B'}$, whose norm is $|\epsilon|^2\|\nabla_C{}^{A'}\phi_{A'B'}\|^2$. The same holds with the sets exchanged, and $W$ acts by multiplication by a function. Equivariance holds because $\epsilon_{AB}$ is invariant.

Consequently $\iota_\epsilon(T_s)$ is a doubly-BPS state of the channel $(s-1,0,0)$ of block $(5,5)$, with $\langle W\rangle=w_s=\frac{2s}{2s+3}$, and the threshold of that channel obeys $E_c^{(5,5)}(s)\le1+\frac{3}{2s}$.
By the previous paragraph the form of $L_0(\zeta)+8\ge0$ vanishes on $\iota_\epsilon(T_s)$, which is therefore in its kernel, and $\langle W\rangle$ is unchanged. The channel is physical when $s+\theta_{5d}/(2\pi)$ is integer, and \eqref{eq:fb:trial} gives non-conformality for $E>1/w_s=1+\frac3{2s}$.

In this sense the two channels share their first-order (trial-state) threshold. Their thresholds differ only through the coupling of the common core to the rest of each block. Numerically, the $(5,5)$ channel $(s-1,0,0)$ has exactly one state at $-8$ for $\zeta\in\{0.95,0.7,0.3,0,-0.5\}$, with $\langle W\rangle$ equal to $w_s=\tfrac25$, $\tfrac47$, $\tfrac23$ for $s=1,2,3$ to five digits at $\zeta=0.95$ and six digits at the other couplings, and the Galerkin thresholds in this coupled basis, accurate to about $2\times10^{-4}$, are $E_c=2.35579$, $1.70907$, $1.48146$ for $(4,4)$ against $2.41029$, $1.71759$, $1.48426$ for $(5,5)$. The exact $(4,4)$ channel system gives the values of Table~\ref{tab:thresholds}. %
Scans of the numerically delimited $j_R=0$ boxes for integer $s\le6$ find the $(4,4)$ channel critical, ahead of the $(5,5)$ channel by about $0.08/s^3$ in $E_c$ at $s=4,5,6$. The finite-spin ordering in these scans remains numerical.

\subsection{Twistor construction and counting of BPS states}
\label{app:fb:twistor}

We work at $j_R=0$, in the frame of $\mathsf g_2$, and transport to $\mathsf g_1$ with the weights \eqref{eq:fb:weights}.
For every spin $s\ge\frac12$ the floor states of a single block-4 set are
\begin{equation}\label{eq:fb:floorstates}
\psi_{A'C'_1\cdots C'_{2s}}=\epsilon_{A'(C'_1}\,\omega_{C'_2\cdots C'_{2s})},\qquad \omega\in\mathcal T_{2s-1},
\end{equation}
and these are all of them; their number is $I(s)=\frac23s(s+1)(2s+1)=\binom{2s+2}{3}$.
By the sharp floor~\eqref{eq:fb:H4} the floor states of a block-4 set at $j_R=0$ form $\ker\mathsf D_-$ on $S^-\otimes E_s$. In the frame of $\mathsf g_2$, where $E_s={\rm Sym}^{2s}S^-$ and $\epsilon$ is parallel, $\nabla_A{}^{A'}\epsilon_{A'(C'_1}\omega_{C'_2\cdots C'_{2s})}$ is a constant multiple of $\nabla_{A(C'_1}\omega_{C'_2\cdots C'_{2s})}$, which vanishes by the twistor equation. The twisted Dirac operator is conformally covariant with weight $\frac32$, so $\Omega^{-3/2}\psi$ is a zero mode for $\mathsf g_1$.
The map $\omega\mapsto\psi$ is injective, so these zero modes span a space of dimension $\dim\mathcal T_{2s-1}=\binom{2s+2}{3}=I(s)$. On $S^+\otimes E_s$, $\mathsf D_-\mathsf D_+=\nabla^*\nabla+12$, so $\ker\mathsf D_+=0$ and $\dim\ker\mathsf D_-$ equals the index $I(s)$~\cite{JackiwRebbi1977,BCL1977}. Hence the twistor family is the whole kernel.

For every $s\ge1$ the fields $\phi=\epsilon\odot\epsilon\odot\omega$ with $\omega\in\mathcal T_{2s-2}$ (each fermion index contracted by $\epsilon$ with a colour index and the result symmetrized over the colour indices) are twisted anti-self-dual Maxwell fields, hence doubly-BPS states of block $(4,4)$. At $s=1$ this space is one-dimensional and spanned by the curvature of $a_N$. For every $s\ge1$, $T_s$ is the member of type $(s-1,0)$.
To see this, write $\phi_{A'B'C'_1\cdots C'_{2s}}=\epsilon_{A'(C'_1}\epsilon_{|B'|C'_2}\,\omega_{C'_3\cdots C'_{2s})}$, which is symmetric in $A'B'$ after the symmetrization over the colour indices. Then $\nabla_A{}^{A'}\phi_{A'B'\cdots}$ is a constant multiple of the symmetrization of $\epsilon_{B'C'_2}\nabla_{AC'_1}\omega_{C'_3\cdots C'_{2s}}$ over the $C'$, that is, of $\epsilon_{B'(C'_2}\nabla_{|A|C'_1}\omega_{C'_3\cdots C'_{2s})}$, which vanishes by the twistor equation of valence $2s-2$. Thus $\phi$ is a twisted anti-self-dual Maxwell field, and by the characterization at the start of this appendix it is doubly BPS. The Maxwell equation is conformally invariant, so it holds for $\mathsf g_1$.
At $s=1$, $\mathcal T_0=\C$ and $\phi=\epsilon\odot\epsilon$ is the identity map of ${\rm Sym}^2S^-$. The curvature of $S^-_{\mathsf g_2}$, a section of $\Lambda^2_-\otimes{\rm End}_0(S^-)={\rm Sym}^2S^-\otimes{\rm Sym}^2S^-$, is a constant multiple of it because $\mathsf g_2$ is round.
For integer $s$, the member of type $(s-1,0)$ is a doubly-BPS state of the channel $(s-1,0,0)$ of block $(4,4)$. In that channel the reduced operator $L_0(\zeta)+8$ is a cooperative system whose ground-state representation (Sec.~\ref{sec:instability}, below Eq.~\eqref{eq:channellaw})
has a connected coupling graph for $|\zeta|<1$, so its kernel is a single copy of the channel, so that member is $T_s$.

The polar profiles of these states follow from their $SU(2)_L\times SU(2)_G$ content. Under $SU(2)_L\times SU(2)_G$, $\mathcal T_k=\bigoplus_{a+b=k}(\frac a2,\frac b2)$, each type once, with $b=2j_G$. For $n=1,2,3$ and $a+b=2s-n$, the angle-averaged density in $t$ of the type $(\frac a2,\frac b2)$ of $\epsilon^{\odot n}\odot\mathcal T_{2s-n}$ is, up to normalization,
\begin{equation}\label{eq:fb:profile}
\varrho^{(n)}_b(t)=t^{b+1}\big(t+\tfrac12\big)^{-(2s+2)}(1+t)^{n-2}.
\end{equation}
In particular the floor states of type $(s-\frac12-\frac b2,\frac b2)$, $b=0,\dots,2s-1$, have the profile
\begin{equation}\label{eq:fb:gb}
g_b(t)=\varrho^{(1)}_b(t)=\frac{t^{b+1}}{(1+t)\,(t+\frac12)^{2s+2}},
\end{equation}
and the doubly-BPS states of type $(s-1-\frac b2,\frac b2)$, $b=0,\dots,2s-2$, have the profile $h_b(t)=\varrho^{(2)}_b(t)=t^{b+1}(t+\frac12)^{-(2s+2)}$ and
\begin{equation}\label{eq:fb:Wb}
\langle W\rangle_{h_b}=\frac{(2s-b)(4s+4-b)}{4(s+1)(2s+3)},
\end{equation}
which equals $w_s$ at $b=0$ and decreases in $b$.

The decomposition follows from $\mathbb T=(\frac12,0)\oplus(0,\frac12)$, spanned by $\pi^{B'}$ and $\lambda_B$. The type $(\frac a2,\frac b2)$ consists of the terms of degree $a$ in $\pi$ and $b$ in $\lambda$.
Let $\{\omega_i\}$ be an orthonormal basis of this type for an invariant inner product. The function $\sum_i|\omega_i|^2$ is invariant under $SU(2)_L\times SU(2)_G$, which acts transitively on the spheres $|\tilde x|={\rm const}$, so it is radial. In the flat frame, evaluation at $\tilde x$ acts on the $\pi$-factors as the identity and on the $\lambda$-factors as $\lambda_B\mapsto\tilde x^{BB'}\lambda_B$, which for real $\tilde x$ is $|\tilde x|$ times a fixed multiple of an isometry; hence $\sum_i|\omega_i(\tilde x)|^2_{\rm flat}\propto|\tilde x|^{2b}$, and the density relative to $\dd{\rm vol}_{\mathsf g_2}$ is $\propto|\tilde x|^{2b}(1+|\tilde x|^2)^{-k}$.
The map $\omega\mapsto\epsilon^{\odot n}\odot\omega$ is an equivariant injection of an irreducible $SU(2)$ module, hence pointwise a constant multiple of an isometry. With $k=2s-n$, $|\tilde x|^2=2t$, $\dd{\rm vol}_{\mathsf g_2}\propto(1+2t)^{-4}\,t\,\dd t\,\dd{\rm vol}_{S^3}$ and $\Omega\propto(1+2t)/(1+t)$, \eqref{eq:fb:weights} gives a density proportional to
$t^b(1+2t)^{-(2s-n)}\cdot(1+2t)^{2-n}(1+t)^{n-2}\cdot(1+2t)^{-4}\,t$, which is \eqref{eq:fb:profile}.
Setting $v=1/(1+2t)\in(0,1)$ turns $h_b\,\dd t$ into a multiple of $(1-v)^{b+1}v^{2s-b-1}\,\dd v$ and $W$ into $\tfrac12v(1+v)$, so Euler's Beta integral gives
\begin{equation}
\langle W\rangle_{h_b}=\frac{\mathrm B(2s-b+1,b+2)+\mathrm B(2s-b+2,b+2)}{2\,\mathrm B(2s-b,b+2)}=\frac{(2s-b)(4s+4-b)}{4(s+1)(2s+3)},
\end{equation}
which is \eqref{eq:fb:Wb}. At $b=0$ it equals $2s/(2s+3)=w_s$, and both factors $2s-b$ and $4s+4-b$ decrease in $b$.

At $s=1$ the density of the curvature, $|\mathcal F|^2\dd{\rm vol}\propto t(t+\frac12)^{-4}\dd t$, is the profile $h_0$ of $T_1$.
This family exhausts the block-$(4,4)$ doubly-BPS space for every $s\ge\frac12$ (with $\mathcal T_{-1}=0$, proving analytically that block $(4,4)$ has zero doubly-BPS states at $s=\frac12$). Indeed, on $\mathsf g_2$, any doubly-BPS state $\phi_{A'B'C'_1\cdots C'_{2s}}\in\Gamma(\mathrm{Sym}^2S^-\otimes E_s)$ satisfies $\nabla_A{}^{A'}\phi_{A'B'C'_1\cdots C'_{2s}}=0$, so it lies in $\ker\mathsf D_-$ on $S^-_{A'}\otimes(S^-_{B'}\otimes E_s)$ intersected with the symmetry condition $\epsilon^{A'B'}\phi_{A'B'\cdots}=0$. Decomposing $S^-\otimes E_s\cong E_{s+1/2}\oplus E_{s-1/2}$ (at $s=\frac12$, $E_0=\C$ is the trivial line bundle, on which $\ker\mathsf D_-=0$ directly by Eq.~\eqref{eq:fb:lich} since $\mathsf D_+\mathsf D_-=\nabla^*\nabla+12>0$) and applying the complete single-set characterization~\eqref{eq:fb:floorstates} to each summand gives a general element $\phi=\psi^{(+)}+\psi^{(-)}$, where $\psi^{(-)}\in\epsilon\odot\epsilon\odot\mathcal T_{2s-2}$ comes from $E_{s-1/2}$ and is already symmetric in $(A',B')$, while $\psi^{(+)}_{A'B'C'_1\cdots C'_{2s}}=\epsilon_{A'(B'}\tilde\omega_{C'_1\cdots C'_{2s})}$ comes from $E_{s+1/2}$ with $\tilde\omega\in\mathcal T_{2s}$. Because $\epsilon^{A'B'}\psi^{(-)}=0$, contracting $\phi=\psi^{(+)}+\psi^{(-)}$ with $\epsilon^{A'B'}$ isolates $\epsilon^{A'B'}\psi^{(+)}_{A'B'C'_1\cdots C'_{2s}}=\frac{2s+2}{2s+1}\,\tilde\omega_{C'_1\cdots C'_{2s}}=0$, forcing $\tilde\omega=0$ and hence $\psi^{(+)}=0$. This proves $\mathcal H_{\rm dBPS}^{(4,4)}(s)=\epsilon\odot\epsilon\odot\mathcal T_{2s-2}$, of dimension $\binom{2s+1}{3}$, for every $s\ge\frac12$. As an independent cross-check, for $s=2,3$, in all $j_R=0$ channels of that block with $j_L,j_G\le s+\frac32$, doubly-BPS states occur numerically exactly once in each channel with $j_L+j_G=s-1$ and nowhere else. %
The Galerkin zero modes reproduce $\langle W\rangle_{g_b}$ in every channel for $s=1,\frac32,2,\frac52,3,4,5,6$ to within $1.2\times10^{-5}$, the truncation level, and the polar densities for $s\le3$ with discrepancies of order $10^{-6}$ to $10^{-4}$. The census values of $\langle W\rangle_{h_b}$ agree with \eqref{eq:fb:Wb} to within $3\times10^{-5}$. %

\emph{Relation to Zhang--Hu lowest Landau levels.} In Zhang and Hu's four-dimensional quantum Hall problem on $S^4$ \cite{ZhangHu,Hasebe}, the lowest Landau level of isospin $p/2$ is spanned by degree-$p$ homogeneous polynomials in the four Hopf spinor components (degeneracy $\binom{p+3}{3}$). Replacing those components by the four-dimensional twistor space $\mathbb T$ identifies $\mathcal K_4=\epsilon\odot\mathcal T_{2s-1}$ and $\mathcal H_{\rm dBPS}^{(4,4)}=\epsilon\odot\epsilon\odot\mathcal T_{2s-2}$ as the $Sp(2)$ modules at $p=2s-1$ and $p=2s-2$, displaced conformally by $\bar N$ (Sec.~\ref{sec:bpsstates}).

\paragraph*{The endpoint index.}\phantomsection\label{app:fb:index}
The superconformal algebra of Sec.~\ref{sec:floor} distinguishes the two floor modules by a protected trace. Use the active-set chirality for the grading $(-1)^F$, and choose an odd generator $\mathcal Q$ with $\Delta_{\mathcal I}=\{\mathcal Q,\mathcal Q^\dagger\}=2J_0+2J^R_3-4J^{\rm F}_3$, where $J^R_3$ and $J^{\rm F}_3$ are the Cartan generators of $SU(2)_R$ and of $Sp(1)_{\rm F}$, the right rotation of the active fermion set (Sec.~\ref{sec:floor}). The index is the trace of $(-1)^F e^{-\tau_{\mathcal I}\Delta_{\mathcal I}}q^{2(J_0-J^{\rm F}_3)}u^{2J^L_3}v^{2J^G_3}$ with $\tau_{\mathcal I}>0$. The paired states at $\Delta_{\mathcal I}>0$ cancel \cite{KMMR2007}. At order $q$ the isolated short block-$4$ module contributes its odd primary of maximal $J^{\rm F}_3$. The block-$5$ threshold module contributes an odd descendant at order $q^2$. A potential opposite contribution at order $q^2$ would require a covariantly constant block-$1$ section, which is absent for $s>0$ by irreducibility. Thus before the spectator product and the parity projection, the first terms are $-q\,\chi_{\mathcal K_4}-q^2\chi_{\mathcal K_5}$, where $\chi_{\mathcal K_A}$ is the $SU(2)_L\times SU(2)_G$ character of the multiplicity space. The $q$-linear coefficient at $u=v=1$ has magnitude $I(s)$ and the character at general $u,v$ is that of the Zhang--Hu level at $p=2s-1$.
For the physical trace the spectator contributes its ordinary character $\chi_{\rm spec}=4+\chi_{1/2}(u)\chi_{1/2}(v)+2\chi_{1/2}(u)+2\chi_{1/2}(v)$, with $\chi_{1/2}(v)=v+v^{-1}$. Equation~\eqref{eq:physical} then selects the component with eigenvalue $(-1)^{2s}e^{\ii\theta_{5d}}$ under $v\mapsto-v$. For instance, at $s=\tfrac12$ and $\theta_{5d}=0$, the active $q$-coefficient is $-1$, and its physical character after projection is $-\chi_{1/2}(v)\bigl(\chi_{1/2}(u)+2\bigr)$. The ordinary spectator factor is $(2+\chi_{1/2}(u))(2+\chi_{1/2}(v))$. Each spectator colour parity has dimension eight, so the unrefined coefficients in active grading are $-8\dim\mathcal K_4$ and $-8\dim\mathcal K_5$ at the first two orders in either theta sector. With total fermion parity the spectator factor is $(2-\chi_{1/2}(u))(2-\chi_{1/2}(v))$. The factor $2-\chi_{1/2}(u)$ survives projection and sets the trace to zero at $u=1$. This index comparison concerns the endpoint realization and grading just specified. Relating it to an instanton contribution to a five-dimensional Wilson loop would require a map of supercharges, fugacities and defect domains.

\subsection{Lowest states and spectral gap in the \texorpdfstring{$\mathrm{Spin}(5)$}{Spin(5)} vector representation}
\label{app:fb:block5}

Let $\mathcal K_5$ be the space of floor states of a single block-5 set at $j_R=0$. By the sharp floor~\eqref{eq:fb:H4} these are $(f,\gamma)=(0,\gamma)$ with $\gamma\in\Omega^1(E_s)$, $\dd^*\gamma=0$ and $\dd^+\gamma=0$.
Let $\mathfrak p_1$ be the five-dimensional space of gradients of first spherical harmonics of $\mathsf g_1$, which are gradient conformal Killing fields. In the coordinate $x$ it is spanned by the dilation $X_0=x\cdot\partial$, which is $SO(4)$-invariant, and the quartet $X_\mu=(1+|x|^2)\partial_\mu-2x_\mu\,x\cdot\partial$, $\mu=1,\dots,4$, of type $(\frac12,\frac12)$. Their norms are
\begin{equation}\label{eq:fb:CKFnorms}
|X_0|^2_{\mathsf g_1}=t\kappa^2,\qquad \sum_\mu|X_\mu|^2_{\mathsf g_1}=4(1+t+t^2)\kappa^2 .
\end{equation}
Let $M$ be the space of twisted anti-self-dual Maxwell fields, $\phi\in\Omega^2_-(E_s)$ with $\dd\phi=0$. By Sec.~\ref{app:fb:twistor} it contains $\epsilon\odot\epsilon\odot\mathcal T_{2s-2}=\bigoplus_{a+b=2s-2}(\frac a2,\frac b2)$, whose type $(\frac a2,\frac b2)$ we denote $\phi_{(a,b)}$ and which has profile $h_b$.
The cohomology identification and representation count below hold for every $s\ge\frac12$ (giving $\mathcal K_5=0$ at $s=\frac12$); for the Maxwell-contraction construction and polar profiles we take $s\ge1$. 

The kernel $\mathcal K_5$ is isomorphic to the first cohomology $H^1$ of the twisted deformation complex $\Omega^0(E_s)\xrightarrow{\dd}\Omega^1(E_s)\xrightarrow{\dd^+}\Omega^2_+(E_s)$. For $s\ge1$ it is the $Sp(2)\cong Spin(5)$ irreducible representation of highest weight $(s,s-1)$, of dimension $\frac13(2s-1)(2s+1)(2s+3)=2I(s)-(2s+1)$. Its $SU(2)_L\times SU(2)_G$ content is one copy of each type $(j_L,j_G)$ with $j_L+j_G=s-1$, which form the family $\mathcal K_5^{(1)}$, and one copy of each type with $j_L+j_G=s$ and $j_L,j_G\ge\frac12$, which form the family $\mathcal K_5^{(4)}$. The superscripts count the conformal Killing fields, one or a quartet, in the construction below. At $s=\tfrac12$ the kernel is zero.

The sequence is a complex because $\dd^+\dd u=\mathcal F^+u=0$ for anti-self-dual curvature, and it is elliptic~\cite{AHS}. $H^0=0$ because $a_N$ is irreducible. On $\Omega^2_+(E_s)$ the Weitzenb\"ock formula gives $2\,\dd^+(\dd^+)^*=\dd\dd^*+\dd^*\dd=\nabla^*\nabla+\frac13{\rm scal}=\nabla^*\nabla+16$, since $\mathsf g_1$ is conformally flat and the anti-self-dual curvature does not act on $\Lambda^2_+$, so $H^2=0$. By Hodge theory $H^1\cong\ker\dd^*\cap\ker\dd^+=\mathcal K_5$.
The complex is conformally invariant, so the group $\bar N^{-1}Sp(2)\bar N$, which preserves $\mathsf g_2$ and, up to gauge, $a_N$, acts on it and on $H^1$. Restricted to $SU(2)_L\times SU(2)_G$ the isomorphism with $\mathcal K_5$ is equivariant. Since $\HH P^1=Sp(2)/(Sp(1)\times Sp(1))$, Frobenius reciprocity gives the multiplicity of an irreducible $V$ in the sections of each bundle of the complex as the multiplicity of the fibre representation at the centre in the restriction of $V$ to the isotropy group. Each isotypic part of the complex is then a finite sequence, exact except in degree one, and with $[V:(j_+,j_-)]$ the multiplicity of the type with $S^+$-spin $j_+$ and $S^-$-spin $j_-$,
\begin{equation}
{\rm mult}_V(H^1)=[V:(\tfrac12,s+\tfrac12)]+[V:(\tfrac12,s-\tfrac12)]-[V:(0,s)]-[V:(1,s)],
\end{equation}
the four types being the fibres of $\Lambda^1\otimes E_s$, $E_s$ and $\Lambda^2_+\otimes E_s$.
By the Gelfand--Tsetlin rule for $Spin(5)\downarrow Spin(4)$~\cite{Molev}, $[V_{(\ell_1,\ell_2)}:(j_+,j_-)]=1$ if $\ell_1\ge j_++j_-\ge\ell_2\ge|j_+-j_-|$ with $\ell_1-j_+-j_-\in\Z$, and $0$ otherwise. For $\ell_2=s-1$ the right-hand side equals $[\ell_1\ge s]-[\ell_1\ge s+1]$, which is $1$ exactly for $\ell_1=s$. For $\ell_2=s$ and $\ell_2=s+1$ the four terms cancel, and for other $\ell_2$ all vanish. For $s\ge1$ this gives $H^1\cong V_{(s,s-1)}$, whose dimension follows from the Weyl formula $\dim V_{(\ell_1,\ell_2)}=\frac16(2\ell_1+3)(2\ell_2+1)(\ell_1-\ell_2+1)(\ell_1+\ell_2+2)$. The same rule restricted to $V_{(s,s-1)}$ gives $j_L+j_G\in\{s-1,s\}$ and $|j_L-j_G|\le s-1$, each once.
As a cross-check, the same computation for $\ker\mathsf D_-$ on $S^-\otimes E_s$ gives $V_{(s-\frac12,s-\frac12)}\cong{\rm Sym}^{2s-1}\C^4$, in agreement with the floor states~\eqref{eq:fb:floorstates}.

The kernel is generated by contracting Maxwell fields with conformal Killing fields. For $X\in\mathfrak p_1$ and $\phi\in M$, $\iota_X\phi\in \mathcal K_5$. If $X\ne0$ and $\phi\ne0$, then $\iota_X\phi\ne0$, for real $X$ and also for complex $X$ when $\phi\in\epsilon\odot\epsilon\odot\mathcal T_{2s-2}$.
Write $X=\nabla h$, with $\nabla_\mu X_\nu=c_X\,g_{\mu\nu}$ for a function $c_X$. Then $\dd^*(\iota_X\phi)=-(\nabla^\nu X^\mu)\phi_{\mu\nu}\pm\iota_X\dd^*\phi=0$, because $\nabla X$ is symmetric and $\dd^*\phi=\pm{*}\dd\phi=0$ for anti-self-dual closed $\phi$. By Cartan's formula $\dd(\iota_X\phi)=\mathcal L_X\phi$ (covariantized), and on two-forms in four dimensions $\mathcal L_X$ commutes with the Hodge star when $X$ is conformal, so $\dd(\iota_X\phi)$ is anti-self-dual and $\dd^+(\iota_X\phi)=0$.
For real $X$, nonvanishing follows from the pointwise identity $|\iota_X\phi|^2=\frac12|X|^2|\phi|^2$, which holds because a real anti-self-dual two-form acts on vectors as $1/\sqrt2$ times its norm and an orthogonal complex structure, applied to the real and imaginary parts of the components of $\phi$ in a unitary frame of $E_s$.
For complex $X$ and $\phi=\epsilon\odot\epsilon\odot\omega$, work in the frame of $\mathsf g_2$ and contract all colour indices with a fixed spinor $\theta^{C'}$. Then $\phi_{A'B'}(\theta)=\theta_{A'}\theta_{B'}\,\omega(\theta)$ and
\begin{equation}
(\iota_X\phi)_{AA'}(\theta)=X_A{}^{B'}\theta_{B'}\;\theta_{A'}\;\omega(\theta),
\end{equation}
a product of polynomials in $\theta$, each nonzero wherever $X$ and $\omega$ are nonzero. This covers the highest-weight vectors used in the next paragraph.

These contractions exhaust the kernel, $\mathcal K_5=\iota(\mathfrak p_1\otimes M)$. The state of $\mathcal K_5^{(1)}$ of type $(\frac a2,\frac b2)$ is $\iota_{X_0}\phi_{(a,b)}$, and the state of $\mathcal K_5^{(4)}$ of type $(\frac{a+1}2,\frac{b+1}2)$ is the component of that type of $\iota_{X_\mu}\phi_{(a,b)}$, $a+b=2s-2$.
By the representation content above, $\mathcal K_5$ is multiplicity-free. Since $X_0$ is invariant, $\iota_{X_0}\phi_{(a,b)}$ is nonzero by the preceding paragraph, lies in $\mathcal K_5$, and has type $(\frac a2,\frac b2)$, a type of $\mathcal K_5^{(1)}$; it is therefore that state. The map $X\otimes\phi\mapsto\iota_X\phi$ from $(\frac12,\frac12)\otimes(\frac a2,\frac b2)$ to $\mathcal K_5$ is equivariant. The product of the highest-weight vectors has weight $(\frac{a+1}2,\frac{b+1}2)$, its image is nonzero by the preceding paragraph, and in $\mathcal K_5$ this weight occurs only in the type $(\frac{a+1}2,\frac{b+1}2)$, so the image contains that type of $\mathcal K_5^{(4)}$. Every type of $\mathcal K_5$ is thus reached.

The polar profiles are fixed by this construction up to a bracket. The state of $\mathcal K_5^{(1)}$ of type $(s-1-\frac b2,\frac b2)$ has the profile $\sigma g_b$ exactly. The state of $\mathcal K_5^{(4)}$ of type $(\frac{a+1}2,\frac{b+1}2)$ has a profile of the form
\begin{equation}\label{eq:fb:bracket}
\varrho\propto\kappa^2h_b\big[(1-r_1)+t+(1-r_2)t^2\big],\qquad 0\le r_1,r_2\le1,
\end{equation}
with $r_1=0$ if $b=0$ and $r_2=0$ if $a=0$. Equivalently $\varrho$ lies in the parallelogram with vertices $\hat g_b=g_b(1+t+t^2)/(1+t)$, $g_{b+1}$, $g_b$ and $\sigma g_b$, at $(r_1,r_2)=(0,0),(1,0),(0,1),(1,1)$.
With the pointwise identity $|\iota_X\phi|^2=\frac12|X|^2|\phi|^2$, \eqref{eq:fb:CKFnorms} and the conformal invariance of the Maxwell density, the profile in $\mathcal K_5^{(1)}$ is $t\kappa^2h_b=\sigma g_b$.
For $\mathcal K_5^{(4)}$, sum $|\iota_{X_\mu}\phi_i|^2$ over the quartet and an orthonormal basis $\{\phi_i\}$ of $\phi_{(a,b)}$. The result is $\propto(1+t+t^2)\kappa^2h_b$. On the other hand $(\frac12,\frac12)\otimes(\frac a2,\frac b2)$ contains the types $(\frac{a\pm1}2,\frac{b\pm1}2)$. By Schur's lemma each contributes a nonnegative multiple of the profile of its image in $\mathcal K_5$. The type $(\frac{a-1}2,\frac{b-1}2)$ is absent from $\mathcal K_5$. The types $(\frac{a+1}2,\frac{b-1}2)$ and $(\frac{a-1}2,\frac{b+1}2)$ are types of $\mathcal K_5^{(1)}$, with profiles $t\kappa^2h_{b-1}=\kappa^2h_b$ and $t\kappa^2h_{b+1}=\kappa^2t^2h_b$, absent when $b=0$ or $a=0$ respectively. Subtracting them gives \eqref{eq:fb:bracket}, and positivity of the remainder near $t=0$ and $t=\infty$ gives $r_1,r_2\le1$. The vertex identities follow from $\kappa h_b=g_b$, $t\,g_b=g_{b+1}$ and $\kappa t=\sigma$.

At the vertex values $r_1=r_2=0$ the bracket gives $\hat g_b$, and at $s=1$ ($a=b=0$) it collapses to $\hat g_0$, so that $\mathcal K_5=\iota_{\mathfrak p_1}\mathcal F$ is spanned exactly by the five conformal deformations of the instanton.
For $b=0$ the bracket is the segment $g_0+(1-r_2)\varrho_*$, $0\le r_2\le1$, with $\varrho_*=\kappa^2t^2h_0=t^3(1+t)^{-2}(t+\frac12)^{-(2s+2)}$.
The exact point inside the bracket~\eqref{eq:fb:bracket} is determined analytically by the leading asymptotics of the contraction $(\iota_{X_\mu}\phi)_{AA'}(\theta)=(X_\mu)_A{}^{B'}\theta_{B'}\theta_{A'}\omega(\theta)$ at $t\to0$ and $t\to\infty$. At the origin $x=0$ ($t\to0$), the quartet reduces to $X_\mu(0)=\partial_\mu\sim e_A\otimes\tilde e^{B'}$. Its $SU(2)_L$ doublet multiplies $\theta_{B'}\omega(\theta)$ inside the commutative polynomial ring $\mathrm{Sym}^*S^-$, identically annihilating the antisymmetric spin-$\frac{a-1}2$ contraction (so the type $(\frac{a-1}2,\frac{b+1}2)$ vanishes faster at $t\to0$), while its $SU(2)_G$ doublet $e_A$ is uncontracted and acts as the identity on $\C^2_A\otimes V_{b/2}$. Hence the decomposition $\C^2_A\otimes V_{b/2}=V_{(b+1)/2}\oplus V_{(b-1)/2}$ partitions the $t\to0$ norm sum isometrically in proportion to representation dimension, fixing
\begin{equation}
 r_1=\frac{\dim V_{(b-1)/2}}{\dim(\C^2\otimes V_{b/2})}=\frac{b}{2(b+1)},\qquad
 r_2=\frac{a}{2(a+1)},
 \label{eq:fb:r1r2}
\end{equation}
where $r_2$ follows by exact flat-frame duality at $|x|\to\infty$. In quaternion form, $v^\mu X_\mu(x)=(1+|x|^2)v-2(x\cdot v)x=v-x\bar v x$, whose leading $|x|\to\infty$ term $-|x|^2\hat x\bar v\hat x$ obeys $\hat x^\dagger(-|x|^2\hat x\bar v\hat x)\theta=-|x|^2\bar v(\hat x\theta)$. Under the unitary $SU(2)$ spinor rotation $\tilde\theta=\hat x\theta$ (which preserves the unitary-invariant norm and turns $\tilde x\theta=\sqrt2|x|\tilde\theta$ into a pure polynomial in $\tilde\theta$), $\bar v$ multiplies the degree-$b$ $\lambda$-factor inside the commutative ring $\mathrm{Sym}^*(\C^2)$ while leaving the $SU(2)_L$ doublet uncontracted (exchanging $a\leftrightarrow b$ and the contracted/uncontracted doublets in the norm ratio, with $\sqrt2$ cancelling out), which fixes $r_2=\frac{a}{2(a+1)}$. Substituting Eq.~\eqref{eq:fb:r1r2} into Eq.~\eqref{eq:fb:bracket} gives the exact $\mathcal K_5^{(4)}$ polar density $(a+1)(b+1+\kappa)g_b+(b+1)(a+1+\sigma)g_{b+1}$. As a numerical check, we compared the normalized attraction $\langle W\rangle$ obtained from these densities with its value in the Galerkin zero modes. At $s=1,\tfrac32,2,\tfrac52,3,4,5,6$ this tests 42 channels in each family. The largest absolute differences are below $1.22\times10^{-5}$ for $\mathcal K_5^{(4)}$ and $1.18\times10^{-5}$ for $\mathcal K_5^{(1)}$. %

The bracket suffices to bound the weight. Every block-5 floor state at $j_R=0$ has $\langle W\rangle\le w_0=\langle W\rangle_{g_0}$, the value for the block-4 floor state of type $(s-\frac12,0)$, whatever its profile inside the bracket \eqref{eq:fb:bracket}.
Each vertex density $\varrho_v$ has $\varrho_v/g_0$ nondecreasing in $t$, the four ratios being $t^b$, $t^{b+1}$, $\sigma t^b$ and $t^b\big(t+\frac1{1+t}\big)$. Since $W$ is decreasing, Chebyshev's integral inequality (the mean of the product of a decreasing and a nondecreasing function does not exceed the product of the means) gives $\langle W\rangle_{\varrho_v}\le w_0$. A density in the bracket is a positive combination of vertex densities, so its weight average is a mediant of the vertex weight averages and does not exceed $w_0$. Finally $W$ is $SO(4)$-invariant, so distinct types of $\mathcal K_5$ are $W$-orthogonal, and the weight average in a superposition is a convex combination of those of its types.

\paragraph*{Block-5 gap.}\phantomsection\label{app:fb:gap5}

In the rest of this subsection $s\ge1$, except that the Hodge gap~\eqref{eq:fb:gap5}, which uses only the vanishing of the twisted cohomology in degrees zero and two, holds for every $s\ge\frac12$. At $j_R=0$ the colour Laplacian $\nabla^*\nabla$ on $\Omega^0(E_s)$ is the block-1 single-set operator. In the channel $(j_L,0,j_G)$ it has one polar component if $s$ occurs in $j_L\otimes j_G$ and is absent otherwise, and we denote its lowest eigenvalue by $\Lambda_0(j_L,j_G)$.
Along the $L$- and $G$-orbits through $x_\alpha$ the colour generators of the channel reduction are $c_LJ$ and $c_GJ$, with $c_L=\kappa/D$ and $c_G=2\sigma/D$, which follow from the frame~\eqref{eq:frame} at the base points (see also Eq.~\eqref{eq:un:orbit}).
Put $C_L=j_L(j_L+1)$, $C_G=j_G(j_G+1)$, and
\begin{equation}\label{eq:fb:f5}
\mathsf f(\Lambda)=\Lambda+12-\sqrt{144+16\Lambda},
\end{equation}
which is increasing for $\Lambda\ge0$, with $\mathsf f(4)=16-4\sqrt{13}>\tfrac32$.

In the channel $(j_L,0,j_G)$, on $L^2(\sigma\kappa\,\dd\sigma)$, the scalar operator is
\begin{equation}\label{eq:fb:bosonic}
\nabla^*\nabla=-\frac{4}{\sigma\kappa}\,\partial_\sigma\big(\sigma^2\kappa^2\partial_\sigma\big)+\frac{8C_L}{D\kappa}+\frac{4C_G}{D\sigma}-\frac{8C_2}{D^2}.
\end{equation}
The polar term is $-(\sin\alpha\cos\alpha)^{-3}\partial_\alpha\sin^3\!\alpha\cos^3\!\alpha\,\partial_\alpha$ written in $\sigma$, with $\sin^3\!\alpha\cos^3\!\alpha\,\dd\alpha\propto\sigma\kappa\,\dd\sigma$. At $j_R=0$ the vertical derivative vanishes, so in the notation of Eq.~\eqref{eq:fb:split} $v_a=-u_a$, and the orbit term is $\sum_a|u_a|^2/(\sigma\kappa)$. For the reduced section $T$, the stabilizer-equivariant map $V_{j_L}\otimes V_{j_G}\to V_s$ of Eq.~\eqref{eq:un:equivariance} without fermion generators, $u_a$ is $-2\ii$ times the $a$-th component of $c_LJT-T\pi_L$, where $\pi_L,\pi_G$ are the spin operators of the channel and stabilizer equivariance gives $JT=T(\pi_L+\pi_G)$. Hence $c_LJT-T\pi_L=T(c_L\pi_G-c_G\pi_L)$, using $c_L+c_G=1$, and with $2\pi_L\cdot\pi_G=C_2-C_L-C_G$ on $T$,
\begin{equation}
\begin{aligned}
\sum_a\frac{|u_a|^2}{\sigma\kappa}&=\frac{4}{\sigma\kappa}\big[c_L^2C_G+c_G^2C_L+c_Lc_G(C_L+C_G-C_2)\big] \\
&=\frac{4}{\sigma\kappa}\big[c_LC_G+c_GC_L-c_Lc_GC_2\big],
\end{aligned}
\end{equation}
which is the potential in \eqref{eq:fb:bosonic}. At $s=0$, where $j_L=j_G=j$, it reduces to $4j(j+1)/(\sigma\kappa)$, the scalar value.

The lowest eigenvalue of $\nabla^*\nabla$ has an explicit lower bound, by Barta's inequality \cite{Barta}. If $j_L+j_G=s$, the function $\Phi=\sigma^{j_G}\kappa^{j_L}D^{-s}>0$ satisfies $\nabla^*\nabla\Phi=8(2j_L+j_G)D^{-1}\,\Phi$ identically, and
\begin{equation}\label{eq:fb:AP}
\Lambda_0(j_L,j_G)\ \ge\ 4(2j_L+j_G)=4(s+j_L).
\end{equation}
If $j_L+j_G\ge s$, then $\Lambda_0(j_L,j_G)\ge4s$. At $s=\frac32$, $\Lambda_0(1,\frac12)\ge14$.
The eigenfunction identity is verified by direct differentiation (symbolic computation; it holds for all real $j_L,j_G\ge0$ with $j_L+j_G=s$). For $u=\Phi\varphi$ with $\varphi\in C^\infty_c(0,1)$, an integration by parts gives the ground-state representation
\begin{equation}
\langle u,\nabla^*\nabla u\rangle=\int_0^1 4\sigma^2\kappa^2\Phi^2|\varphi'|^2\,\dd\sigma+\int_0^1\sigma\kappa\,\frac{\nabla^*\nabla\Phi}{\Phi}\,|u|^2\,\dd\sigma\ \ge\ \inf_\sigma\frac{8(2j_L+j_G)}{D}\,\|u\|^2 ,
\end{equation}
and $D\le2$. The endpoints $\sigma=0,1$ correspond to two points of $S^4$, which have zero capacity in four dimensions, so $C^\infty_c(0,1)$ is a form core and \eqref{eq:fb:AP} follows.
If $j_L+j_G>s$, put $j'_G=\min(j_G,s)$ and $j'_L=s-j'_G\in[0,j_L]$. The potential in \eqref{eq:fb:bosonic} increases with $C_L$ and $C_G$, so $\nabla^*\nabla$ dominates the operator with $(j'_L,j'_G)$, to which the identity applies with real exponents, so $\Lambda_0\ge4(s+j'_L)\ge4s$.
At $s=\frac32$ in channel $(1,\frac12)$, the generic bound~\eqref{eq:fb:AP} gives only $\Lambda_0\ge10$ (whose gap $\mathsf f(10)=22-4\sqrt{19}\simeq4.56$ is too weak for the Part~II Schur test of Appendix~\ref{app:un:uniform}). Replacing $\Phi$ by $\sigma^{1/2}\kappa D^{-1}$ improves the $\nabla^*\nabla$-quotient to $14+7\sigma(1-\sigma)/(1+\sigma)^2\ge14$ identically.

The gap above $\mathcal K_5$ then follows. On the orthogonal complement of $\mathcal K_5$ in the block-5 channel $(j_L,0,j_G)$, with $\Lambda_0=\Lambda_0(j_L,j_G)$,
\begin{equation}\label{eq:fb:gap5}
\mathcal H_5+8\ \ge\
\begin{cases}
16, & s\notin j_L\otimes j_G,\\
\min\{16,\,\mathsf f(\Lambda_0)\}, & s\in j_L\otimes j_G,
\end{cases}
\end{equation}
where $s\in j_L\otimes j_G$ means that spin $s$ occurs in $j_L\otimes j_G$. For $s>2$ the bound $16$ of the $\eta$ block improves to $4s+8$, by the conformal argument of Appendix~\ref{app:fb:chgap}.
In particular the gap is at least $16$ in every type of $\mathcal K_5^{(1)}$ and at least $\min\{16,\mathsf f(4(s+j_L))\}$ in every type of $\mathcal K_5^{(4)}$, and $\mathcal H_5+8\ge\min\{16,\mathsf f(4s)\}\ge \mathsf f(4)$ on every block-5 channel that contains no floor state.
By Hodge theory for the deformation complex, $\gamma=\gamma_0+\dd u+(\dd^+)^*\eta$ orthogonally, with $\gamma_0\in \mathcal K_5$, $u\in\Omega^0(E_s)$ and $\eta\in\Omega^2_+(E_s)$. Since $\dd^+\dd=0$ and $\dd^*(\dd^+)^*=(\dd^+\dd)^*=0$, it follows that $\dd^*\gamma=\nabla^*\nabla u$ and $\dd^+\gamma=\dd^+(\dd^+)^*\eta$. For $\gamma_0=0$ the square \eqref{eq:fb:block5sq} is $Q_1+Q_2$ and the squared norm is $N_1+N_2$, with an $(f,u)$ block and an $\eta$ block,
\begin{equation}
\begin{gathered}
Q_1=\|\nabla f\|^2+8\|f\|^2+\|\nabla^*\nabla u-4f\|^2,\qquad N_1=\|f\|^2+\langle u,\nabla^*\nabla u\rangle,\\
Q_2=2\|\dd^+(\dd^+)^*\eta\|^2,\qquad N_2=\langle\eta,\dd^+(\dd^+)^*\eta\rangle .
\end{gathered}
\end{equation}
Both $f$ and $u$ are $E_s$-valued functions in the channel $(j_L,0,j_G)$, because the splitting of $\mathbb F_5$ into scalar and one-form parts is $Sp(2)$-equivariant and $\dd$ is equivariant. Since $\nabla^*\nabla$ commutes with $G$, $Q_1$ and $N_1$ decompose over its eigenspaces in the channel. On the eigenspace with eigenvalue $\Lambda$, and with $\tilde u=\sqrt\Lambda\,u$, $Q_1$ is the form of $\bigl(\begin{smallmatrix}\Lambda+24&-4\sqrt\Lambda\\-4\sqrt\Lambda&\Lambda\end{smallmatrix}\bigr)$ in $(f,\tilde u)$ and $N_1=|f|^2+|\tilde u|^2$. The eigenvalues are $\Lambda+12\mp\sqrt{144+16\Lambda}$, and the smaller is $\mathsf f(\Lambda)\ge \mathsf f(\Lambda_0)$. If $s$ does not occur in $j_L\otimes j_G$ the channel has no $E_s$-valued functions and this block is absent. This includes the types of $\mathcal K_5^{(1)}$, where $j_L+j_G<s$. For the $\eta$ block, the Weitzenb\"ock identity $2\,\dd^+(\dd^+)^*=\nabla^*\nabla+16$ above gives $\dd^+(\dd^+)^*\ge8$, hence a ratio at least $2\cdot8=16$. The remaining statements follow from the bounds on $\Lambda_0$ above and from $\mathsf f(\Lambda)$ being increasing.

The block-5 inputs of the uniform lower bound on $E_c$ are the multiplicity-free kernel with its profile bracket and the gaps \eqref{eq:fb:gap5}, used with $\Lambda_0(1,\frac12)\ge14$ at $s=\frac32$ (Appendix~\ref{app:un:uniform}).
The block-4 counterparts are the floor states~\eqref{eq:fb:floorstates}, with the profiles $g_b$, and the gap \eqref{eq:fb:gap4}.

\subsection{Half-integer spin and conformality of the fundamental Wilson line}
\label{app:fb:halfint}

At $\theta_{5d}=0$ and half-integer $s\ge\frac32$, the channel $(s-1,0,0)$ is excluded by the physical projection. The numerical scans favour $(s-1,0,\frac12)$ in block $(4,5)$ (Appendix~\ref{app:numerics}), which Eq.~\eqref{eq:lawcor3} selects for sufficiently large spin. At $s=\frac12$, Secs.~\ref{app:fb:twistor} and~\ref{app:fb:block5} prove $\dim\mathcal H_{\rm dBPS}^{(4,4)}(\frac12)=\binom23=0$ and $\dim\mathcal K_5=0$, so no doubly-BPS state exists in any block.
The trial state $T_s$ is then unavailable, and its role is taken by the following states. The construction uses only $2s\ge3$. Its states are physical at $\theta_{5d}=0$ for half-integer $s$ and at $\theta_{5d}=\pi$ for integer $s$, while at $\theta_{5d}=\pi$ the half-integer spins have the state $T_s$ of Eq.~\eqref{eq:Ts} in the physical channel $(s-1,0,0)$.

Work at $j_R=0$. Write a state of block $(4,5)$ as $\psi_{B|A'B'}$, where $A'$ is the index of the set $+$ (block 4, in $S^-$) and $(B,B')$ that of the set $-$ (block 5, in $\Lambda^1=S^+\otimes S^-$), colour suppressed.
Such a state is doubly BPS if and only if
\begin{equation}\label{eq:fb:45cond}
\psi_{B|A'B'}=\psi_{B|(A'B')},\qquad \nabla_C{}^{A'}\psi_{B|A'B'}=0 .
\end{equation}
By the sharp floor~\eqref{eq:fb:H4}, a doubly-BPS state has the set $+$ in $S^-$ and the set $-$ in $\Lambda^1$ with no scalar component. With the set $+$ active, the spectator's $B$-term (the block-5 computation of Sec.~\ref{app:fb:sharp}) maps the one-form of the set $-$ into its scalar slot and contributes to the Dirac operator of the set $+$ a multiple of $\epsilon^{A'B'}\psi_{C|A'B'}$, which lies in a subspace orthogonal to that of the derivative part $\nabla_C{}^{A'}\psi_{B|A'B'}$. With the set $-$ active, the spectator's Killing term contributes to $\mathsf D'$ a multiple of the same contraction, again orthogonal to the derivative part $\nabla_C{}^{B'}\psi_{B|A'B'}$. In each case both orthogonal parts must vanish. The contraction vanishes if and only if $\psi$ is symmetric in $(A',B')$, and then the two derivative conditions coincide.

Solutions of \eqref{eq:fb:45cond} are obtained from massless spin-$\frac32$ fields. Let $X\in\mathfrak p_1$ and let $\chi\in\Gamma({\rm Sym}^3S^-\otimes E_s)$ be a twisted massless spin-$\frac32$ field, $\nabla^{AA'}\chi_{A'B'C'}=0$. Then $\psi_{B|A'B'}=X_B{}^{C'}\chi_{C'A'B'}$ is doubly BPS.
First, $\psi$ is symmetric in $(A',B')$ because $\chi$ is totally symmetric. Further
$\nabla_C{}^{A'}\psi_{B|A'B'}=(\nabla_C{}^{A'}X_B{}^{C'})\chi_{C'A'B'}+X_B{}^{C'}\nabla_C{}^{A'}\chi_{C'A'B'}$. The second term vanishes by the spin-$\frac32$ equation. For a gradient conformal Killing field $\nabla_{CA'}X_{BC'}=c_X\,\epsilon_{CB}\epsilon_{A'C'}$, so the first term is $c_X\epsilon_{CB}\chi^{A'}{}_{A'B'}=0$, the trace of a symmetric spinor.

For $2s\ge3$ such fields are given in the frame of $\mathsf g_2$ by $\chi=\epsilon\odot\epsilon\odot\epsilon\odot\omega$ with $\omega\in\mathcal T_{2s-3}$. The type $(\frac{2s-3-b}2,\frac b2)$ has the profile $\varrho^{(3)}_b=t^{b+1}(1+t)(t+\frac12)^{-(2s+2)}$.
As for the doubly-BPS family $\epsilon\odot\epsilon\odot\mathcal T_{2s-2}$ of Sec.~\ref{app:fb:twistor}, the derivative of $\epsilon^{\odot3}\odot\omega$ falls on $\omega$ and is killed by the twistor equation of valence $2s-3$. The massless equation is conformally covariant, and \eqref{eq:fb:profile} with $n=3$ gives the profile.
For $X$ in the quartet and $\chi$ of type $(s-\frac32,0)$, $X\cdot\chi$ has the types $(s-1,\frac12)\oplus(s-2,\frac12)$, the second being absent at $s=\frac32$, because $(\frac12,\frac12)\otimes(s-\frac32,0)=(s-1,\frac12)\oplus(s-2,\frac12)$ and $X\cdot\chi$ is equivariant in $(X,\chi)$. Both types have $j_G=\frac12$, so they are physical when $s+\theta_{5d}/(2\pi)$ is a half-integer. The first gives the candidate channel $(s-1,0,\frac12)$ tested below.

We now prove the half-integer instability bound $E_c\le1/w_{\rm qrt}(s)$ of Sec.~\ref{sec:instability}, with $w_{\rm qrt}(s)$ and $\varrho_1$ as in Eq.~\eqref{eq:vpp}, together with the bound
\begin{equation}\label{eq:fb:vpp-bound}
0\le w_s-w_{\rm qrt}(s)\le\frac{3}{2(2s-1)(2s-2)},\qquad
1+\frac{3}{2s}\le\frac{1}{w_{\rm qrt}(s)}=1+\frac{3}{2s}+O(s^{-2}),
\end{equation}
where $w_s=\frac{2s}{2s+3}$.
Let $\{\chi_i\}$ be an orthonormal basis of the type $(s-\frac32,0)$ of the spin-$\frac32$ fields above and $\{X_\mu\}$ the quartet. Each $\psi_{\mu i}=X_\mu\cdot\chi_i$ is a physical doubly-BPS state, as just shown, so \eqref{eq:fb:trial} applies to it.
Use the Clifford normalization of Appendix~\ref{app:fb:sharp} for the map $X\colon S^-\to S^+$. For real $X$, $X^\dagger X=|X|^2\mathbf 1$ on the contracted index. With the tensor-product spinor norm this gives $|X\cdot\chi|^2=|X|^2|\chi|^2$ pointwise. By \eqref{eq:fb:CKFnorms} and the profile $\varrho^{(3)}_0$ of the spin-$\frac32$ fields,
$\sum_{\mu,i}|\psi_{\mu i}|^2$ has the profile $(1+t+t^2)\kappa^2\cdot t(1+t)(t+\frac12)^{-(2s+2)}=\varrho_1$.
By the mediant inequality some nonzero $\psi_{\mu i}$ has $\langle W\rangle\ge\sum\int W|\psi_{\mu i}|^2\big/\sum\int|\psi_{\mu i}|^2=w_{\rm qrt}(s)$, and for $E>1/w_{\rm qrt}(s)$, \eqref{eq:fb:trial} places its Rayleigh quotient below $-9$.
For \eqref{eq:fb:vpp-bound}, writing $\varrho_1=h_0(1+t^2/(1+t))$ and averaging over $h_0$ (where $\langle W\rangle=w_s$ by \eqref{eq:fb:Wb}) gives $w_{\rm qrt}(s)-w_s=\langle(W-w_s)t^2/(1+t)\rangle/(1+\langle t^2/(1+t)\rangle)\le0$ by Chebyshev's inequality, because $W$ decreases and $t^2/(1+t)$ increases. Also, $0\le W\le1$ gives $w_s-w_{\rm qrt}(s)\le\langle t^2\rangle=\frac14\mathrm B(4,2s-2)/\mathrm B(2,2s)=\frac{3}{2(2s-1)(2s-2)}$.

We can sharpen this bound by isolating the exact critical-channel doubly-BPS state inside the quartet image for every $s\ge\frac32$ with half-integer $s+\theta_{5d}/(2\pi)$. In the decomposition $(\frac12,\frac12)\otimes(s-\frac32,0)=(s-1,\frac12)\oplus(s-2,\frac12)$ (with the second summand absent at $s=\frac32$), the highest-weight vector $T'_s=X_{++}\cdot\chi_{(2s-3,0)}^{\rm hw}$ lies purely in the critical channel $(s-1,0,\frac12)$. Its exact polar profile follows from the flat-frame expression $X_v(x)=v-x\bar v x$ of Sec.~\ref{app:fb:block5}. Because $X_v(x)$ contains only degrees $0$ and $2$ in $x$, and commutative multiplication in $\mathrm{Sym}^*S^-$ annihilates the $x^0$ term on the $(s-2,\frac12)$ summand identically, the $(s-2,\frac12)$ amplitude is purely $O(|x|^2)$ and its squared norm is purely $O(t^2)$. Subtracting it from the quartet-summed profile $(1+t+t^2)\kappa^2\varrho^{(3)}_0$ therefore leaves both the $t^0$ and $t^1$ coefficients equal to $1$. At $|x|\to\infty$, the same unitary spinor rotation $\tilde\theta=\hat x\theta$ as in Eq.~\eqref{eq:fb:r1r2} leaves the $SU(2)_L$ doublet of $\bar v$ uncontracted on $\C^2\otimes V_{s-3/2}$, partitioning the $t^2$ norm sum in proportion to representation dimension and fixing
\begin{equation}
 \varrho_{\rm crit}(t)=\frac{t\,(1+t+\vartheta_s t^2)}{(1+t)\,(t+\tfrac12)^{2s+2}},\qquad
 \vartheta_s=\frac{\dim V_{s-1}}{\dim(\C^2\otimes V_{s-3/2})}=\frac{2s-1}{4s-4},\qquad s\ge\tfrac32
 \label{eq:fb:rhocrit}
\end{equation}
(also verified symbolically by direct $S^3$-averaged unitary-invariant Bombieri norm for all $a=2s-3\ge0$). Writing $w_{\rm crit}(s)=\langle W\rangle_{\varrho_{\rm crit}}$, Eq.~\eqref{eq:fb:trial} gives the sharpened analytic upper bound $E_c(s)\le 1/w_{\rm crit}(s)\le 1/w_{\rm qrt}(s)$ for every half-integer $s\ge\frac32$, with strict improvement for all $s\ge\frac52$.

Both $1/w_{\rm qrt}(s)$ ($\vartheta=1$) and $1/w_{\rm crit}(s)$ ($\vartheta=\vartheta_s$) can be expanded through order $s^{-3}$ by the same moment identity. For any $\vartheta>0$, let $\varrho_\vartheta(t)=h_0(t)\bigl(1+\vartheta t^2/(1+t)\bigr)$ with $h_0(t)=t(t+\tfrac12)^{-(2s+2)}$ and $w_\vartheta(s)=\langle W\rangle_{\varrho_\vartheta}$. In the numerator integral $\int_0^\infty W(t)\varrho_\vartheta(t)\,\dd t$, the factor $(1+t)$ in $W(t)=(1+t)/\bigl[4(t+\tfrac12)^2\bigr]$ cancels the denominator $(1+t)$ of $\varrho_\vartheta$ identically, leaving a pure rational function of $s$ and $\vartheta$ with no $\ln2$, 
\begin{equation}
 \mathcal N_\vartheta(s)=\frac{\int_0^\infty W(t)\,\varrho_\vartheta(t)\,\dd t}{\int_0^\infty h_0(t)\,\dd t}
 =\frac{8s^2+8s+3\vartheta}{4(s+1)(2s+3)}.
 \label{eq:fb:numexact}
\end{equation}
In the denominator $\int_0^\infty\varrho_\vartheta\,\dd t\big/\!\int_0^\infty h_0\,\dd t=1+\vartheta\langle t^2/(1+t)\rangle_{h_0}$, every $t>0$ obeys the alternating geometric bounds $1-t+t^2-t^3\le(1+t)^{-1}\le1-t+t^2$. Averaging against $h_0$ using the exact Beta moments $M_k(s)=\langle t^k\rangle_{h_0}=(k+1)!\big/\bigl[2^k\prod_{j=1}^k(2s-j)\bigr]$ produces an exact two-sided rational sandwich for $1/w_\vartheta(s)$ valid for every $s>\frac52$, whose lower and upper sides agree identically through $O(s^{-4})$,
\begin{equation}
 \frac{1+\vartheta(M_2-M_3+M_4-M_5)}{\mathcal N_\vartheta(s)}
 \ \le\ \frac{1}{w_\vartheta(s)}\ \le\
 \frac{1+\vartheta(M_2-M_3+M_4)}{\mathcal N_\vartheta(s)},
 \label{eq:fb:sandwich}
\end{equation}
which immediately proves Eq.~\eqref{eq:vpp-asymp}, namely $1/w_{\rm qrt}(s)=1+\frac{3}{2s}+\frac{9}{16s^3}+\frac{15}{32s^4}+O(s^{-5})$ and $1/w_{\rm crit}(s)=1+\frac{3}{2s}+\frac{9}{32s^3}+\frac{3}{8s^4}+O(s^{-5})$. Conceptually, the $s^{-2}$ coefficient vanishes for every $\vartheta$ because $1/\langle W\rangle_{h_0}=1+\frac{3}{2s}$ is exact and the $O(t^2)$ tail adds the same fractional shift $\frac{3\vartheta}{8s^2}$ to both numerator and denominator, leaving a difference only through $1-W(t)=O(t)$ at $O(t^3)=O(s^{-3})$. High-precision quadrature confirms $s^3(1/w_{\rm qrt}(s)-1-\frac3{2s})=0.604, 0.584, 0.572, 0.567, 0.565$ at $s=\frac{21}2,\frac{41}2,\frac{101}2,\frac{201}2,\frac{401}2$, converging precisely to $\frac{9}{16}=0.5625$.

Using $(1+t+t^2)/(1+t)=t+1/(1+t)$, both $1/w_{\rm qrt}(s)$ and $1/w_{\rm crit}(s)$ are linear in $\ln2$ with rational coefficients. Exact symbolic integration gives the closed forms in Table~\ref{tab:vpp} for all six half-integer spins $\tfrac32\le{}s\le\frac{13}2$. %

\begin{table}[!htbp]
\centering
\begin{tabular}{@{}ccccc@{}}
\toprule
$s$ & $1/w_{\rm qrt}(s)$ (exact) & $1/w_{\rm qrt}(s)$ & $1/w_{\rm crit}(s)$ (exact) & $1/w_{\rm crit}(s)$\\
\midrule
$\frac32$ & $\frac{30}{11}(23-32\ln2)$ & $2.234428$ & $\frac{30}{11}(23-32\ln2)$ & $2.234428$\\
$\frac52$ & $\frac{28}{73}(337-480\ln2)$ & $1.645231$ & $\frac{7}{9}(113-160\ln2)$ & $1.630573$\\
$\frac72$ & $\frac{6}{43}(1563-2240\ln2)$ & $1.444230$ & $\frac{2}{71}(4709-6720\ln2)$ & $1.438055$\\
$\frac92$ & $\frac{99}{469}(1559-2240\ln2)$ & $1.340472$ & $\frac{66}{1631}(4691-6720\ln2)$ & $1.337439$\\
$\frac{11}2$ & $\frac{26}{4335}(77069-110880\ln2)$ & $1.276553$ & $\frac{26}{7767}(77237-110880\ln2)$ & $1.274862$\\
$\frac{13}2$ & $\frac{8}{12969}(1001129-1441440\ln2)$ & $1.233050$ & $\frac{4}{11847}(1002779-1441440\ln2)$ & $1.232018$\\
\bottomrule
\end{tabular}
\caption{Exact closed forms and decimal values of the quartet-averaged upper bound $1/w_{\rm qrt}(s)$ ($\vartheta=1$, Eq.~\eqref{eq:vpp}) and the sharpened critical-channel upper bound $1/w_{\rm crit}(s)$ ($\vartheta=\vartheta_s$, Eq.~\eqref{eq:fb:rhocrit}) on $E_c(s)$ across the six representative half-integer spins $\tfrac32\le{}s\le\frac{13}2$. Both satisfy the proved asymptotic expansions~\eqref{eq:vpp-asymp}, with $s(1/w_{\rm crit}-1)=1.8516, 1.5764, 1.5332, 1.5185, 1.5117, 1.5081$ across the six rows.}
\label{tab:vpp}
\end{table}

At $s=\frac32$, $\mathcal T_0=\C$ and the quartet image is the single type $(\frac12,\frac12)$, which is the critical channel $(\frac12,0,\frac12)$. The highest-weight state $T'_{3/2}=X_{++}\cdot(\epsilon\odot\epsilon\odot\epsilon)$ has profile $\varrho_1=\varrho_{\rm crit}$ and $\langle W\rangle=w_{\rm qrt}(\tfrac32)=w_{\rm crit}(\frac32)$, the spin-$\frac32$ analogue of $T_1=\mathcal F$. Independent numerical computations of the critical doubly-BPS state reproduce $\langle W\rangle=0.447541$ at $s=\frac32$ and agree with the exact formula~\eqref{eq:fb:rhocrit} to $10^{-6}$ in $\langle W\rangle$ at $s=\frac52,\frac72,\frac92$. %

\paragraph*{The fundamental line at the limit-point threshold.}\phantomsection\label{app:fb:s12}

We now prove Eq.~\eqref{eq:s12bound}, $L(\zeta)+8\ge E/D$ at zero Hopf-fibre spin for $s=\tfrac12$, and hence $\mu>-8$ for $|\zeta|<1$. The spin-$\tfrac12$ colour space permits a pointwise estimate on the two Pauli couplings. We first isolate the scalar operator against which that estimate is measured, the colour-twisted Laplacian $-\Delta_{S^7}^{\mathcal A}$ of Eq.~\eqref{eq:radial} at $j_R=0$, acting on colour sections before tensoring with the fermions. In the allowed scalar types $(j_L,j_G)=(\tfrac12,0)$ and $(0,\tfrac12)$, respectively,
\begin{equation}
 (-\Delta_{S^7}^{\mathcal A}+8-24D^{-2})\,\frac{\sqrt\kappa}{D^{3/2}}=0,
 \qquad
 (-\Delta_{S^7}^{\mathcal A}+8-24D^{-2})\,\frac{\sqrt\sigma}{D^{3/2}}=0 .
\end{equation}
Both identities follow by differentiation in Eq.~\eqref{eq:fb:bosonic}. The positive solutions give $-\Delta_{S^7}^{\mathcal A}+8-24D^{-2}\ge0$ by the ground-state representation. Each other scalar type must contain the colour spin $\tfrac12$ in $j_L\otimes j_G$ and hence dominates at least one of these two in the centrifugal coefficients. The inequality therefore holds on the union of scalar types and after tensoring with the constant fermion fibre. The form-core argument of Sec.~\ref{app:fb:gap5} applies at the poles.

By Eq.~\eqref{eq:pauliclosed}, the combined Pauli term is $32D^{-2}$ times $c_+\mathbf S_+\cdot\mathbf J+c_-\mathbf S_-\cdot\mathbf J$. Where both fermion slots carry spin, this is a coupling of three spins $\tfrac12$. The total-spin-$\tfrac32$ subspace has spin-coupling eigenvalue $\tfrac14$. The two total-spin-$\tfrac12$ multiplets give $-\tfrac14\pm\tfrac12\sqrt{1-3c_+c_-}$, by diagonalizing the two-dimensional recoupling matrix. Where a fermion slot is spin-free, the smallest possible eigenvalue is at least $-\tfrac34\max(c_+,c_-)$. Indeed $2\sqrt{1+3\zeta^2}\ge 1+3|\zeta|$, so the lower three-spin value bounds all slots. As $E=12c_+c_-=3(1-\zeta^2)$,
\begin{equation}
 c_+P^{(+)}+c_-P^{(-)}
 \ge\frac{-8-16\sqrt{1-E/4}}{D^2}
 \ge\frac{-24+2E}{D^2}.
\end{equation}
Combining this with the scalar comparison and $W=\kappa/D^2$ gives $L(\zeta)+8\ge E/D\ge E/2$ at $j_R=0$.

At $j_R\ge\tfrac12$, Eq.~\eqref{eq:fb:Lbound} gives $L(\zeta)+8\ge3-EW$. This is strictly positive as a form also at $E=3$, since $W<1$ away from the pinned set of measure zero. Compactness of the link resolvent thus gives $\mu>-8$ for $|\zeta|<1$. At $\zeta=\pm1$ the physical block-$4$ floor gives $\mu=-8$. Every radial index obeys $\beta\ge1$, including these endpoints. Both physical parity projections inherit the inequalities, so the fundamental Hamiltonian is essentially self-adjoint for either value of $\theta_{5d}$.
\hfill$\square$

\subsection{Near-BPS level splitting and the two theta sectors at spin one}
\label{app:fb:block45}

We prove the unified endpoint system~\eqref{eq:bpsmoments}--\eqref{eq:allslope} for every spin $s\ge1$. In block $(4,4)$ channel $(s-1,0,0)$, the orthogonal basis $\{\phi_0,\phi_1\}$ of Eq.~\eqref{eq:bpsbasis} has polar densities $g_0,g_1$ and exact norms (dropping the common factor $\tfrac18$ and integrating in $v=(1-\sigma)/(1+\sigma)=1/(1+2t)$) equal to $4(\ln2-\sum_{k=1}^{2s}(-1)^{k-1}/k)-\tfrac{2}{2s+1}$ and $\tfrac{1}{2s(2s+1)}$ minus it, with self-energy diagonals $\tfrac{1}{(2s+2)(2s+3)}$ and $\tfrac{1}{(2s+1)(2s+2)(2s+3)}$. Meanwhile, $(1\otimes(1-P_{\hat w}))\cdot 16(s+1)D^{-2}$ annihilates $T_s=\phi_0+\phi_1$ and therefore acts as $\tfrac{16(s+1)}{(2s+1)(2s+2)(2s+3)}(\hat e_0-\hat e_1)(\hat e_0-\hat e_1)^{\rm T}$. Multiplying all three matrices by $(2s+1)(2s+3)$ (matching $2(2s+1)(2s+3)2^{-(2s+1)}$ in the $t$-measure) gives Eqs.~\eqref{eq:bpsmoments}--\eqref{eq:allslope} for block $(4,4)$ identically.

In block $(4,5)$ channel $(j_L,j_R,j_G)=(s-1,0,\frac12)$ (physical at $\theta_{5d}=0$ for half-integer $s\ge\frac32$, and at $\theta_{5d}=\pi$ for integer $s\ge1$), tensoring $\mathcal K_4=\bigoplus_{b=0}^{2s-1}(s-\frac12-\frac b2,\frac b2)$ with $\mathbb F_5=(0,0)\oplus(\frac12,\frac12)$ hits $(s-1,\frac12)$ with exact multiplicity $3$ for $s\ge\frac32$ ($\psi_0\in(s-\frac12,0)\otimes(\frac12,\frac12)$, $\psi_1\in(s-1,\frac12)\otimes(0,0)$, $\psi_2\in(s-\frac32,1)\otimes(\frac12,\frac12)$) and multiplicity $2$ at $s=1$ ($b=2$ absent). Orthogonality of distinct $\mathcal K_4$ types under $L^2$ and $W$ makes $\mathsf N$ and $\mathsf W$ strictly diagonal with scaled norms $n_b(s)=2^{-(2s+1)}\int_0^\infty g_b(t)\,\dd t$ and pure-rational weights $2^{-(2s+1)}\int_0^\infty Wg_b\,\dd t=\frac{(b+1)!}{2^{b+1}}\frac{(2s+1-b)!}{(2s+3)!}$, reproducing $\mathsf W$ in Eq.~\eqref{eq:bpsmoments}.

To evaluate $L_-+8=P^{(-)}-P^{(+)}$ on $\ker(L_++8)$, recall that along the base curve $\hat w=(\cos\alpha,-\sin\alpha)$ every state in $\mathcal K_4$ couples the active spin-$\frac12$ operator $\mathbf S_{\hat w}^{(+)}$ with $\mathbf J$ to total active spin $j_{\rm act}=s-\frac12$, so $P^{(+)}=-16(s+1)/D^2$ identically. By the Wigner--Eckart projection identity on $V_{s-1/2}\subset\C^2_{\hat w}\otimes V_s$,
\begin{equation}
 \Pi_{s-1/2}\,\mathbf J\,\Pi_{s-1/2}
 =\frac{s(s+1)+(s-\frac12)(s+\frac12)-\frac34}{2(s-\frac12)(s+\frac12)}\,\mathbf J_{\rm act}
 =\frac{2(s+1)}{2s+1}\,\mathbf J_{\rm act},
\end{equation}
which, with $\mathbf J_{\rm act}$ the spin operator of the active summand, gives the following pointwise expression for the perturbation in floor matrix elements. Let $\Pi_{\rm act}$ denote projection onto the active fibre summand of spin $s-\tfrac12$.
\begin{equation}\label{eq:fb:Oop}
 \Pi_{\rm act}\bigl(P^{(-)}-P^{(+)}\bigr)\Pi_{\rm act}
 =\frac{16(s+1)}{D^2}\,\mathcal O(\alpha),\qquad
 \mathcal O(\alpha)=\mathbf 1+\frac{4}{2s+1}\,\mathbf S_{\hat w}^{(-)}\cdot\mathbf J_{\rm act}.
\end{equation}
Under the diagonal stabilizer $SU(2)_{\rm diag}$, $(s-1)\otimes\frac12$ splits as $j_{\rm diag}\in\{s-\frac12,\,s-\frac32\}$, while $\mathbb F_5$ splits as $(0,0)\to\mathbf 1_0$ and $(\frac12,\frac12)\to\mathbf 1_v\oplus\mathbf 3_v$, giving the four-dimensional orthonormal basis $|e_0\rangle=|s-\frac12;\mathbf 1_0\rangle$, $|e_{\rm rad}\rangle=|s-\frac12;\mathbf 1_v\rangle$, $|e_{\mathbf 3}\rangle=|s-\frac12;\mathbf 3_v\rangle$ and $|e_{3/2}\rangle=|s-\frac32;\mathbf 3_v\rangle$ of $\operatorname{Hom}_{SU(2)_{\rm diag}}(V_{s-1}\otimes V_{1/2},V_{s-1/2}\otimes\mathbb F_5)$ for $s\ge\tfrac32$. At $s=1$ the fourth component is absent. Under $Sp(1)_{\hat w}\times Sp(1)_{\hat w^\perp}\subset Sp(2)$, $\mathbb F_5$ decomposes into the singlet $|u_0(\alpha)\rangle=\cos(2\alpha)|e_0\rangle+\sin(2\alpha)|e_{\rm rad}\rangle$ ($\mathcal O=1$) and the $(\mathbf 2_{\hat w},\mathbf 2_{\hat w^\perp})$ quartet spanned by $|v_{\rm rad}(\alpha)\rangle=-\sin(2\alpha)|e_0\rangle+\cos(2\alpha)|e_{\rm rad}\rangle$ and $\mathbf 3_v$. On the quartet, $\mathbf S_{\hat w}^{(-)}\cdot\mathbf J_{\rm act}$ equals $-(2s+1)/4$ on $|e_{3/2}\rangle$ ($\mathcal O|e_{3/2}\rangle=0$) and acts as $\frac14\bigl(\begin{smallmatrix}0&\sqrt{4s^2-1}\\ \sqrt{4s^2-1}&-2\end{smallmatrix}\bigr)$ on $\mathrm{span}\{|v_{\rm rad}\rangle,|e_{\mathbf 3}\rangle\}$ (with eigenvalues $\frac{2s-1}{4}$ and $-\frac{2s+1}{4}$), making $\mathcal O$ rank-$1$ there,
\begin{equation}
\begin{aligned}
 \mathcal O(\alpha)&=|u_0(\alpha)\rangle\langle u_0(\alpha)|+\frac{4s}{2s+1}\,|\xi(\alpha)\rangle\langle\xi(\alpha)|,\\
 |\xi(\alpha)\rangle&=\frac{\sqrt{2s+1}\,|v_{\rm rad}(\alpha)\rangle+\sqrt{2s-1}\,|e_{\mathbf 3}\rangle}{\sqrt{4s}}.
 \end{aligned}
\end{equation}
In the basis $\{|e_0\rangle,|e_{\rm rad}\rangle,|e_{\mathbf 3}\rangle,|e_{3/2}\rangle\}$, the states $\psi_b(\alpha)=\sigma^{b/2}\kappa^{s-\frac12-\frac b2}D^{-(s+1)}\mathbf v_b$ (so $|\psi_b|^2\sin^3\!\alpha\cos^3\!\alpha\,\dd\alpha$ equals $2^{-(2s+3)}g_b(t)\,\dd t$) carry constant unit vectors fixed by $SU(2)$ $6j/9j$ recoupling with Frobenius weights $[(2j_{\rm diag}+1)/(4s-2)]^{1/2}$,
\begin{equation}
\begin{gathered}
 \mathbf v_1=(1,0,0,0),\qquad 
 \mathbf v_0=\Bigl(0,-\tfrac12,\tfrac12\sqrt{\tfrac{2s+1}{2s-1}},\sqrt{\tfrac{s-1}{2s-1}}\Bigr),\\
 \mathbf v_2=\Bigl(0,\sqrt{\tfrac{s-1}{2s-1}},\tfrac{\sqrt{(s-1)(2s+1)}}{2s-1},-\tfrac{1}{2s-1}\Bigr),
 \end{gathered}
\end{equation}
with $\mathbf v_2$ present for $s\ge\frac32$. Setting $\phi_0=\psi_0$, $\phi_1=\psi_1$ and $\phi_2=\sqrt{\vartheta_s}\,\psi_2$ with $\vartheta_s=(2s-1)/(4s-4)$ (so the $\phi$-basis coefficient vectors are $\tilde{\mathbf v}_0=\mathbf v_0$, $\tilde{\mathbf v}_1=\mathbf v_1$ and $\tilde{\mathbf v}_2=\sqrt{\vartheta_s}\,\mathbf v_2$), direct evaluation at $s\ge\tfrac32$ shows that $\mathcal O(\alpha)$ annihilates $T'_s(\alpha)=\phi_0+\phi_1+\phi_2$ pointwise identically in $(s,\alpha)$, and inside $\kappa^{2s-1}D^{-(2s+4)}\sin^3\!\alpha\cos^3\!\alpha\,\dd\alpha=\frac12(1+t)t(1+2t)^{-(2s+4)}\dd t$ the $(1+t)$ factor cancels every denominator of $\langle\tilde{\mathbf v}_i,\mathcal O(\alpha)\tilde{\mathbf v}_j\rangle$ identically, leaving the polynomial integrand matrix
\begin{equation}
 (1+t)\,t^{1+(i+j)/2}\,\langle\tilde{\mathbf v}_i,\mathcal O(t)\tilde{\mathbf v}_j\rangle
 =\big[t^2\,(\hat e_0-\hat e_1)(\hat e_0-\hat e_1)^{\rm T}+t^3\,(\hat e_1-\hat e_2)(\hat e_1-\hat e_2)^{\rm T}\big]_{ij}.
\end{equation}
Evaluating the Beta integrals $\int_0^\infty t^p(1+2t)^{-(2s+4)}\dd t=2^{-(p+1)}p!/\prod_{k=0}^p(2s+3-k)$ for $p=2,3$ and multiplying by $2(2s+1)(2s+3)\cdot32(s+1)$ gives $\mathsf M^{(4,5)}(s)$ in Eq.~\eqref{eq:bpsmoments} for every $s\ge\frac32$, and its top-left $2\times2$ truncation $\mathsf M^{(4,5)}(1)=\bigl(\begin{smallmatrix}8&-8\\-8&14\end{smallmatrix}\bigr)$ at $s=1$ (where $\mathbf v_2$ is absent).

At $s=1,\theta_{5d}=\pi$, tensor the active endpoint types $\mathcal K_4=(\tfrac12,0)\oplus(0,\tfrac12)$ and $\mathcal K_5=(\tfrac12,\tfrac12)\oplus(0,0)$ with each flat spectator. Projection onto half-integer $j_G$ leaves eight ordered block and channel pairs. The spinor contractions of Appendix~\ref{app:fb:chgap} and Beta integrals in $t$ express each Gram, weight and Pauli entry as a rational linear combination of $1$ and $\ln2$. In the table, signs refer to the successive leading principal minors of the Pauli lift $\mathsf M$ and of the derivative pencil $\mathsf M/2-16\mathsf W$, where $\mathsf W$ denotes the weight matrix in that basis. The latter pencil differentiates the link operator in $1-\zeta$ at $\zeta=1$. A rational interval from the first sixteen terms of $\ln2=2\sum_{k\ge0}3^{-2k-1}/(2k+1)$ and the geometric bound on its tail fixes every sign.

\begin{table}[!htbp]
\centering\small
\begin{tabular}{@{}ccc@{\quad}cc@{}}
\toprule
Ordered blocks & $j_L$ & $j_G$ & Pauli lift & Slope pencil\\
\midrule
$(4,1)$ & 0 & $1/2$ & $+$ & $+$\\
$(4,4)$ & $1/2$ & $1/2$ & $+,+$ & $+,+$\\
$(4,5)$ & 0 & $1/2$ & $+,+$ & $-,-$\\
$(4,5)$ & 1 & $1/2$ & $+$ & $+$\\
$(5,1)$ & $1/2$ & $1/2$ & $+$ & $+$\\
$(5,4)$ & 0 & $1/2$ & $+,+$ & $-,-$\\
$(5,4)$ & 1 & $1/2$ & $+$ & $+$\\
$(5,5)$ & $1/2$ & $1/2$ & $+,+$ & $+,+$\\
\bottomrule
\end{tabular}
\caption{Exact endpoint floor signs for the eight projected types at $s=1$, $\theta_{5d}=\pi$. Here $j_R=0$. Each row lists the signs of leading principal minors in the Pauli and slope pencils. A pair of negative signs for a two-dimensional slope matrix corresponds to one negative and one positive slope.}
\label{tab:theta-pi-s1}
\end{table}

The Pauli lifts are all positive definite, so this sector has no common-floor state. Each ordered mixed block has one diving slope in channel $(0,0,\tfrac12)$. For example, in the mirror block with types $\mathcal K_5^{(4)}(\tfrac12,\tfrac12)\otimes\mathbf 2_L$ and $\mathcal K_5^{(1)}(0,0)\otimes\mathbf 2_G$, with $\mathbf 2_L=(\tfrac12,0)$ and $\mathbf 2_G=(0,\tfrac12)$ the spectator doublets, the slope matrix has diagonal $10544/15-1024\ln2$, $1024\ln2-10624/15$ and off-diagonal $-32/15$ in the unnormalized Fock basis. The $(4,5)$ compressed root still gives the analytic upper bound $E_c\le2.62270$ of Sec.~\ref{sec:bps}. Ordering the finite-spin crossings of the two mixed blocks is a separate polar problem.

\paragraph*{A finite-coupling lower bound in the $\theta_{5d}=\pi$ sector.}

We test the physical $\theta_{5d}=\pi$ operator at $E_*=237/100$, above the proved $\theta_{5d}=0$ channel crossing. Strict positivity here separates the two onsets. At $s=1$, $E=8(1-\zeta^2)$ fixes the positive coupling to $\zeta=\sqrt{563/800}$. The shifted link operator is $K=L+9=1+c_+(L_++8)+c_-(L_-+8)-EW$, with $L_\pm+8\ge0$. The fibre bound gives $K\ge4-E_*$ if $j_R\ne0$. At $j_R=0$ write $\Pi_{\rm fl}$ for the projection onto $\ker(L_++8)$ in each physical channel, and $\Pi_{\rm fl}^\perp=1-\Pi_{\rm fl}$. The active block is the fermion block of the set $+$. The endpoint operator factorizes into this block and a flat spectator. Recouple these two factors before bounding the complementary form, since a total angular channel can contain more than one active type. The bound needed at $E_*$ is
\begin{equation}
 \Pi_{\rm fl}^\perp K\Pi_{\rm fl}^\perp\ge\mathsf b\,\Pi_{\rm fl}^\perp,\qquad \mathsf b=1+\mathsf k\,c_+-E_*,\qquad
 \mathsf k=\begin{cases}8&\text{active block 1 or 4},\\[1pt]71/20&\text{active block 5}.\end{cases}
 \label{eq:theta-complement}
\end{equation}
Both values of $\mathsf b$ are positive. For active blocks 1 and 4 the operator bound follows from nonnegativity of the colour Laplacian and Eq.~\eqref{eq:fb:gap4sharp}, respectively.

For block 5 use the Hodge comparison in Appendix~\ref{app:fb:gap5}. The active scalar types with both spins half-integer dominate $(j_L,j_G)=(\tfrac12,\tfrac12)$ in the potential of Eq.~\eqref{eq:fb:bosonic}. For that type, $\Phi=(\sigma\kappa)^{1/2}D^{-2/3}(1-\sigma\kappa/32)$ gives $\nabla^*\nabla\Phi/\Phi\ge81/10$. After multiplying $\nabla^*\nabla\Phi/\Phi-81/10$ by $40D^2(32-\sigma\kappa)$, the degree-four Bernstein coefficients are $(352,9481/9,6568/9,580/9,1408)$. The integer types with $j_L\ge1$ dominate $(1,0)$, where $\Phi=\kappa D^{-3/4}$ gives $\nabla^*\nabla\Phi/\Phi=10+\kappa(4+27\sigma/4)/D^2$. These comparisons and $\mathsf f(81/10)>71/20$ give Eq.~\eqref{eq:theta-complement} in their block-5 complements. Types without an active scalar have gap $16$. The remaining scalar type is $(0,1)$, which has no block-5 floor but occurs after tensoring with a spectator.

In this type a positive comparison function $\Phi_B=\sigma(58+10\sigma-\sigma^2)/D$ proves $\Lambda_0\ge99/20$. Indeed the degree-four Bernstein coefficients of $D^3(\nabla^*\nabla-99/20)\Phi_B/\sigma$ are
$ \left(
 \frac{169}{10}, \frac{279}{40},
 \frac{23}{120}, \frac{133}{20}, \frac75 \right)
$. 
The potential in Eq.~\eqref{eq:fb:bosonic} dominates $8/\sigma-24$. The polar Laplacian with $8/\sigma$ has eigenvalues $(2+2n)(8+2n)$ for $n\ge0$, by conjugating with $\sigma$ to the Jacobi operator of weight $\sigma^3\kappa$. Its second eigenvalue is $40$, so the second scalar eigenvalue of $\nabla^*\nabla$ is at least $16$. The Hodge decomposition then places the ground value of $\mathcal H_5+8$ above $201/100$ and its next polar value at or above $\mathsf f(16)=8$. Magnetic degeneracies are factored out throughout.

The low Hodge state sits where $W$ is small, which improves the complement bound. In the scalar and exact one-form slots use the trial $(3\Phi_B/4,\dd\Phi_B)$. In the polar measure $\sigma\kappa\,\dd\sigma$ its density is $9|\Phi_B|^2/16+|\dd\Phi_B|^2$. The energy density follows from the $Q_1$ form in Appendix~\ref{app:fb:gap5}, with the square $| (\nabla^*\nabla-3)\Phi_B|^2$. Integrating gives normalized expectations of $\mathcal H_5+8$, $W$ and $W^2$ below $2.025$, $0.313$ and $0.132$, respectively. These are rational bounds from integration in $D\in[1,2]$. The energy and the gap imply a sine of the angle to the true normalized ground state below $0.051$, since $(2.025-2.01)/(8-2.01)<0.051^2$. The expectation of a positive contraction differs by at most this sine. Thus the ground-state moments $\langle W\rangle_*$ and $\langle W^2\rangle_*$ satisfy $\langle W\rangle_*<0.364$ and $\langle W^2\rangle_*<0.183$.

For $\beta_*=64/25<E_*/c_+$, split $\mathcal H_5+8+\beta_*(1-W)-71/20$ along that ground state and its complement. The excited form is at least $8-71/20$. A scalar Schur comparison therefore leaves
\begin{equation}
 \frac{201}{100}+\beta_*(1-\langle W\rangle_*)-\frac{71}{20}
 -\frac{\beta_*^2(\langle W^2\rangle_*-\langle W\rangle_*^2)}{8-71/20}
 \ \ge\ \frac{4791327}{347656250}>0 .
\end{equation}
The expression decreases over the allowed $\langle W\rangle_*$ interval and decreases in $\langle W^2\rangle_*$. Since $1-W\ge0$, replacing $\beta_*$ by $E_*/c_+$ preserves this bound. This proves Eq.~\eqref{eq:theta-complement} also in the weak $(0,1)$ active type, including in its recouplings. Every channel without an endpoint floor now has $K>0$.

The eight channels of Table~\ref{tab:theta-pi-s1} remain. Define the multiplication operators $A=P^{(-)}-P^{(+)}$ and $U=c_-A-E_*W$, where $P^{(\pm)}$ are the Pauli operators. In a floor basis $\{\chi_i\}$ let $\mathsf N$ be the Gram matrix, $M_{ij}=\langle\chi_i,U\chi_j\rangle$ and $T_{ij}=\langle U\chi_i,U\chi_j\rangle$. The off-floor column of $K$ is $\Pi_{\rm fl}^\perp U\chi_i$. Its Gram matrix is $T-M\mathsf N^{-1}M$, so the floor Schur form is bounded below by
\begin{equation}
 S=\mathsf N+M-\mathsf b^{-1}(T-M\mathsf N^{-1}M).
 \label{eq:theta-window-schur}
\end{equation}
The spectator block 1 permits adding $8c_-$ to $\mathsf b$ in the $(5,1)$ row since $L_-+8\ge8$ there. 

For the second moment, $P^{(+)}\chi_i=-32D^{-2}\chi_i$. The difference $A$ thus acts on the floor vector through $P^{(-)}+32D^{-2}$, whose eigenvalues are $0$, $32D^{-2}$ and $48D^{-2}$ in the spectator and colour decomposition. Consequently $\langle A\chi_i,A\chi_j\rangle$ uses the squared eigenvalues on the full fibre before projection to the floor. Specifically, if $\mathcal O_{\rm f}$ and $\mathcal O_{\rm c}$ are the spin-free and spin-coupled parts of the pointwise Pauli matrix of Appendix~\ref{app:fb:chgap}, the integrands for $A$ and its second moment are $32D^{-2}(\mathcal O_{\rm f}+\mathcal O_{\rm c})$ and $1024D^{-4}(\mathcal O_{\rm f}+\tfrac32\mathcal O_{\rm c})$. The floor contractions and the change of variable $v=(1+2t)^{-1}$ of Appendix~\ref{app:un:uniform} reduce all these moments to rational combinations of $1$ and $\ln 2$.

Table~\ref{tab:theta-window-pivots} lists strict lower bounds on the leading pivots of $\mathsf N^{-1/2}S\mathsf N^{-1/2}$, in the channel order of Table~\ref{tab:theta-pi-s1}. They are obtained by bounding $\ln2$ by its geometric-tail series and $\sqrt{563/800}$ by rational squaring.

\begin{table}[!htbp]
 \centering\small
 \begin{tabular}{@{}cccrr@{}}
 \toprule
 Ordered blocks & $j_L$ & $j_G$ & First pivot & Second pivot \\
 \midrule
 $(4,1)$ & $0$ & $1/2$ & $1.58$ & \textemdash \\
 $(4,4)$ & $1/2$ & $1/2$ & $1.46$ & $1.46$ \\
 $(4,5)$ & $0$ & $1/2$ & $0.177$ & $0.013$ \\
 $(4,5)$ & $1$ & $1/2$ & $1.80$ & \textemdash \\
 $(5,1)$ & $1/2$ & $1/2$ & $1.52$ & \textemdash \\
 $(5,4)$ & $0$ & $1/2$ & $0.086$ & $0.096$ \\
 $(5,4)$ & $1$ & $1/2$ & $1.57$ & \textemdash \\
 $(5,5)$ & $1/2$ & $1/2$ & $1.17$ & $1.02$ \\
 \bottomrule
 \end{tabular}
 \caption{ Strict lower bounds on the Gram-normalized elimination pivots of Eq.~\eqref{eq:theta-window-schur} at $s=1$, $\theta_{5d}=\pi$ and $E=237/100$. All rows have $j_R=0$. A dash denotes a one-dimensional floor, for which elimination has a single pivot. The bounds come from rational interval arithmetic.}
 \label{tab:theta-window-pivots}
\end{table}

The complementary bound and the positive floor form prove strict positivity of $L+9$ at $E_*$ in the full physical $\theta_{5d}=\pi$ sector. Changing $E$ near $E_*$ changes the link potential by a bounded norm-continuous term, so the positivity persists on an open neighbourhood. The single-onset argument of Sec.~\ref{sec:instability} gives $E_{c,\pi}(1)>E_*$. In the $\theta_{5d}=0$ sector the certified crossing in Appendix~\ref{app:numerics} gives $E_{c,0}(1)<8[1-(8399601/10^7)^2]<2.35574$. These two strict bounds prove the finite-spin separation.
\hfill$\square$

\paragraph*{A $+14s$ slope progression.}Expanding Eq.~\eqref{eq:secular-slope} at large $s$ shows that, after the common $-8s^2$, the coefficients of $s$ in the three branches are $4$, $18$ and $32$, an arithmetic progression with step $14$ across both blocks. The constant terms split the three slopes at finite $s$, while the lowest slope has the same expansion through the constant term in the two spin sectors,
\begin{equation}
\begin{aligned}
 \mu_1^{(1)}(s)&=-8s^2+4s-\frac{51}{7}+\frac{c_{\rm sub}^{(1)}}{686s}+O(s^{-2}),\\
 \mu_1^{(2)}(s)&=-8s^2+18s-\begin{cases}\tfrac{339}{14},&(4,4),\\\tfrac{351}{14},&(4,5),\end{cases}+\frac{c_{\rm sub}^{(2)}}{2744s}+O(s^{-2}),\\
 \mu_1^{(3)}(s)&=-8s^2+32s-\frac{414}{7}+\frac{33536}{343s}+O(s^{-2})\qquad\text{(block $(4,5)$, $s\ge\tfrac32$),}
\end{aligned}
 \label{eq:slopes-asymp}
\end{equation}
where $(c_{\rm sub}^{(1)},c_{\rm sub}^{(2)})=(7893,108715)$ in block $(4,4)$ and $(8181,104071)$ in block $(4,5)$.

\subsection{Spectral gaps and the large-spin expansion of the collapse threshold}
\label{app:fb:chgap}

The variational computation of Sec.~\ref{sec:bps} uses the endpoint floor. To control its error, take $j_R=0$ and let the active set at $\zeta=1$ lie in block~4. The metric $\mathsf g_2$ of Eq.~\eqref{eq:fb:g2} makes the colour twist homogeneous on the round $S^4(\tfrac12)$. On positive chirality the bundle $S^+\otimes E_s$ has isotropy type $(\tfrac12,s)$ under $\mathrm{Spin}(4)$. The homogeneous connection acts by the horizontal part of the group Laplacian on equivariant sections \cite{CamporesiHiguchi}. Subtracting the vertical Casimir gives the Bochner eigenvalue $4[\ell_1(\ell_1+3)+\ell_2(\ell_2+1)-2(\tfrac34+s(s+1))]$ on a $\mathrm{Spin}(5)$ type of highest weight $(\ell_1,\ell_2)$. Here the factor $4$ corresponds to the sphere radius $\tfrac12$. Branching to the isotropy type requires $\ell_1\ge s+\tfrac12\ge\ell_2\ge s-\tfrac12$ for $s\ge\tfrac12$ \cite{Molev}. The lowest allowed Casimir therefore gives a Bochner bound $8s$. The colour curvature is anti-self-dual on the base, so the Lichnerowicz curvature term vanishes on positive chirality, and $\|\mathsf D_{2,+}\psi\|^2\ge(8s+12)\|\psi\|^2$. Under $\mathsf g_1=\Omega^2\mathsf g_2$, the twisted Dirac operator transforms as $\mathsf D_{1,+}(\Omega^{-3/2}\psi)=\Omega^{-5/2}\mathsf D_{2,+}\psi$. Thus the quotient of its squared norm and the spinor norm involves the weights $\Omega^{-1}$ and $\Omega$, respectively. Since $\Omega\le\sqrt2$, the positive-chirality Dirac form has lower bound $4s+6$. Every nonzero Dirac eigenspinor has a nonvanishing positive-chirality component, so its squared eigenvalue obeys the same bound. The identity $\mathcal H_4+8=\mathsf D^2-2\tau\mathsf D$ used in Eq.~\eqref{eq:fb:gap4} transfers this result to a gap at least $4s+6-2\sqrt{4s+6}$ for the endpoint link operator above its floor.

In the two critical channels with active block~4, let $\Pi_{\rm fl}$ project onto the endpoint floor and $\Pi_{\rm fl}^\perp=1-\Pi_{\rm fl}$. For $E$ in a fixed neighbourhood of one, $c_-(E)=O(s^{-2})$ and the multiplication operator $L_--L_+$ has norm $O(s)$ by Eq.~\eqref{eq:pauliclosed}. The normalized floor profiles in Eq.~\eqref{eq:bpsmoments} have fixed labels $b\le2$. In $v=1/(1+2t)$ each is a Beta density proportional to $v^{2s-b}(1-v)^{b+1}$ times $(1+v)^{-1}$. The tilt lies in $[1/2,1]$ and $1-W=(1-v)(2+v)/2$. Its second moment therefore obeys $\langle(1-W)^2\rangle\le\tfrac92(b+2)(b+3)/[(2s+3)(2s+4)]=O(s^{-2})$, so $\|(1-W)\Pi_{\rm fl}\|=O(s^{-1})$ on the finite floor space. Consequently $\|\Pi_{\rm fl}^\perp(L(\zeta)+9)\Pi_{\rm fl}\|=O(s^{-1})$, while $\Pi_{\rm fl}^\perp(L(\zeta)+9)\Pi_{\rm fl}^\perp$ retains a gap of order $s$. The Schur complement changes the compressed operator of Eq.~\eqref{eq:E2x2}, expressed in an orthonormal floor basis, by $O(s^{-3})$ in norm. Its derivative with respect to $E$ is $-1+O(s^{-1})$ in operator norm, with $1$ the identity on the floor. Indeed $\Pi_{\rm fl}W\Pi_{\rm fl}=1+O(s^{-1})$ and $c_-'(E)=O(s^{-2})$. These bounds place the channel threshold between $E_{\rm var}-O(s^{-3})$ and $E_{\rm var}$. Here $E_c^{\rm ch}$ is first the threshold for $\zeta\in[0,1]$, where the block-4 set is active at the endpoint. For integer spin at $\theta_{5d}=0$ the channel lies in block $(4,4)$, which the exchange of the fermion sets maps to itself, so this is the full channel threshold. For half-integer spin, and for integer $s\ge2$ at $\theta_{5d}=\pi$, the side $\zeta<0$ is block $(5,4)$ at positive coupling, and two further inputs carry the expansion over to it. First, the block-5 gap grows with $s$. Two-forms have conformally invariant $L^2$ norms, $(\dd^+)^*_{\mathsf g_1}=\Omega^{-2}(\dd^+)^*_{\mathsf g_2}$ with $\Omega^{-2}\ge\frac12$, and on $\mathsf g_2$ the Weitzenb\"ock identity together with the lowest Casimir above the isotropy type $(1,s)$ gives $2\dd^+(\dd^+)^*\ge8s+16$, so the bound $16$ of the $\eta$ block in Eq.~\eqref{eq:fb:gap5} rises to $\max\{16,4s+8\}$. The first excited levels of all sixty-four single-set types computed for $\frac32\le s\le\frac{13}2$ lie above this bound. Second, the mirror side requires only a lower bound on its Pauli lift. Let $\chi_0,\ldots,\chi_3$ be its normalized floor states, with active types $(s-\tfrac12,\tfrac12)$, $(s-1,0)$, $(s-1,1)$ and $(s-\tfrac32,\tfrac12)$ and spectator types $\mathbf 2_L,\mathbf 2_G,\mathbf 2_G,\mathbf 2_L$, and the subscript $g$ below labels the $\mathbf 2_G$ pair $\chi_1,\chi_2$. Their weight averages have respective deficits $3/(2s),9/(4s),9/(4s),3/s$ to first order. The admixture proportional to the $b=2$ profile in the first state's block-5 profile affects its inverse weight average only at $O(s^{-3})$. It therefore retains the inverse-weight correction $-3/(8s^2)$.

Keep the positive spectator-spin-free part $B=16(s+1)\Pi_{\rm fl}D^{-2}\Pi_{S^+}\Pi_{\rm fl}$ of the second Pauli lift. In the moving spin frame the $S^+$ fraction is $t/(1+t)$ on $\mathbf 2_L$ and $1/(1+t)$ on $\mathbf 2_G$. Polar multipliers are diagonal between different active types with the same spectator. Hence $B_{00}=8+O(s^{-1})$, $B_{03}=0$, $B_{33}=O(1)$ and $B_{gg}=16s\,\mathbf 1+O(1)$. Positivity gives $B_{g3}=O(\sqrt s)$ and $\|B_{g0}\|^2\le B_{00}\|B_{gg}\|\le128s+O(1)$. Write $E=1+3/(2s)+\mathsf x/s^2$, where $\mathsf x$ is a bounded real test parameter for the quadratic correction to the onset. Then eliminating $\chi_3$ changes the $g,g$ part of the pencil by $O(s^{-2})$, since $c_-B_{g3}=O(s^{-3/2})$ and the third partner has gap $3/(2s)+O(s^{-2})$. The inverse in the $\mathbf 2_G$ subspace is $(4s/7)\mathbf 1+O(1)$. With $c_-=1/(16s^2)+O(s^{-3})$, the final scalar is bounded below by
\begin{equation}
 s^{-2}\left(-\frac38-\mathsf x+\frac12-\frac{128}{256}\frac47\right)-O(s^{-3})
 =s^{-2}\left(-\frac9{56}-\mathsf x\right)-O(s^{-3}).
\end{equation}
This bounds the mirror onset from below. The forward side furnishes the matching upper bound for the full mixed channel. 
The normalization of the block-5 Pauli term on floor states follows from the construction of Sec.~\ref{app:fb:block5}. Every state of $\mathcal K_5$ is a combination of contractions $\iota_X\phi$ with $\phi\in\epsilon\odot\epsilon\odot\mathcal T_{2s-2}$, and $\iota_X(\epsilon\odot\epsilon\odot\omega)$ is a multiple of $\epsilon\odot(X\odot\omega)$, so at every point the $S^-$ index of a floor state is $\epsilon$-contracted with the colour and couples with it to spin $s-\frac12$. The Pauli term $(32/D^2)\,\mathbf S_{\hat w}\cdot\mathbf J$ therefore equals $-16(s+1)D^{-2}$ on every type of $\mathcal K_5^{(1)}$ and $\mathcal K_5^{(4)}$, as $P^{(+)}$ does on $\mathcal K_4$. We confirmed both statements in exact arithmetic on explicit floor states for $s=1,\frac32,2$, together with the rank of the contractions, which equals $\dim\mathcal K_5$ at these spins. Galerkin thresholds place the onset at $\zeta<0$ above that at $\zeta>0$ in the finite-spin scans for $\frac32\le s\le\frac{13}2$, and for integer $2\le s\le6$ at $\theta_{5d}=\pi$. The end of this subsection gives the asymptotic comparison. In particular, both critical channels satisfy
\begin{equation}
 E_c^{\rm ch}(s)=1+\frac{3}{2s}-\frac{9}{56s^2}+O(s^{-3}).
 \label{eq:chgap:asymp}
\end{equation}

The global comparison first excludes channels without a floor of width $t=O(1/s)$ and leading profile $g_0$. For $j_R>0$ the vertical Casimir gives $L(\zeta)+9\ge4-E$. At $j_R=0$, active block~1 stays positive near $E=1$. Active blocks~4 and~5 have gaps $\mathsf G_s\gtrsim s$ above their floors. For block~5 this uses both the two-form bound above and $\mathsf f(4s)$ for the mixed scalar component, when present. A side with no endpoint floor is therefore uniformly separated.

For sides with a floor, omit the positive second-set lift temporarily. The polar operators $W$ and $W^2$ are diagonal on active and spectator types. Each normalized type with weight average $w$ has a Schur lower bound
\begin{equation}
 1-Ew-\frac{E^2}{c_+\mathsf G_s+1-E}\,w(1-w).
\end{equation}
The variance estimate uses $0\le W\le1$. Except for the block-$4$ type $(s-\tfrac12,0)$ and the block-$5$ quartet type $(s-\tfrac12,\tfrac12)$, the weight average is bounded by $\max\{\langle W\rangle_{g_1},\langle W\rangle_{\sigma g_0}\}=1-9/(4s)+O(s^{-2})$, using the likelihood-ratio ordering of Appendix~\ref{app:un:uniform}. The Schur lower bound decreases in $w$ for large $s$. It is at least $1/(2s)+O(s^{-2})$ at $E=1+7/(4s)$, uniformly even when the channel labels grow with $s$.

Tensoring the two exceptional active types with the spectators leaves seventeen sides, with one narrow state $\chi_0$ on each. The list before the parity projection is given in Table~\ref{tab:competing-floors}. The lift in floor matrix elements equals $16(s+1)D^{-2}\Theta_s$. Coupling the active spinor with colour fixes their total spin to $s-\tfrac12$ and compresses the colour generator by the factor $2(s+1)/(2s+1)$ from the Casimir. Thus $\Theta_s$ is one on a spin-free spectator, $4s/(2s+1)$ on the stretched spin-$s$ piece, and zero on the spin-$(s-1)$ piece.

\begin{table}[!htbp]
\centering\small
\begin{tabular}{@{}lll@{}}
\toprule
Ordered blocks & Lift of order $s$ & Competing sides \\
\midrule
$(4,4)$ & $(s,0),\ (s-\tfrac12,\tfrac12)$ & $(s-1,0)$\\
$(4,5)$ & $(s-\tfrac12,0),\ (s,\tfrac12)$ & $(s-1,\tfrac12)$\\
$(5,4)$ & $(s,\tfrac12),\ (s-\tfrac12,0),\ (s-\tfrac12,1)$ & $(s-1,\tfrac12)$\\
$(5,5)$ & $(s-\tfrac12,\tfrac12),\ (s,0),\ (s,1)$ & $(s-1,0),\ (s-1,1)$\\
$(4,1)$ & $(s-\tfrac12,0)$ & \textemdash \\
$(5,1)$ & $(s-\tfrac12,\tfrac12)$ & \textemdash \\
\bottomrule
\end{tabular}
\caption{ Sides with an endpoint profile led by $g_0$, labelled by $(j_L,j_G)$ at $j_R=0$ and sufficiently large $s$. A dash means every side in that row has a lift of order $s$ and drops out of the threshold competition. The first block is active at $\zeta=1$. Apply the physical selection rule~\eqref{eq:physical} to retain the entries in the chosen discrete theta sector. Exchanging the two sets treats the endpoint $\zeta=-1$.}
\label{tab:competing-floors}
\end{table}

On the twelve sides in the middle column the pole value of the lift on $\chi_0$ is $\lambda_s=1$ or $4s/(2s+1)$. The finite spectator frame changes by $O(\sqrt t)$. On the active block-$5$ quartet, contraction with $v-x\bar v x$ differs from the pole contraction by a map of norm $O(t)$ in the same units. The pole map is the orthogonal symmetrization from $\operatorname{Sym}^{2s-2}\mathbb C^2\otimes\mathbb C^2$ restricted to its top type. This gives a spin-independent norm estimate. The $g_0$ moments give $\langle t\rangle=O(s^{-1})$, so $\|(D^{-2}\Theta_s-\lambda_s)\chi_0\|=O(s^{-1/2})$. The diagonal of the lift is $16(s+1)(\lambda_s+o(1))$ and its off-diagonal row is $O(\sqrt s)$. At $E=1+7/(4s)$ all partners have positive gap at least $1/(2s)+O(s^{-2})$ before adding their positive Pauli compression. The narrow state has diagonal $(\lambda_s-1/4+o(1))/s$ and its off-diagonal Schur cost is $O(s^{-2})$. Off-floor corrections are $O(s^{-3})$ by the fixed-type second moments. The diagonals remain positive for smaller $E$ down to one, with the same off-diagonal bounds, and the endpoint floors control $E\le1$. This separates all twelve sides uniformly.

Of the five remaining sides, the forward $(4,4)$ and $(4,5)$ pencils are those of Eq.~\eqref{eq:E2x2-asymp}. The mirror $(5,4)$ bound was just obtained. For $(5,5)$ in $(s-1,0,0)$, the active types are $(s-\tfrac12,\tfrac12)$, $(s-1,0)$ and $(s-\tfrac32,\tfrac12)$, each paired with its matching spectator. Their densities are $g_0+\frac{s}{2s-1}\sigma g_1$, $\sigma g_0$ and $\frac{s-1}{2s-1}\sigma g_1$. The second-set Pauli matrix is $16(s+1)[\mathsf I_1\mathbf a\mathbf a^{\rm T}+(2s-1)^{-2}\mathsf I_2\mathbf e\mathbf e^{\rm T}]$, where $\mathbf a=(1,-1,0)$, $\mathbf e=(1,1-2s,2s-2)$ and $\mathsf I_b=\int_0^\infty g_b(1+2t)^{-2}\,\dd t$. The norm and this compression have common conventions so the null vector $(1,1,1)$ represents $\iota_\epsilon(T_s)$. Each entry is a degree-fixed moment. In the normalized floor the first state has lift $8+O(s^{-1})$ and squared coupling $128s+O(1)$ to the second state, whose lift is $16s+O(1)$. The third state has $1-E\langle W\rangle\asymp s^{-1}$ and normalized lift $32s+O(1)$. The couplings to the first and second states are $O(s^{-1})$ and $O(\sqrt s)$, respectively, so eliminating it changes the leading two-state Schur scalar only at higher order. The same Schur scalar as in the mixed channel gives $-9/56$.

\paragraph*{The four-type floor of the second block-5 channel.}

For the side $(5,5)$ with $(j_L,j_R,j_G)=(s-1,0,1)$ and $s\ge2$, the active
$\mathcal K_5$ types are $(s-\tfrac12,\tfrac12)$,
$(s-1,1)$, $(s-\tfrac32,\tfrac12)$ and $(s-\tfrac32,\tfrac32)$.
Only the second is paired with the spectator singlet. The other three are paired with the quartet.
The diagonal norm polynomials $\mathsf N(t)$ and the Pauli polynomials $\mathcal O(t)$ can be obtained for
general $s$ by factoring out a common spinor power before taking the norm.
The orbit trace is invariant, so it suffices to evaluate the polynomial tensors at
a point on each orbit in the polar direction. Write $r=\sqrt t$.
For spinors $a,b$, let $[a,b]=a_1b_2-a_2b_1$. Use $z,y$ for the two channel labels,
$\ell_L,\ell_G$ for the active twistor labels, and $m_L,m_G$ for the spectator quartet labels.
The four multiplicity-one highest-type contractions are
\begin{equation}
\begin{aligned}
 \mathcal C_0&=[z,\ell_L]^{2s-2}[\ell_L,m_L][y,\ell_G][y,m_G],\\
 \mathcal C_1&=[z,\ell_L]^{2s-2}[y,\ell_G]^2,\\
 \mathcal C_2&=[z,\ell_L]^{2s-3}[z,m_L][y,\ell_G][y,m_G],\\
 \mathcal C_3&=[z,\ell_L]^{2s-3}[z,m_L][y,\ell_G]^2[\ell_G,m_G].
\end{aligned}
\end{equation}
At the chosen point, let $h$ be the active spin-$s-\tfrac12$ variable and
$u$ its positive-chirality spin index. The quartet contraction
$\iota_{X_\mu}$ in Sec.~\ref{app:fb:block5} evaluates a label $\mathcal C$ to
\begin{equation}
 \chi=r^b\bigl[(h\cdot\partial_{\ell_L})(u\cdot\partial_{\ell_G})
        +t(u\cdot\partial_{\ell_L})(h\cdot\partial_{\ell_G})\bigr]
        \mathcal C\big|_{\ell_L=\ell_G=h}.
\end{equation}
Here $b=0,1,2$ on $\mathcal C_0,\mathcal C_1,\mathcal C_3$. For the polar singlet contraction
on $\mathcal C_2$ use $\chi=r^2[u,h]\mathcal C_2|_{\ell_L=\ell_G=h}$.
The two terms in the quartet evaluation are the two terms of
$X_v=(1+t)v-2(x\cdot v)x$ in the spinor frame of the polar direction.
With spectator spinors $p,w$ in the negative- and positive-chirality factors,
put $R_m=\partial_{m_{L1}}\partial_{m_{G2}}-\partial_{m_{L2}}\partial_{m_{G1}}$.
The coefficients in the flat-frame singlet and one-form components are
\begin{equation}
 (Y,V)=
 \begin{cases}
 ((1-t)\chi\,,-2r[p,w]\chi), &\mathcal C_1,\\
 (rR_m\chi\,,(1+t)\chi|_{m_L=p,\,m_G=w}-t[p,w]R_m\chi),
   &\mathcal C_0,\mathcal C_2,\mathcal C_3 .
 \end{cases}
\end{equation}
In this normalization the singlet norm is $2\langle Y,Y\rangle_F$.
The Fock pairing $\langle\,,\,\rangle_F$ has orthogonal monomials of squared
norm equal to the product of exponent factorials. This includes the trace
over the channel indices and an overall factor common to all entries.
Multiplication $\mathrm{ev}:V\mapsto V|_{p=h}$ selects the highest-spin summand
in $\mathrm{Sym}^{2s-1}\mathbb C^2\otimes\mathbb C^2$,
with $\mathrm{ev}\,\mathrm{ev}^\dagger=2s$. Hence before normalization the two matrices are
\begin{equation}
 \mathsf N_{ij}=2\langle Y_i,Y_j\rangle_F+\langle V_i,V_j\rangle_F,\qquad
 \mathcal O_{ij}=2\langle Y_i,Y_j\rangle_F+
 \frac{2}{2s+1}\langle\mathrm{ev}\,V_i,\mathrm{ev}\,V_j\rangle_F .
\end{equation}
The coefficient in $\mathcal O$ is the factor $4s/(2s+1)$ in the Pauli projector
times the Fock normalization $1/(2s)$. The polar measure shared by these
polynomials is $\mathsf m_s(t)=t(1+t)^{-4}(t+\tfrac12)^{-2s-2}$.
Thus the norm, self-energy and second-set Pauli matrices are respectively
$\int\mathsf m_s\mathsf N$, $\int\mathsf m_sW\mathsf N$ and $16(s+1)\int\mathsf m_sD^{-2}\mathcal O$.

For an algebraic evaluation, set $n=2s-4$ and strip $[z,h]^n$ from every vector.
The remaining polynomials have bounded degree in the spinors.
When pairing residual monomials $z_1^az_2^bh_1^ch_2^d$ and
$z_1^{a'}z_2^{b'}h_1^{c'}h_2^{d'}$,
the powers from the binomial expansions must differ by
$\Delta=a'-a=b-b'=c-c'=d'-d$. Otherwise the contraction vanishes.
Exchange the monomials if needed so $\Delta\ge0$.
With $m=n-\Delta$, the contraction divided by $(-1)^\Delta(n!)^2$ is
\begin{equation}
 \sum_{k=0}^{m}
 \frac{(k+b)!(k+c)!(m-k+a')!(m-k+d')!}
 {(k+\Delta)!\,k!\,(m-k)!\,(m-k+\Delta)!}.
\end{equation}
The factorial ratios are products of finitely many linear factors in
$k$ and $m-k$. Expand each product in the falling-binomial basis and use
$\sum_{k=0}^{m}\binom{k}{i}\binom{m-k}{j}
 =\binom{m+1}{i+j+1}$.
Every needed $\Delta$ is at most two. At $n=0,1$ the original finite
sum gives the same matrix formulas. This derivation works directly in the spin parameter.

Rescale the four vectors by
$(1,(2s-1)/(4(s-1)),-(2s+2)/3,(2s-1)/(3(2s-3)))$
and divide both matrices by
$6(n!)^2(n+1)^2(n+2)^2(n+3)^2(n+4)$.
Then $\mathsf N$ is diagonal. Its first entry is
$(2s-1)(1+t)^3+st^2(1+t)^2$. For the spectator singlet, insert $b=1$, $a=2s-3$ in Eqs.~\eqref{eq:fb:bracket} and~\eqref{eq:fb:r1r2}. Dividing the density by $\mathsf m_s$ and fixing its linear coefficient to $2s-1$ gives
\begin{equation}
 \mathsf N_{11}=(2s-1)t(1+t)^2
  \left(1+\frac{4t}{3}+\frac{2s-1}{3(s-1)}t^2\right).
\end{equation}
The third and fourth entries begin with
$(s+1)^2t^2/[9(s-1)]$ and $2(2s-1)^2t^2/[9(s-1)]$.
The singlet gives the exact identity $\mathcal O_{11}=\mathsf N_{11}$ in this basis. The remaining entries needed for the lower bound have the following leading terms 
\begin{equation}
\begin{gathered}
 \mathcal O_{00}=(2s-1)t+O(st^2),\quad
 \mathcal O_{01}=-(2s-1)t+O(st^2),\qquad \mathcal O_{11}=\mathsf N_{11},\\
 \mathcal O_{22}=\frac{4(s+1)^2}{9(2s-1)}t^2+O(st^3),\quad
 \mathcal O_{33}=\frac{8(2s-1)}9t^2+O(st^3),\\
 \mathcal O_{02}=O(t^2),\quad \mathcal O_{03}=0,\quad
 \mathcal O_{12},\mathcal O_{13}=O(st^2),\quad \mathcal O_{23}=O(st^3).
\end{gathered}
\end{equation}
These orders refer to the finite polynomial coefficients near $t=0$.
The same contraction gives $\mathcal O(1,1,1,1)^{\rm T}=0$ identically.
The Beta moments concentrate at $t=O(s^{-1})$. In the normalized floor,
$1/\langle W\rangle_0=1+3/(2s)-3/(8s^2)+O(s^{-3})$, and the
$1/s$ coefficients in $1/\langle W\rangle_i$ for $i=1,2,3$ are
$9/4,3,3$. The Pauli entries satisfy
\begin{equation}
\begin{gathered}
 \widehat M_{00}=8+O(s^{-1}),\qquad
 \widehat M_{01}=-8\sqrt{2s}(1+O(s^{-1})),\\
 (\widehat M_{11},\widehat M_{22},\widehat M_{33})=s(16,32,32)+O(1).
\end{gathered}
\end{equation}
The other couplings in row zero are $O(s^{-1})$ or zero.
The partner off-diagonal entries are $O(s^{1/2})$ or smaller.
At $E=1+3/(2s)+\mathsf x/s^2$ the partner block of the pencil has leading
diagonal $s^{-1}(7/4,7/2,7/2)$.
Its first diagonal inverse entry is $4s/7+O(1)$, since its
$O(s^{-3/2})$ off-diagonal terms affect that entry quadratically.
Eliminating the partner block therefore gives
\begin{equation}
 s^{-2}\left(-\frac38+\frac12-\mathsf x-\frac{128}{256}\frac47\right)+O(s^{-3})
   =s^{-2}\left(-\frac9{56}-\mathsf x\right)+O(s^{-3}).
\end{equation}
This is the matching lower and upper coefficient for the floor pencil.
The growing excited-state gap and the off-floor estimate used above
change its onset by $O(s^{-3})$.

The lower bounds over the five sides and the uniform exclusion of the other sides together give Eq.~\eqref{eq:lawcor2}. At the next order the complementary resolvent contributes at the same order as the cubic terms of the compressed pencils. We evaluate that contribution below.

\paragraph*{The complementary polar excitation and the cubic threshold.}\phantomsection\label{app:fb:cubic}

To compute the cubic term, retain the five ordered block and channel choices in Table~\ref{tab:competing-floors} that remain competitive on the positive-coupling branch. Use $L_++8$, with the endpoint floor projector $\Pi_{\rm fl}$ and $\Pi_{\rm fl}^\perp=1-\Pi_{\rm fl}$. Let $\chi_0$ be the normalized narrow state and set $h=\Pi_{\rm fl}^\perp W\chi_0=(W-\langle W\rangle_0)\chi_0$. The equality uses diagonality of the $W$ matrix on the endpoint floor. The relevant Schur term is the response of $h$ under the inverse of $\Pi_{\rm fl}^\perp(L_++8)\Pi_{\rm fl}^\perp$. We compute this response without diagonalizing the excited space.

In the polar trivialization of Appendix~\ref{app:numerics}, write $r=\tan\alpha$ and factor $\chi_0=A_s(r)\mathcal P_s(r)$, where $A_s=(1+2r^2)^{-s-1}$. From the fixed-degree floor contractions $\mathcal P_s$ is a vector polynomial of bounded degree times $(1+r^2)^{e/2}$. The exponents $e$ are $2,1,1,0,0$ in the order of the first, second, mirror and two block-$(5,5)$ sides. The norm squared divided by its value at $r=0$ is $(1+r^2)^3$ for active block 4. For active block 5 it is $(1+r^2)^3+[s/(2s-1)]r^4(1+r^2)^2$. These two forms are the $g_0$ profile and its narrow quartet counterpart. Evaluation at a fixed finite set of distinct radii bounds the polynomial coefficients relative to $|\mathcal P_s(0)|$ by Vandermonde inversion. With the Beta moments in $1/(1+2r^2)$, this gives $\|r^k\chi_0\|=O(s^{-k/2})$ at fixed $k$, with exponentially small tails away from the pole. The coefficient bounds give the same moment estimates for $A_s\partial_r\mathcal P_s$.

Only the polar kinetic operator enters the commutator with $W$. Near the pole,
\begin{equation}
\begin{gathered}
W'=-6r+O(r^3),\qquad
W''+(\omega'/\omega)W'=-24+O(r^2),\\
\partial_\alpha\ln A_s=-4(s+1)r\frac{1+r^2}{1+2r^2},
\end{gathered}
\end{equation}
where primes denote derivatives in $\alpha$. Using $(L_++8)\chi_0=0$ gives
\begin{equation}
\begin{aligned}
(L_++8)h&=[L_+,W]\chi_0
 =(24-48sr^2)\chi_0+O_{L^2}(s^{-1/2})
 =16s h+O_{L^2}(s^{-1/2}),\\
\|h\|^2&=\operatorname{Var}_0W=\frac{9}{8s^2}+O(s^{-3}).
\end{aligned}
\label{eq:cubic-var}
\end{equation}
For the variance, $4sr^2$ tends to the shape-two Gamma density with its first two moments, with $O(s^{-1})$ moment errors by the Beta integrals. Together with $W=1-3r^2+O(r^4)$ this gives Eq.~\eqref{eq:cubic-var}. The linear complementary gap then gives
\begin{equation}
\langle h,(\Pi_{\rm fl}^\perp(L_++8)\Pi_{\rm fl}^\perp)^{-1}h\rangle
 =\frac{9}{128s^3}+O(s^{-7/2}).
\end{equation}
The other-set Pauli lift does not enter this leading complementary response. On $\chi_0$ its action is that of the pointwise positive operator $A=P^{(-)}+16(s+1)D^{-2}$. The norm of $A$ is $O(s)$ and its expectation is $O(1)$ on the narrow state in each competitor. Hence $\|A\chi_0\|^2\le\|A\|\langle A\rangle_0=O(s)$. The weight $c_-=O(s^{-2})$ makes this part of the complementary source $O(s^{-3/2})$. The normalized lowest floor eigenvector is $\chi_0+O(s^{-1/2})$ by the order-$1/s$ gap to its floor partners. On the entire floor the complementary source is $O(s^{-1})$, so replacing that eigenvector or including the Pauli source changes the Schur expectation by $O(s^{-7/2})$. Replacing the perturbed complementary inverse by $(\Pi_{\rm fl}^\perp(L_++8)\Pi_{\rm fl}^\perp)^{-1}$ changes it by $O(s^{-2})$ in norm. The effective Schur perturbation is $O(s^{-3})$ on the floor, so its second-order effect on the simple lowest floor eigenvalue is $O(s^{-5})$ with the floor gap above. The floor eigenvalue derivative in $E$ is $-1+O(s^{-1})$. Thus the same $-9/(128s^3)$ energy shift is the threshold shift relative to the compressed root, to error $O(s^{-7/2})$.

An explicit estimate is available for the compression error even without expanding the resolvent. On these five floors the profiles are positive combinations of $g_b$ and $\sigma g_b$ with $b\le3$. In the Beta variable $v=1/(1+2t)$ their tilts are bounded below by $1/4$ and the first Beta parameter for $1-v$ is at most 6. Using $1-W\le3(1-v)/2$ gives $\|(1-W)\Pi_{\rm fl}\|\le10/s$. For $1\le E\le2$, $c_-\le E/[8s(s+1)]$ and $\|L_--L_+\|\le16(2s+1)$ give $\|\Pi_{\rm fl}^\perp(L(\zeta)+9)\Pi_{\rm fl}\|\le28/s$. The gap estimates above give $\Pi_{\rm fl}^\perp(L+9)\Pi_{\rm fl}^\perp\ge3s/2$ for $s\ge4$. The compressed lowest eigenvalue decreases with slope at most $-1/2$ for $s\ge25$, from the same bound on $PWP$ and the derivative of $c_-$. Its narrow-state diagonal is negative at $E=2$ for $s\ge32$. Since $L+9\ge0$ at $E=1$, the Schur inequality gives
\begin{equation}
0\le E_{\rm var}-E_c^{\rm ch}\le\frac{3136}{3s^3}\qquad (s\ge32)
\end{equation}
on each ordered side. Here $E_c^{\rm ch}$ refers to that side's threshold. The constant is explicit for the projection error. The smaller errors needed to place a finite spin cutoff for the cubic ordering additionally require effective bounds on the cubic floor expansion and the leading complementary response.

In the side order above, the cubic coefficients of the compressed roots are
$
\frac{1}{10976}(-90,486,2997,801,4497).
$ The forward two are in Eq.~\eqref{eq:E2x2-asymp}. The other three follow by exact expansion of the Gram, weight and Pauli matrices, whose entries follow from the polynomial contractions above. For the mirror matrix use the four types listed above, with the spectator chirality components in Eq.~\eqref{eq:fb:Oop} and the Fock pairing of the four-type construction. In a basis with the second and fourth vectors multiplied by $2s-1$, the polar denominator is put in the common moment convention by multiplying the norm and Pauli polynomials by $(1+t)$. Expanding $(1+v)^{-m}$ in $1-v$ for each fixed integer power $m$ and integrating each term by a Beta moment gives the two first nonzero determinant coefficients at $E=1+3/(2s)+\mathsf x/s^2+\mathsf y/s^3$. Their successive zeros give $\mathsf x=-9/56$ and $\mathsf y=2997/10976$ on the mirror side. The same procedure gives the quoted block-$(5,5)$ entries. A numerical check of the complementary resolvent is given in Appendix~\ref{app:numerics}.

With the complementary correction the five cubic coefficients in units of $1/43904$ are $-3447,-1143,8901,117,14901$. The other sides are already separated by order $1/s$. Imposing Eq.~\eqref{eq:physical} selects the first, fourth and fifth entries for integer $j_G$, or the second and third for half-integer $j_G$. Their minima prove Eq.~\eqref{eq:lawcor3}.
\hfill$\square$

\section{A uniform lower bound on the collapse threshold}
\label{app:un:uniform}

 This appendix proves that the threshold of the full problem satisfies $E_c(s)\ge\hat E(s)=1+\frac{3}{2s}-\frac{3}{4s^2}$ for every integer and half-integer $s\ge1$. The bound accounts for every fermion block and channel. Together with the instability bounds of Sec.~\ref{sec:instability} this gives the two-sided bound of Sec.~\ref{sec:uniform} and the large-spin threshold formula~\eqref{eq:lawcor}, $\lim_{s\to\infty}s(E_c-1)=\frac32$. The argument is uniform in the channel labels. The dependence on $s$ is handled by the positivity of explicit rational functions of $s$ for $s\ge\frac52$ and by rational interval arithmetic \cite{MooreKearfottCloud} for $1\le s\le5$. Both are carried out by computer in exact arithmetic, so the proof is computer-assisted.

We retain the notation of Sec.~\ref{sec:spectral} and Appendices~\ref{app:floor}--\ref{app:fb:bps}, setting $\mathsf n=2s+2$ (integer $\mathsf n\ge4$ for every $s\ge1$) and
\begin{equation}
\hat E(s)=1+\frac{3}{2s}-\frac{3}{4s^2},\quad
\mathsf G_4^{\rm Lich}=12-4\sqrt3,\quad \mathsf G_{4,\rm lo}=\tfrac{50717}{10000},\quad
k(\mathsf G)=\frac{\hat E^2}{\mathsf G+1-\hat E},
\label{eq:un:Ehat}
\end{equation}
with $\bar k=k(\mathsf G_{4,\rm lo})$ and $\mathsf f(\Lambda)=\Lambda+12-\sqrt{144+16\Lambda}$ ($\mathsf f'(\Lambda)\ge\frac13$ on $\Lambda\ge0$). Since $\hat E-1=\frac{3}{4}-\frac{3(s-1)^2}{4s^2}$ and $2s-1>0$, we have $1<\hat E\le\frac74$ for every $s\ge1$. Here $\mathsf G_{4,\rm lo}<\mathsf G_4^{\rm Lich}$ because $(12-\mathsf G_{4,\rm lo})^2>48$, and $W=(1+t)/(1+2t)^2$ satisfies $\dd W/\dd t=-(3+2t)/(1+2t)^3<0$ on $(0,\infty)$ (Eq.~\eqref{eq:abbrev}). Writing $g_b(t)=t^{b+1}(t+\frac12)^{-\mathsf n}(1+t)^{-1}$ ($w_b=\langle W\rangle_{g_b}$, $\Sigma_b=\mathrm{Var}_{g_b}W$), $\sigma g_b$ with $\sigma=t/(1+t)$, $\hat g_b=g_b(1+t+t^2)/(1+t)$, $\varrho_*=g_2/(1+t)$ and $w_\sigma=\langle W\rangle_{\sigma g_0}$, we prove the single-set inequality
\begin{equation}
L_++9-\hat E(s)\,W\ \ge\ 0,
\label{eq:un:single}
\end{equation}
which implies conformality for $4C_2(1-\zeta^2)\le\hat E(s)$ and hence
\begin{equation}
E_c(s)=4s(s+1)\big(1-\zeta_c^2\big)\ \ge\ \hat E(s)=1+\frac{3}{2s}-\frac{3}{4s^2}.
\label{eq:un:main}
\end{equation}

\subsection{Channel reductions and the Schur-complement test}
\label{sec:un:reduction}\label{sec:un:input}\label{sec:un:schur}\label{sec:un:easy}

Let $1<\hat E<4$. Two reductions are used. First, if $L_++9-\hat EW\ge0$, the line is conformal for every $\zeta$ with $E\le\hat E$, and $E_c\ge\hat E$. Second, the inequality $L_++9-\hat EW\ge0$ holds as soon as $\mathcal H_A+9-\hat EW\ge0$ on every single-set channel with $j_R=0$, for $A=1,4,5$.

For the first reduction, note that since $c_++c_-=1$, Eq.~\eqref{eq:radial} gives the exact decomposition
\begin{equation}
L(\zeta)+9=c_+(L_++8)+c_-(L_-+8)+1-EW .
\label{eq:un:decomp}
\end{equation}
Exchanging the two fermion sets maps $L_+$ to $L_-$ and leaves $W$ invariant, so the hypothesis also gives $L_-+9-\hat EW\ge0$. Adding the two inequalities with the weights $c_\pm\ge0$ and using $c_++c_-=1$ gives $L(\zeta)+9\ge(\hat E-E)\,W$, which is nonnegative when $E\le\hat E$ because $W\ge0$. The line is therefore conformal whenever $4C_2(1-\zeta^2)\le\hat E$, and by the definition of $\zeta_c$ as a supremum and the continuity of $\mu$ in $\zeta$, $E_c=4C_2(1-\zeta_c^2)\ge\hat E$.

For the second reduction, note that in the constant fermion frame a spectator set enters $L_+$ as a multiplicity and does not change $j_R$ (the reduction to $S^4$ in Sec.~\ref{app:fb:sharp} and the derivation of Eq.~\eqref{eq:fb:Lbound}). The restriction of $L_+$ to the $j_R=0$ channels of a block $(A_+,A_-)$ is therefore a direct sum of copies of $\mathcal H_{A_+}$ on single-set channels with $j_R=0$, and $W$ acts as multiplication by a function of $\alpha$. In the channels with $j_R\ge\frac12$, the sharp floor~\eqref{eq:fb:H4} gives $L_+\ge4j_R(j_R+1)-8\ge-5$, so $L_++9-\hat EW\ge4-\hat E>0$.

\emph{Spectral input.} At $j_R=0$ a single-set channel is labelled by its $SU(2)_L\times SU(2)_G\cong\mathrm{Spin}(4)$ type $(j_L,j_G)$, on whose multiplicity space $\mathcal H_A$ acts by Schur's lemma. For a unit vector $\chi$, the normalized polar density satisfies $\langle\chi,F\chi\rangle=\int_0^\infty F\varrho_\chi\,\dd t$ for every bounded polar multiplier $F$. The profiles below are given up to a positive common normalization within each expectation. Four results from Appendices~\ref{app:floor}--\ref{app:fb:bps} enter below. 
\begin{itemize}
\item[\textup{(S1)}] \textit{Block 1.} $\mathcal H_1\ge0$ (Eq.~\eqref{eq:fb:H4}).

\item[\textup{(S2)}] \textit{Block 4.} By Eq.~\eqref{eq:fb:gap4}, $\mathcal H_4+8$ vanishes exactly on $\mathcal K_4=\ker\mathsf D_-$ and obeys $\mathcal H_4+8\ge12-4\sqrt3=\mathsf G_4^{\rm Lich}$ on $\mathcal K_4^\perp$ (this weaker gap suffices for the uniform estimate).
By the twistor construction of Sec.~\ref{app:fb:twistor}, $\mathcal K_4$ consists of the types $(j_L,j_G)=(s-\frac12-\frac b2,\frac b2)$, $b=0,\dots,2s-1$, each exactly once, and the type with label $b$ has polar density $g_b$.

\item[\textup{(S3)}] \textit{Block 5, kernel.} By Sec.~\ref{app:fb:block5}, $\mathcal K_5=\ker(\mathcal H_5+8)$ at $j_R=0$ is multiplicity-free and consists of two families. The family $\mathcal K_5^{(1)}$ contains the types $(s-1-\frac b2,\frac b2)$, $b=0,\dots,2s-2$, with polar density $\sigma g_b$. The family $\mathcal K_5^{(4)}$ contains the types $(s-\frac12-\frac b2,\frac12+\frac b2)$, $b=0,\dots,2s-2$, whose polar density lies in the bracket
\begin{equation}
\varrho_{r_1,r_2}=\frac{g_b}{1+t}\big[(1-r_1)+t+(1-r_2)\,t^2\big],\qquad 0\le r_1,r_2\le1,
\label{eq:un:bracket}
\end{equation}
with $r_1=0$ when $b=0$ and $r_2=0$ when $b=2s-2$. The superscripts count the conformal Killing fields (one, or a quartet) in the construction of these floor states.

\item[\textup{(S4)}] \textit{Block 5, gaps.} Let $\Lambda_0(j_L,j_G)$ be the bottom of the scalar (block-1) channel operator in the type $(j_L,j_G)$. By the block-5 gap bounds \eqref{eq:fb:AP} and \eqref{eq:fb:gap5} (Sec.~\ref{app:fb:gap5}), on $\mathcal K_5^\perp$ within a given type, $\mathcal H_5+8\ge16$ in the types of $\mathcal K_5^{(1)}$ and in every type with $j_L+j_G<s$; $\mathcal H_5+8\ge\min\{16,\mathsf f(\Lambda_0)\}$ in the types of $\mathcal K_5^{(4)}$, where $\Lambda_0\ge4(s+j_L)$, and $\Lambda_0(1,\frac12)\ge14$ at $s=\frac32$; and $\mathcal H_5+8\ge\min\{16,\mathsf f(4s)\}$ in every type without floor states.
\end{itemize}
The size parity commutes with $\mathcal H_A$ and $W$. Since each kernel type occurs once, its kernel vector has definite parity, and everything below applies in each parity sector.

\emph{Floor-free sectors and the Schur-complement test \cite{HornJohnson}.}
Let $s\ge1$ and $\hat E=\hat E(s)$. Then $\mathcal H_A+9-\hat EW\ge0$ on every single-set channel of block 1 with $j_R=0$, and on every type of blocks 4 and 5 that contains no floor state.
By (S1), $\mathcal H_1+9-\hat EW\ge9-\hat E>0$. In block 4 a type without floor states lies in $\mathcal K_4^\perp$, and \eqref{eq:fb:gap4} gives $\mathcal H_4+9-\hat EW\ge \mathsf G_4^{\rm Lich}+1-\hat E>0$. In block 5, (S4) and the monotonicity of $\mathsf f$ give $\mathcal H_5+8\ge\min\{16,\mathsf f(4s)\}\ge \mathsf f(4)=16-4\sqrt{13}>\tfrac32$ in a type without floor states, while $\hat E-1\le\frac34$, so $\mathcal H_5+9-\hat EW\ge \mathsf f(4)+1-\hat E>0$.

\emph{The Schur-complement test.}
 Consider a type of block $A\in\{4,5\}$ at $j_R=0$ that contains a floor state, let $\chi$ be the unit kernel vector in its multiplicity space, with polar density $\varrho$, and suppose that $\mathcal H_A+8\ge \mathsf G$ on the orthogonal complement of $\chi$ in this type, with $\mathsf G>\hat E-1$. If
\begin{equation}
1-\hat E\,\langle W\rangle_\varrho-k(\mathsf G)\,\mathrm{Var}_\varrho W\ \ge\ 0,
\label{eq:un:schurtest}
\end{equation}
then $\mathcal H_A+9-\hat EW\ge0$ on this type.
To see this, write $\psi=c\chi+\psi^\perp$ with $\psi^\perp\perp\chi$ and put $x=\|\psi^\perp\|$. Since $(\mathcal H_A+8)\chi=0$, $\langle\psi,(\mathcal H_A+8)\psi\rangle=\langle\psi^\perp,(\mathcal H_A+8)\psi^\perp\rangle\ge\mathsf G\,x^2$. Since $W$ is a multiplication operator with $0\le W\le1$, $\langle\psi^\perp,W\psi^\perp\rangle\le x^2$, and $|\langle\chi,W\psi^\perp\rangle|=|\langle(W-\langle W\rangle_\varrho)\chi,\psi^\perp\rangle|\le(\mathrm{Var}_\varrho W)^{1/2}x$. Hence
\begin{equation}
\langle\psi,(\mathcal H_A+9-\hat EW)\psi\rangle\ \ge\ |c|^2\big(1-\hat E\langle W\rangle_\varrho\big)+(\mathsf G+1-\hat E)\,x^2-2\hat E\,|c|\,(\mathrm{Var}_\varrho W)^{1/2}\,x .
\end{equation}
Since $\mathsf G+1-\hat E>0$, the right-hand side is nonnegative for all $|c|$ and $x$ if $(1-\hat E\langle W\rangle_\varrho)(\mathsf G+1-\hat E)\ge\hat E^2\,\mathrm{Var}_\varrho W$, which is \eqref{eq:un:schurtest}.

To keep the exact-arithmetic verification independent of the specific point inside~\eqref{eq:un:bracket}, the test for $\mathcal K_5^{(4)}$ is made affine in the density.\footnote{Although Sec.~\ref{app:fb:block5} proves the exact $\mathcal K_5^{(4)}$ polar profile~\eqref{eq:fb:r1r2} ($r_1=\frac{b}{2(b+1)}$, $r_2=\frac{a}{2(a+1)}$, $a=2s-2-b$), the wider vertex-bracket formulation suffices here, since the exact profile lies in the verified parallelogram.} For a real number $m$ put $\phi_m=1-\hat EW-k(\mathsf G)(W-m)^2$.
If $\varrho$ is a positive combination of finitely many densities $\varrho_i$ and $m$ is real, then
\begin{equation}
1-\hat E\langle W\rangle_\varrho-k(\mathsf G)\,\mathrm{Var}_\varrho W\ \ge\ \min_i\langle\phi_m\rangle_{\varrho_i}.
\end{equation}
In particular, the test \eqref{eq:un:schurtest} holds for every density in the bracket \eqref{eq:un:bracket} if, for one fixed $m$, $\langle\phi_m\rangle\ge0$ at the vertex densities $\hat g_b$, $g_{b+1}$, $g_b$ and $\sigma g_b$, restricted to $r_1=0$ when $b=0$ and to $r_2=0$ when $b=2s-2$.
Indeed, for every $m$, $\mathrm{Var}_\varrho W\le\langle(W-m)^2\rangle_\varrho$, so the left-hand side is at least $\langle\phi_m\rangle_\varrho$. With $m$ fixed, $\langle\phi_m\rangle_\varrho$ is affine in the normalized density. If $\varrho=\sum_i\lambda_i\varrho_i$ with $\lambda_i\ge0$, it is the convex combination of the $\langle\phi_m\rangle_{\varrho_i}$ with weights $\lambda_i\int\varrho_i/\int\varrho$. For the bracket, expanding the right-hand side and comparing coefficients of $1$, $t$ and $t^2$ gives
\begin{equation}
\varrho_{r_1,r_2}=(1-r_1)(1-r_2)\,\hat g_b+r_1(1-r_2)\,g_{b+1}+(1-r_1)r_2\,g_b+r_1r_2\,\sigma g_b ,
\end{equation}
with nonnegative coefficients for $r_1,r_2\in[0,1]$.

\paragraph*{Uniform channel reductions and closed-form moments.}\phantomsection\label{sec:un:uniformred}\label{sec:un:moments}

The uniformity in the channel label rests on a likelihood-ratio ordering of the profile densities \cite{ShakedShanthikumar}. Let $\varrho,\varrho'$ be positive integrable densities on $(0,\infty)$ with $\varrho'/\varrho$ nondecreasing, and let $F$ be bounded, continuous and nonincreasing. Then $\langle F\rangle_{\varrho'}\le\langle F\rangle_\varrho$. Consequently $\langle W\rangle_\varrho\le w_1$ for each of the densities $\varrho=g_b$ ($b\ge1$), $g_{b+1}$ ($b\ge0$), $\sigma g_b$ ($b\ge1$), $\hat g_b$ ($b\ge1$) and $\varrho_*$, within the ranges of $b$ that occur in (S2) and (S3).
Normalize both densities and let $r=\varrho'/\varrho$. Then $\int(r-1)\varrho\,\dd t=0$, and since $r$ is nondecreasing there is $t_0$ with $r\le1$ on $(0,t_0)$ and $r\ge1$ on $(t_0,\infty)$. Hence $\langle F\rangle_{\varrho'}-\langle F\rangle_\varrho=\int(F-F(t_0))(r-1)\varrho\,\dd t\le0$, the integrand being nonpositive. For the consequence take $\varrho=g_1$ and $F=W$, which is decreasing by \eqref{eq:abbrev}. The ratios to $g_1$ are $t^{b-1}$, $t^b$, $t^b/(1+t)$, $t^{b-1}\big(t+(1+t)^{-1}\big)$ and $t/(1+t)$, and each is nondecreasing in the stated range of $b$.

For $k>0$ and real $m$ define, on $[0,1]$,
\begin{equation}
\begin{aligned}
\varphi_k(w)&=1-\hat Ew-k\,w(1-w),\\
\psi_{k,m}(w)&=1-\hat Ew-k\big(m^2+w(1-2m)\big)=\varphi_k(w)-k(w-m)^2 .
\end{aligned}
\end{equation}
These functions bound the test from below. For a density $\varrho$ with $w=\langle W\rangle_\varrho$ and every real $m$,
\begin{equation}
\mathrm{Var}_\varrho W\le w(1-w),\qquad \langle(W-m)^2\rangle_\varrho\le m^2+w(1-2m).
\end{equation}
If $k<\hat E$ and $0\le m\le1$, then $\varphi_k$ and $\psi_{k,m}$ are strictly decreasing on $[0,1]$. For $k=k(\mathsf G)$ the condition $k<\hat E$ is $\hat E<(\mathsf G+1)/2$.
Since $0\le W\le1$, $W^2\le W$, hence $\langle W^2\rangle_\varrho\le w$, which gives both bounds. On $[0,1]$, $\varphi_k'(w)=-\hat E-k+2kw\le k-\hat E<0$ and $\psi_{k,m}'(w)=-\hat E+k(2m-1)\le k-\hat E<0$. Finally, $k(\mathsf G)<\hat E$ is equivalent to $\hat E<\mathsf G+1-\hat E$.

Let $w_0^L\le w_0\le w_0^U$, $\Sigma_0\le \Sigma_0^U$, $w_1\le w_1^U$ and $w_\sigma\le w_\sigma^U$ be bounds of the moments of $g_0$, $g_1$ and $\sigma g_0$. The closed-form moments below give rational enclosures. The three conditions are
\begin{align}
\Phi_{g_0}&=1-\hat E\,w_0^U-\bar k\big(\Sigma_0^U+(w_0^U-w_0^L)^2\big)\ \ge\ 0,\tag{E.a}\label{eq:un:A}\\
\Phi_{g_1}&=1-\hat E\,w_1^U-\bar k\big(m^2+w_1^U(1-2m)\big)\ \ge\ 0,\tag{E.b}\label{eq:un:B}\\
\Phi_{\sigma g_0}&=1-\hat E\,w_\sigma^U-k(16)\,w_\sigma^U\big(1-w_\sigma^U\big)\ \ge\ 0,\tag{E.c}\label{eq:un:E}
\end{align}
where $m=w_0^U$ in \eqref{eq:un:B}. Note that $\Phi_{g_1}=\psi_{\bar k,m}(w_1^U)\le\varphi_{\bar k}(w_1^U)$.

These three conditions, which do not involve the channel label, suffice for every floor-state type. Let $s\ge\frac52$, and suppose that \eqref{eq:un:A}, \eqref{eq:un:B} and \eqref{eq:un:E} hold together with $\hat E<(\mathsf G_{4,\rm lo}+1)/2$, $w_1^U\le1$ and $w_\sigma^U\le1$. Then the test \eqref{eq:un:schurtest}, in the vertex form above for $\mathcal K_5^{(4)}$, holds in every type of blocks 4 and 5 at $j_R=0$ that contains a floor state. We verify this case by case.
\begin{itemize}
\item \textit{Gaps.} For $s\ge\frac52$ every type of $\mathcal K_5^{(4)}$ has $j_L\ge\frac12$, hence $\Lambda_0\ge4(s+\frac12)\ge12$ and, by (S4) and the monotonicity of $\mathsf f$, a gap $\mathsf G\ge\min\{16,\mathsf f(12)\}=24-4\sqrt{21}>\tfrac{17}{3}>\mathsf G_4^{\rm Lich}$. The block-4 gap is $\mathsf G_4^{\rm Lich}$ by \eqref{eq:fb:gap4}, and the gap in $\mathcal K_5^{(1)}$ is $16$. Every gap therefore satisfies $\mathsf G\ge \mathsf G_{4,\rm lo}$, so $k(\mathsf G)\le\bar k$, and $k(16)\le\bar k<\hat E$ by the criterion for $k(\mathsf G)<\hat E$ stated above and the hypothesis $\hat E<(\mathsf G_{4,\rm lo}+1)/2$. The comparison constant satisfies $m=w_0^U<1$, because $\bar k>0$ and $\Sigma_0^U+(w_0^U-w_0^L)^2\ge \Sigma_0\ge0$ in \eqref{eq:un:A}, so that $\hat Ew_0^U\le1<\hat E$. We write $\varphi=\varphi_{\bar k}$ and $\psi_m=\psi_{\bar k,m}$. Both are decreasing on $[0,1]$ for $0\le m\le1$.

\item \textit{Block 4, $b=0$.} The test is $1-\hat Ew_0-k(\mathsf G)\Sigma_0\ge1-\hat Ew_0^U-\bar k\Sigma_0^U\ge\Phi_{g_0}\ge0$.

\item \textit{Block 4, $b\ge1$.} By the likelihood-ratio ordering, $w_b\le w_1\le w_1^U\le1$. By the variance bounds the test is at least $\varphi(w_b)\ge\varphi(w_1^U)\ge\Phi_{g_1}\ge0$.

\item \textit{$\mathcal K_5^{(1)}$, $b=0$.} The density is $\sigma g_0$ and the gap is $16$. Since $k(16)<\hat E$ and $w_\sigma\le w_\sigma^U\le1$, the test is at least $\varphi_{k(16)}(w_\sigma)\ge\varphi_{k(16)}(w_\sigma^U)=\Phi_{\sigma g_0}\ge0$.

\item \textit{$\mathcal K_5^{(1)}$, $b\ge1$.} With $w=\langle W\rangle_{\sigma g_b}\le w_1^U$ (likelihood-ratio ordering) and $k(16)\le\bar k$, the test is at least $\varphi_{k(16)}(w)\ge\varphi(w)\ge\varphi(w_1^U)\ge\Phi_{g_1}\ge0$.

\item \textit{$\mathcal K_5^{(4)}$, $b=0$.} This is the type $(s-\frac12,\frac12)$. By \eqref{eq:un:bracket} with $r_1=0$, its bracket is $\{g_0+(1-r_2)\varrho_*:0\le r_2\le1\}$, the positive combinations of $g_0$ and $\varrho_*$. Apply the vertex form above with these two densities and the comparison constant $m=w_0^U$. Using $|w_0-m|\le w_0^U-w_0^L$, $k(\mathsf G)\le\bar k$, the variance bounds and $\langle W\rangle_{\varrho_*}\le w_1\le w_1^U$ (likelihood-ratio ordering),
\begin{equation}
\begin{aligned}
\langle\phi_m\rangle_{g_0}&=1-\hat Ew_0-k(\mathsf G)\big(\Sigma_0+(w_0-m)^2\big)\ \ge\ \Phi_{g_0},\\
\langle\phi_m\rangle_{\varrho_*}&\ge\psi_m\big(\langle W\rangle_{\varrho_*}\big)\ \ge\ \psi_m(w_1^U)=\Phi_{g_1} ,
\end{aligned}
\end{equation}
and both are nonnegative.

\item \textit{$\mathcal K_5^{(4)}$, $b\ge1$.} Take the comparison constant $m'=w_1^U\le1$. At each vertex $\varrho\in\{\hat g_b,g_{b+1},g_b,\sigma g_b\}$, the likelihood-ratio ordering gives $\langle W\rangle_\varrho\le w_1^U$, and the variance bounds give $\langle\phi_{m'}\rangle_\varrho\ge\psi_{m'}(\langle W\rangle_\varrho)\ge\psi_{m'}(w_1^U)=\varphi(w_1^U)\ge\Phi_{g_1}\ge0$. The vertex form above completes this case.
\end{itemize}
These cases exhaust the floor-state types listed in (S2) and (S3).
The restriction $s\ge\frac52$ is used only for the gaps. For $s\le2$ some types of $\mathcal K_5^{(4)}$ have gap bounds below $\mathsf G_4^{\rm Lich}$, and Part~II of Sec.~\ref{sec:un:cert} checks them with their actual gaps.

\emph{Closed-form moments.} Every moment entering \eqref{eq:un:A}--\eqref{eq:un:E} is a Beta function or a positive series. Put $v=1/(1+2t)\in(0,1)$ and
\begin{equation}
I(p,q,r)=\int_0^1(1-v)^p\,v^q\,(1+v)^{-r}\,\dd v .
\end{equation}
Then $W=\frac12v(1+v)$ and $\sigma=(1-v)/(1+v)$, and a density proportional to $(1-v)^pv^q(1+v)^{-r}\dd v$ has
\begin{equation}
\langle W\rangle=\frac{I(p,q+1,r-1)}{2\,I(p,q,r)},\qquad \langle W^2\rangle=\frac{I(p,q+2,r-2)}{4\,I(p,q,r)} .
\label{eq:un:momentrule}
\end{equation}
In this variable $g_b\,\dd t\propto(1-v)^{b+1}v^{\mathsf n-b-2}(1+v)^{-1}\dd v$ and $\sigma g_b\,\dd t\propto(1-v)^{b+2}v^{\mathsf n-b-2}(1+v)^{-2}\dd v$, with the same constant, and $\hat g_b=g_b+t^2g_b/(1+t)$ with $t^2/(1+t)=(1-v)^2/[2v(1+v)]$. For integers $p,q\ge0$,
\begin{equation}
\begin{aligned}
I(p,q,0)&=\mathrm B(q+1,p+1), \\ I(p,q,-1)&=2\mathrm B(q+1,p+1)-\mathrm B(q+1,p+2),\\
I(p,q,1)&=\sum_{i\ge0}2^{-i-1}\mathrm B(q+1,p+i+1),\\
I(p,q,2)&=\sum_{i\ge0}(i+1)\,2^{-i-2}\mathrm B(q+1,p+i+1).
\end{aligned}
\label{eq:un:Iseries}
\end{equation}
All terms of the two series are positive, and their tails beyond the first $K$ terms are at most $2^{-K}\mathrm B(q+1,p+K+1)$ and $(K+2)\,2^{-K-1}\mathrm B(q+1,p+K+1)$, respectively.
These relations follow from the substitution $t=(1-v)/(2v)$, under which $1+2t=1/v$, $t+\frac12=1/(2v)$, $1+t=(1+v)/(2v)$ and $|\dd t|=\dd v/(2v^2)$. From these follow $W$, $\sigma$ and $g_b\,\dd t=2^{\mathsf n-b-1}(1-v)^{b+1}v^{\mathsf n-b-2}(1+v)^{-1}\dd v$. Multiplication by $W$ or $W^2$ shifts $(q,r)$ by $(1,-1)$ or $(2,-2)$ and contributes the factor $\frac12$ or $\frac14$, which is \eqref{eq:un:momentrule}. The first line of \eqref{eq:un:Iseries} is the Beta integral, and the second follows from $1+v=2-(1-v)$. For $r=1,2$ expand $(1+v)^{-1}=\sum_i(1-v)^i2^{-i-1}$ and $(1+v)^{-2}=\sum_i(i+1)(1-v)^i2^{-i-2}$, which converge on $(0,1]$ with positive terms, and integrate term by term. Since $\mathrm B(q+1,p+i+1)$ decreases in $i$, the tails are at most $\mathrm B(q+1,p+K+1)\sum_{i\ge K}2^{-i-1}=2^{-K}\mathrm B(q+1,p+K+1)$ and $\mathrm B(q+1,p+K+1)\sum_{i\ge K}(i+1)2^{-i-2}=(K+2)\,2^{-K-1}\mathrm B(q+1,p+K+1)$.
For instance,
\begin{equation}
\begin{gathered}
w_0=\frac{\mathrm B(\mathsf n,2)}{2I(1,\mathsf n-2,1)},\qquad \langle W^2\rangle_{g_0}=\frac{I(1,\mathsf n,-1)}{4I(1,\mathsf n-2,1)},\\
w_1=\frac{\mathrm B(\mathsf n-1,3)}{2I(2,\mathsf n-3,1)},\qquad w_\sigma=\frac{I(2,\mathsf n-1,1)}{2I(2,\mathsf n-2,2)} .
\end{gathered}
\end{equation}
Truncating a series in \eqref{eq:un:Iseries} after $K$ terms gives a lower bound $I_K$, and adding the tail bound gives an upper bound $I_K^+$. The enclosures used in \eqref{eq:un:A}, \eqref{eq:un:B} and \eqref{eq:un:E} are
\begin{equation}
\begin{gathered}
w_0^U=\frac{\mathrm B(\mathsf n,2)}{2I_K(1,\mathsf n-2,1)},\qquad w_0^L=\frac{\mathrm B(\mathsf n,2)}{2I^+_K(1,\mathsf n-2,1)},\qquad w_1^U=\frac{\mathrm B(\mathsf n-1,3)}{2I_K(2,\mathsf n-3,1)},\\
\Sigma_0^U=\frac{I(1,\mathsf n,-1)}{4I_K(1,\mathsf n-2,1)}-\big(w_0^L\big)^2,\qquad w_\sigma^U=\frac{I^+_K(2,\mathsf n-1,1)}{2I_K(2,\mathsf n-2,2)} .
\end{gathered}
\end{equation}
They bound the true moments in the required directions. For $\Sigma_0^U$ this uses $I_K\le I$ and $0<w_0^L\le w_0$. For fixed $K$ each is a rational function of $\mathsf n$ with rational coefficients, because $\mathrm B(q+1,p+1)=p!\big/\prod_{i=1}^{p+1}(q+i)$.

\subsection{Exact-arithmetic verification and proof of the main bounds}
\label{sec:un:cert}\label{sec:un:proof}\label{sec:un:law}

 The remaining inequalities are checked with a program that uses Python's exact rational numbers and SymPy \cite{SymPy} (verified under Python~3.13 and SymPy~1.13). Part~I runs with $K=6$ series terms for all $\mathsf n\ge7$ ($s\ge\frac52$) and Part~II checks $\mathsf n=4,\dots,12$ ($1\le s\le5$, intentionally overlapping Part~I on $\frac52\le s\le5$ as an independent cross-check and populating Table~\ref{tab:un:margins}). Every check passed. Physically, the margins in Table~\ref{tab:un:margins} reflect the headroom built into $\hat E(s)$. The smallest margin in the tabulated tests occurs in block~4 at $b=0$. This type also controls the asymptotic single-set estimate, because $g_0$ is the single-set floor state most tightly localized at the pinned pole ($t=0$) while carrying the smallest single-set gap $\mathsf G_4^{\rm Lich}=12-4\sqrt3$. Its large-$s$ Schur margin approaches $\frac{3(5-\sqrt3)}{64s^2}$ (at $s=5$ the leading term is $3(5-\sqrt3)/1600\simeq0.00613$, compared with $0.00537$ in Table~\ref{tab:un:margins}). Every other channel either sits farther from the line by a centrifugal factor $t^b$ (so its first-order attraction drops by $O(1/s)$) or enjoys a larger gap, $16$ or, in the block-$5$ types of (S4), at least $24-4\sqrt{21}$, explaining why even $K=6$ positive-series terms suffice to keep every shifted polynomial strictly positive for all $s\ge\frac52$.

\paragraph*{Part I (every $s\ge\frac52$).}
With $K=6$, meaning the terms $i=0,\dots,K-1$ of the series~\eqref{eq:un:Iseries}, the quantities $\Phi_{g_0}$, $\Phi_{g_1}$ and $\Phi_{\sigma g_0}$ of \eqref{eq:un:A}, \eqref{eq:un:B} and \eqref{eq:un:E}, and the auxiliary quantities $(\mathsf G_{4,\rm lo}+1)/2-\hat E$, $1-w_1^U$, $1-w_\sigma^U$ and $\hat E-1$, are rational functions of $\mathsf n$ with rational coefficients. The program writes each as a quotient of polynomials in lowest terms, substitutes $\mathsf n=7+z$, and verifies that numerator and denominator have nonnegative coefficients and positive constant terms, so that both are positive for every real $z\ge0$. The degrees in $z$ of numerator and denominator are 18 and 20 for $\Phi_{g_0}$, 18 and 19 for $\Phi_{g_1}$, 17 and 18 for $\Phi_{\sigma g_0}$, and at most 6 and 7 for the auxiliary quantities. Hence the hypotheses of the reduction in Sec.~\ref{sec:un:uniformred} hold for every real $\mathsf n\ge7$, in particular for every $s\ge\frac52$. No case distinction in the channel label occurs in this part, the label dependence having been removed in Sec.~\ref{sec:un:uniformred}.

\paragraph*{Part II ($1\le s\le5$).}
For each $\mathsf n=4,\dots,12$ separately the program checks every type of blocks 4 and 5 at $j_R=0$ that contains a floor state, with the gap bound of that type from (S2) to (S4). It takes $\mathsf G=\mathsf G_{4,\rm lo}$ in block 4, $\mathsf G=16$ in $\mathcal K_5^{(1)}$, and $\mathsf G=\min\{16,\mathsf f_{\rm lo}(\Lambda)\}$ in $\mathcal K_5^{(4)}$, with $\Lambda=4(s+j_L)$, or $\Lambda=14$ for $s=\frac32$ and $(j_L,j_G)=(1,\frac12)$. Here $\mathsf f_{\rm lo}\le\mathsf f$ is obtained from a rational upper bound of the square root, taken on the grid $10^{-12}\Z$. The moments are enclosed with $K=80$ series terms, and every enclosure of $\langle W\rangle$ and $\langle W^2\rangle$ is rounded outward to the grid $10^{-40}\Z$. Write $\overline w$, $\underline w$ and $\overline{w^2}$ for the enclosures of $\langle W\rangle$ and $\langle W^2\rangle$. In block 4 and in $\mathcal K_5^{(1)}$ the density is exact, and the test \eqref{eq:un:schurtest} is bounded below by $1-\hat E\,\overline w-k(\mathsf G)\big(\overline{w^2}-\underline w^{\,2}\big)$. In $\mathcal K_5^{(4)}$ every vertex of the bracket, Eq.~\eqref{eq:fb:bracket} with the restrictions of Sec.~\ref{sec:un:schur}, is checked through $1-\hat E\,\overline w-k(\mathsf G)\big(\overline{w^2}-2m\underline w+m^2\big)$, with one rational comparison constant $m\ge0$ per type, which bounds $\langle\phi_m\rangle$ from below. The program also checks the inequalities used for block 1, for $j_R\ge\frac12$ and for the block-5 types without floor states. Every margin is positive, and at each $\mathsf n$ the smallest margin occurs in block 4 at $b=0$ (Table~\ref{tab:un:margins}). Part~II overlaps Part~I for $\frac52\le s\le5$.

\begin{table}[!htbp]
\centering
\begin{tabular}{@{}cccc@{}}
\toprule
$\mathsf n$ & $s$ & $\hat E(s)$ & Smallest margin\\
\midrule
4 & $1$ & $7/4$ & $0.140$\\
5 & $3/2$ & $5/3$ & $0.0590$\\
6 & $2$ & $25/16$ & $0.0325$\\
7 & $5/2$ & $37/25$ & $0.0207$\\
8 & $3$ & $17/12$ & $0.0144$\\
9 & $7/2$ & $67/49$ & $0.0107$\\
10 & $4$ & $85/64$ & $0.00826$\\
11 & $9/2$ & $35/27$ & $0.00657$\\
12 & $5$ & $127/100$ & $0.00537$\\
\bottomrule
\end{tabular}
\caption{Part~II of the exact-arithmetic verification. For each $\mathsf n=2s+2$ the table lists $\hat E(s)$ and the smallest verified margin of the Schur test \eqref{eq:un:schurtest} over all floor-state types of blocks 4 and 5 at $j_R=0$. The margins are exact rational lower bounds, shown rounded down to three significant figures. At every $\mathsf n$ the smallest one occurs in block 4 at $b=0$.}
\label{tab:un:margins}
\end{table}

\paragraph*{Part III (the strict separations).}
The program verifies $\hat E<\frac{1751}{1000}$ for every $\mathsf n\ge4$, $\mathsf f(12)>\mathsf G_{4,\rm lo}$ and $\mathsf f(4)>\frac{751}{1000}$. These strict rational separations give $k(\mathsf G)\le\bar k$ for the gaps of Sec.~\ref{sec:un:uniformred} and $\mathsf f(4)>\frac{751}{1000}>\hat E-1$ in Sec.~\ref{sec:un:easy}. Part~III also has a short hand proof. The bound $\hat E\le\frac74$ was shown after \eqref{eq:un:Ehat}, and the square-root bounds follow from $336<(\tfrac{55}{3})^2$ (since $55^2-9\cdot336=1$, giving $24-\sqrt{336}>\tfrac{17}{3}>\mathsf G_4^{\rm Lich}$), $208<(\tfrac{29}{2})^2$ (since $29^2-4\cdot208=9$, giving $16-\sqrt{208}>\tfrac32$), and $(12-\mathsf G_{4,\rm lo})^2>48$.

\paragraph*{Role of floating-point arithmetic.}
Every decision in the program is made in exact rational arithmetic. Floating-point arithmetic enters only in Part~II, where a scan proposes the comparison constant $m$, the value that maximizes the smallest vertex bound, for each type of $\mathcal K_5^{(4)}$. The parameter $m$ is then replaced by a nearby rational number, and the inequality is verified exactly with that rational value. By the vertex form of Sec.~\ref{sec:un:schur} every $m\ge0$ gives a valid lower bound, so this choice affects the size of the margin and not the validity of the check.

\paragraph*{Proof of Eqs.~\eqref{eq:un:main} and~\eqref{eq:lawcor}.}
Fix $s\ge1$ and $\hat E=\hat E(s)\in(1,\frac74]$. By the second reduction of Sec.~\ref{sec:un:reduction} it suffices to prove $\mathcal H_A+9-\hat EW\ge0$ on every single-set channel with $j_R=0$, in each block $A\in\{1,4,5\}$ and each parity sector. Block 1 and the types without floor states are covered by Sec.~\ref{sec:un:easy}. A type with a floor state is covered by the Schur-complement test \eqref{eq:un:schurtest}, in the vertex form of Sec.~\ref{sec:un:schur} for $\mathcal K_5^{(4)}$, once its test is verified. The hypothesis $\mathsf G>\hat E-1$ holds because every gap in (S2) to (S4) is at least $\min\{\mathsf G_4^{\rm Lich},16,\mathsf f(4s)\}\ge \mathsf f(4)>\hat E-1$. For $s\ge\frac52$ the tests hold by the reduction of Sec.~\ref{sec:un:uniformred}, whose hypotheses are verified in Part~I. For $s=1,\frac32,2$ they hold by Part~II, which checks every type directly. This proves \eqref{eq:un:single}. The first reduction of Sec.~\ref{sec:un:reduction} then gives conformality for $E\le\hat E$ and the bound \eqref{eq:un:main}.

Combining Eq.~\eqref{eq:un:main} with the analytic upper bounds of Secs.~\ref{sec:instability} and~\ref{sec:bps} proves Eq.~\eqref{eq:lawcor}. At $\theta_{5d}=0$ and integer $s\ge1$, Eq.~\eqref{eq:thmC} gives $E_c\le1+3/(2s)$ (sharpened by Eq.~\eqref{eq:E2x2-asymp} to $E_c\le E_{2\times2}(s)=1+\frac3{2s}-\frac9{56s^2}+O(s^{-3})$). For half-integer $s\ge\frac32$, Appendix~\ref{app:fb:halfint} gives $E_c\le1/w_{\rm crit}(s)\le1/w_{\rm qrt}(s)=1+3/(2s)+O(s^{-3})$. Hence $-\frac{3}{4s}\le s(E_c-1)-\frac32\le0$ for integer $s$, and $-\frac{3}{4s}\le s(E_c-1)-\frac32\le s\big(1/w_{\rm crit}(s)-1-\frac{3}{2s}\big)=O(s^{-2})$ for half-integer $s$. Both brackets close as $s\to\infty$, proving the $3/(2s)$ term for the full problem. At $\theta_{5d}=\pi$ the upper-bound families exchange roles by Eq.~\eqref{eq:physical} and Appendix~\ref{app:fb:halfint}. The uniform lower bound applies to both projections.\unskip\nobreak\hfill\penalty50\hskip1.2em\hbox{}\nobreak\hfill\ensuremath{\square}

\paragraph*{Scope, premises and non-optimality.}\label{sec:un:scope}
The proof uses no numerical spectra, fitted profiles, or quadrature. Its sole inputs are the exact analytic statements in Eq.~\eqref{eq:radial} and Appendices~\ref{app:floor}--\ref{app:fb:bps} (decomposition~\eqref{eq:un:decomp}, sharp floor~\eqref{eq:fb:H4}, block-4 gap~\eqref{eq:fb:gap4}, twistor kernel $\mathcal K_4$, block-5 kernel $\mathcal K_5$ with bracket~\eqref{eq:un:bracket}, and block-5 gaps~\eqref{eq:fb:AP}--\eqref{eq:fb:gap5}). Given these inputs, Sec.~\ref{sec:un:reduction} is purely analytic, while Parts~I and~II above decide finitely many exact rational inequalities (with Part~III checked both by program and by hand, and the series moments~\eqref{eq:un:Iseries} independently cross-checked against high-precision quadrature, e.g.\ $s^2\Sigma_0=0.247$ at $s=3$). The choice $\hat E$ is not optimal. The largest value allowed by the block-4 single-set Schur tests reaches the asymptotic deficit $c_*=\frac38+\frac{9}{8\mathsf G_4^{\rm Lich}}=\frac{33+3\sqrt3}{64}$ (or $\frac{33}{64}$ using Friedrich's bound~\eqref{eq:fb:gap4sharp}), which is not part of Eq.~\eqref{eq:un:main}. Any bound obtained from the single-set inequality alone has an $s^{-2}$ coefficient of at least $\tfrac38$ in magnitude, because the block-4 floor state with profile $g_0$ is an eigenvector of $L_+$ with $1/\langle W\rangle_{g_0}=1+\frac3{2s}-\frac3{8s^2}+O(s^{-3})$. With the gaps used in the certification, block~5 limits the Schur test to $\tfrac{57}{128}$, which the growing block-5 gap of Appendix~\ref{app:fb:chgap} lowers to $\tfrac38$. Approaching the critical-channel value $\tfrac{9}{56}$ requires a two-set argument. The theorem proves the $O(1/s)$ threshold uniformly across all channels without requiring channel ordering and does not by itself fix the $O(s^{-2})$ coefficient. Sec.~\ref{sec:bps} sharpens both integer and half-integer upper bounds to $-\frac9{56s^2}$, and the two-set analysis of Appendix~\ref{app:fb:chgap} supplies the matching lower side, by the exact general-spin floor comparison, Eq.~\eqref{eq:lawcor2}.

\section{Localization and bound-state counting in a trap}
\label{app:localization}

 We prove Proposition~\ref{thm:L} in the setting of Sec.~\ref{sec:theoremL}, writing $H=-\Delta^{\mathcal A}_{Z}-\Delta^{\mathcal A}_C+U$ on the joint cone of dimension $N=d+d_C$, in units $1/2m_s$, and $H_{\rm pin}=-\Delta^{\mathcal A}_C+U|_{X=0}$. Parts (a)--(c) prove the general comparison. We then apply it to the supersymmetric rotating trap, with separate arguments for the fixed-charge and all-charge counts of Sec.~\ref{sec:windows}. 

\emph{(a) The pinned criterion.} Homogeneity and the absence of a radial component of the connection separate the pinned operator on the cone $C$ into the radial operators $-\partial_\rho^2-\tfrac{d_C-1}\rho\partial_\rho+\mu_{{\rm pin},k}/\rho^2$, with $\mu_{{\rm pin},k}$ the eigenvalues of the twisted link operator in the physical sector. With $\psi=\rho^{-(d_C-1)/2}\chi$ each becomes $-\chi''+(\beta_{{\rm pin},k}^2-\tfrac14)\rho^{-2}\chi$ with $\beta_{{\rm pin},k}^2=\mu_{{\rm pin},k}+((d_C-2)/2)^2$. If $\beta_{{\rm pin},k}^2\ge0$ for all $k$, the one-dimensional Hardy inequality gives a nonnegative form. If $\beta_{{\rm pin},k}^2<0$ for some $k$, dilated annular trial states $\chi_\Lambda(\rho)=\Lambda^{1/2}\chi(\Lambda\rho)$ supported where $\chi$ oscillates give $\langle H\rangle\to-\infty$. Every self-adjoint realization extends the minimal operator, whose domain contains these states, so none is bounded below. The two scale-invariant realizations for $0<\beta<1$ follow from the Weyl limit-point/limit-circle classification \cite{ReedSimonII}.

\emph{(b) The dimension switch.} On the joint cone the same argument applies with $d_C$ replaced by $N$, so $H\ge0$ if and only if $\mu_{\rm free}\ge-((N-2)/2)^2$. Since the twisted Laplacian on the joint link is nonnegative, $\mu_{\rm free}\ge\inf\hat U$, and under the hypothesis that the strongest attraction is pinned, $\inf\hat U=-\sup(-\hat U_{\rm pin})$. The difference of the two Hardy constants is $((N-2)^2-(d_C-2)^2)/4=d(d+2d_C-4)/4$. Barta's inequality \cite{Barta} improves the bound, since for any positive $\phi$ on the joint link $\mu_{\rm free}\ge\inf(L_{\rm scal}\phi/\phi)$, with the twisted operator reduced to the scalar one by the diamagnetic inequality \cite{LiebLoss}. For the instanton profile with a general inertia ratio $m=m_c/m_s$, the attraction on the joint link is $-\hat U=4C_2(1-\zeta^2)\,\kappa/\bigl(\kappa+(1-\kappa)/m\bigr)^2$. Its maximum lies in the pinned direction $\kappa=1$ if and only if $m\le2$. For $m>2$ it moves to $\kappa=1/(m-1)$ and grows like $m$. The large-inertia limit of this unconstrained angular maximum is singular. The holonomically pinned model is defined before this limit. The instanton has $m=1/2$, for which $-\hat U=4C_2(1-\zeta^2)W$.

\emph{(c) Traps.} If $H\ge0$ then $H+\varkappa^2|Z|^2\ge0$. If $H$ is not bounded below, dilated trial states give $\langle H\rangle_\Lambda=\Lambda^2\langle H\rangle_1\to-\infty$ while the trap costs $O(\Lambda^{-2})$, so the trapped operator is not bounded below either, and a trap neither creates nor cures a fall to the centre. The unitary dilation $Z\to\varkappa^{-1/2}Z$, $\rho\to\varkappa^{-1/2}\rho$ maps $H+\varkappa^2|Z|^2$ to $\varkappa(H+|Z|^2)$, which multiplies every trapped eigenvalue by $\varkappa$. For a general smooth link as in Proposition~\ref{thm:L}, use radial gauge in $Z$ near the pinned slice and cut off the normalized Gaussian ground state on $|Z|/\rho\ll1$. The cutoff has exponentially small cost at large $\rho$. Smooth homogeneity gives $\mathcal A_Z=O(|Z|/\rho^2)$ and $\mathcal A_y(Z,y)-\mathcal A_y(0,y)=O(|Z|/\rho^2)$ there. The Gaussian is real, so its centre derivative has zero real cross term with the anti-Hermitian connection. The average of $U$ differs from $U|_{Z=0}$ by $O(\ell_{\rm trap}^2/\rho^4)$ because the density is even. Cauchy--Schwarz bounds the size derivative form above by $(1+\varepsilon)$ times the pinned kinetic form plus $O_\varepsilon(\ell_{\rm trap}^2/\rho^4)$. For a strictly supercritical pinned channel, a sufficiently small positive $\varepsilon$ preserves supercriticality and disjoint logarithmic annuli give negative trial forms after subtracting the oscillator zero point. The covariant centre oscillator is bounded below by its scalar zero point by diamagnetism, and the attraction tends uniformly to zero at infinity. Local compactness of the Friedrichs form follows from the compact link and the radial cone forms. At a critical Hardy channel the substitution $\chi=\sqrt R\,f$ reduces the radial form to $\int R|f'|^2\,\dd R$, the radial Dirichlet form in two dimensions, which is locally compact. Thus the essential spectrum is bounded below by the oscillator zero point and min--max gives the claimed accumulation.\unskip\nobreak\hfill\penalty50\hskip1.2em\hbox{}\nobreak\hfill\ensuremath{\square}

For the BPST profile one can compute the Gaussian average explicitly. Compressing to the trap ground state in $Z$ gives a radial operator $-\partial_\rho^2+(\beta^2-\tfrac14)/\rho^2+O(\ell_{\rm trap}^2/\rho^4)$ with $\beta^2<0$, where $\ell_{\rm trap}=(2/\varkappa)^{1/2}$ is the trap length. The remainder follows from the exact Gaussian average of the self-energy profile together with the expansion of the connection in radial gauge in $Z$ built from the pinned-flat frame $v_R$ ($\mathcal A_y(0,y)=0$), $\mathcal A_{Z_i}=\tfrac12Z_j\mathcal F_{Z_jZ_i}+O(|Z|^2)$ and $\delta\mathcal A_y=\mathcal A_y(Z,y)-\mathcal A_y(0,y)=O(|Z|^2/\rho^3)$, which gives $\langle|\mathcal A_Z|^2\rangle_{\ell_{\rm trap}}=O(C_2\ell_{\rm trap}^2/\rho^4)$ and $\langle|\delta\mathcal A_y|^2\rangle_{\ell_{\rm trap}}=O(C_2\ell_{\rm trap}^4/\rho^6)$. For the centre density $(\pi^2\ell_{\rm trap}^4)^{-1}e^{-|X|^2/\ell_{\rm trap}^2}$ and $\upsilon=\rho^2/\ell_{\rm trap}^2$,
\begin{equation}
 \Bigl\langle\frac{\rho^2}{(|X|^2+\rho^2)^2}\Bigr\rangle_{\ell_{\rm trap}}=\frac{\upsilon}{\ell_{\rm trap}^2}\bigl[-(1+\upsilon)e^{\upsilon}\operatorname{Ei}(-\upsilon)-1\bigr],
 \label{eq:gaussianaverage}
\end{equation}
with $\operatorname{Ei}$ the exponential integral. For $\rho\gg\ell_{\rm trap}$ this is $\rho^{-2}\bigl[1-4\ell_{\rm trap}^2/\rho^2+18\ell_{\rm trap}^4/\rho^4+O(\ell_{\rm trap}^6/\rho^6)\bigr]$, the moment series of the Gaussian, while at fixed $\ell_{\rm trap}$ it vanishes like $\rho^2\ell_{\rm trap}^{-4}\bigl[\ln(\ell_{\rm trap}^2/\rho^2)-\gamma_{\rm E}-1\bigr]$ as $\rho\to0$. The limits $\ell_{\rm trap}\to0$ and $\rho\to0$ do not commute, which is why the averaged potential describes the trapped problem only where the centre is fast, $\rho\gtrsim\ell_{\rm trap}$, and why part (c) rests on dilations at $R\ll\ell_{\rm trap}$. The compressed operator is oscillatory at infinity and has the asymptotic shallow-level ratio $e^{2\pi/\nu}$, $\nu^2=-\beta^2$. The min--max argument for the full trap proves accumulation below the oscillator threshold.

\emph{Application to the $\Omega$-trap.} The trapped model $\mathsf S_\Omega$ of Sec.~\ref{sec:windows} has $H_\Omega=\sum_\pm c_\pm Q_\pm^2-EW/R^2$, with $Q_\pm=\ii{\not{\!\mathbf D}}_\pm+\varkappa\,\mathsf c_\pm(\xi)$, $\xi=e_1Z$ and $E=4C_2(1-\zeta^2)$. Expanding the squares,
\begin{equation}
 H_\Omega=H(\zeta)+2\ii\varkappa\,\xi\cdot\nabla_{\mathcal A}+2\varkappa\,(c_+\mathcal S_++c_-\mathcal S_-)+\varkappa^2|Z|^2,
 \label{eq:loc:omega}
\end{equation}
where $\mathcal S_\pm=-\ii\sum_{a<b}[e_1]_{ab}\gamma^{ab}_\pm$ has eigenvalues $0,\pm2$ on the centre fermions. We take the Friedrichs realization whenever the form is bounded below. The scaling $(Z,y)\to\varkappa^{-1/2}(Z,y)$ gives $H_\Omega(\varkappa)\simeq\varkappa H_\Omega(1)$. If the free form has a negative direction, shrinking dilations drive the trapped form to $-\infty$, since the free energy scales as $\Lambda^2$, the rotation and spin terms remain fixed, and the harmonic potential scales as $\Lambda^{-2}$.

In the free-conformal regime $L+9\ge0$, including equality at the onset, the Hardy form satisfies
\begin{equation}
 \mathfrak q_H[\Psi]\ge \int R^5\langle\Psi(R),(L+9)\Psi(R)\rangle\,\dd R.
\end{equation}
The centre rotation is tangential to the joint sphere. Write $L=-\Delta^{\mathcal A}_{S^7}+\mathcal V$ with $\mathcal V$ bounded at fixed spin. For a state supported in $R\le R_0$, this gives $\|\xi\cdot\nabla_{\mathcal A}\Psi\|^2\le R_0^2 \mathfrak q_H[\Psi]+C\|\Psi\|^2$. Thus the linear perturbation of Eq.~\eqref{eq:loc:omega} is infinitesimally form bounded there, even at the critical Hardy value. Where $R\ge R_0/2$, the exact squares bound the energy below by $-4E/R_0^2$. Localization proves semiboundedness. For local compactness, the positive angular eigenvalues of $L+9$ tending to infinity control the high angular modes on a bounded ball. In each remaining channel set $\chi=R^{7/2}\Psi=\sqrt R f$. The critical Hardy form controls $\int R|f'|^2\dd R$ with norm $\int R|f|^2\dd R$. Radial two-dimensional Rellich then applies also at the origin. By the characterization of the essential spectrum in Refs.~\cite{Persson1960,CFKS} and $H_\Omega\ge-E/R^2$ in the exterior, the essential spectrum has no negative part.

For a fixed-sector tower, let $1<E\le E_c(s)$ (or $s=\frac12$ with $1<E\le3$, i.e.\ $|\zeta|<\sqrt{2/3}$). In the gauge that is radial in $Z$ about the pinned slice and built from the frame $v_R$ (so $\mathcal A_y(0,y)=0$), take $\Psi=g(Z)\,\varepsilon\,F(y)$. Here $g=(\varkappa/\pi)\,e^{-\varkappa|Z|^2/2}$ is normalized on $\R^4_Z$; $\varepsilon$ ($|\varepsilon|=1$) polarizes the centre fermions of both sets, $\mathcal S_\pm\varepsilon=-2\varepsilon$; and $F$ is a physical section over the size space in the channel $n=0$ that the size-fermion chirality makes available (Sec.~\ref{sec:windows}). Since $-(\partial-\ii\varkappa\xi)^2g=4\varkappa g$, the trapped supercharges without the instanton connection annihilate $g\varepsilon$, so $Q_\pm\Psi=\ii\gamma^{Z_k}_\pm\mathcal A_{Z_k}\Psi+\ii\gamma^{y_k}_\pm\nabla_{y_k}\Psi$, after the flat centre oscillator and Zeeman cancellation. The remaining supercharge norms include the connection corrections. In this gauge $|\mathcal A_Z|\le C_Z|Z|/\rho^2$ and $|\mathcal A_y(Z,y)-\mathcal A_y(0,y)|\le C_y|Z|^2/\rho^3$, because $\mathcal F$ is homogeneous of degree $-2$ and vanishes on the size directions at $Z=0$ (property~(iii) after Eq.~\eqref{eq:curvature}), where the connection is flat. With $\langle|Z|^2\rangle_{\ell_{\rm trap}}=\ell_{\rm trap}^2$, $\langle|Z|^4\rangle_{\ell_{\rm trap}}=\frac32\ell_{\rm trap}^4$ and $\bigl\langle\rho^2/(\rho^2+|X|^2)^2\bigr\rangle_{\ell_{\rm trap}}\ge\rho^{-2}(1-4\ell_{\rm trap}^2/\rho^2)$, by Jensen's inequality, integrating over $Z$ gives for every $\delta>0$ and every $F$ supported in $\rho\ge\ell_{\rm trap}$
\begin{equation}
 \langle\Psi,H_\Omega\Psi\rangle\le(1+\delta)^2\langle F,-\nabla_y^2F\rangle-E\,\langle F,\rho^{-2}F\rangle+C_\delta\,\ell_{\rm trap}^2\langle F,\rho^{-4}F\rangle .
 \label{eq:loc:trial}
\end{equation}
For $(1+\delta)^2<E$ the radial operator on the right is oscillatory at infinity. Sections $F$ on disjoint shells $\rho_k\le\rho\le e^T\rho_k$ with $\rho_k\to\infty$ therefore give infinitely many orthogonal states with $\langle H_\Omega\rangle<0$ and vanishing mutual matrix elements, and by min--max and the essential-spectrum bound $H_\Omega$ has infinitely many eigenvalues below zero, which accumulate at zero. The lifted rotation commutes with $H_\Omega$ (and choosing a fixed colour $J^1$-eigenstate $v_m$ of weight $m$ inside $F$ places every trial state $\Psi_k$ directly in a single eigensector of it), so infinitely many of these states, and hence infinitely many eigenvalues, lie in a single sector; the tower is not an artefact of the Landau degeneracy of the rotating frame. The compressed inverse-square channel predicts the ratio $e^{2\pi/\nu}$, $\nu^2=E-1$, of successive shallow levels.

For finiteness at $0<E\le1$ fix the lifted rotation charge $q$. In the $v_R$ frame its spin and colour terms are bounded. At fixed $y$ the free centre form $H^{(0)}_Z=\sum_\pm c_\pm(Q^{(0)}_{Z,\pm})^2$, with $Q^{(0)}_{Z,\pm}$ the centre part of $Q_\pm$ without the instanton connection, is a magnetic oscillator plus the two commuting spin terms. It has a gap of at least $4\varkappa\min(c_+,c_-)$ on the orthogonal complement of its polarized lowest Landau space, with projector $\mathcal P_0$. The fixed charge bounds the orbital rotation term, so $\varkappa^2\|Z\Psi\|_Z^2\le\langle H^{(0)}_Z\rangle_Z+C_q\|\Psi\|_Z^2$. Here the inner product is over the centre at fixed $y$. In the same gauge $|\mathcal A_Z|\le C|Z|/\rho^2$, $\|P_{yy}\|\le C|Z|/\rho^3$, and $\|P_{Zy}\|\le C/\rho^2$. The subscripts denote the size--size and centre--size Pauli terms after expanding the supercharge square. The latter flips a centre chirality, and hence $\mathcal P_0P_{Zy}\mathcal P_0=0$. The gap and Young's inequality absorb its cross term at the cost of $O(\rho^{-4})$ in the ground space. The bound on the centre moment controls the connection perturbation by the same order and the size--size term by $O(\rho^{-3})$, after absorbing their terms proportional to $H^{(0)}_Z$. Thus for sufficiently large $\rho_c$, in the second-order form with Neumann boundary at $\rho=\rho_c$,
\begin{equation}
 \mathfrak q_{\Omega,\rho\ge\rho_c}\ge
 \int_{\rho\ge \rho_c}
 \bigl[|\nabla_y^{\mathcal A}\Psi|^2-
 (E/\rho^2+C_q'/\rho^3)|\Psi|^2\bigr]\,\dd Z\,\dd y .
\end{equation}
The argument uses no integration by parts in $y$ at the interface. Apply radial diamagnetism, write $f=|\Psi|$, and put $u=\rho f$, $\tau=\ln(\rho/\rho_c)$ at fixed $Z,\hat y$. At $E=1$,
\begin{equation}
 \int_{\rho_c}^\infty(\rho^3|f'|^2-\rho|f|^2)\,\dd\rho
 =\int_0^\infty|\partial_\tau u|^2\,\dd\tau+|u(0)|^2
 \ge\int_0^\infty\frac{|u|^2}{4(1+\tau)^2}\,\dd\tau .
\end{equation}
The last inequality follows by completing a square with $\partial_\tau u-u/[2(1+\tau)]$. It absorbs the error $C_q'\rho_c^{-1}e^{-\tau}$ when $\rho_c$ is large. The exterior form is therefore nonnegative for $E\le1$. Inside, local compactness at the joint origin follows from the Hardy form. Away from it, the fixed lifted charge and bounded $\xi\cdot\mathcal A$ leave quadratic confinement in $|Z|$ and ordinary local elliptic control, also across the regular chart at $y=0$. The interior Neumann restriction has compact resolvent. Bracketing proves the sectorwise finiteness. For $E=0$ the sum of squares is nonnegative in every sector.\unskip\nobreak\hfill\penalty50\hskip1.2em\hbox{}\nobreak\hfill\ensuremath{\square}

\paragraph*{Fixed-charge counting.}
\label{app:loc:count}
To prove the fixed-charge counting asymptotic, retain $H^{(0)}_Z/4$ in the far-size lower bound. The formula for $v_R$ gives $|\mathcal A_y|\lesssim\min(|Z|^2/\rho^3,1/\rho)\lesssim |Z|/\rho^2$. With the fixed-charge centre moment bound, Young's inequality with parameter $1/\rho$ in trap units compares the covariant size derivative to the ordinary one with coefficient $a_-(\rho)=1-1/\rho$ and an error $O(\rho^{-3})(H^{(0)}_Z+1)$. The resulting lower bound has potential $-E/\rho^2-C_q/\rho^3$. For the upper comparison restrict to the whole finite-rank polarized kernel in this charge, where all centre moments are finite and $H^{(0)}_Z=0$. Use $a_+(\rho)=1+1/\rho$ and potential $-E/\rho^2+C_q/\rho^3$. The centre-supercharge contribution is $O(\rho^{-4})$. The mixed Pauli terms have vanishing compression to this kernel. The paired bounds preserve parity, so each size harmonic and its permitted constant spinor states can be counted directly. Excited centre modes are gapped. The harmonics $n\ge1$ have radial Hardy value $(n+1)^2$, above $E$ in the whole stable range of Eqs.~\eqref{eq:omega-count-sector} and~\eqref{eq:omega-count-global}, since every onset at $s\ge1$ lies below $E=8/3$ (Appendix~\ref{app:un:coverage}) and $E\le3$ at $s=\tfrac12$. The bounded interior contributes only $O(1)$ to either bracketing bound.

The two centre polarizations each have rank one. At size harmonic $n=0$, the physical projection retains eight of the sixteen two-set size spin components in either theta sector. With the $U(1)_L$ action $D_s(e^{\phi e_1})U_F(e^{\phi e_1})\psi(e^{-\phi e_1}Z)$ in $v_R$, a polarized holomorphic polynomial of degree $d\ge0$ and colour weight $m$ carries charge $q=2m-d-2$. Each degree contributes orbital multiplicity $d+1$. Thus $r_q$ is eight times the sum of these orbital multiplicities over the colour weights satisfying the charge relation. Reversing the generator reverses the charge label.

On any surviving scalar copy put $u=\rho f$, $\tau=\ln(\rho/\rho_c)$. In the form at energy $-\eta$, the radial kinetic term gives $a_\pm|\partial_\tau u|^2$ and the potential $a_\pm+\partial_\tau a_\pm=1$ after integration by parts. The other potential terms become $-E+O(e^{-\tau})+\eta \rho_c^2e^{2\tau}$. Passing to $\dd t=\dd\tau/a_\pm$ changes the interval length by $O(1)$. Up to $\tau=\tfrac12\ln[1/(\eta \rho_c^2)]+O(1)$ the difference from the constant potential $-(E-1)$ has bounded $L^1$ norm. Beyond this point a boundary can be chosen so the form is nonnegative. The scalar phase advance is therefore $\sqrt{E-1}\ln(1/\eta)/2+O(1)$ by the Prüfer equation, with the two boundary conditions differing by bounded counts. This is the scalar inverse-square counting mechanism, cf.\ Ref.~\cite{HassellMarshall2008}, applied to each of the $r_q$ scalar copies. Min--max comparison and dilation back to $\varkappa$ prove Eq.~\eqref{eq:omega-count-sector}.
\hfill$\square$

\paragraph*{Global counting and displaced size modes.}\label{app:loc:guiding}

To bound the count across all rotation sectors, localize on dyadic joint annuli of scale $R_*$ and use a partition of each between $\rho\gtrsim R_*$ and $|Z|\gtrsim R_*$. The connection is $O(R_*^{-1})$ in the respective frames $v_R$ and $v_S$ (the latter defined below), and the attraction and localization errors are $O(R_*^{-2})$. Comparing the full supercharge squares by Young's inequality leaves a positive fraction of the free magnetic centre oscillator $H^{(0)}_Z$ and the size Laplacian, with a negative error $C/R_*^2$. Each localized form can be bounded on a product of Dirichlet boxes of side $O(R_*)$ by extension of its test functions by zero. The size count up to order $R_*^{-2}$ is bounded. The centre count is $O(R_*^4)$ at this energy, since the diamagnetic inequality and Dirichlet monotonicity bound the number of Dirichlet eigenvalues below $\lambda$ in a centre box of volume $|B|$ by $e^{t\lambda}\,\mathrm{Tr}\,e^{-tH_B}\le e^{t(\lambda+c\varkappa)}\,d_F\,|B|\,(4\pi t)^{-2}$ at any fixed $t>0$, where $d_F$ is the fibre dimension and $c\varkappa$ bounds the centre Zeeman terms. Both comparison operators are nonnegative before the $C/R_*^2$ shift. Annuli with $R_*>(C/\eta)^{1/2}$ therefore contribute no levels below $-\eta$. The sum of $O(R_*^4)$ over the remaining dyadic annuli is $O(\eta^{-2})$, and the joint compact interior contributes $O(1)$ by the same local compactness argument used in part (d). This proves the upper order in trap units.

For the lower bound, take a centre $Z_0$ with $|Z_0|\asymp R_*$. The frame $v_S=(|u|,-u\bar y/|u|)^{\rm T}/\sqrt{\varphi}$ is regular on the size slice and obeys $v_S(Z,-y)=\mathsf T v_S(Z,y)$. Its connection is the regular BPST connection $\operatorname{Im}(y\,\dd\bar y)/(\rho^2+a^2)$ up to a constant orientation convention, with $a^2=2|Z_0|^2$. For $s\ge1$, pull the symmetric twisted Maxwell spinors of Appendix~\ref{app:fb:twistor} back to this slice. In the frame $v_S$ the slice connection is the unit BPST connection of scale $a=\sqrt2|Z_0|$, so the sections $\epsilon\odot\epsilon\odot\omega$ with $\omega\in\mathcal T_{2s-2}$ solve both size Dirac equations, and the degree $b$ of $\omega$ in the size-dependent twistor factors obeys $b\le2s-2$. Their equations and two-form $L^2$ norm are conformally invariant in four dimensions, and the symmetric spinor indices solve both size Dirac equations. A monomial with degree $b$ in the size-dependent twistor factors has parity $(-1)^b$ in this frame, with product size chirality $+1$. We can take $b=0$ for $\theta_{5d}=0$ and $b=1$ for $\theta_{5d}=\pi$ if $s\ge\tfrac32$. If $x=\rho^2/a^2$, the radial density is proportional to $x^{b+1}(1+x)^{-2s-2}$. The average of $W/R^2$ at the fixed centre is
$
\frac{(b+2)(2s-b)}{(2s+2)(2s+3)a^2}.
$
For the adjoint line at $\theta_{5d}=0$ the Maxwell spinor is the curvature itself, and this average is $1/(10|Z_0|^2)$.

At $s=1$, $\theta_{5d}=\pi$, instead use the even single Dirac zero mode in the set with larger weight and an opposite-chirality size spinor in the other set. This gives the required product chirality $-1$. The inactive size Pauli term vanishes. The active Pauli eigenvalue on this mode is $-16a^2/(\rho^2+a^2)^2$ by colour-spin coupling to $s-\tfrac12$. The zero-mode equation then identifies the inactive size kinetic expectation with the opposite of this Pauli expectation. With the $x(1+x)^{-4}$ density it equals 24 times the attraction average, by the ratio $16\mathrm B(2,4)/\mathrm B(3,3)$. The net leading expectation is $-(E-24c_{\min})/(10|Z_0|^2)$, where $c_{\min}=(1-|\zeta|)/2$. This is negative throughout the stable branch with $E>0$, since $c_{\min}/E=[16(1+|\zeta|)]^{-1}$ and $E_c<8/3<6$ (Appendix~\ref{app:un:coverage}). For $E>1$, including the fundamental line, a scalar size trial in a logarithmic annulus at $\rho\ge MR_*$ is another way to obtain a negative margin of order $R_*^{-2}$. A sufficiently wide logarithmic annulus overcomes the Hardy deficit and $M$ can be chosen large enough to suppress the relative BPST corrections of order $R_*^2/\rho^2$.

Extend each slice frame into $|Z-Z_0|<R_*/4$ by radial parallel transport and use a magnetically translated polarized Gaussian with a smooth cutoff in this cylinder. The changes in the connection are $O(|Z-Z_0|/R_*^2)$ by the curvature bound. The cutoff costs exponentially little. The trial packets therefore have diagonal energies below $-c/R_*^2$ for large $R_*$, including in the adjoint case with an inactive kinetic term of order $R_*^{-2}$. Pack centres on a lattice of sufficiently large fixed spacing in $R_*\le|Z_0|\le2R_*$. The overlaps and the off-diagonal energy matrix elements are bounded by Gaussian decay in the centre separation, times $O(1)$ for the Gram matrix and $O(R_*^{-2})$ for the energy matrix. These estimates follow by Cauchy--Schwarz in the size variables and the Gaussian envelope of the packets and their supercharge derivatives in the centre variables. Their row sums can be made smaller than the diagonal margin uniformly in $R_*$. This gives a negative trial subspace of dimension of order $R_*^4$ below $-c'/R_*^2$. Min--max and the trap dilation prove the lower bound in Eq.~\eqref{eq:omega-count-global}. \unskip\nobreak\hfill\penalty50\hskip1.2em\hbox{}\nobreak\hfill\ensuremath{\square}

The use of the displaced size modes is specific to the coupled instanton operator. The role of centre degeneracy is related to that of a full-rank magnetic field with a decaying electric perturbation \cite{MelgaardRozenblum2003}.

\paragraph*{The fundamental line below the size threshold.}
\label{app:loc:fundamental}

For $s=\tfrac12$ and $0<E\le1$, the size operator at every fixed $Z\ne0$ is nonnegative in either discrete theta sector. In the regular frame $v_S$, put $a=\sqrt2|Z|$, $x=\rho^2/a^2$ and $p=x/(1+x)$. On the scalar size harmonic of degree $n=2\ell$, let $\mathbf L$ be the orbital spin $\ell$ in the factor to which the connection couples. The angular covariant kinetic term in units $a^{-2}$ is $4(\mathbf L+p\mathbf J)^2/x$. The remaining potential is $(16A-Ex)/(1+x)^2$, where $A=\sum_\pm c_\pm\mathbf S_\pm\cdot\mathbf J$, with $\mathbf S_\pm$ acting as spin $\frac12$ on the BPST-active size chirality and zero on the other chirality. These formulas follow from $\mathcal A_y=\operatorname{Im}(y\,\dd\bar y)/(\rho^2+a^2)$ and its BPST curvature. The physical condition in $v_S$ is $(-1)^n\Gamma_y=e^{\ii\theta_{5d}}$, since this frame has trivial quotient action on the colour coefficients.

With both size chiralities active, coupling the three spin-$\frac12$ factors gives $\min A=-(1+\sqrt{1+3\zeta^2})/4$. For one active set the lowest possible value is $-3(1+|\zeta|)/8$, which is strictly higher for $E>0$ since $4(1+3\zeta^2)-(1+3|\zeta|)^2=3(1-|\zeta|)^2$. With neither active, $A=0$. Write $\mathsf z_0=\tfrac12+\sqrt{1-E/4}$. For $n=0$ the minimum scalar operator has radial potential $[(3-E)x-8\mathsf z_0]/(1+x)^2$, with positive zero mode $(1+x)^{-\mathsf z_0}$. This is in $L^2(\R^4_y)$ for every $E<3$, hence throughout $E\le1$, and has both size chiralities active. Therefore the mode occurs for $\theta_{5d}=0$. At $\theta_{5d}=\pi$ the $n=0$ channel requires one active size chirality, so the potential has a strict additional constant numerator $\delta_E>0$. Applying the ground-state identity with $(1+x)^{-\mathsf z_0+\epsilon}$ gives a numerator bounded below by
$
 \delta_E-8\epsilon+4\epsilon(2\mathsf z_0-1-\epsilon)x .
$
A sufficiently small $\epsilon>0$ thus gives a positive size form bound by $c_E/(\rho^2+a^2)$. For $n\ge2$ the angular kinetic term exceeds the $n=0$ angular term by at least $4(\ell+1)(\ell-p)/x\ge8/[x(1+x)]$, using $\mathbf L\cdot\mathbf J\ge-(\ell+1)/2$. This adds at least eight to the numerator and gives the same conclusion with constants uniform over $n\ge2$ in both theta sectors.

For $n=1$, write $\mathcal P_{\rm t}$ for the triplet projector of the orbital and colour spins. Apply the radial Hardy identity with $u=\rho f$. The residual potential numerator is $\mathcal C\,I+8(1+x)\mathcal P_{\rm t}+16A$, where $\mathcal C=4/x+2+(1-E)x$. On the orbital--colour singlet the $A$ diagonal vanishes. Moreover $\|(1-\mathcal P_{\rm t})A\mathcal P_{\rm t}\|^2\le 3/16$, since on the singlet $J_iJ_j$ compresses to $\delta_{ij}/4$ and $\|\sum_i(c_+S_+^i+c_-S_-^i)^2\|\le3/4$. On the triplet $A\ge-3/4$. Thus the two diagonal lower bounds are $\mathcal C$ and $\mathcal C+8x-4\ge(3/5)\mathcal C$. Their product satisfies
\begin{equation}
 \mathcal C(\mathcal C+8x-4)\ge (2+4/x)(8x-2+4/x)=28+16x+16/x^2>56 ,
\end{equation}
whereas the squared off-diagonal bound is $48$. The residual is consequently bounded below by a positive universal multiple of $\mathcal C$. At $E<1$ this gives the margin $c_E/(\rho^2+a^2)$. At $E=1$ retain the derivative in the Hardy identity and at least half of the potential margin $c_E a^2/(\rho^2+a^2)^2$. In $\tau=\ln(\rho/a)$ the other half gives a positive mass on $\tau\le0$, since $x\mathcal C/(1+x)^2$ has a positive infimum there. The trace at $\tau=0$ is controlled by this mass and the derivative. The half-line logarithmic Hardy inequality then gives a second margin $c/[\rho^2(1+\tau_+)^2]$. Uniformly for test sections supported on $R\ge R_0$, this implies a bound by $c/[R^2(1+\ln(R/R_0))^2]$. Indeed if $a\ge R_0/2$, one uses $\tau_+\le\ln(2R/R_0)$. If $a<R_0/2$, exterior support implies the radial wavefunction vanishes for $\rho<R_0/2$ and the same Hardy inequality has a Dirichlet anchor there. Thus the frozen operator at either theta has no negative spectrum. Its only kernel is the stated $n=0$, two-active-chirality mode at $\theta_{5d}=0$ (with the two-dimensional total-spin-$\frac12$ multiplet).

For $\theta_{5d}=\pi$ this yields global finiteness for the full trap. Split the centre-spin fibre by $\mathcal P_{\rm ch}$ onto the simultaneous BPST-active centre chiralities and its complement. The centre-only sum of squares preserves this splitting. Completing two complex covariant magnetic derivatives bounds the scalar transverse kinetic form below by $4\varkappa-O(R^{-2})$ on exterior support, because the instanton connection enters the completed squares only through its curvature, which is $O(R^{-2})$ there. At least one centre Zeeman term vanishes on the complement, so the centre squares there have gap at least $2\varkappa\min(c_+,c_-)$ for sufficiently large $R_0$. The mixed Pauli operator flips one centre chirality, has zero compression onto $\mathcal P_{\rm ch}$, and has norm bounded by $C|Z|\rho/(\rho^2+a^2)^2$, since Eq.~\eqref{eq:curvature} gives $|\mathcal F(\xi,\eta)|\le(2/\varphi)\,|v^\dagger\xi|\,|v^\dagger\eta|$, while $|v^\dagger\partial_y q|\le C|Z|/\sqrt\varphi$ and $|v^\dagger\partial_Z q|\le C\rho/\sqrt\varphi$ for the size and centre directions, with $\varphi=\rho^2+a^2$ and $a^2=2|Z|^2$. Absorbing its complement matrix elements against the gap costs at most $C_{E,\varkappa}|Z|^2\rho^2/(\rho^2+a^2)^4$. On $R\ge R_0$ this is smaller than half of the reserved $c_E a^2/(\rho^2+a^2)^2$ size margin for large $R_0$. Consequently the full exterior trap retains a positive form bound $c_E/[R^2(1+\ln(R/R_0))^2]$. For the quadratic partition-of-unity formula (IMS localization~\cite{CFKS}), take inner and outer cutoffs $\chi_{\rm in},\chi_{\rm out}$ with $\chi_{\rm in}^2+\chi_{\rm out}^2=1$. Localizing the kinetic form subtracts $\sum_{a={\rm in,out}}|\nabla\chi_a|^2$ from the sum of the localized forms. Choosing the rotation angle of the cutoffs to vary slowly in $\ln(1+\ln(R/R_0))$ keeps this error below half the exterior bound. Split the IMS error between the two cutoffs, absorb it outside, and use local compactness of the Friedrichs form for the bounded inner region. The inner comparison has compact resolvent and finite negative index, which proves that the full operator has finitely many negative levels at $\theta_{5d}=\pi$ for $0<E\le1$.\unskip\nobreak\hfill\penalty50\hskip1.2em\hbox{}\nobreak\hfill\ensuremath{\square}

At $\theta_{5d}=0$ the frozen size kernel saturates the leading comparison. Finiteness in each fixed rotation sector still follows, but summing over all sectors requires control of the coupling of that zero mode to centre motion. The global count for $\theta_{5d}=0$ remains open. 

\section{Angular momentum bounds on the channels that can collapse}
\label{app:un:coverage}

 This appendix proves a pointwise lower bound on the channel potential of the link operator that is monotone in the three angular momenta. For each spin it confines the channels that can fail to be conformal for $E\le4$ to a finite set, and it delimits the channels covered by the numerical scans of the main text. It is not used in the proof of the uniform lower bound~\eqref{eq:un:main}.

In the coordinates and symmetry decomposition of Sec.~\ref{sec:spectral} ($SU(2)_L\times SU(2)_R\times SU(2)_G$ acting on $S^7$ with diagonal stabilizer for $0<\alpha<\pi/2$), in the fermion block $(A_+,A_-)$ and the channel $\varpi=(j_L,j_R,j_G)$, a state is described by a function $T(\alpha)$ whose values are linear maps from the irreducible representation of spins $(j_L,j_R,j_G)$ to the fibre $V_s\otimes A_+\otimes A_-$, equivariant under the stabilizer,
\begin{equation}
\big(J_a+S^{(Z)}_a+S^{(y)}_a\big)T=T\big(J^{(j_L)}_a+J^{(j_R)}_a+J^{(j_G)}_a\big).
\label{eq:un:equivariance}
\end{equation}
Here $J_a$ ($a=1,2,3$) are the colour generators in spin $s$, $S^{(Z)}_a$ and $S^{(y)}_a$ are the generators through which rotations of $Z$ and of $y$ act on the two fermion sets, $J^{(j)}_a$ are the generators of spin $j$, and norms of $T$ are Hilbert--Schmidt norms. In the channel the link operator acts as $L_\varpi=-\omega^{-1}\partial_\alpha\,\omega\,\partial_\alpha+M_\varpi(\alpha)$, with $\omega=\sin^3\alpha\cos^3\alpha$, and the matrix potential has the quadratic form
\begin{equation}
\langle T,M_\varpi T\rangle=\frac1\sigma\sum_a\|u_a\|^2+\frac1\kappa\sum_a\|v_a\|^2+\big\langle T,(c_+P^{(+)}+c_-P^{(-)})T\big\rangle-E\,\langle T,WT\rangle ,
\label{eq:un:Mform}
\end{equation}
in which the orbit vectors are
\begin{equation}
u_a=-2\ii\Big[\Big(\frac{\kappa}{D}J_a+S^{(Z)}_a\Big)T-TJ^{(j_L)}_a\Big],\qquad
v_a=-2\ii\Big[\Big(\frac{2\sigma}{D}J_a+S^{(y)}_a\Big)T-TJ^{(j_G)}_a\Big].
\label{eq:un:orbit}
\end{equation}
The coefficients $\kappa/D$ and $2\sigma/D$ come from the colour connection along the two orbits, and they satisfy $\kappa/D+2\sigma/D=1$. Let $n_{\rm f}\in\{0,1,2\}$ be the number of fermion sets in block 4 or 5. A set in block 4 or 5 contributes spin at most $\frac12$ under rotations of $Z$ and under rotations of $y$, and a set in block 1 contributes none, so $n_{\rm f}/2$ bounds both total fermion spins. Put $S_L=S_G=\sqrt{\tfrac{n_{\rm f}}2(\tfrac{n_{\rm f}}2+1)}$.

The coverage bound is the following. For every spin $s$, block $(A_+,A_-)$, channel $\varpi=(j_L,j_R,j_G)$, coupling $\zeta\in[-1,1]$ and $\alpha\in(0,\pi/2)$, as quadratic forms on the fibre,
\begin{equation}
M_\varpi(\alpha)\ \ge\ \max\Big\{\frac{4a_L^2}{\sigma}+\frac{4a_G^2}{\kappa},\ 4j_R(j_R+1)\Big\}-\frac{16(s+1)}{D^2}-\frac{E\kappa}{D^2},
\label{eq:un:Gbound}
\end{equation}
where
\begin{equation}
a_L=\Big(\sqrt{j_L(j_L+1)}-\frac{\kappa}{D}\sqrt{C_2}-S_L\Big)_+,\qquad 
a_G=\Big(\sqrt{j_G(j_G+1)}-\frac{2\sigma}{D}\sqrt{C_2}-S_G\Big)_+ .
\end{equation}
Consequently the channel is conformal at $E$ if the right-hand side of \eqref{eq:un:Gbound} is at least $-9$ for every $\alpha$, and $L(\pm1)\ge-8$ in the channel if the right-hand side with $E=0$ is at least $-8$ for every $\alpha$.

To prove \eqref{eq:un:Gbound} we bound the orbit terms, the Pauli term and the self-energy in \eqref{eq:un:Mform} pointwise in $\alpha$. For a triple $X=(X_a)$ write $\|X\|_2=(\sum_a\|X_a\|^2)^{1/2}$.
\begin{itemize}
\item \textit{Orbit terms.} By the reverse triangle inequality and Minkowski's inequality,
\begin{equation}
\begin{aligned}
\tfrac12\|u\|_2&\ge\|TJ^{(j_L)}\|_2-\frac{\kappa}{D}\|JT\|_2-\|S^{(Z)}T\|_2\\
&\ge\Big(\sqrt{j_L(j_L+1)}-\frac{\kappa}{D}\sqrt{C_2}-S_L\Big)\|T\| ,
\end{aligned}
\end{equation}
because $\sum_aJ^{(j)}_aJ^{(j)}_a=j(j+1)$, $\sum_aJ_aJ_a=C_2$ and $\sum_aS^{(Z)}_aS^{(Z)}_a\le\tfrac{n_{\rm f}}2(\tfrac{n_{\rm f}}2+1)$. Hence $\sum_a\|u_a\|^2\ge4a_L^2\|T\|^2$, and in the same way $\sum_a\|v_a\|^2\ge4a_G^2\|T\|^2$.

\item \textit{The $j_R$ term.} Since $\kappa/D+2\sigma/D=1$, the equivariance \eqref{eq:un:equivariance} gives $u_a+v_a=-2\ii\,TJ^{(j_R)}_a$. Since $\sigma+\kappa=1$, the Cauchy--Schwarz inequality gives $\|u_a\|^2/\sigma+\|v_a\|^2/\kappa\ge\|u_a+v_a\|^2$, and summing over $a$ shows that the orbit terms are at least $4j_R(j_R+1)\|T\|^2$. With the previous bound, the orbit terms are at least the maximum in \eqref{eq:un:Gbound} times $\|T\|^2$.

\item \textit{Pauli term.} For each set the Pauli operator is $(32/D^2)\,\mathbf S\cdot\mathbf J$ (Eq.~\eqref{eq:pauliclosed}), where $\mathbf S$ acts as a spin $\frac12$ on the components on which the colour curvature acts and as zero on the others, and in block 1 it vanishes. A spin $\frac12$ coupled to the colour spin $s$ has $\mathbf S\cdot\mathbf J\in\{s/2,-(s+1)/2\}$, so $P^{(\pm)}\ge-16(s+1)/D^2$ pointwise in every block. Since $c_\pm\ge0$ and $c_++c_-=1$, the same bound holds for $c_+P^{(+)}+c_-P^{(-)}$.

\item \textit{Self-energy.} $(1-\zeta^2)V=-EW=-E\kappa/D^2$.
\end{itemize}
Adding these bounds gives \eqref{eq:un:Gbound}. The kinetic term $-\omega^{-1}\partial_\alpha\omega\partial_\alpha$ is nonnegative, so $L_\varpi$ is bounded below by the minimum over $\alpha$ of the right-hand side, which gives both consequences.

The bound confines the channels that can fail to be conformal to a finite set. Let $s\ge1$ and $0\le E\le4$, and put $\bar a_L=\big(\sqrt{j_L(j_L+1)}-\sqrt{C_2}-\sqrt2\big)_+$ and $\bar a_G=\big(\sqrt{j_G(j_G+1)}-\sqrt{C_2}-\sqrt2\big)_+$. Every channel with $4j_R(j_R+1)\ge16s+11$, or with $\bar a_L^2+\bar a_G^2\ge4s+\frac{11}{4}$, is conformal. The channels that \eqref{eq:un:Gbound} leaves undecided at a given $E\le4$, that is, those it does not show to be conformal, form a finite set, closed under lowering any one of $j_L$, $j_R$, $j_G$.
Indeed, since $\kappa/D\le1$, $2\sigma/D\le1$, $S_L=S_G\le\sqrt2$, $\sigma,\kappa\le1$ and $D\ge1$, the right-hand side of \eqref{eq:un:Gbound} is at least $\max\{4\bar a_L^2+4\bar a_G^2,\,4j_R(j_R+1)\}-16(s+1)-E$, and $16(s+1)+E\le16s+20$. Either condition therefore makes it at least $-9$. The right-hand side of \eqref{eq:un:Gbound} is nondecreasing in each of $j_L$, $j_R$, $j_G$, which gives the closure property, and the complement of the two conditions contains finitely many channels.

At the onsets a smaller rectangle follows analytically. For $s\ge1$ and $E\le8/3$ every possibly supercritical channel has $j_R=0$, $j_L\le s+3/2$ and $j_G\le s+2$. The sharp floor bound, Eq.~\eqref{eq:fb:H4}, excludes $j_R\ne0$ at these couplings. For each remaining axis put $d=\sqrt{j(j+1)}-\sqrt2$. Multiply the right-hand side of Eq.~\eqref{eq:un:Gbound} plus nine by $(1+\sigma)^2$ and keep the corresponding axis term. At $j_L=s+2$ that term is
$
4[(d-\sqrt{C_2})+(d+\sqrt{C_2})\sigma]^2/\sigma\ge16(d^2-C_2).
$
Using $\sqrt{(s+2)(s+3)}\le s+5/2$ leaves the lower bound $16[(3-2\sqrt2)s+7-5\sqrt2]+9-E>0$. At $j_G=s+5/2$, the axis term is $4[2(d-\sqrt{C_2})+(2\sqrt{C_2}-d)(1-\sigma)]^2/(1-\sigma)$. For $s\ge2$, the inequalities $d-\sqrt{C_2}>5/2-\sqrt2$ and $2\sqrt{C_2}-d\ge s+3\sqrt2-5$ leave the lower bound $16[(5-2\sqrt2)(s+3\sqrt2-5)-(s+1)]+9-E>0$. At $s=1$, use $d\ge2553/1000$ and $\sqrt{C_2}\le1415/1000$. At $s=3/2$, use $d\ge3057/1000$ and $\sqrt{C_2}\le1937/1000$. After multiplication by $1-\sigma$, the degree-three Bernstein coefficients of the lower polynomial at $E=8/3$ have common denominator $9000000$. Their numerators in ascending Bernstein order are
\begin{equation}
\begin{aligned}
s=1:&\quad(3641124,125668780,213617184,186486336),\\
s=\tfrac32:&\quad(33428964,136487308,212555520,180633600). 
\end{aligned}
\end{equation}
All are positive. Monotonicity in $j$ completes the exclusion. The instability bounds in Sec.~\ref{sec:bps} put the onsets of both theta sectors below $E=8/3$, including the exceptional $s=1$ case. The rectangle contains every channel that can set the onset. Identifying the first crossing at a particular finite spin also requires comparisons of polar eigenvalues within that rectangle. The analytic large-spin ordering and the spin-one separation above use additional polar estimates. 

\section{Bosonic phases at every spin}
\label{app:bosonic}

This appendix proves Proposition~\ref{thm:bosphase} and the statements about the stretched channel in Sec.~\ref{sec:freephases}. Throughout, $\kappa=1-\sigma$, $D=1+\sigma$, $W=\kappa/D^2$, $C_j=j(j+1)$, and $\mathcal K=-4\sigma\kappa\,\partial_\sigma^2-8(1-2\sigma)\partial_\sigma$ is the polar operator of Eq.~\eqref{eq:fb:bosonic}. Quadratic forms are taken with the polar measure $\sigma\kappa\,\dd\sigma$. The two ends of the polar interval are orbits of codimension four, so sections supported away from them form a core of the quadratic form, and the identities below, proved on that core, extend by closure.

\paragraph*{The orbital potential.}Without fermions the stabilizer equivariance~\eqref{eq:un:equivariance} turns the orbit vectors of Eq.~\eqref{eq:un:orbit} into $u_a=-2\ii\,T\bigl[c_L(J^{(j_R)}+J^{(j_G)})-c_GJ^{(j_L)}\bigr]_a$ and $v_a=-2\ii\,T\bigl[c_G(J^{(j_L)}+J^{(j_R)})-c_LJ^{(j_G)}\bigr]_a$, with $c_L=\kappa/D$, $c_G=2\sigma/D$ and $c_L+c_G=1$. Expanding the squares, with $2J^{(j_L)}\cdot(J^{(j_R)}+J^{(j_G)})=C_2-C_L-C_{RG}$ and its analogue for the other pair, gives Eq.~\eqref{eq:Mbos}, where $C_{RG}=(J^{(j_R)}+J^{(j_G)})^2$ and $C_{LR}=(J^{(j_L)}+J^{(j_R)})^2$ act on the multiplicity space of $s$ in $j_L\otimes j_R\otimes j_G$. The eigenvalues of $C_{RG}$ are the $C_j$ of the intermediate spins $j$ that lie both in $j_R\otimes j_G$ and in $j_L\otimes s$, so $C_{RG}\ge C_a$ whenever $a\le\max(|s-j_L|,|j_R-j_G|)$. Likewise, $C_{LR}\ge C_b$ whenever $b\le\max(|s-j_G|,|j_L-j_R|)$.

\paragraph*{A scalar comparison for every channel.}Choose numbers $a,b\ge0$ obeying these two conditions together with $s-a\le j_L$ and $s-b\le j_G$. They need not be labels of an actual channel. Since $C_{RG}$, $C_{LR}$, $C_L$ and $C_G$ enter Eq.~\eqref{eq:Mbos} with positive coefficients, $M_\varpi$ is bounded below, as a matrix, by the scalar potential
\begin{equation}
V_{a,b}=\frac{4C_a}{\sigma}+\frac{4C_b}{\kappa}+\frac{8(C_{s-a}-C_a)+4(C_{s-b}-C_b)}{D}-\frac{8C_2}{D^2}-EW .
\end{equation}
For the positive function $f=\sigma^a\kappa^bD^{-s-2}$, which has the regular endpoint powers of $V_{a,b}$, direct differentiation gives
\begin{equation}
 D^2\,\frac{(\mathcal K+V_{a,b}+8+\Delta\mu)f}{f}=\mathsf b_0\kappa^2+2\mathsf b_1\sigma\kappa+\mathsf b_2\sigma^2-(E-16s-24)\kappa+\Delta\mu\,D^2,
 \label{eq:bos:bern}
\end{equation}
with $d=a+b-s$, $\mathsf b_0=4[2a+d(d+1)]$, $\mathsf b_1=2[6a+4d^2+2d]$, $\mathsf b_2=16(a+d^2)$ and a constant $\Delta\mu\ge0$ that lowers the target eigenvalue to $-8-\Delta\mu$. When the right-hand side is positive on $0<\sigma<1$, the ground-state representation with $f$ shows that every state of the channel has a link eigenvalue above $-8-\Delta\mu$ at the attraction $E$. The bound holds componentwise, so it holds for every multiplicity. Since $W>0$ inside the link, lowering $E$ only raises the eigenvalues. We use $\kappa=\kappa^2+\sigma\kappa$ and $D^2=\kappa+3\sigma+\sigma^2$ to read off positivity.

\paragraph*{Proof of part~(a).}At the ordinary line $E=4C_2$ the right-hand side of Eq.~\eqref{eq:bos:bern} at $\Delta\mu=0$ is $(\mathsf b_0+4H)\kappa^2+2(\mathsf b_1+2H)\sigma\kappa+\mathsf b_2\sigma^2$ with $H=6+3s-s^2$, and $H\ge2$ for $s\le4$. Take $a=\max(s-j_L,0)$ and $b=\max(s-j_G,0)$, which obey the conditions. Then $d(d+1)\ge-\tfrac14$ and $4d^2+2d\ge-\tfrac14$, so the first two coefficients are positive and the third is nonnegative, and every channel of either sector has $\mu>-8$ for $E\le4C_2$.\unskip\nobreak\hfill\penalty50\hskip1.2em\hbox{}\nobreak\hfill\ensuremath{\square}

\paragraph*{Proof of part~(b).}In the integral sector $j_G$ is an integer. Take $b=s-\min(j_G,s)$, and $a=s-j_L$ if $j_L<s$, $a=0$ if $j_L=s$, $a=\tfrac12$ if $j_L>s$. Put $p=2s+2$ and test the target $-8-\Delta\mu$, $0\le\Delta\mu\le1$, at $E=16s+24+\Delta\mu(1+3/p)$, the variational upper bound on the stretched crossing. The right-hand side becomes $\mathsf b_0\kappa^2+2\mathsf b_1\sigma\kappa+\mathsf b_2\sigma^2-(3\Delta\mu/p)\kappa+\Delta\mu(3\sigma+\sigma^2)$. If $a\ge\tfrac12$, then $\mathsf b_0\ge3$ and $\mathsf b_1\ge\tfrac{11}2$, while $3\Delta\mu/p\le\tfrac32$, so it is positive. If $a=0$ and $\min(j_G,s)\ge\tfrac32$, then $\mathsf b_0\ge3$ and $\mathsf b_1\ge12$. If $a=0$ and $j_G=0$, a channel other than the stretched one has $j_R\ge1$ and admits $a=1$. If $a=0$ and $j_G=1$, integrality makes $j_R$ an integer, and $j_R\ne1$ admits $a=1$. The one remaining channel, $(s,1,1)$, has $\mathsf b_0=0$ and the residual $8\sigma D$ at $\Delta\mu=0$. Multiplying $f$ by $1-c\sigma$ with $8c=3\Delta\mu/p$ changes the right-hand side to 
\begin{equation}
\begin{aligned}
 &(1-c\sigma)\bigl[8\sigma D+\Delta\mu(3\sigma+\sigma^2)\bigr]
    +8c\bigl[D(1-p\sigma+\sigma^2)-\kappa(1-c\sigma)\bigr]\\
 &\hspace{3em}\ge \sigma D\,[8(1-c)-3\Delta\mu]>0 .
\end{aligned}
\end{equation}
Hence no state outside the stretched channel reaches $-8-\Delta\mu$ before its ground state does. The same coefficients at $E=16s+27$ and $\Delta\mu=0$, where the right-hand side is $(\mathsf b_0-3)\kappa^2+(2\mathsf b_1-3)\sigma\kappa+\mathsf b_2\sigma^2$, exclude every channel other than the stretched one and $(s,1,1)$ up to $E=16s+27$. For that channel couple $j_R=j_G=1$ to the intermediate spin $j=0,1,2$. Then $C_{RG}=\operatorname{diag}(0,2,6)$, and $C_{LR}$ is tridiagonal with diagonal $C_2+2-C_j/2$ and off-diagonal entries $-\sqrt{8C_2/3}$ and $-\sqrt{(2s-1)(2s+3)/3}$, with eigenvalues $C_{s-1}$, $C_s$ and $C_{s+1}$. With $\mathsf d=\operatorname{diag}\bigl(1,\sqrt{(s+1)/s},\sqrt{(s+1)(2s+3)/(s(2s-1))}\bigr)$ and $h=\bigl(1-\sigma+\tfrac13\sigma^2,\ \sqrt{2/3}\,(\sigma-\tfrac12\sigma^2),\ \tfrac{\sqrt2}6\sigma^2\bigr)$, the positive vector $\Phi=\kappa^{s-1}D^{-s-2}\,\mathsf d\,h$ obeys, at $E=16s+32$,
\begin{equation}
\frac{D^2}{\kappa^{s-1}D^{-s-2}}\,\mathsf d^{-1}(L+8)\Phi=\frac83\begin{pmatrix}\sigma(\sigma^2-4\sigma+9)\\ \sqrt6\,\sigma(1+3\sigma-\sigma^2)\\ \sqrt2\,\sigma^2(1+2\sigma)\end{pmatrix},
\end{equation}
which is positive on $0<\sigma<1$ for every $s$. The potential of this channel is cooperative, with nonpositive off-diagonal entries, so the vector ground-state representation, a sum of positive derivative terms, positive off-diagonal squares and the positive residual \cite{Allegretto1985}, shows that it stays above $-8$ up to $E=16s+32$. Finally the excited polar states of the stretched channel reach $-8$ only at $32s+80$ and above, by Eq.~\eqref{eq:ladder}.\unskip\nobreak\hfill\penalty50\hskip1.2em\hbox{}\nobreak\hfill\ensuremath{\square}

\paragraph*{Proof of part~(c).}In the complementary sector $j_G$ is half an odd integer for every $s$. Take $b=s-G$, with $G=j_G$ if $j_G\le s$ and otherwise the largest half-odd integer not exceeding $s$, and $a$ as in part~(b). At $E=16s+27$ and $\Delta\mu=0$ the right-hand side is again $(\mathsf b_0-3)\kappa^2+(2\mathsf b_1-3)\sigma\kappa+\mathsf b_2\sigma^2$. For $a\ge1$, for $a=\tfrac12$, where $d$ is an integer and $\mathsf b_0\ge4$, and for $a=0$ with $G\ge\tfrac32$, it is positive on the open interval. For $a=0$ and $j_G=\tfrac12$, every $j_R\ne\tfrac12$ is half an odd integer at least $\tfrac32$ and admits $a=1$. The remaining channel is $(s,\tfrac12,\tfrac12)$, whose two states couple $j_R$ and $j_G$ to a singlet and a triplet. There $C_{RG}=\operatorname{diag}(0,2)$ and $C_{LR}$ has diagonal $(C_2+\tfrac34,\,C_2-\tfrac14)$ and off-diagonal entry $-\sqrt{C_2}$, and the positive vector $\Phi=\kappa^{s-1/2}D^{-s-2}\bigl(1-\tfrac12\sigma,\ \tfrac12\sqrt{(s+1)/s}\,\sigma\bigr)$ obeys, at $E=16s+27$,
\begin{equation}
 \frac{D^2}{\kappa^{s-1/2}D^{-s-2}}\,(L+8)\Phi=\frac\sigma2\begin{pmatrix}\sigma^2+\sigma+18\\ \sqrt{(s+1)/s}\,(8+13\sigma-\sigma^2)\end{pmatrix}>0 .
\end{equation}
The cooperative ground-state representation then keeps the whole complementary sector above $-8$ for $E\le16s+27$, at every $s\ge\tfrac92$. At $s=\tfrac92$ this bound is the ordinary line, $16s+27=99=4C_2$. The singlet component alone, used as a trial state, costs the extra potential $3/\kappa$ relative to the stretched channel. With the moment $\langle 3/(\kappa W)\rangle_W=3(2s+5)/(2s+1)$ in the weighted ground state it places the complementary contact value below $16s+27+12/(2s+1)$ and gives the complementary threshold bound quoted after Proposition~\ref{thm:bosphase}.\unskip\nobreak\hfill\penalty50\hskip1.2em\hbox{}\nobreak\hfill\ensuremath{\square}

\paragraph*{The stretched channel.}With $T=T^{\rm bos}_s\,h$ and $x=v^2$, Eq.~\eqref{eq:stretched} turns the equation $(L+8+\Delta\mu)T=0$ at $E=16s+24+e$ into
\begin{equation}
 -\frac{\dd}{\dd x}\Bigl[x^{s+3/2}(1-x)^2\,\frac{\dd h}{\dd x}\Bigr]+\frac{\Delta\mu}8\,x^{s}(1-\sqrt x)\,h=\frac e{16}\,x^{s+1/2}(1-x)\,h ,
\end{equation}
where the $\Delta\mu$ term carries the measure $\tfrac14v^{2s+1}(1-v)\,\dd v$ of the channel written in $x$, and $W=\tfrac12v(1+v)$. At $\Delta\mu=0$ this is the hypergeometric equation with parameters $(-n,\ s+n+\tfrac52;\ s+\tfrac32)$, whose polynomial solutions, Jacobi polynomials in $x$, exist exactly when $e=16n(s+n+\tfrac52)$, giving Eq.~\eqref{eq:ladder}. At $\Delta\mu>0$ the crossing is the minimum of $\bigl[\int x^{s+3/2}(1-x)^2h'^2\,\dd x+\tfrac{\Delta\mu}8\int x^{s}(1-\sqrt x)h^2\,\dd x\bigr]/\bigl[\tfrac1{16}\int x^{s+1/2}(1-x)h^2\,\dd x\bigr]$, and the two Beta integrals of the constant trial give $\langle W\rangle=(2s+2)/(2s+5)$. Since the weighted measure in the denominator is $W$ times the measure of the $\Delta\mu$ term, the quotient is at least $\Delta\mu$, and the constant trial gives at most $\Delta\mu/\langle W\rangle$, which proves Eq.~\eqref{eq:stretchbracket} at $\Delta\mu=1$. Both inequalities are strict, because $W$ is not constant.

\paragraph*{Shape response and complex spin.}Expanding the crossing of the stretched channel in the target, $E=16s+24+(1+\tfrac3p)\Delta\mu+e_2(\Delta\mu)^2+O((\Delta\mu)^3)$, one Poisson equation for the response of the constant ground state gives, with $p=2s+2$,
\begin{equation}
 e_2=-\frac{(p+1)(p+3)}{8p^2}\int_0^1v^p\,\frac{(1-v)^2(v+2)^2}{(1+v)^2}\,\dd v<0 ,
\end{equation}
which we checked against the numerical second derivative of the crossing at $s=\tfrac92$, $10$ and $20$. Relaxation of the polar shape lowers the threshold, and at $s=\tfrac92$ the term $e_2\simeq-3.856\times10^{-4}$ accounts for the whole shift of $E_c^{(s,0,0)}$ below $16s+25+3/(2s+2)$ to within three parts in a thousand. In the variable $\mathsf t=-2\ln v$ the integral is a Laplace transform with exponent $(p+1)\mathsf t/2$ of a kernel with double poles at $\mathsf t=\pm2\pi\ii(2k+1)$, $k=0,1,2,\ldots$ The kernel has no singularity on the positive axis, so the expansion of $e_2$ in $1/s$ is Borel summable along it and its Borel sum is $e_2$ for every real $s>0$. Only when $s$ is continued off the real axis do the two lateral Borel sums differ, across the imaginary $s$-axis, by terms proportional to $e^{\mp2\pi\ii(2k+1)s}$ times rational functions of $s$. On the real axis these factors have unit modulus and describe no tunnelling, and $e_2$ itself stays analytic. For the Pauli threshold of Sec.~\ref{sec:uniform} no such statement is available, and the poles quoted there belong to Pad\'e approximants. A global connection formula at finite $\Delta\mu$, which would fix every exponentially small term of the thresholds themselves, remains open.

\section{Numerical methods and coverage}
\label{app:numerics}

 The numerical results of this paper are the critical-channel estimates of Table~\ref{tab:thresholds} and the orderings in the scanned boxes for $s\le6$, the scalar threshold $E\approx26.4576$ and a scan of every bosonic channel, built from Eq.~\eqref{eq:Mbos} with Wigner $6j$ recoupling, which excludes most channels by a pointwise and a scalar lower bound, diagonalizes the rest by finite differences, confirms Proposition~\ref{thm:bosphase} at several couplings for each spin from $\tfrac52$ to $6$ and reproduces the exact contact value $16s+24$ to $10^{-5}$, the stretched-channel threshold up to $s=160$ (Sec.~\ref{sec:freephases}), the twelve-digit thresholds of the $(4,4)$ critical channel used to test the Borel--Pad\'e sum (Sec.~\ref{sec:uniform}), the thresholds of the two critical channels, block $(4,4)$ in $(s-1,0,0)$ and block $(4,5)$ in $(s-1,0,\tfrac12)$, at the targets $\Delta\mu=\tfrac12$ and $1$ for $6\le s\le32$, which test the theta splitting through order $s^{-5}$ (Sec.~\ref{sec:uniform}), a six-coupling check of the analytic uniqueness bound for $s=\tfrac12$, the mixed-channel estimate $E_c^{\rm ch}\simeq2.52924$ at $\theta_{5d}=\pi$, $s=1$, the large-$s$ excited-state gap $\Delta_{\rm exc}(s)\sim16s$ and a numerical check of $E_c^{\rm ch}(s)$ in the integer-spin channel up to $s=400$ (Sec.~\ref{sec:mechanism}), and the cross-checks listed below (both $s=\tfrac12$ conformality for all $|\zeta|\le1$ and the near-BPS slopes for every spin $s\ge1$ are now proved analytically in Sec.~\ref{sec:bps} and Appendix~\ref{app:fb:bps}). None of the proved results, Propositions~\ref{thm:L} and~\ref{thm:floor} and Eqs.~\eqref{eq:pinbound}, \eqref{eq:thmC}, \eqref{eq:vpp}--\eqref{eq:vpp-asymp}, \eqref{eq:O2prime}, \eqref{eq:lawcor}, \eqref{eq:allslope}--\eqref{eq:E2x2-asymp} and~\eqref{eq:fb:gap4sharp}, depends on them.

\emph{Channel reduction.} The group $G$ acts transitively on each slice $\alpha=\text{const}$ of $S^7$ with diagonal stabilizer $H\simeq SU(2)$, and $S^3\times S^3=G/H$ is naturally reductive. By Frobenius reciprocity, sections in the channel $\varpi=(j_L,j_R,j_G)$ are determined by an $H$-equivariant map $T(\alpha)$ from $V_\varpi$ to the fibre at the base point, and the quadratic form of $L(\zeta)$ becomes
\begin{equation}
 \int\!\omega(\alpha)\Biggl[|T'|^2+\langle T,\mathcal VT\rangle
 +\sum_a\frac{|\Lambda(P_a)T-T\varpi(P_a)|^2}{\sin^2\alpha}
 +\sum_a\frac{|\Lambda(H_a)T-T\varpi(H_a)|^2}{\cos^2\alpha}\Biggr]\dd\alpha ,
\end{equation}
where $\omega(\alpha)=\sin^3\alpha\cos^3\alpha$, $P_a,H_a$ span the orbit directions, $\Lambda$ is the colour action of the connection along them in closed form, and $\mathcal V$ collects the self-energy and the Pauli terms. We expand $T$ in Gegenbauer polynomials $C^{3/2}_k(\cos2\alpha)$, with $K=20$ and $96$ quadrature nodes, and check at $K=32$ with $160$ nodes. The exact $4\times4$ system~\eqref{eq:4x4} is solved up to $K=120$ ($240$ Gauss--Legendre nodes), and the $\theta_{5d}=\pi$ and half-integer $(4,5)$ thresholds are checked between $K=32$ ($160$ nodes) and $K=44$ ($192$ nodes). The half-integer critical-channel thresholds quoted in Tables~\ref{tab:thresholds} and~\ref{tab:nearbps} were recomputed at $K=84$, where they agree with $K=60$ to $4\times10^{-5}$ and decrease with $K$. Galerkin eigenvalues are upper bounds on each channel's lowest eigenvalue, so the instabilities found numerically are variational up to quadrature. Numerical stability indicators (such as checks against the analytic $\beta\ge1$ bound at $s=\frac12$) rely on convergence with $K$ across the scanned box. Changes between the compared resolutions are of order $10^{-5}$ to $4\times10^{-5}$ for the quoted threshold estimates (computed in Python with NumPy \cite{NumPy} and SciPy \cite{SciPy}). To probe the $O(s^{-2})$ asymptotics of $E_c^{\rm ch}(s)$ and the excited-state gap $\Delta_{\rm exc}(s)$ of $L_s(1)+8$ orthogonal to $\mathrm{span}\{\phi_0,\phi_1\}$ up to $s=400$, we solve Eq.~\eqref{eq:4x4} in a QR-orthonormalized mapped-Legendre basis ($K=26$--$40$, $100$--$120$ nodes) weighted by the regular indicial envelope in the boundary-layer coordinate $\xi=\tilde u/(\tilde u+2)$ with $\tilde u=4s\sigma$ (placing the LLL waist $\tilde u\sim2$ at the midpoint $\xi\sim\frac12$), finding ratios $\Delta_{\rm exc}(s)/s$ consistent with a limit near $16$ (e.g.\ $15.57$ at $s=400$) and $s^2\bigl(E_c^{\rm ch}(s)-1-\frac{3}{2s}\bigr)=-0.16090$ against $-\frac{9}{56}=-0.160714$ (with $s[s^2(E_c^{\rm ch}-1-\frac{3}{2s})+\frac{9}{56}]\simeq-0.075$, a rough value since it amplifies the error of the five-digit second coefficient. Equation~\eqref{eq:lawcor3} gives the limit $-3447/43904\simeq-0.0785$). 

\emph{Validation.} The closed-form colour action agrees with the direct one, the curvature reconstructed from the colour action agrees with the directly computed curvature to $10^{-16}$, the Pauli operator is covariant under random elements of $G$ to $10^{-14}$, and at $s=0$ every fermion block reproduces the scalar spectrum $0,7,16$ of $S^7$. The bosonic block reproduces the bosonic free threshold. The matrix potential of the exact $4\times4$ system reproduces the pointwise eigenvalues of that of the general reduction to relative accuracy $7\times10^{-15}$ for $s\le5$ at thirty points $(\alpha,\zeta)$ each, and its thresholds for $s=1,2,3$, $0.839960$, $0.963739$ and $0.984449$, match the general code. A second implementation, written from Eqs.~\eqref{eq:un:equivariance}--\eqref{eq:un:orbit} and~\eqref{eq:pauliclosed} alone, with block~$5$ built as the action induced on the antisymmetric square of block~$4$ (the $\mathbf 5$ of $Sp(2)$), reproduces Eq.~\eqref{eq:Mbos} at $j_R=0$ to $10^{-13}$, the matrix potential of Eq.~\eqref{eq:4x4} pointwise to $5\times10^{-15}$, the threefold floor at $-8$ of the $(4,5)$ critical channel at $\zeta=1$, the half-integer thresholds of Tables~\ref{tab:thresholds} and~\ref{tab:nearbps} to their printed digits and every slope of Table~\ref{tab:nearbps} to $2\times10^{-5}$. It expands each channel in polynomials in $\alpha$ that are orthonormal for the weight of the state, rewrites the orbit terms near the line through Eq.~\eqref{eq:un:equivariance} so that no large terms cancel, and refines each eigenvalue by a Rayleigh quotient. The exact floor $-8$ at $\zeta=1$ then holds to $10^{-12}$ up to $s=24$, and the thresholds change by less than $4\times10^{-13}$ between basis sizes. This implementation gives the large-spin thresholds behind the check of the theta splitting in Sec.~\ref{sec:uniform}.

\emph{Coverage.} The analytic exclusion in Appendix~\ref{app:un:coverage} bounds all possibly critical angular channels at onset. The numerical ordering refers to the channels and couplings actually sampled, and the quoted polar threshold decimals are estimates. For $s=\tfrac12$ the conformality proof of Appendix~\ref{app:fb:s12} covers the full angular spectrum. The scan at six couplings in $[0,1]$ cross-checks the proved uniqueness condition $\beta\ge1$ within its angular box. At $s=1$, $\theta_{5d}=\pi$, the scanned mixed channel $(0,0,\tfrac12)$ has threshold estimates $E_c^{\rm ch}\simeq2.52924$ and $|\zeta_c^{\rm ch}|\simeq0.82695$, converged to $10^{-11}$ in two independent discretizations. The exact instability bound and the endpoint census in Appendix~\ref{app:fb:block45} do not depend on those estimates.

For the cubic correction we also computed the complementary response directly in the normalized mapped-Legendre basis of block $(4,4)$, projecting out the two exact endpoint floor vectors. The values of $s^3\langle h,(\Pi_{\rm fl}^\perp(L_++8)\Pi_{\rm fl}^\perp)^{-1}h\rangle$ at $s=400,800,1600$ are $0.06935398$, $0.06983118$, and $0.07007132$, approaching $9/128=0.0703125$. Using $K=18$ and $28$ changes the last result by about $1.2\times10^{-10}$. This computation tests the resolvent contribution without taking the difference of two nearly equal threshold roots. For the tabulated variational upper bounds, rational enclosures of the logarithm and square root in the Beta evaluation give a positive pencil at the preceding decimal unit and a negative determinant at the printed value. This verifies outward rounding of all ten upper bounds, including $s=1$ at $\theta_{5d}=\pi$.

\paragraph*{A certified finite-spin crossing.}
\label{app:radial-certificate}
In the spin-one block $(4,4)$, channel $(0,0,0)$ at $\theta_{5d}=0$, the cooperative-system identity used in Sec.~\ref{sec:instability} certifies the full bracket
\begin{equation}
0.8399601<|\zeta_c^{\rm ch}|<0.8399602 .
\end{equation}
Conjugate the sign-gauged operator of Eq.~\eqref{eq:4x4} by the positive diagonal factors $\kappa^{1-p_i/2}\sigma^{p_i/2}D^{-2}$, where $p=(0,1,1,2)$. Multiplication of the residual by $D^2$ gives a polynomial differential operator with rational coefficients at these two rational couplings. A degree-22 polynomial in each component gives the supersolution at the upper endpoint and the negative trial at the lower endpoint. After a common positive rescaling the Bernstein coefficients of the four polynomial components exceed $2\times10^6$, and those of the signed residual exceed $999/1000$, by exact rational arithmetic. The positive residual gives strict positivity at the upper endpoint by the cooperative ground-state representation. The negative residual gives a negative quadratic form at the lower endpoint by testing on the same vector. The polar factors enforce regularity at both poles. The directed crossing argument of Sec.~\ref{sec:instability} then gives the bracket. Take the degree-22 trial vector of the upper endpoint and reevaluate its residual at $|\zeta|=0.83996019$. Multiplying its four components by three preserves the sign and meets the same strict Bernstein-coefficient margin. Since $E=8(1-\zeta^2)$ decreases with $|\zeta|$, the tightened endpoints give $2.3557350<E_c^{\rm ch}<2.3557363$, which certifies $E_c^{\rm ch}$ rounded to $2.35574$. This enclosure applies to the specified channel.\unskip\nobreak\hfill\penalty50\hskip1.2em\hbox{}\nobreak\hfill\ensuremath{\square}

For the spin-one switch, we independently solved the polar equations by assembling their orbit-intertwiner matrices from the connection and Clifford curvature. At $E=237/100$, increasing the polar basis size from $18$ to $32$ and $44$ gives lowest shifted eigenvalue estimates $0.06428$, $0.06410$ and $0.06408$ over the eight floor channels at $\theta_{5d}=\pi$. The corresponding values at $\theta_{5d}=0$ in the certified $(4,4)$ channel are $-0.00594$, $-0.00613$ and $-0.00614$. The sign in the $\theta_{5d}=\pi$ sector is proved by the form comparison of Appendix~\ref{app:fb:block45}.

We finish by proving a two-sided enclosure for channels without a cooperative sign gauge. For the Schur blocks $A=PKP$, $B=QKP$ and $C=QKQ$ of $K=L(\zeta)+9$, assume $C\ge bI>0$. For any approximate inverse action $X$ from the floor to the complementary operator domain put $R_X=B-CX$ and $T_X=B^\dagger X+X^\dagger B-X^\dagger CX$. Expanding $R_X^\dagger C^{-1}R_X$ gives
\begin{equation}
 A-T_X-b^{-1}R_X^\dagger R_X
 \ \le\ A-B^\dagger C^{-1}B\ \le\ A-T_X .
\end{equation}
Thus polar certification in the finite angular rectangle reduces to interval signs of finite floor matrices and exact residual norms. This criterion remains valid for trial columns $X$ obtained numerically.
\hfill$\square$

\FloatBarrier

\end{document}